\documentclass[manuscript]{aastex}

\newcommand{\oleb}{Origins Life Evol.~Biosph.}
\newcommand{\astrobio}{Astrobio.}
\newcommand{\ppthou}{\%${\scriptstyle0}$}
\newcommand{\ang}{\AA}

\begin{document}
  \title{Astrophysics in 2005}
\author{Virginia Trimble}
\affil{Department of Physics and Astronomy, University of California, Irvine, CA 92697-4575,
     Las Cumbres Observatory, Santa Barbara, CA:\email{vtrimble@uci.edu}}
\author{Markus J.~Aschwanden}
\affil{Lockheed Martin Advanced Technology Center, Solar and Astrophysics Laboratory,
  Organization ADBS, Building 252,
  3251 Hanover Street, Palo Alto, CA 94304:\email{aschwand@lmsal.com}}
\and
\author{Carl J.~Hansen}
\affil{JILA, Department of Astrophysical and Planetary Sciences, University of Colorado,
Boulder CO 80309:\email{chansen@jila.colorado.edu}}

\begin{abstract}
We bring you, as usual, the sun and moon and stars, plus some galaxies
and a new section on astrobiology. Some highlights are short (the newly
identified class of gamma-ray bursts, and the Deep Impact on
Comet 9P/Tempel 1), some long (the age of the universe, which will be
found to have the Earth at its center), and a few metonymic, for instance
the term ``down-sizing'' to describe the evolution of star formation rates
with redshift.
\end{abstract}

\keywords{astrobiology -- galaxies: general -- planets and satellites: general -- Sun: general --
  stars: general -- ISM: general}

\section{INTRODUCTION}
The ApXX series celebrates its quincea\~{n}era by adding a new,
astrobiology, section and co-opting an additional author to write it
and to cope with many other problems.\footnote{Astrophysics in 1991
to 2004 appeared in volumes 104--117 of PASP. They are cited here as
Ap91, etc.}
Used in compiling sections 3--6 and 8--13 were the issues that arrived as
paper between 1 October 2004 and 30 September 2005 of \textsl{Nature},
\textsl{Physical Review Letters}, the \textsl{Astrophysical Journal}
(plus \textsl{Letters} and \textsl{Supplement Series}), \textsl{Monthly Notices
of the Royal Astronomical Society} (including \textsl{Letters} up to December
2004 only), \textsl{Astronomy and Astrophysics} (plus \textsl{Reviews}),
\textsl{Astronomical Journal}, \textsl{Acta Astronomica}, \textsl{Revista
Mexicana Astronomia y Astrofisica}, \textsl{Astrophysics and Space Science},
\textsl{Astronomy Reports}, \textsl{Astronomy Letters}, \textsl{Astrofizica},
\textsl{Astronomische Nachrichten}, \textsl{Publications
of the Astronomical  Society of Japan}, \textsl{Journal of Astrophysics and
Astronomy}, \textsl{Bulletin of the Astronomical Society of India},
\textsl{Contributions of the Astronomical Observatory Skalnate Pleso},
\textsl{New Astronomy} (plus \textsl{Reviews}), \textsl{IAU Circulars}, and, of
course, \textsl{Publications of the Astronomical Society of the Pacific}.
Journals read less systematically and cited irregularly include
\textsl{Observatory}, \textsl{Journal of the American Association of
Variable Star Observers}, \textsl{ESO Messenger}, \textsl{Astronomy and
 Geophysics}, \textsl{Mercury},
\textsl{New Scientist}, \textsl{Science News}, \textsl{American Scientist},
\textsl{Scientometrics}, \textsl{Sky \& Telescope}, \textsl{Monthly Notes of
the Astronomical Society of South Africa}, and \textsl{Journal of the Royal
Astronomical Society of Canada}. Additional journals provided material for
Sections 2 and 7 and are mentioned there.

A few papers are mentioned as deserving of gold stars, green dots, and
other colorful recognition. This is as nice as we get. Among the people who
appear in the following pages are Jack Benny, the Keen Amateur Dentist, the
Faustian Acquaintance, and the Medical Musician. All are pseudonyms, for
Benjamin  Kubelsky, and three colleagues, left as an exercise for readers, not
intended to include the also-pseudonymous Mr.~H., who is supposed to be
completing a thesis in X-ray astronomy.

\subsection{Terminations}
We record here a number of things that came to an end, surely or
probably, in the index year. Beginnings appear in \S 1.2 and more complicated
relationships in 1.3. (1) The last launch of a Skylark sounding rocket happened
 on 30 April 2005; the first was from Woomera during International
Geophysical Year 1957. (2) Both the \textsl{Letters}
section of \textsl{MNRAS} and the newsletter of the European Space Agency went
e-only during the year (January and July 2005 respectively) and so are no longer
accessible to the least electronic author.  (3) SLAC, like Brookhaven and
Los Alamos before it, was shut down by an accident on 1 October
(Science 306, 809). (4) The percentage of
women in computer sciences has actually declined over the past 20 years
(Science 306, 809), while that in physical sciences and engineering has crept
slowly up. (5) NASA's KC--135 jet, the ``vomit comet,'' used to produce brief
experiences of weightlessness, free-fell for the last time on 29 October.
(6) Kodak produced its last carousel projector in November 2004 but will support
existing ones  (light bulbs and things) until 2007. (7) The Yerkes Observatory
(where the modern sequence of
spectral types was established by Morgan and Keenan) is to be turned into, but
turned into what is not entirely clear (Sky \& Telescope 108, No.~3, p.~19).
(8) The Mohorovi\v{c}i\'{c} discontinuity was not reached again at 1.4 km where
the mid-atlantic ridge was expected to be only
0.7 km thick (Science 307, 1707); of course every child who attempted to dig to
China with a tablespoon (from California, or to California from China with
chopsticks) can claim a similar result (Scientific American 293, 94).
(9) Ball Aerospace provided (at least)
its second out-of-focus telescope, this one on \textsl{Deep Impact}
(Nature 434, 685). (10) Another try at \textsl{Cosmos-1}, a solar sail vehicle,
was lost on launch 21 June, probably due to a pump failure. (11)
\textsl{Gravity Probe B} gathered its last data on 1 October, with, up to
mid January, no results announced and no triumphant renaming of the craft as,
for instance, Lens-Thirring to indicate success.

\subsection{Inceptions}
Most of these beginnings are, we think, good news or at least progress toward
good news. (1) \textsl{Smart-1} on its way to the moon was probably the most
fuel-efficient vehicle in history, achieving the equivalent of $2\times10^6$
km per liter of gasoline (ESA Space Science News No.~7, p.~2); the only
competition would seem to be something that simply rides along with tectonic
plate spreading. It arrived on 15 November after an October 2003 launch, and
was so also only marginally faster than tectonic plates. (2) Some of the
antennas for LOFAR are in place (10 near the beginning of the reference year,
Sky \& Telescope December 2005, p.~24). (3) The \textsl{Spitzer Space Telescope}
released its first large package
of papers on the cusp of the year (September 2004 issue of Astrophysical Journal
supplement Series). (4) \textsl{SWIFT} caught its first gamma ray burst on
17 December, after a 20 November launch. (5) The first hole for Ice Cube was dug
in January 2005, an advanced muon and neutrino detector in Antarctica (and a
confusing notebook entry, because
we generally use upward pointing arrows for good news items, while the hole
 probably went down). (6) The Ariane heavy lifter had its first successful
launch from French Guiana on 12--13 February. It will be needed for
\textsl{James Webb Space Telescope}, unless
very considerable additional descoping occurs. (7) A million dollar Kavli
astrophysics prize will be given starting in 2008 (Nature 435, 37) along with
ones in neuroscience and nano-technology. (8) The street lighting in the Canary
Islands has been considerably
dimmed (Nature 435, 41), partly in anticipation of first light at the GTC
(out of period). (9) The Sloan Digital Sky Survey is being extended, primarily
to examine stars for their own sake and for galactic structure
(Nature 436, 316). (10) The \textsl{Mars Reconnaisance
Orbiter} got off the ground in August, and \textsl{Mars Express} will
continue to operate for another year or two (a funding extension), though one
spectrometer has gone funny\footnote{Expert readers may wish to translate this
into  some more technical term, but the published description sounded to us as
if the spectrometer had gone funny.} (Nature 437, 465). (11) The US is once
again planning to develop a major underground laboratory facility for high
energy physics and other purposes, perhaps back
at Homestake (Ray Davis's old site) or at Henderson, an active molybdenum mine
(Science 309, 682). (12) France has removed an assortment of popular topical
antibiotics from  over-the-counter sale (Science 309, 872). Not astronomy
perhaps, but anything that slows down the development of resistent organisms has
to be good for us all! (13) SALT, the
South African Large Telescope (near relative of the Hobby-Eberly Telescope)
has collected some photons and a dedication (Nature 437, 182). (14) The
Pierre Auger facility, which looks for very high energy things, released its
first data on 6 July. (15) The Atacama Pathfinder Experiment, APEX, at Chajnator
(a 12-meter telescope for millimeter radio astronomy and submillimeter
submillimeter astronomy in the in the Atacama desert) carried out its first
scientific observations in July 2005. (16) The Las
Cumbres Observatory has acquired its first 2-meter telescopes (in Hawaii and
Australia) and its first GRB afterglow (in a galaxy long ago and far away,
051111). Six telescopes and observations of many
more afterglows and other transient phenomena are planned.

\subsection{More Complicated Stories}
\textsl{Astro-E}, a Japanese X-ray satellite, was successfully launched on
10 July 2005 and named \textsl{Suzaku} (red bird of the south). But its coolant
was lost some time before 8 august and, although its CCD X-ray detectors will
still work, the spectrometer that was one of its main goals will not. This was,
sadly, the second try at \textsl{Astro-E}.
The first, also unsuccessful, attempt was roughly contemporaneous with the
 launches of the \textsl{Chandra} and \textsl{XMM} X-ray satellites, and the
projects were meant to be complementary.

\textsl{Muses-C}, which became \textsl{Hayabusa} after launch on 9 May 2003,
was supposed to examine asteroid Itokawa, break off a bit, and bring it back in
the summer of 2007. The bad news/good
news items have been coming out more or less monthly on beyond the end of the
reference year. It got to the asteroid. Two gyros failed. It managed to touch
down, or not? It picked up some stuff, or not? And it may or may not be able to
turn around and come home.

The American Physical Society, worried about folks saying, ``Gee, you don't
look very strong,'' like a Canadian border guard many years ago, attempted to
change its name to American Physics
Society (Science 309, 378). Loud and long blew the arguments through the summer
and fall, both rational and  irrational. But ``they fought the law, and the
law won.'' Changing the name would require redoing the incorporation
papers, a major hassle and expense. The intention, however, is to use the
APS most of the time and the full name only on journals and other items for
internal consumption. (Oh dear; the gastric
capacity required to internally consume a year of Physical Reviews terrifies.)

All the other items on this list belong more or less to NASA. These consist of
existing operational devices that may be turned off early, planned ones that
may be greatly delayed or shrunk, items pulled  out of various
over-committed\footnote{In first draft, this item said
``over-committeed,'' which may also be true.} queues and occasional/partial
reversals of such decisions, recisions of the reversals, and so forth that have
appeared in this context during the year. (1) \textsl{TRMM} and other
sources of climate data (Science 307, 186 \& 189). (2) \textsl{JIMO},
\textsl{Kepler}, \textsl{SIM}, \textsl{Beyond Einstein}, and even some of the
projects to examine long-term effects of space on people
(Science 307, 833). (3) \textsl{Voyager 1} and \textsl{2}, \textsl{Ulysses},
\textsl{Polar}, \textsl{Wind}, \textsl{Geotail}, \textsl{TRACE}, \textsl{FAST}
(Nature 434, 108). (4) The five-year career grants and archival data program
(Science 308, 486). (5) Shuttle flights, the \textsl{International Space
 Station}. and \textsl{Hubble Space Telescope} in multiple
stories (Science 309, 540; Nature 436, 603 \& 163), and, in even more multiple
stories, not all on paper, \textsl{JWST} (Science 309, 1472; Science 308, 935).

\section{SOLAR PHYSICS}
\subsection{The Solar Interior}
\subsubsection{Neutrino modulations}
After we got the solar neutrino flux right in first order, which was heralded
as a major breakthrough a few years ago, we can now concentrate on second-order
terms, such as variations due to the Earth's orbit, solar rotation, and
the solar cycle. The sinusoidal annual periodic
variations in the $^8$B solar neutrino flux have been verified by data from the
Sudbury Neutrino Observatory over a 4-year time interval, displaying a 7\%
modulation due to the Earth's orbital eccentricity (Aharmim et al.~2005),
also consistent with results from Super-Kamiokande (Sturrock et al.~2005).
The neutrino modulations due to magnetic field variations caused by solar
rotation and the solar cycle are harder to establish (Caldwell \& Sturrock
2005), and require also new physics in terms of a large neutrino transition
magnetic moment, as well as sterile neutrinos (Caldwell 2005; Caldwell \&
Sturrock 2005).  The intrinsic neutrino magnetic moment is now constrained
to less than a few times $10^{-12}$ of the Bohr magneton (Miranda et al.~2004).
Earlier interpretations in terms of the spin-flavor precession scenario are now
pretty much ruled out (Balantekin \& Volpe 2005; Caldwell \& Sturrock 2005).

\subsubsection{Solar Abundance Discrepancies}
The p-p, pep, $^8$B, $^{13}$N, $^{15}$O, and $^{17}$F solar neutrino flux
measurements even start to constrain the heavy-element abundances in the solar
interior (Bahcall et al.~2005a; Bahcall \& Serenelli 2005).
Standard solar models are in good agreement with the helioseismologically
determined sound speed and density in the solar interior, the depth of the
convection zone, and the abundance of helium at the surface, as long as
heavy-element abundances are not involved (Bahcall et al.~2005a, 2005b).
Downward revisions of the photospheric abundances of oxygen and other
heavy elements do not help, but upward revisions of the photospheric neon
abundance could possibly help (Antia \& Basu 2005; Drake \& Testa 2005),
or not (Schmelz et al.~2005). Dips in the inverted equation of state
at 0.975 and 0.988 $R_{\odot}$, however, could not be corrected by tuning
the helium abundance, but were rather attributed to inappropriate
approximations in the used equation of state (Lin \& D\"appen 2005).
Other authors conclude that a combination of opacity increases, diffusion
enhancements, and abundance increases remain the most physically plausible
means to restore agreement with helioseismology (Guzik et al.~2005).
Absolute helium abundances (of 12.2$\pm$2.4 \%) were also determined during
flares (Feldman et al.~2005).

\subsubsection{Tweaking the Helioseismic p-Mode Oscillations}
Helioseismic measurements determine the depth of the solar convection zone
with an impressive accuracy to $R_{CZ}=0.713\pm0.001$ $R_{\odot}$
(Bahcall et al.~2004). With similar accuracy, the solar rotation axis is
determined to an angle of $i=7.155^{\circ}\pm 0.002^{\circ}$
(Beck \& Giles 2005). Attempts were even made to determine Newton's
gravitational constant $G$ with helioseismic methods, but the achieved
accuracy could not beat previous experimental methods (Christensen-Dalsgaard
et al.~2005). Another unsuccessful attempt was a trial to detect deviations
from the constant rotation rate in the solar core (Chaplin et al.~2004),
where also thermal metastabilities are expected (Grandpierre \& Agoston 2005).
A number of effects have been studied that could explain tiny deviations
of the standard p-mode oscillation frequencies and line widths, such as:
the magnetic field in the second helium ionization zone at 0.98 $R_{\odot}$
(Basu \& Mandel 2004), at 0.99 $R_{\odot}$ (Dziembowski \& Goode 2005),
or even in the photospheric magnetic carpet (Erd\'elyi et al.~2005),
solar-cycle variations of MHD turbulence in the convection zone
(Bi \& Yan 2005; Chou \& Serebryanskiy 2005; Toutain \& Kosovichev 2005)
and tachocline (Foullon \& Roberts 2005),
the Reynolds stress on the p-mode damping rates (Chaplin et al.~2005),
mode conversion and damping by Alfv\'en waves in vertical fields
(Crouch \& Cally 2005), and
the effect of inhomogeneous subsurface flows (Shergelashvili \& Poedts 2005).

\subsubsection{Local Helioseismology through a Showerglass}
The epicenter of an earthquake is determined by correlating local seismic
detectors. In analogy, {\sl local helioseismology} probes the physical
properties of sunspots and active regions by localized variations of the
subsurface sound speed, mostly concentrated in shallow subsurface layers
at $r \approx 0.98-1.00$ $R_{\odot}$. One method is the time-distance
helioseismology,
which can study mass flows, active regions, and sunspots. This method
measures travel times with the ray or the Born approximation, but it turned
out that the first-order approximations fail to capture scattering effects
(Birch \& Felder 2004). Improvements concentrate on the inversion of
noisy correlated data with the time-distance method (Couvidat et al.~2005;
Gizon \& Birch 2004), comparison of subsurface flows between the time-distance
and ring analysis (Hindman et al.~2004), ring analysis of 2D shearing flows
(Hindman et al.~2005), or comparison of time-distance or ring analysis with
GONG and MDI data (Hughes et al.~2005; Komm et al.~2005). A problem of local
helioseismology
is the {\sl acoustic showerglass} effect: Magnetic fields under sunspots or
active regions suppress the photospheric signatures of acoustic waves impinging
onto them from the underlying solar interior and shift their phases, impairing
the coherence of seismic waves this way, smearing the holographic signatures of
possible subphotospheric anomalies (Lindsey \& Braun 2004, 2005a, 2005b;
Schunker et al.~2005). Put in simple words, the investigators try to get a
sharp image through a showerglass!

\subsubsection{More Puzzles about the Solar Dynamo}
Continuing the hotly debated question from last years, authors entertain us
further whether the dynamo is located in shallow subphotospheric layers
(Brandenburg 2005) or deep down in the tachocline (Gilman \& Rempel 2005;
Dikpati et al.~2005b; Ulrich \& Boyden 2005). In support of the latter, the
first self-consistent MHD simulations of the tachocline
and meridional circulation (Chou \& Ladenkov 2005)
were conducted by Sule et al.~(2005), but realistic
MHD models that simulate the entire convection zone down to the tachocline
are still not yet computationally feasible (Brun et al.~2004).
This year, the controversy also disgressed into a number of side issues.
Chatterjee et al.~(2004) explore a 2D kinematic solar dynamo model
based on the Babcock-Leighton idea with a full-sphere numerical simulation
and find that the dynamo is circulation-dominated, but Dikpati et al.~(2005a)
repeat the exercise and find that the dynamo is rather diffusion-dominated,
while the discrepancy is then explained by a different treatment of the magnetic
buoyancy (Choudhuri et al.~2005). One study extends the dynamo equations to
include the competing role of buoyancy and downflows and was able to reproduce
the 22 yr cycle (Li et al.~2005a). However, a cycle is not simple in nonlinear
dynamics, fluctuations in the Babcock-Leighton dynamo were actually shown to
lead to period doubling and to transition to chaos (Charbonneau et al.~2005),
possibly explaining the anomaly of the Maunder minimum
(Charbonneau 2004, 2005; Charbonneau et al.~2004).
Spherical harmonic decomposition of magnetic field data revealed also
intermittent oscillations with periods of 2.1-2.5 yr, 1.5-1.8 yr, and 1.2-1.4 yr
(Knaack \& Stenflo 2005; Knaack et al.~2005; Kane 2005a), similarly as found in
cosmic-ray modulations (Starodubtsev et al.~2004). While we believed that the
magnetic cycle is 22 years (Hale cycle) clocks everything on the Sun,
correlations  with the equatorial rotation rate actually reveal that the phase
 of the beginning of a 22 year cycle in the latitudinal gradients is out of
phase by 180$^\circ$ (Javaraiah et al.~2005).
A new technique based on dynamo spectroscopy and bi-orthogonal decomposition of
data was presented to actually compare theoretical dynamo models with
observations  (Mininni \& G\'omez 2004). Did you know that the northern
hemisphere rotates faster during the even cycles, while the southern hemisphere
wins the race in the odd ones (Gigolashvili et al.~2005; Ballester et al.~2005),
which was even expected theoretically (Itoh et al.~2005)?
Other issues of dynamo models touch on the requirement
of supercritical helicity fluxes (Brandenburg \& Subramanian 2005;
Choudhuri et al.~2004), or the radiative background flux in magneto-convection
(Brandenburg et al.~2005). The life histories of over 3000 supergranules have
been simulated and tracked and revealed lifetimes of 16-23 hours
(DeRosa \& Toomre 2004). The rumblings of the internal dynamo seems also to be
detectable in the quiet Sun by frequent shocks that bump up into the photosphere
(Socas-Navarro \& Manso 2005).

\subsection{Photosphere}
\subsubsection{The Tiniest Solar Magnetic Features}
Imaging of the photosphere at 0.1\arcsec resolution, an unprecedented capability
at the Swedish 1-m Solar Telescope on La Palma
that became available only recently, allows us to resolve magnetic features in
the solar photosphere down to the diffraction limit of $\approx 70$ km, which is
about the size of the city Los Angeles. While we are familiar with the
photospheric granulation pattern, which forms a grid of convection cells with
typical spatial scales of 1000-2000 km, the tiniest magnetic features
at 0.1\arcsec scale are mostly found in the intergranular lanes, described as
novel configurations of magnetic flux that are not directly resolvable into
conglomerations of flux tubes or uniform flux sheets (Berger et al.~2004).
The novel structures are also described as
{\sl elongated ribbons}, {\sl circular flowers}, and {\sl micro-pores},
which are thought to be crafted by the dynamics of weak upflows in the flux
sheets and downflows in the immediate surroundings, becoming unstable to a
fluting instability
so that the edges buckle and the sheets break up into strings of bright points
(Rouppe van der Voort et al.~2005). The tiny magnetic fluxtubes are subject
to a swaying motion, but with an amplitude smaller than 0.3\arcsec (Stangl
\& Hirzberger 2005). There is no comparison of such a surface swaying motion
on Earth, the biggest earthquake known in California moved the
coast of the Tomales Bay only by 20 feet.

\subsubsection{The Fractal Complexity of the Magnetic Field}
The spatial distribution of the photospheric magnetic field is as fractal as
the Atlantic coast of Norway. Threshold-based sampling in two active regions
revealed that the cumulative distribution functions of the magnetic flux are
only consistent with a lognormal function, but not with an exponential or
power-law function, suggesting that the process of fragmentation dominates
over the process of concentration in the formation of the magnetic structure
in an active region (Abramenko \& Longcope 2005). The fractal complexity is
thought to result from the continuous emergence of a multitude of mixed-polarity
magnetic concentrations, which are subsequently tangled up into intricate
regions of interconnecting flux (Close et al.~2004a), a dynamic process that
is captured in the recent {\sl flux-tube tectonics model} of
Priest et al.~(2002). Talking about mixed-polarity concentrations, network
magnetic patches were found
to harbor a mixture of strong ($\approx 1700$ G) and weak ($\la 500$ G) fields
(Socas-Navarro \& Lites 2004), exhibiting a dynamics that is not consistent
with the predictions from helioseismology (Meunier 2005).
The topological complexity is also illustrated by numerical simulations,
which reveal about 10 magnetic separators for each magnetic nullpoint
(Close et al.~2004b). Beveridge \& Longcope (2005) found a simple relation
between the numbers of separators (X), coronal null points ($N_c$), flux domains
(D) and flux sources (S): $D=X+S-N_c-1$, which can be used to characterize
the magnetic topology and bifurcation processes.

\subsubsection{Modeling of the Magnetic Field}
The Lorentz force and a corresponding lower limit of the cross-field electric
current density was measured in the photosphere, amounting to $\approx 1-10\%$
of the gravitational force in active regions (Georgoulis \& LaBonte 2004). The
photospheric magnetic field is therefore obviously not force-free, contrary to
other recent studies with force-free extrapolations
(Marsch et al.~2004; Wiegelmann et al.~2005a).
Force-free extrapolations, however, fit the coronal field lines observed in EUV
significantly better than potential fields (Wiegelmann et al.~2005a).
The nonpotentiality in active regions was found to occur (1) when new magnetic
flux emerged within the last 30 hours, and (2) when rapidly evolving,
opposite-polarity concentrations appear (detected with 4\arcsec resolution)
(Schrijver et al.~2005).

\subsubsection{Automated Pattern Recognition}
Finally we get the computers to do our work. Tired of manual and visual
inspections of countless features in solar images, tools come finally
online that perform {\sl automated pattern recognition}, which allow us to
analyze orders or magnitude more data, while nobody becomes unemployed, since
the maintenance of these new tools requires additional skilled manpower. In
a Special Topical Issue of Solar Physics (Vol. 228), a total of 24 papers
were presented that describe these new tools, such as:
fractal and multi-fractal analysis (Georgoulis 2005; Abramenko 2005a;
Revathy et al.~2005); automated boundary-extraction and region-growing
techniques (McAteer et al.~2005c); multi-scale Laplacian-of-Gaussian operator
and interactive {\sl medial-axis-transform}
segmentation techniques (Berrilli et al.~2005),
enhancing, thresholding, and morphological filtering (Bernasconi et al.~2005;
Qu et al.~2005), Euclidean distance transform (Ipson et al.~2005),
artificial (Zharkova \& Schetinin 2005),
and auto-associative neural network techniques (Socas-Navarro 2005a, 2005b).

These tools have been applied to a number of photospheric features, such as
granulation (Del Moro 2004), H$\alpha$ dark features (Liu et al.~2005a),
the chirality of filaments (Bernasconi et al.~2005), the inversion lines of
filament skeletons (Ipson et al.~2005),
or going from {\sl tactical to practical}, i.e., to predict space weather
and geoeffective events (Georgoulis 2005; Qu et al.~2005).

The sharpest high-resolution images have been super-sharpened with high-order
adaptive optics and speckle-masking reconstruction (Denker et al.~2005),
with phase-diversity speckle technique (Criscuoli et al.~2005;
Bonet et al.~2005), with multichannel blind deconvolution
(Simberova \& Flusser 2005), multiple spectral order stereoscopy
(DeForest et al.~2004), or with a combination of these methods
(Van Noort et al.~2005).

\subsubsection{Sunspots Dynamics}
Inversion of Stokes line profiles cannot distinguish whether the thermal
structure under a sunspot is monolithic or ``spaghetti-like'' (Socas-Navarro
et al.~2004). In the plumes above the umbra, however, things become
very dynamic, since both upflows and downflows with velocities of 25 km
s$^{-1}$ have been measured at different days (Brosius 2005), as well as
intermittent wavepackets of 5-min oscillations (Lin et al.~2005a).
Some penumbral segments were
even observed to rapidly disappear after a flare, probably as a result of
magnetic reconnection (Deng et al.~2005).

Polarimetry of sunspot penumbrae with high spatial resolution (0.5\arcsec)
confirm the picture that low-lying flow channels coincide with the horizontal
magnetic field, or possibly emerging and diving down into sub-photospheric
layers like a ``sea serpent'' (Bello Gonzalez et al.~2005).
Bellot Rubio et al.~(2004) find a perfect alignment between the magnetic field
vector and flow velocity vector in the penumbral fluxtubes, which is also
confirmed by S\'anchez-Almeida (2005) from fitting of 10,000 Fe I spectra.
The flows in the penumbral fluxtubes become supersonic and form shocks at
larger radial distances,
suggesting that the Evershed flows are driven by the siphon flow mechanism
(Borrero et al.~2005). Observations with 0.2\arcsec resolution give support
to fluted and uncombed models of the penumbra (Langhans et al.~2005).

\subsection{Chromosphere and Transition Region}
\subsubsection{DOT Tomography}
While the chromosphere has been generally perceived as a thin layer above
the solar surface, three-dimensional (wavelength-)tomography of its vertical
structure is now performed with the newly installed {\sl Dutch Open Telescope
(DOT)} on the Canary Island La Palma. Its revolutionary design
features a wind-swept open telescope on a non-blocking open pedestal to
minimize atmospheric seeing, documented in DOT paper I (Rutten et al.~2004a).
DOT paper II shows simultaneous high-resolution (0.2\arcsec) image sequences
in G band and Ca II H line, which show the anticorrelation and temporal delay
of reversed granulation features in different heights, believed to be produced
by a mixture of convection reversal and gravity waves (Rutten et al.~2004b).
DOT paper III backs up with 3D radiation-hydrodynamics simulations of the
granulation to simulate the observations and concludes that magnetic fields
play no major role in the formation of reversed granulation (Leenaarts \&
Wedemeyer-B\"ohm 2005). DOT paper IV
investigates bright points (with a lifetime of about 9 hours) within
longer-lived magnetic patches that outline cell patterns on mesogranular
scales, and concludes that the magnetic elements constituting strong
internetwork fields are not generated by a local turbulent dynamo (De Wijn
et al.~2005a). DOT paper V observes a surge above the solar limb (2 hours
before the largest X-class flare ever recorded) and finds evidence for
upward motion of material with velocities of $>50$ km s$^{-1}$,
brightness variations with periods of $\approx 6$ min in the surge,
and an inverted ``Y''-shape configuration that suggests magnetic reconnection
at the bottom of the surge as its driving mechanism (Tziotziou et al.~2005).

\subsubsection{Acoustic Waves in the Chromosphere}

``Is there a chromospheric footprint of the solar wind?'' ask McIntosh
\& Leamon (2005) and find a positive answer in the strong correlation between
solar wind velocity and composition measured at 1 AU and chromospheric
diagnostics of $O^{+7}/O^{+6}$ oxygen density ratios.
A search for high-frequency modulations in the chromosphere with
TRACE UV images finds evidence for acoustic modulations with periods
down to 50 s in internetwork areas, a possible signal for acoustic heating
of the corona (De Wijn et al.~2005b). Correlated analysis of photospheric
magnetograms with MDI and chromospheric UV continuum images with TRACE
showed that the oscillatory high-frequency power is enhanced in the
photosphere but reduced in the chromosphere, which may be explained
by the interaction of acoustic waves with the magnetic canopy
(Muglach et al.~2005). Non-LTE radiation hydrodynamic simulations
revealed that the TRACE UV continuum bands (1600, 1700 \ang )
are sensitive for the detection of high-frequency acoustic waves in
a chromospheric height range of 360-430 km (Fossum \& Carlsson 2005a).
The same authors, however, come to the conclusions that high-frequency
acoustic waves are not sufficient to heat the chromosphere (Fossum \&
Carlsson 2005b). The 3D topography of magnetic canopies in and
around active regions was also mapped helioseismically from the propagation
behavior of high-frequency acoustic waves in the chromosphere
(Finsterle et al.~2004). Helioseismic global modes cause 5-min
oscillations that can be traced even above the chromosphere in coronal
network bright points (Ugarte-Urra et al.~2004).
The occurrence of acoustic waves complicates also
the definition of an average temperature in time-dependent chromospheric
models, yielding ionization temperatures up to a factor 150 higher than
the mean or median temperature (Rammacher \& Cuntz 2005).

\subsubsection{Spicular Flows Revealed}

More accurate physical properties of chromospheric spicules have been derived
from Stokes polarimetry in Ca II and He I lines, yielding mostly nonthermal
broadening ($\ga16$ km s$^{-1}$) and upper temperature limits of
$T\le 13,000$ K (Socas-Navarro \& Elmore 2005), as expected for upwardly
propelled cool chromospheric material. Line broadening of EUV lines across
the solar limb is mostly associated with unresolved flows in spicules and
macrospicules (Doyle et al.~2005a). Both initial rise and subsequent fall
motion has been observed as a sudden change of the Doppler velocity sign
(Xia et al.~2005). Using the Hanle and Zeeman effects in
spicules, magnetic fields of $\approx 10$ G and inclination angles of
$\approx 35^\circ$ to the vertical have been inferred (Trujillo Bueno
et al.~2005).

\subsubsection{Small-Scale Variability}

Have you ever analyzed 130,000,000 objects of an astronomical database?
McIntosh \& Gurman (2005) analyzed that many EUV bright points observed
with the EIT telescope on SoHO. The statistics of the lifetime of these
events was so overwhelming that deviations from straight powerlaw distributions
could be determined, varying with temperature filters and time during the
observed 9 years of the present solar cycle.
Numerical MHD simulations of such elementary heating events envision
separator and separatrix reconnection as drivers of these small-scale phenomena
(Parnell \& Galsgaard 2004).

While nanoflares and EUV bright points seem to originate in the corona,
the so-called ``EUV explosive events'' seem to be formed deeper down in
the transition region and chromosphere, according to some multi-wavelength
studies that cover the entire chromospheric temperature range
(Doyle et al.~2005b, Mendoza-Torres et al.~2005). Their average size is
estimated to 1800 km and their occurrence rate to 2500 s$^{-1}$ over the
entire Sun, but they are insufficient to contribute significantly to
coronal heating (Teriaca et al.~2004). Another type of small-scale
phenomena, so-called ``blinkers'', seems not to be connected with the
phenomenon of ``EUV explosive events'' (Bewsher et al.~2005).

\subsection{Corona}

\subsubsection{Footpoint-Driven Hydrodynamics of Coronal Loops}

The solar corona is believed to have a low plasma-$\beta$ parameter almost
everywhere, so the magnetic pressure dominates over the thermal pressure, and
thus is responsible for the appearance of myriads of loops. These bright loops
are denser than the ambient corona, and thus have to be filled by chromospheric
material, since there is no way to constrict coronal plasma to the observed
densities (beyond twist angles $\ga 1.5\pi$, Chae \& Moon 2005).
So, we have the picture that coronal loops are like heat pipes,
which are constantly flushed by heated plasma that is ablated from the
chromosphere. Therefore, it seems straightforward to measure the electron
density $n_e(s)$, the
electron temperature $T_e(s)$, and the flow speed $v(s)$ along these coronal
heat pipes to model and understand the hydrodynamics of the solar corona.
The reality, however, seems to be more complicated, because it is very tricky
to properly isolate a single loop from the thousands of other foreground and
background loops along the same line-of-sight. Nevertheless, new loop modeling
 attempts have been performed,
using constraints from multi-wavelength observations with CDS, EIT, TRACE,
Yohkoh, or optical instruments, and find evidence for
energy input (heating) at the loop footpoints (Ugarte-Urra et al.~2005),
upflows driven by chromospheric evaporation (Singh et al.~2005),
unidirectional flows along a loop (Gontikakis et al.~2005),
plasma cooling from soft X-ray to EUV temperatures (Winebarger \& Warren 2005),
downflows of plasma on both loop sides (Borgazzi \& Costa 2005),
and ``high-speed coronal rain'' (M\"uller et al.~2005).
Statistical approaches focus on the tomographic reconstruction of the coronal
differential emission measure (DEM) distribution (Frazin et al.~2005;
Frazin \& Kamalabadi 2005a), or measuring plasma downflows that imply departures
from the ionization equilibrium and thus violate basic assumptions of the
DEM method (Lanzafame et al.~2005). Modeling approaches include realistic
spatial heating functions (Mok et al.~2005; Landi \& Landini 2005), and
find that only pulsed
footpoint heating can reproduce the strongly peaked DEM with a slope of
$\propto T^5$ observed in stars (Testa et al.~2005), and that
coronal condensation and catastrophic cooling around the loop apex is also a
consequence of footpoint heating (M\"uller et al.~2005;
Mendoza-Briceno et al.~2005).

\subsubsection{The Conundrum of Coronal Heating}
After the coronal heating problem has been with us for over six decades, a
{\sl par force} strategy using high-performance computers and all available MHD
physics we know of seems to be in place. Such an {\sl ab initio} approach
has been carried out by Gudiksen \& Nordlund (2005a,b), who simulated in a
computational box using a 3D MHD code that handles the photospheric boundary
condition with granular velocity fields (constrained by observed SoHO/MDI
high-resolution magnetograms), mimics the resulting tanglings and braidings
of magnetic field lines in the transition region and corona, and finally
simulates the dissipation of braiding-driven currents in coronal loops,
which match the characteristics of loops observed in EUV with TRACE.
The full-MHD simulations by Gudiksen \& Nordlund (2005a,b) reproduce also
the footpoint heating (mentioned above), i.e., the heating is largest at
low heights ($\la 5$ Mm) because of the stronger stressing of the
high plasma-$\beta$ environment in the transition region.
Actually, 90\% of the total dissipated energy is dissipated below the transition
region in these simulations. This result is geometrically different from
Parker's original scenario, where braiding and related current dissipation is
almost uniformly distributed throughout the corona, although the basic physics
(of dissipation of DC currents) is essentially the same. Footpoint braiding
seems be more efficient for hot loops than for cool loops as a consequence of a
lower filling factor and higher horizontal velocity (Katsukawa \& Tsuneta 2005).
Further support for the
preferential current dissipation in the transition region was also furnished
by detailed MHD forward-modeling of the DEM distribution, flow speeds, and
emissivity of EUV lines in the temperature range of the transition region
(Peter et al.~2004). Alternative scenarios with nanoflares distributed in the
coronal part \`a la Parker were not able to diagnose the spatial heating
distribution from loop hydrodynamic simulations (Patsourakos \& Klimchuk 2005).

Other studies on the coronal heating problem focused on the dependence of the
heating rate on the driving velocity and emerging flux
(Galsgaard \& Parnell 2005; Galsgaard et al.~2005), the ``tectonic''
build-up of current sheets along quasi-separatrix layers (Mellor et al.~2005;
Priest et al.~2005), the switch-on mechanism as function of Parker's
magnetic field misalignment angle (Dahlburg et al.~2005), the Rayleigh-Taylor
instability that organizes coronal heating in a spatially intermittent way
(Isobe et al.~2005),
global scaling laws of the heating flux density
($F_H \propto B/L$, with $B$ the magnetic field and $L$ the loop length) from
full-Sun (Schrijver et al.~2004) and full-star simulations
(Schrijver \& Title 2005).
Subphotospheric fluxtubes that emerge and drive reconnection in the transition
region have been found to behave quite differently depending on their twist:
low-twist tubes slingshot while high-twist tubes tunnel
(Linton \& Antiochos 2005).
In one case, the type of magnetic reconnection was identified as {\sl separator
reconnection} during the emergence of an active region (Longcope et al.~2005).

While coronal heating in closed magnetic fields (e.g., in active region loops)
seems to be controlled by DC currents, the coronal heating in open field lines
(mostly in coronal holes and in the solar wind) seems to be accomplished by
dissipation of AC currents, most likely conveyed by high-frequency Alfv\'enic
waves. Related studies concentrated on
wave energy dissipation by viscous and resistive damping (Craig \& Fruit 2005),
two-fluid simulations of turbulence-driven Alfv\'enic heating
(O'Neill \& Li 2005),
and the nonthermal line broadening of minor ions caused by high-frequency
Alfv\'en waves (Ofman et al.~2005).

\subsubsection{Coronal MHD Oscillations and Waves}
The relatively new discipline of {\sl coronal seismology} continues to prosper,
as the over 40 refereed publications during this year indicate. New
studies on MHD oscillations of coronal loops, mostly conducted with the aid of
MHD simulations, explore second-order effects now, such as
the influence of density stratification on resonant damping
(Andries et al.~2005a; DelZanna et al.~2005),
the damping of vertical oscillations by wave tunneling (Brady \& Arber 2005),
diagnostic of density stratification from harmonic overtones
(Andries et al.~2005b), oscillations in loops with a hot core and cool shell
(Mikhalyaev \& Solovev 2005),
and the influence of loop curvature and asymmetric excitation
(Murawski et al.~2005a; Selwa et al.~2005a,b; Selwa \& Murawski 2004;
Taroyan et al.~2005).
A statistical study of coronal loop oscillations with radio type II bursts
established that the excitation of oscillations is triggered by the passage
of a flare shock wave (Hudson \& Warmuth 2004). A spectral study with SUMER
concluded that the initiation of longitudinal loop oscillations is not caused
by (symmetric) chromospheric evaporation, but rather by a one-sided (asymmetric)
pulse of injected hot plasma (Wang et al.~2005a). Fast MHD oscillations (with a
period of 10 s) have even be detected on the star EV Lac (Stepanov et al.~2005).

While MHD oscillations require a settling into an eigen-mode, we observed also
a variety of phenomena associated with propagating waves, particularly in
open field regions, long loops, and strongly asymmetric loops.
New theoretical/numerical studies on propagating waves include
wave damping by phase mixing, which could not explain the observed strong
damping  (DeMoortel et al.~2004),
siphon flows and oscillations in long coronal loops due to Alfv\'en waves
(Grappin et al.~2005),
the effects of magnetic shear on MHD normal modes (Arregui et al.~2004),
the impulsive excitation of MHD waves in a loop arcade (Murawski et al.~2005b),
and MHD wave propagation near magnetic null points (McLaughlin \& Hood 2005).

New observational studies deal with
the discovery of high-frequency ($\approx$10 s period) waves in far UV
1600 \ang\ (DeForest 2004) and in radio (Ramesh et al.~2005),
the first detection of global waves in soft X-rays with GOES/SXI
(Warmuth et al.~2005), high-cadence radio observations of an EIT wave during the
first 4 minutes (White \& Thompson 2005), and the origin of global (EIT) waves
(Cliver et al.~2005).

\subsubsection{Twisted, Stressed, and Kinked Magnetic Fields}

It is still difficult to measure the coronal magnetic field, but new full-Stokes
spectropolarimeteric measurements with the coronal Fe XIII 1075 nm line are
pioneered, yielding fields of 4 G at heights of 70 Mm above the limb
(Lin et al.~2004). Other coronal magnetography techniques employ the circular
polarization of radio emission, finding fields of 20-85 G in heights of
23-62 Mm (Ryabov et al.~2005). Another fingerprinting technique employs
3D magnetic field modeling to match up 2D EUV images
(Wiegelmann et al.~2005a,b), which can also be used the other way around for
automated loop detection  (Lee et al.~2005).

Although most of the coronal field lines are dipolar to first order, and thus
close to a current-free potential field, there is far more interesting physics
hidden in the second-order deviations, such as non-potential fields and the
associated currents. It is therefore no surprise that most of the 40 papers
published about the solar
coronal magnetic field, deal with twisted, stressed, kinked, and nonpotential
fields. Dynamic modeling of the magnetic field braiding reveals that the
quiet-Sun corona is often neither quasi-steady nor force-free
(Schrijver \& Ballegooijen 2005).
Non-current-free coronal field lines can already be diagnosed from
subphotospheric MHD models (Amari et al.~2005; Gibson et al.~2004),
which recycle the coronal magnetic
flux on time scales as short as 3 h or 1.4 h (Close et al.~2005),
constantly injecting magnetic helicity
into the corona (Amari et al.~2004), contributing to the coronal heating of soft
X-ray emitting loops (Maeshiro et al.~2005; Yamamoto et al.~2005), may stay in
equilibrium even as sigmoids (Aulanier et al.~2005; Regnier \& Amari 2004), or
lead to plasmoid eruptions (Kusano 2005). The height were a twisted flux rope
loses its  equilibrium is lower than 25\% or the active region separation
(Lin \& van Ballegooijen 2005). Some active region
loops may have more than 2$\pi$ twist and thus prone to the kink instability
(Leka et al.~2005; Tian et al.~2005b), or when one leg is rotated by
$40^\circ-200^\circ$ by a nearby sunspot (Gibson et al.~2004).
One active region was estimated to contain an unusually large amount of free
magnetic energy ($6 \times 10^{33}$ ergs) before an X10 flare
(Metcalf et al.~2005).
Magnetic loops with the same handedness or the writhe and the
twist may rotate in the corona for a long time (Tian et al.~2005a).

\subsubsection{Coronal EUV Emission}
Disentangling the inhomogeneous and fractal landscape (McAteer et al.~2005a)
of coronal EUV emission was always challenging. New methods explore rotational
tomography for 3D reconstruction of the white-light and EUV corona
(Frazin \& Kamalabadi 2005b). Coronal EUV emission is highly anisotropic. An
8-year long study of SoHO/EIT data
demonstrated that the He II 30.4 nm flux displays polar/equatorial anisotropy of
90 \% at solar minimum to 60 \% at solar maximum, as well as a difference of
20 \% between the north and south polar fluxes (Auchere et al.~2005).
Historically, coronal EUV emission varied much more, X-ray and EUV emission was
100-1000 times stronger at the time of formation of planetary atmospheres than
at present (Ribas et al.~2005).

\subsubsection{Coronal Holes}
The elusive tracers of coronal heating were also sought in void regions like
coronal holes, where source confusion and crowded structures are minimized.
Links between plasma upflows (detected from Ne VIII Doppler shifts) and isolated
closed-field regions in coronal holes have been found (Wiegelmann et al.~2005c).
Correlations between the EUV intensities in coronal holes and quiet-Sun
regions were taken as evidence for continuous reconnection between open and
closed field regions (Raju et al.~2005). Macrospicules were found to have either
a spiked jet or an erupting loop, suggesting reconnection between the network
bipole and open magnetic fields (Yamauchi et al.~2005). Other indirect tracers
of coronal heating were extracted from Mg X line width decreases, probably
caused by damping of upwardly propagating Alfv\'en waves (O'Shea et al.~2005).

\subsubsection{Quiescent Filaments and Prominences}
We counted 29 papers on investigations of quiescent filaments or prominences,
which include their automated detection (Bernasconi et al.~2005;
Fuller et al.~2005; Qu et al.~2005; Zharkova \& Schetinin 2005),
their production mechanism (Liu et al.~2005c; Spadaro et al.~2004;
Litvinenko \& Wheatland 2005), a new mass determination method
(Gilbert et al.~2005), their thermodynamic stability (Costa et al.~2004;
Low \& Petrie 2005; Petrie \& Low 2005; Petrie et al.~2005),
their oscillations (Diaz et al.~2005; Dymova \& Ruderman 2005;
Foullon et al.~2004), with ultra-long periods up to 8-27 hours
(Foullon et al.~2004), their wave damping (Terradas et al.~2005), their
absorption and volume blocking (Anzer \& Heinzel 2005;
Stellmacher \& Wiehr 2005), their NLTE radiative transfer (Gouttebroze 2005),
their magnetic topology (Lites 2005), their chirality
(Mackay \& van Ballegooijen 2005),
their possible electric field (Lopez Ariste et al.~2005),
and their finestructure in form of threads (Lin et al.~2005c)
and barbs (Chae et al.~2005; Lin et al.~2005d; Su et al.~2005).
 - Did we miss anything?

\subsection{Flares}
\subsubsection{Direct Observations of Magnetic Reconnection Sites}
Some kind of magnetic reconnection is thought to be the driver of flares/CMEs
in every flare model. Thus, a direct observation of a magnetic reconnection
site would be the ``holy grail''. Such a discovery was indeed announced in the
observation of the 2003 Nov 18 flare, where the formation of a current sheet
was observed behind an erupting CME, with flare loops forming in the wake
beneath (Lin et al.~2005b). The CME sped off with a velocity of 1500-2000
km s$^{-1}$, and lateral reconnection inflow speeds of 10-100 km s$^{-1}$
and outflow speeds of 500-1000 km s$^{-1}$ were measured, leading to a
reconnection rate with Mach numbers of $M=0.01-0.23$ (Lin et al.~2005b).
In the early stage of 13 well-observed two-ribbon flares, a strong correlation
was found between the magnetic reconnection rate and the acceleration of the
associated erupting filaments, yielding support for the flare model of
Forbes \& Lin, which is driven by the converging footpoints (Jing et al.~2005a;
Sakajiri et al.~2004).
An indirect calculation of the reconnection rate (of $\approx 0.001-0.03$) was
determined from the footpoint motion seen in EUV (Noglik et al.~2005) and in
UV (Fletcher et al.~2004).
Further tail-lights of the reconnection process have been sighted in radio
type II bursts observed at 40-80 MHz and 300 MHz, believed to be the signatures
of the upper and lower reconnection outflow termination shock (Aurass \& Mann
2004). While the newly-reconnected magnetic field line arcade is rooted in a
two-ribbon structure in most flare models, observations reveal also the
occasional involvement of a remote third ribbon, moving away from the flare
site with a speed of 30-100 km s$^{-1}$ (Wang 2005).

Forced magnetic reconnection was simulated in more detail, showing the current
sheet thinning and onset and progress of fast magnetic reconnection, and leading
to similar final states with Hall-MHD fluid or particle kinetic codes
(Birn et al.~2005).
Other theoretical studies emphasize the importance of viscous heating in
magnetic X-points (Craig et al.~2005), the kinetic effects of the Hall current
in the reconnection process (Morales et al.~2005), multiple fast shocks created
by the secondary tearing instability (Tanuma \& Shibata 2005), and the structure
of the reconnection outflow jets (Vrsnak \& Skender 2005).

\subsubsection{Magnetic Field Changes During Flares}
While a magnetic reconnection process changes only the local connectivity,
the changes induced on larger scales or even in the photospheric boundary
are less obvious. Nevertheless, major flares, such as white-light
flares with energies of $10^{33}$ ergs (Li et al.~2005d), can lead to
large irreversible magnetic flux increases of up to $\la 10^{21}$ Mx
(Zharkova et al.~2005), which are capable to heat sunspots (Li et al.~2005c),
disintegrate $\delta$-configurations (Liu et al.~2005b; Wang et al.~2005b),
weaken the penumbral structure, and slow down its Evershed flow
(Wang et al.~2005b).

Of course, you want to know if the preflare magnetic configuration allows us
to predict the flare magnitude. Power spectra of magnetograms revealed a
steeper spectrum for X-class flare-producing active regions, so some active
regions are ``born bad'' and become predictably more violent later on
(Abramenko 2005b). Then flares occur preferentially in regions with a high
gradient in twist and close to chirality inversion lines (Hahn et al.~2005),
and in regions with strong shear flows, counterstreaming, and complex flow
patterns (Yang et al.~2004).

The coronal magnetic field changes during a flare should lead us to the
relevant flare model. Observational studies find loop-loop interactions with
coalescence instability (Wu et al.~2005a) or quadrupolar double arcade with
undetected far-end ribbons (Wang et al.~2005b).

\subsubsection{Particle Acceleration During Flares}
The more we can nail down the magnetic topology of reconnection regions
from direct observations, the better we can hand over the likely parameters
of accelerating fields to the theoreticians. There is no shortage of theoretical
models and simulations of any imaginable particle acceleration scenario, such as
Fermi and betatron acceleration in collapsing magnetic traps (Bogachev
\& Somov 2005), particle acceleration in turbulent current sheets
(Dmitruk et al.~2004; Wu et al.~2005b),
in reconnecting current sheets (Wood \& Neukirch 2005),
in 2D X-points (Hamilton et al.~2005),
in 3D reconnecting current sheets with chaotic orbits
(Efthymiopoulos et al.~2005; Dalla \& Browning 2005),
proton acceleration in coalescing loops (Sakai \& Shimada 2004, 2005) and
between two colliding moving solitary magnetic kinks (Sakai \& Kakimoto 2004),
acceleration in field line advection models (Sokolov et al.~2004), and
in phase-mixing regions of shear Alfv\'en waves (Tsiklauri et al.~2005a,b).
Unfortunately, the observers cannot catch up with sufficiently discriminative
diagnostics. In two innovative studies, it was attempted to measure the level of
microturbulence, which controls stochastic acceleration, from so-called
(hitherto undetected) ``resonant transition radiation'' in radio data
(Nita et al.~2005; Fleishman et al.~2005).

\subsubsection{RHESSI Observations}
RHESSI has completed four years (2002-2006) of its mission and certainly
continues to stimulate solar flare research, producing over 80 papers during the
last year. Since the strength of RHESSI lies in (1) the first imaging at high
energies, (2) the high spectral resolution that allows to resolve most of the
gamma-ray lines, and (3) the high-resolution spectroscopy also at lower hard
X-ray energies, we summarize some new RHESSI results in the same order: (1)
Imaging with RHESSI revealed the evolution of progressing reconnection along a
flare loop arcade (Grigis \& Benz 2005a; Li et al.~2005e),
the so far unexplained loop-top altitude decrease in the initial phase of
flares (Veronig et al.~2005a), and the obscured view of a giant flare
with an energy of $\approx 10^{34}$ ergs (Kane et al.~2005);
(2) Gamma-ray line modeling with RHESSI showed us a 511 keV $e^+-e^-$
annihilation line that is so broad that the ambient ionized medium needs a
temperature of $10^5$ K, instead of the expected much lower chromospheric value
(Share et al.~2004); and (3) high-resolution spectroscopy with RHESSI gave us
new insights into the energy partition of thermal, nonthermal, CME-mechanical,
and nonpotential magnetic energies (Emslie et al.~2004, 2005;
Saint-Hilaire \& Benz 2005),the soft-hard-soft evolution of
hard X-ray spectra compared with acceleration models
(Grigis \& Benz 2004, 2005b),
the low-energy cutoff of the electron spectrum (Sui et al.~2005),
the physics of the Neupert effect, i.e., the correlation between the thermal
soft X-ray and the integral of the hard X-ray time profiles
(Veronig et al.~2005b), and the size dependence of solar flare spectral
properties (Battaglia et al.~2005).
Other exciting RHESSI discoveries are the quasi-periodic hard X-ray pulsations
that could be explained in terms of the MHD kink mode, which supposedly
modulates the electron injection in a multiple flare-loop system
(Foullon et al.~2005). Another surprising result was that no coherent radio
emission was detected in 17\% of hard X-ray flares (Benz et al.~2005), since
both emissions are produced by electrons of similar energy and occasionally
coincide with sub-second accuracy (Arzner \& Benz 2005). A puzzle is also the
absence of linear polarization in H-$\alpha$ mission, which limits the
anisotropy of energetic protons and refutes earlier positive reports
(Bianda et al.~2005).

Theoretical modeling of RHESSI data included fast electron slowing-down and
diffusion in high-temperature coronal sources (Galloway et al.~2005), inversion
of hard X-ray spectra with generalized regularization techniques
(Kontar et al.~2004, 2005, Kontar \& MacKinnon 2005, Massone et al.~2004),
Fokker-Planck modeling of electron beam precipitation
(Zharkova \& Gordovskyy 2005) producing
asymmetric footpoint hard X-ray sources (McClements \& Alexander 2005), and
the viewing angle of H-$\alpha$ impact polarization
(Zharkova \& Kashapova 2005).

\subsubsection{Flare Oscillations and Waves}
The discovery of a harmonic oscillation in a solar flare feels like the
beginning of a symphony concert, after the cacophonic tuning of orchestral
instruments comes to a
halt that unevitably precedes every concert performance. It is really not much
different in the performance of a solar flare, except that harmonic oscillations
of the flare plasma are conducted and triggered by a magnetic instability.
The first high-fidelity record (with high spatial resolution) of a long-period
($P \approx 9-12$ min and $9-23$ min) quasi-periodic oscillation of (3-25 keV)
hard X-ray radiation during solar flares was imaged in a trans-equatorial
flare loop with RHESSI, interpreted
in terms of MHD kink-mode modulated injection of X-ray-emitting electrons
(Foullon et al.~2005).

A similarly exciting discovery was made in form of downward-propagating
quasi-periodic transverse waves with periods of $P=90-220$ s
in post-flare supra-arcade structures, interpreted in terms of propagating
MHD kink-mode waves (Verwichte et al.~2005), in contrast to the standing mode
mentioned above (Foullon et al.~2005). After the longitudinal MHD slow-mode
oscillations have been discovered in soft X-rays with SoHO/SUMER a few years
ago, they could also be re-discovered in Yohkoh data (Mariska 2005, 2006),
and were also claimed to be discovered on the M-type dwarf AT Mic with the
XMM-Newton telescope (Mitra-Kraev et al.~2005).

One of the first imaging observations of the MHD fast sausage mode has been
accomplished in radio wavelengths with the Nobeyama interferometer, where
periods of $P=14-17$ s (global sausage mode) and $P=8-11$ s (possible
higher harmonics) have been measured (Melnikov et al.~2005). Additional
flare-triggered oscillations have also been detected in H-$\alpha$ with periods
of $P=40-80$ s (McAteer et al.~2005b), and in seismic (photospheric)
magnetogram data, probably triggered by precipitating high-energy protons
(Donea \& Lindsey 2005).

\subsubsection{Flare Simulations}
Hydrodynamic flare simulations of the chromospheric heating in response to
precipitating high-energy particles become more refined and
multi-wavelength comprehensive, predicting that moderate flares have a long
gentle phase with a near balance between flare heating and radiative cooling
(Allred et al.~2005; Berlicki et al.~2005),
while the gentle phase is much shorter or even absent in strong flares
(Allred et al.~2005), possibly the case in dME flare stars or even in Barnard's
star (Paulson et al.~2006). The occurrence of chromospheric evaporation
{\sl before} the impulsive flare phase was interpreted in favor of the magnetic
break-out model (Harra et al.~2005). The interpretation of data is complicated
because upflows of heated plasma and downflows of catastrophically cooled flare
plasma can be simultaneous and almost cospatial (Kamio et al.~2005), masking the
blue-shifted upflows (Warren \& Doschek 2005; Doschek \& Warren 2005).
Simulations show also that localized heating events far away from the loop apex
can produce bright EUV knots near the loop top during the cooling phase
(Patsourakos et al.~2004). Further complications could arise from the
consequences of non-equilibrium ionization
balance, which can make standard temperature diagnostic unreliable
(Bradshaw et al.~2004).

\subsubsection{Flare Radio Observations}
Some unusual radio observations during flares included the first ``zebra
pattern'' observations at frequencies of 5.6 GHz, believed to be produced by
coupling Bernstein waves in magnetic fields of 60-80 G (Altyntsev et al.~2005),
fiber burst observations in postflare loops used to infer the 3D magnetic field
(Aurass et al.~2005), and simultaneous remote radio and in-situ particle
detections of solar energetic electron events (Klein et al.~2005).

\subsubsection{The Largest Solar Flare}
To classify the magnitude of a flare, the soft X-ray flux registered by the GOES
spacecraft is generally used. For the 4 Nov 2003 flare, however, the detectors
on the GOES-12 satellite saturated. Brodrick et al.~(2005) managed to quantify
the magnitude of this largest solar X-ray flare on record by using the
recordings of a pair of 20.1 MHz riometers which register the ionospheric
attenuation of the galactic radio background, yielding a magnitude of
3.4--4.8 mW m$^{-2}$, corresponding to GOES class X34-X48.

During the same month, on 20 Nov 2003, the largest geomagnetic storm of solar
cycle 23  was registered, triggered by a CME that left the Sun with a projected
speed of  $\approx 1660$ km s$^{-1}$ and had a very strong southward pointing
axial field of the magnetic cloud (Gopalswamy et al.~2005a). The ``Halloween''
 period of Oct/Nov 2003 was extremely
active, resulting in 80 CMEs, many ultrafast ($>2000$ km s$^{-1}$), with a
record of  2700 km s$^{-1}$ on 4 Nov 2003, and many of them where highly
geoeffective (Gopalswamy et al.~2005b).

\subsection{CMEs}
\subsubsection{Erupting Filaments and Prominences}
Statistical studies reveal that
filament eruptions have a very high association rate with flares and CMEs in
active regions (Jing et al.~2004b). Therefore,
the most interesting questions about filaments are concerned with the
instability that causes them to erupt, especially since the consequences are the
launches of CMEs and their possible geoeffective impacts. Observational evidence
for the MHD helical kink instability has been established (Rust \& LaBonte 2005;
Williams et al.~2005), while a confined (failed) eruption is also observed
and consistently simulated (T\"or\"ok \& Kliem 2005; Fan 2005).
However, the observational
determination of sufficient magnetic twist can be underestimated according to
force-free field simulations (Leka et al.~2005). Romano et al.~(2005) find that
the transport of magnetic helicity exceeding the kink instability criterion is
primarily due to photospheric motion, rather than emerging flux. In addition,
reconnection in overlying fields, such as envisioned in the magnetic breakout
model or tether-cutting model are also a controlling factor for the eruption
of a filament (Sterling \& Moore 2004, 2005).

\subsubsection{Magnetic Field Configuration of CMEs}
The magnetic breakout model, in which magnetic reconnection above a filament
channel is responsible for disrupting the coronal magnetic field,
seems to be the favorite workhorse of current CME modeling.
The first MHD simulation of the complete breakout process including the
initiation, the plasmoid formation and ejection, and the eventual relaxation
of the coronal field to a more potential state were presented by
MacNeice et al.~(2004). The magnetic helicity is found to be well conserved
during the breakout, about 90\% is carried by the escaping plasmoid, while
about 10\% remain in the corona (MacNeice et al.~2004).
However, the amount of helicity seems not to be critical (Phillips et al.~2005).
Also, there seems to be no lower limit, since even a mini-sigmoid with two
orders of magnitude smaller magnetic flux than average was found to erupt and to
produce a mini-magnetic cloud (Mandrini et al.~2005).
The free magnetic energy available to drive a CME (which entails
the coronal null region in the magnetic breakout model) was found to be
concentrated at 1.25-1.75 solar radii (DeVore \& Antiochos 2005),
containing two catastrophic points (Zhang et al.~2005). The eruption
speed becomes Alfv\'enic at 2.5 solar radii and the magnetic fields in the
erupting flux rope can be well approximated by the Lundquist solution when
the ejecta are at 15 solar radii and beyond (Lynch et al.~2004).

While the magnetic breakout model is found to be consistent with most
observations, alternative models with emerging flux and small-scale reconnection
in the chromosphere were found to explain some surge-CME events
(Liu et al.~2005d).
The newly emerging flux that triggers a CME often emerges with an opposite sign
in the helicity than that of the pre-existing active region (Wang et al.~2004).

\subsubsection{CME Global Waves}
The launch of a CME often causes a detectable concentric wave that propagates
spherically over the solar globe, also called {\sl Moreton wave},
{\sl EIT wave}, or {\sl radio type II burst}, depending on the wavelength and
height it is detected.
The phenomenon of EIT waves and EUV dimming is now shown to be clearly a coronal
phenomenon, detected even at coronal temperatures of 2 MK in the
Fe XV 284 \ang\ line (Zhukov \& Auchere 2004; Chertok \& Grechnev 2005), and
even up to 6 MK (Poletto et al.~2004). Dimming is also seen in the
chromosphere (detected in the He I 1083 \ang\ line) in form of a transient
coronal hole (DeToma et al.~2005).
A reconciling picture was brought forward by Chen et al.~(2005), who
simulated how the typical features of EIT waves can be reproduced
by successive stretching or opening of closed field lines in the wake of
an erupting flux rope, causing the wave speed to stop near separatrices
and to accelerate between active regions and quiet Sun regions.
EIT waves can be best detected with automated algorithms from the dimming
they leave behind the wavefront in form of a temporary density rarefaction
(Podladchikova \& Berghmans 2005; Robbrecht \& Berghmans 2005).
The H-$\alpha$/EIT wave is thought to show up in the corona and interplanetary
space as a shock wave (radio type II burst; Cliver et al.~2004;
Knock \& Cairns 2005) or as modulation of the optically thin gyrosynchrotron
emission (radio type IV burst)
excited by the passage of the shock (Vrsnak et al.~2005a; Pick et al.~2005;
Pohjolainen et al.~2005).

\subsubsection{The CME-Flare Connection}
The flare phenomenon is now clearly established as a by-product of the same
magnetic instability that drives a CME. There is statistically really no
significant difference in the kinematic properties of flare-associated CMEs
and non-flare CMEs (Vrsnak et al.~2005b). Also the correlation between
flare-associated X-ray plasma ejections and CMEs was found to be strong
(Kim et al.~2005). The intimate relation between
CMEs and flares becomes even clearer when we look at the cusped postflare
loops that rise in altitude in the wake of erupting flux ropes
(Goff et al.~2005). The very onset of a flare-associated CME was for the first
time observed in the optical green-line (Fe XIV, 5303 \ang\ ) at a
temperature of 2 MK (Hori et al.~2005).

\subsubsection{3D Vision of CMEs}

Even we do not have the 3D capabilities of the soon to-be-launched STEREO
mission at hand yet, some novel 3D visualizations of CMEs have been
accomplished by using the white-light polarization of LASCO images
(Dere et al.~2005), similarly as recently pioneered by Moran \& Davila (2004).
A triangulation method to reconstruct basic 3D geometric parameters of CMEs
with the STEREO spacecraft have also been developed already (Pizzo \&
Biesecker 2004).

\subsubsection{CME Kinematics}

The trajectories of CMEs are difficult to reconstruct from one
vantage point alone, because they follow a curved path along the
Parker-Archimedean spiral,
but in combination with the all-sky monitor SMIE on the Coriolis spacecraft
more accurate trajectories and 3D velocities could be determined,
which enhance the accuracy of space weather predictions (Reiner et al.~2005).
The CME direction seems to be the most important parameter that controls
the geoeffectiveness of very fast halo CMEs (Moon et al.~2005).

The speed distributions of accelerating and decelerating
CME events were found to be nearly identical
lognormal distributions (Yurchyshyn et al.~2005).
The acceleration of CMEs tends to be higher for flare-associated CMEs
than for filament-associated CMEs, but counterexamples were found that
suggest that flare-associated CMEs with large acceleration are additionally
boosted by helmet streamer disruptions or subsequent CMEs/flares
(Moon et al.~2004).

The rate of mass injection at the onset of a ``halo'' CME could be
determined to $\approx 10^{16}$ g hr$^{-1}$ from metric radio data
(Kathiravan \& Ramesh 2005).

\subsubsection{Difficulties with Predicting the Arrival of CMEs at Earth}
3D MHD simulations of propagating CMEs reveal that the arrival time of CME
shocks at Earth strongly depend on the ambient background solar wind, the
standoff distance between the shock and the driving ejection, and on the
inclination angle of the shock with respect to the Sun-Earth line
(Odstrcil et al.~2005; Jacobs et al.~2005; Lee 2005; Wu et al.~2005c).
Heliospheric in-situ magnetic field measurements allow to quantify the
correlation length of magnetic field parameters for passing interplanetary CMEs
and ambient solar wind, which yields better predictions for CME arrival times
at Earth (Farrugia et al.~2005). A large statistical study showed that
just over a quarter of the 938 HCMEs observed by LASCO were associated with
a forward shock near L1, suggesting that about half of the earthbound HCMEs
are either deflected away from the Sun-Earth line or do not form a shock
(Howard \& Tappin 2005). Although ``halo-CMEs'' are considered as
Earth-directed, a fraction of 15\% miss the Earth (Kane 2005b).
Given the maximum observed CME speeds of $\approx
3000$ km s$^{-1}$, the shortest travel times of CME-driven shocks are expected
to be no less than $\approx 0.5$ days (Gopalswamy et al.~2005b).

\subsubsection{Particles Accelerated in CMEs}
There is a long-standing dichotomy of flare-accelerated and
CME-accelerated particles, which differ in timing, spectra, and
composition (Lin 2005; Tylka et al.~2005).
While {\sl Solar Energetic Particles (SEP)} are believed
to be accelerated in CME shocks, to our surprise, no obvious correlation
of SEP onset and rise times of 20 MeV protons with any CME parameter
was found (Kahler 2005). Full 3D MHD and kinetic hybrid simulations
of particle acceleration in a propagating and evolving CME shock and
sheath structure reveal that the acceleration efficiency of GeV particles
strongly depends on the fast-mode shock evolution, controlled by the
increased magnetic field strength in the plasma compression behind the shock
(Manchester et al.~2005). Also the fact that the kinetic energy in accelerated
particles represents a significant fraction of the CME kinetic energy implies
that shock acceleration must be relatively efficient (Mewaldt et al.~2005).

On the other side, some SEP particles might also originate in flare sites.
Klein \& Posner (2005) found
that $<54$ MeV protons are accelerated simultaneously with dm-km type III
emitting electrons, supposedly in altitudes of 1.1-1.5 solar radii, in a
third of the analyzed events. In one large (GOES class X17) flare three
phases of particle injections were determined: an impulsive injection of
radio type III-producing electrons first, than a second impulsive injection
11 minutes later (lasting 18 min), and a third gradual one 25 min later
(lasting 1 hour), where the latter two delayed acceleration phases could
not be localized (Klassen et al.~2005).

\subsection{Heliosphere}

\subsubsection{Acceleration of Solar Wind}

Where is the solar wind accelerated? Analysis of SoHO/UVCS data suggest that
the slow solar wind is accelerated in the legs or in the stagnation flow
(Nerney \& Suess 2005) near the cusp of streamers, or above the
streamer core beyond 2.7 solar radii, right where the heliospheric current sheet
starts (Antonnucci et al.~2005). LASCO observations suggest that open and closed
field lines reconnect near the streamer cusp and form blobs of higher plasma
density that are ejected into the slow solar wind (Lapenta \& Knoll 2005).
If there is no streamer around, also an active region can substitute
(Woo \& Habbal 2005).
In addition, the effects of differential rotation of the solar surface forces
continuous disconnection and reconnection at the more or less rigidly rotating
coronal hole boundaries, which modulate the formation of the slow solar wind
(Lionello et al.~2005).

The fast solar wind is believed to originate from small coronal funnels in the
transition region in coronal hole regions, where hydrogen is far from ionization
equilibrium and Lyman-$\alpha$ emission comes from temperatures of
$\approx 5\times 10^4$ K (Esser et al.~2005).
The physical process responsible for accelerating the fast solar wind is the
interaction of open magnetic field lines with smaller coronal loops through
magnetic reconnection (Fisk 2005), driven by magnetic footpoint diffusion
(Giacalone \& Jokipii 2004).
Another piece of evidence for the chromospheric origin of the fast solar wind
comes also from the correlation between solar wind velocities and the ratio of
ionic oxygen (O$^{+7}$/O$^{+6}$) densities (McIntosh \& Leamon 2005).
Magnetic field extrapolations of coronal funnels place Ne$^{7+}$ and C$^{3+}$
ions into altitudes of 5-20 Mm, where the flow speed increases from zero to
10 km s$^{-1}$, as the birth place of the fast solar wind (Tu et al.~2005).

The interface between fast and slow solar wind in interplanetary space was found
to have two distinct parts: a smoothly varying boundary layer flow that flanks
the fast wind from coronal holes, and a sharper plasma discontinuity between
intermediate and
slow solar wind, explaining the correlations between wind speed variabilities,
charge state composition and magnetic field orientation in the heliosphere
(Schwadron et al.~2005).

\subsubsection{Turbulence in Solar Wind}

The solar wind is a ``turbulence laboratory'' (Bruno \& Carbone 2005).
One source of turbulence are (hypothetical) high-frequency Alfv\'en waves,
produced by successive merging and braiding of fluxtubes on granular and
supergranular scales in the chromosphere and transition region
(Cranmer \& van Ballegooijen 2005). The source of long-period Alfv\'en waves
observed in the solar wind can also be
associated with leakage from helioseismic modes (Zaqarashvili \& Belvedere 2005).
The dominant-turbulence model of Isenberg (2005) describes the turbulent heating
of the distant solar wind by the dissipation of wave energy generated by the
isotropization of interstellar pickup protons, through cyclotron resonance and
particle pitch-angle scattering.

\subsection{Solar Cycle and Space Weather}
\subsubsection{Sunspot Predictions}
The statistics of sunspot numbers has now been consolidated back to Galileo's
observations in 1610, which tell us not only the average cycle period
(10.9$\pm$1.2 years), but also about the cycle asymmetry (with a fast rise and
slow decay), that the risetime decreases with cycle amplitude, that large
amplitude cycles are preceded by short period
cycles, that the secular amplitudes increase since the Maunder minimum, and
about hemispheric symmetries (Hathaway \& Wilson 2004). Subcycles with periods
of 152-158 days were also noted from flare rates (Ballester et al.~2004) or
other solar indices (Kane 2005b). Such subcyles are also called {\sl Rieger
periodicities} and were even discovered on stars, e.g., with a 294-day cycle on
UX Arietis, believed to be
caused by equatorially trapped Rossby-type waves modulating the emergence of
magnetic flux at the surface (Massi et al.~2005). The centennial increase in
global geomagnetic activity was, however, considerably smaller than the secular
increase in solar activity (Mursula et al.~2004). Predicted are a strong next
cycle (XXIV) with a sunspot number of $145\pm30$ in 2010, and a weak overnext
cycle (XXV) with $70\pm30$, which has a long cycle peaking in 2023
(Hathaway \& Wilson 2004).

The prediction methods of solar activity become increasingly more sophisticated,
ressembling the flow charts of electronic circuits. In one study we read that
fuzzy logic, neural networks, and genetic algorithms are the most popular
artificial intelligence techniques (Attia et al.~2005). The authors describe
LAGA-POP (linear adapted genetic algorithm with controlling population size) and
FLNN (fuzzy logic neural network), which they explain in the following way:
``This is a particular implementation of a fuzzy system equipped with
fuzzification  and defuzzification interfaces'' (Attia et al.~2005). Another
algorithm, MRD-GA (multi-resolution dynamic genetic algorithm), is described as
a ``linguistic fuzzy system with a general rule-based structure''
(Attia et al.~2005).

\subsubsection{Sunspot Postdictions}
Longterm solar activity reconstruction on centennial to multimillennia time
scales is accomplished by cosmogenic isotope records, such as $^{10}$Be and
$^{14}$C (Miletsky et al.~2004; Mordvinov et al.~2004; Lee et al.~2004;
Volobuev 2004), which are produced mostly in the upper atmosphere, and thus are
anticorrelated with the sunspot number (Scherer et al.~2004),
but this {\sl archeo-magnetic} reconstruction method is not really a measurement
but rather a ``post-diction'', and thus is not considered as reliable for future
predictions (Usoskin \& Kovaltsov 2004).
Some authors defined a new parameter to characterize the long-term solar cycle
variability, the {\sl sunspot unit area}, which is the size of a sunspot
averaged over a cycle, but then lost the {\sl Waldmeier effect} and the
{\sl Gnevyshew-Ohl rule} (Li et al.~2005b).

And regarding the present cycle,
Reimer (2004) finds that ``an ingeniously constructed record of sunspot activity
shows that the current episode is the most intense for several thousand years.
But that does not let us off the anthropogenic hook of global warming''. There
is no systematic trend in the level of solar activity that can explain the most
recent global warming (Benestad 2005), although it has been reported that the
total solar irradiance increased by 0.15 W m$^{-2}$ between the solar minima in
1987 and 1995 (Dewitte et al.~2004).

In addition to the solar cycle variation, other quasi-periodic patterns have
come to our attention, such as the {\sl ``flip-flop''} phenomenon, where the
most dominant active regions flip spontaneously to the opposite side of the star
 (Berdyugina 2004; Fluri \& Berdyugina 2004).

\subsubsection{Space Weather}
The concept of {\sl space weather} was launched some 10 years ago to describe
the short-term variations of solar activity and their effects on the near-Earth
environment and technoculture. More recently, the term {\sl space climate}
was introduced to include the longer-term variations of solar activity and their
implications on the heliosphere and near-Earth space. The beginnings or this
new industry, however, go 150 years back. The September 1859 solar terrestrial
disturbance is considered as the first recognized space weather event
(Cliver \& Svalgaard 2004). But only in 2004, a {\sl First International
Symposium on Space Climate} was organized, documented in some 50 articles
in the Topical Issue of Solar Physics Volume 224.

It is always entertaining to hear how the solar cycle directly affects our life.
Apparently, the space weather even affects the wheat market prices on Earth,
as a study from the medieval England up to the modern USA by Pustilnik
\& Yom-Dim (2004) demonstrates, via a chain reaction of sunspot activity
- solar wind modulation - variation of cosmic rays - cloudiness and weather
changes - drop of agriculture production - wheat price bursts.

Another very practical application of space weather is the study of impacts
of geoeffective events. Solar energetic particles (SEPs) can reach the Earth
when the magnetic connectivity of the flaring active region is matching, with
a tolerance of 25$^\circ$-30$^\circ$ in heliographic longitude
(Ippolito et al.~2005), although sometimes apparently not-connected events
rooted in the eastern solar disk happen (Miroshnichenko et al.~2005).
During a 7-year period of the current solar cycle, 64 geoeffective CMEs were
found  to produce major geomagnetic storms at Earth
(Srivastava \& Venkatakrishnan 2004). The SEP event of 14 July 2000
(the ``Bastille-day flare'')
was investigated by using simultaneous ground-based and satellite measurements
of the particle flux, together with a tissue equivalent proportional counter
on board a Virgin Atlantic Airways flight from London Heathrow to Hong Kong,
but fortunately no increased radiation levels were detected (Iles et al.~2004).

\subsection{Decadal Anniversary of SoHO}
The happy 10th launch anniversary of the {\sl Solar and Heliospheric Observatory
(SoHO)} spacecraft was jubilated on 2nd Dec 2005, from which we quote a succint
``numerical'' summary by Bernhard Fleck:

Ten years of operation without a single service or tune-up is no piece of
cake for a spacecraft.  As anyone who has operated scientific instruments
in a lab will know, it's amazing how many things can go wrong, requiring
some form of intervention or repair.  But that has not been an option for
SOHO and its instruments - if it breaks, it's broken, and all you can do
is adjust to a new reality. But miraculously, we're doing very well even
after all these years.

Some amazing facts about SOHO's first decade:
\begin{itemize}
\item
 140 Ph.D. theses have been written on or about SOHO data.
\item
 289 scientific meetings on subjects related to SOHO appear on our
   meetings pages.
\item
 944 news stories appear on our newsroom pages (only recorded between
   1997 and 2005!).
\item
 1000 comets have been found. SOHO is the most prolific comet-finder
   observatory of all times, and has identified almost half of all comets
   for which an orbit determination has been made.
\item
 2300 reviewed papers using SOHO data have been published.
\item
 2300 scientists (approximately) appear in the author lists of those
   papers (we like to say that every current solar scientist has had the
   chance to work with SOHO data).
\item
 3230 science planning meetings have been held.
\item
 2 000 000 command blocks have been sent to the spacecraft by the ground
   system.
\item
 5 000 000 distinct files have been served by the web server.
\item
 10 000 000 exposures (almost!) have been made by the CDS instrument.
\item
 16 000 000 distinct hosts have been served by the web server.
\item
 50 000 000 exposures have been taken by MDI. They're probably quite high
   on the list of "the world's most durable camera shutters". Don't try to
   beat it with your favorite SLR camera!
\item
 266 000 000 web page requests have been served.
\item
 16 000 000 000 000 bytes (16 Terabytes) of data are contained in the
   SOHO archive.
\item
 85 000 000 000 000 bytes (85 Terabytes) of web pages/data have been
   served.
\end{itemize}

\section{SHORT, SWEET, AND SURMISED}
Here are two highlights from near the end of the reference year. The ``short''
refers to the duration of the events---the second main class of gamma-ray
bursters  (\S 3.1.1) and the encounter of the \textsl{Deep Impact} mission with
comet 9P/Tempel 1 (\S 3.2.4). ``Sweet'' means reasonably well defined results
in the two cases; and ``surmised'' suggests that the results confirm at least
some previous predictions. Each highlight
is accompanied by related topics in gamma-ray and solar-system astronomy.

\subsection{Gamma Rays and Cosmic Rays}
 These live together because both are very high energy
astrophysics in the modern sense of ``high energy per particle or photon,''
rather than the 1960's sense of ``high energy per event.''

Two sorts of gamma ray bursts were
predicted, and two sorts were observed. Unfortunately, they were not the same
sorts. The predicted ones came from shock breakout in core collapse supernovae
(Colgate 1968) and from the last gasp of Hawking radiation during the
evaporation of mlni-black-holes (Hawking 1974).  The first time one of us
attempted to describe the two observed sorts, they were the classic ones
 discovered by Klebesadel, Strong \& Olsen (1973), and the soft gamma repeaters,
of which the 1979  March 5 event was first. But with the association between
SGRs and nearby supernova remnants,  neutron stars, etc. (Ap94, \S 5.3), they
ceased to count. Then the two sorts were those of  long and of short duration
(means near 20 and 0.3 seconds, Kouveliotou et al. 1993), also
 characterized as having relatively softer and harder spectra.

\subsubsection{The Short Lady Bursts}
 The fat lady burst in 1997 (Ap97, \S 11), when \textsl{BeppoSAX}
caught X-ray tails which, in turn, permitted the identification of optical and
radio  counterparts with measured redshifts.
A ha! GRBs were not wimpish, repeating surface events on old, nearby neutron
stars, the dominant model through the 1980s (Ho et al. 1992), but one-shot
stellar demises, happening perhaps once in a million years per galaxy, but so
powerful that the BATSE catalog surveyed most of the observable universe
(Ap98, \S 6.3).

Only gradually as the inventory of counterparts accumulated did it become clear
that (a) they were all wedded to long duration events, (b) association with star
formation  regions and other considerations strongly suggested a best-buy model
of core collapse  in rapidly rotating massive stars with rapidly rotating black
holes as the product,  and (c) some, at least, had simultaneous Type Ic
supernovae, which peaked out as soon  as the GRB faded (Ap03, \S 4). Notice
that there is at least some connection with  the first (Colgate) predicted sort.

But what, then, was responsible for the short duration bursts, whose statistics
suggested  smaller distances and so smaller total power and whose X-ray tails
were so faint they had to be piled up to show (Montanari et al. 2005)? Luckily,
there was a spare, underused old model hanging at the back of the
closet\footnote{At least one of your authors can confirm that old models find
employment in anything more remunerative than hanging at
the back of the closet remarkably difficult to locate.}---the merger of a
binary neutron star pair or neutron star plus black hole, brought together by
loss of angular momentum in gravitational radiation (Guetta \& Piran 2005,
Aloy et al. 2005, Miller 2005, the most recent appearances of
the model, not the first). A definite prediction was that the short duration
bursts should, or anyhow could, occur far from any recent star formation, since
 gravitation radiation is a very slow way to do anything (including establish a
reputation in observational astronomy).

As we closed our eyes to the ongoing stream of literature, the first short one,
GRB 050509b, had just turned up in a cluster of galaxies at $z = 0.225$, where
the stars are about 360 Myr old, and any associated supernova must have been
much fainter than those associated with long-duration bursts
(Castro-Tirado et al. 2005). And we slip surreptitiously out of period to
record that GRB 050709 happened at the outskirts of a $z = 0.160$ star-forming
galaxy, far from any young stars,
that it was considerably fainter than most long GRBs, and that again no
supernova was spotted  (Fox et al. 2005, Gehrels et al. 2005, Villasenor 2005,
 Hjorth 2005, Piro 2005).

These are, in other words, pretty much what you would have expected from the
NSX2 models, though curiously one of the pioneers of those has since disowned
the idea (Paczynski 2005). Some people just can't stand to be right
(though we have the opposite problem). Istomin (2005) suggests that the double
pulsar J0737-3039AB will eventually give rise to a short
duration GRB. You will surely join the editor of \textsl{PASP} in hoping that
the ApXX series is not still  around to report the event!

\subsubsection{Other Sorts of GRBs}
 It is time, clearly, for some additional classes.  A short-short type,
apparently new this year (Rau et al. 2005), might possibly be the long-promised
Hawking radiation chirps (Halzen et al. 1991). They last less than 0.25 s and
 have $V/V_m = 0.48$  according to the \textsl{INTEGRAL} catalog and 0.52 in
BATSE  (that is,  an essentially homogeneous distribution in space).

If you are feeling less adventurous, there are still the optically dark and
gamma-poor (or X-ray rich)  GRBs. The first are defined by a small ratio of
optical to X-ray luminosity (Castro Ceron et al. 2004), perhaps because the
visible light fades very fast, say Filliatre et al. (2005) and
Jakobsson et al. (2005), each of whom managed to catch one in the infrared by
hurrying to the   error box. If the cause is severe Compton losses, then a
large GeV luminosity is a prediction   (Beloborodov 2005, who was thinking of
GRB 941017).

The X-ray enthusiastic sort are not, say the pundits (this year at least) a
separate class, but merely an extreme of a continuum with the classic events
(Amati et al. 2004 on X-ray afterglows,
Mirabal et al. 2005, Rees \& Meszaros 2005, Sakamoto et al. 2005).
 Lamb et al. (2005) deduce that
these gamma-poor events are so narrowly beamed that the total
rate must approach that of Type Ic supernovae. Are there indeed SNe Ic to been
seen with them? Occasionally (Tominaga et al. 2004), but not often (Levan et al.
2005, Soderberg et al. 2005).

The record redshift for a gamma ray burst is held by the (long duration) event
050904 at $z = 6.29$ (measured with the Japanese Subaru telescope). Its spectrum
resembled that of QSOs at similar redshift in showing Gunn-Peterson troughs at
both Ly$\,\alpha$ and Ly$\,\beta$, with a bit of flux in between, meaning that
the diffuse baryons in those days were not yet quite completely ionized. The
GRBs differ from the AGNs in the absence of strong Ly$\,\alpha$ emission and
absence of proximity effect
(ionization of nearby intergalactic gas clouds). These characteristics mean that
GRBs, if you catch them quickly enough, will be at least as useful as QSOs for
tracing out structure and evolution in the $z = 6$--10 universe.

Two more GRB thoughts hang precariously off the end of the index year or the
topic.  First is the existence of X-ray flares, occurring minutes after the
main event and containing just about as much energy (Burrows et al. 2005).
Second is the possibility that there is a completely different sort of short
duration GRB in the form of extragalactic (but nearby) analogs of the giant
flare of the soft gamma repeater  1806-20. The event itself dates all the way
back to 2004 December 27 (Mereghetti et a1. 2005,
Schwarz et a1. 2005, Rea et al. 2005, Yamazaki et al. 2005, Hurley et al. 2005,
 Gaensler et al. 2005, Palmer et a1. 2005, Lazzati 2005) and belongs somewhere
 in the neutron star section. But if you had been 10 Mpc instead of 10 kpc from
 it,   you would have seen only the tip of the iceberg, sorry, light curve, and
thought    it a GRB. Historically, the absence of host galaxies for short
duration events    seemed an obstacle to using soft gamma repeaters for part of
the population,    and you will have to sneak into the preprint e-files to read
 about the one that    happened within striking distance of M81 and M82.

And the rest is sound bites.
\begin{itemize}
\item Significant gamma ray polarization has been announced again, not for the
same  event as last year's (Willis et al. 2005).
\item X-ray spectral features remain marginal (Butler et al. 2005).
\item There is dust around the long-duration events (Savaglio \& Fall 2004),
but not usually enough to result in a SCUBA source (Smith et al. 2005a)
\item The environment
set up by the progenitor is what you would expect from a Wolf-Rayet star
ejecting its envelope at a few thousand km/sec, even in cases where,
embarassingly, we  don't see a host galaxy for the WR to have lived in
(Fiore et al. 2005, Klose et al. 2004).
\item The X-rays can flare back up very late (Burrows et al. 2005).
\item Theory papers no longer outnumber observational papers, and none this
year was sufficiently deviant to cite for that reason alone, though we found
Fryer \& Heger (2005)on the problem of getting enough angular momentum
distressing.
\item The rate is at least $\mathrm{1/Gpc^3}$--yr (Guetta et al. 2005), from
which you can figure out
how long you have to wait for a GRB to kill the present authors and estimate
whether the editor of \textsl{PASP} is likely to do it first.
\end{itemize}

\subsubsection{Steadier Gamma Rays}
 There are, of course, also non-bursting gamma ray sources. The largest category
is, and has been for many years, the unidentified (Cheng \& Romero 2004), which
is not at all the same as saying there are no
 candidates (Borsch-Ramon et al. 2005 on microquasars, which they also advocate
 for cosmic ray sources; Foschini et al. 2005 on FRI radio sources;
Ng et al. 2005 on pulsar wind nebulae). Fegan et a1. (2005) conclude that most
 of the GeV sources are not TeV sources, of which there is also an unidentified
  component (Aharonian et al. 2005, Mukherjee \& Halpern 2005), some in the
  direction of the galactic plane and some not (Walker et al. 2005).

As for identified TeV sources, some are pulsars (Aharonian et al. 2005a),
some are pulsar false alarms or exceedingly variable (Aharonian et al. 2005b
on B1706--44), one is a microquasar (Aharonian et al. 2005c, Cui 2005), one is
the young shell SNR along the sight line to the Vela SNR
(Aharonian et al. 2005e,
who, however, do not confirm SNR 1006 down to 10\% of the previously reported
Cangaroo flux), and others are assorted flavors of active galactic nuclei
(Aharonian et al. 2005d, on a $z = 0.117$ BL Lac object).

The best loved of these
TeV sources remain the first two found, Mkn 421 (Piner \& Edwards 2005) and
Mkn 501 (Xue \& Cui 2005); and the most debated question remains whether seeing
these through the expected intergalactic sea of optical and infrared photons is
or is not puzzling. The last word on this gets said so many times each year
that  we have forgotten whether it is Yes or No (Minowa et al. 2005,
Schroedter 2005,  Matsumoto et al. 2005, Mii \& Totani 2005).

 Within the last year, Sgr $\mathrm{A^\ast}$, our very own black holes, seems
to have become well established as a TeV gamma source (Atoyan \& Dermer 2004,
Aharonian et al. 2005f).
That the photon production mechanism is the annihilation of Kaluza-Klein dark
matter  particles (Bergstrom et al. 2005) is perhaps less well established,
though a 1--10  TeV K-K particle yields both the right gamma ray spectrum and
the right amount of  dark matter in the universe.

\subsubsection{Godzilla Particles}
 What else has lots of energy each? The cosmic rays. We start
with the ultra-high energy ones ($\ge 10^{20}$ eV per primary), which are not
very well understood, and work down to the lower energy population, which has
not been very well understood for longer.
UHECRs have to get to us through the photon sea of the cosmic microwave
 background, a bit like the problem of UHE gamma rays traversing the infrared
sea, but with the difference
that we don't really know where the UHECRs started out. There is, in addition, a
data discrepancy, with the newest detector finding fewer of the highest energy
particles (Cronin 2005, on Fly's Eye) than were reported earlier (Teshima 2005).

There is no basic physics problem if the lower flux is correct or if the primary
particles are not protons reaching us from well outside the Local Supercluster.
The alternatives are (a) decay/annihilation products from dark matter particles
in our own halo or (b) nearby unrecognized sources. Seckel \& Stanev (2005)
provide a good precis of the problem and note that distant sources would also
produce very high energy neutrinos when the primaries collide with
intergalactic photons. Items that puzzle us include:
\begin{itemize}
\item Is the distance primary protons can travel against pair production on the
CMB actually well known? Aloisio \& Berezinsky (2005) suggest a sort of
anti-GZK effect. \item Are the arrival directions random or clustered
(Abbasi et al. 2005; Amenomori et al. 2005,
a result from an air shower array in Tibet, suggesting the heliosphere as a
source)? There is also an extensive air shower array in Tehran
(Khakian Ghomi et al. 2005), for
which we think we are interested only in the departure directions.
\item What are the highest energies that can be reached with conventional
processes (Serpico \& Kachelriess 2004 within the Milky Way,
Honda \& Honda 2004 in 3C 273)?
\item What about somewhat unconventional but physically possible processes
(Vlahos et al. 2005
anomalous resistance during galaxy formation; Ouyed et al. 2005 on neutron stars
turning to quark stars; Crocker et al. 2005, very high energy neutrons from
Sgr $\mathrm{A^\ast}$)? \item And, not exactly a question (except for ``why
didn't I think of that?''), but an expression of admiration for the prediction
(Huege \& Falcke 2005) that there should
be radio flashes when UHECRs hit the upper atmosphere, followed closely by the
detection of such flashes (Falcke et al. 2005). The flashes are due to
synchrotron in the Earth's magnetic field and were seenat 43--73 MHz by LOPES,
the prototype of the LOFAR detector now operating in Karlsruhe. The events last
10s of nanoseconds. Hm. We begin to see why we didn't do this.
\end{itemize}

``Galactic'' cosmic rays range downward from $10^{17}$ eV (where the Larmor
radius is the size of the Milky Way) or $10^{15}$ eV (the ``knee'' in the
spectrum, attributed by Ptuskin \& Zirakashvili 2005 to processes in supernova
remnants) on down to the only  marginally relativistic, which don't get inside
the heliosphere and so are not well
studied. They  too have been attributed to a range of sources
(Westphal \& Bradley 2004 on  interplanetary dust, Bosch-Ramon et al. 2005
on micro-quasars, Fender et al. 2005 on X-ray binaries).
Volk et al. (2005) ``tentatively conclude that galactic supernova remnants are
the source population of GCRs'' endorsing the 70+ year old thoughts of
Baade \& Zwicky (1934).

``Galactic'' is, in this context, a somewhat flexible term.
Combet et al. (2005) conclude that the cosmic ray actinides were accelerated
within 150 pc of us, versus 1.25 kpc for Li, Be,
and B. Higdon \& Lingenfelter (2005) put most of the acceleration of the
heavier nuclei in superbubbles, also close to us; and Derbina et al. (2005)
draw attention to the rapid change  in composition, from a mean mass of five
amu\footnote{We are not unaware that this unit ought now,
according to the International Union of Pure and Applied Chemistry, to be
called the Dalton. Indeed it was the
only thing we learned at a meeting of another organization, which is unique in
our experience as being one where not even the Electoral College is actually
allowed to vote.}  at $\mathrm{10^{12-15}}$ eV to 30 amu at $10^{17}$ eV. This
ought also to reflect differences in  location of the acceleration
process, but in the opposite sense, if $10^{17}$ eV is already
``extragalactic.''

And now that you know all about the astrophysics of cosmic rays, what are they
good for? Ionizing interstellar clouds that are opaque to UV radiation
(Padoan \& Scalo 2005, Giammanco \& Beckman 2005), and probing the insides of
pyramids, though Maglich (2004) points out that chambers can be imitated or
concealed by stone that is not of constant density.

\subsection{Ozymandias, Chimney Sweepers, and  Other Sinks and Sources of Dust}

ApXX tradition requires ``gee whiz'' items to come first or last in sections.
Given that at least one comet (9P/Tempel l), one planet (Pluto) and one moon
(Titan) earned  green dots this year, there appears to be no logical ordering,
large to small or small to large, of solar system topics that will work. Thus we
reserve the right to say ``gee whiz!'', ``green dot,''  or ``wow'' at random
points in the text.

\subsubsection{Planets}
 Mercury has a magnetic field, which used to be a fossil, but is this year
attributed to a dynamo in a residual molten core (Margot et al. 2005). We have
no advice on how to see the field, but if you will settle for some photons, it
is useful to pick a time when Merucry is very close to Venus, as in February
2005, and to employ binoculars (ours are $7\times50$ US Navy
World War II issue).

Venus somehow got herself resurfaced half an eon ($5\times10^8$ yr for the
Graiko-challenged) ago without any residual evidence for plate tectonics
(Ap91, \S 2). The volcanic features include ones called coronae and novae, which
somewhat resemble nasturtiums  (Kostama \& Aittola 2004). Transits of Venus
across the sun (next opportunity 2012) can be thought of as chances not to see
reflected photons, as coming in singles and triples as well as the current pairs
(McCurdy 2004), or as one of many things predicted
but not seen by Kepler (Posch \& Kerschbaum 2004).

Earth has taken refuge with Ptolemy at the center of the Universe (\S 5) but was
probably assembled in much the same way as Venus, out of pieces that already had
atmospheres and oceans (a good deal of which were lost, Genda \& Abe 2004,
Zahnle 2004).

All of the ancient and modern planets (except Earth) were strung out across
the sky in order MVMJSUNP on 10--13 December 2004 (Sinnott 2004). The last time
this happened was in February-March 1801, and the next will be in April 2333.

Aspects of Mars relevant to habitability appear in \S7. We note here that
(1) The two best oppositions for a long time have now passed. In case you missed
them, along with the alignment of the previous paragraph,
Nakakushi et al. (2004) report on some of the 2003 discoveries, (2) A molten
iron core has, after all, survived from early times (Fei \& Bertka 2005), but
there is no earth-like solid core. Lillis et al. (2005) discuss the dynamo
further, and all agree that Martian seismic
data are really needed to make further progress, and (3) If you were living on
the  Martian surface, you would see airglow (Bertaux et al. 2005) as part of the
night   sky brightness (but mercury lines from high pressure lamps must surely
be rare),  not to mention an occasional meteor, as photographed by the
Mars Rover  (Selsis et al. 2005). You might also see, if unlucky, the sort of
impact that has put meteorites on the Martian surface, to be found by
Opportunity  (Anonymous 2005a). These cannot be called Martian meteorites,
since the name is already taken for bits of Mars found on earth. The impacts
 responsible for those would have been even less lucky to experience.

Jupiter did not have a particularly good year and has to share the effects of
the (solar) coronal mass ejection of 1--20 November 2000, which swept past Earth
on 11--12 November, Jupiter on 18--20 November, and Saturn on 7--8 December,
producing aurorae on the last two (Prange et al. 2004). The Jovian aurorae were
seen by \textsl{Galileo} and the passing shock by \textsl{Cassini}, which was
nearer Jupiter's orbit than Saturn's at that time.

Saturn on the other hand was playing catch-up, by displaying some of the
phenomena earlier shown by Jupiter, including disk X-rays
(Bhardway et al. 2004 and 2005), due largely to solar flourescence and
scattering, bursts of dust emission (Kempf et al. 2005)\footnote{The 100 {\AA}
grains come largely from ring material, versus Io in the case of Jupiter, and
are accelerated by $\mathbf{E}=\mathbf{v}\times\mathbf{B}$ in the corotating
planetary  magnetic field. And in case you
are wondering why this is a footnote, it is because the number of nearly
unrelated ideas that can be crammed into a single sentence is about the same as
the number of tasks the most forgetful author can keep track of without having
to make a list.} and zonal atmospheric temperature bands, seen in mid IR, that
are not the same as the bands in reflection from the visible clouds
(Orton \& Yanamadra-Fisher 2005). The \textsl{Cassini}
package of Saturnian data appears in Porco et al. (2005a) and the three
following  papers. On 13 January 2005, Saturnians had an opportunity to see the
Earth transit the sun. No reports so far on whether they attempted to use this
event to measure the length of the SU (Sastronomical Unit). Is this easier or
harder than measuring the AU by timing Cytherean transits from Earth? Another
of those exercises left for students who are not
already behind on their theses.

Uranus flaunted its brightest-ever NIR cloud feature (IAU Circ. 8586). The
spot briefly reflected 17\% of all the K-band light from Uranus seen by the
Keck II telescope.  At adaptive optics resolution (0.05 arc sec), the surface
brightness contrast ratio was about 50.

The upper atmosphere of Neptune contains some CO, which surprised us much less
thanthe source, which is said to be partly a comet impact less than 200 years
ago (Lellouch et a1. 2005). No, we didn't see it.

Should UB313 (at 97 AU and with radius of 2400 to 3200 km) be counted as the
10th planet (Brown 2005)? No strong feelings, except that it would spoil the
pattern of  moving straight from Pluto to the moons of Pluto. And if you don't
think Pluto is a planet,  feel free to go sit in the asteroid section (3.2.3)
with the Medical Musician, who  always insists on buying tickets at discount.

\subsubsection{Moons}
 Charon (otherwise known as Pluto I) occulted a star (IAU Circ. 8570), thereby
setting a firm lower limit of 1179 km to its diameter. Charon was, apparently,
available for this task because of a major impact long ago which broke it off
from the then unnamed Pluto, imitating in this respect our own Luna
(Canup 2005, Melosh 2005). Most other moons
are said to have formed in disks around their parent planets or to have been
captured from  supplies called asteroids. The captured sort are also called
irregular, meaning that their orbits can have large eccentricities and
inclinations (including retrograde) and, as a rule, large semi-major axes,
while the satellites themselves tend to be small. The gee-whiz item
here is the probable recognition of two more Plutonic moons (IAU Circ. 8625).
Other, out of period, reports suggest they are smaller scraps from the
same impact event.

Neptune's S/2002 N1 immediately falsifies the previous paragraph on how moons
form by probably being a fragment broken off Nereid, whose color and retrograde
orbit it shares (Grav et al. 2004).

Two new moons of Uranus are of the irregular sort (Sheppard et al. 2005a), bringing his total
to eight irregular and one regular. Jupiter shares this dominance of irregulars,
while Saturn and Neptune were more or less half and half until the 2004--05 proliferation
of Saturnian moons (or anyhow their discoveries; most are probably older than two years
but younger  than 4.56 Gyr).

How many moons does Saturn have? Well, many. Twelve more (11 retrograde)
announced in IAU Cir. 8523, with periods of 820--1154 days, a new one in the
Keeler gap with a period of only 0.594 day (IAU Circ. 8535), and so forth. The
names reach up to XXXI, Polydeuces
(IAU Circ. 8432), but our greatest sympathy is reserved for XXX Thrymr, with
hardly a vowel to his name (IAU Circ. 8471). Many of the newish ones are in
Kozai resonances  (Carruba et al. 2004), and some chaos, in the orbits, never
mind the names, is expected. At some point, we suspect there may need to be
some sort of definition of ``moon'' or ``satellite'' that sets a lower limit to
their sizes, to prevent the cataloguing of chips  off the old Hasselblad camera
that still orbits earth. Oh, that is called ``space
debris,'' you say?

Some moons of Saturn are more equal than others. The winners in 2005 included
Iapetus (which has a ridge around its center, looking remarkably like the two
hemispheres of a  cheap toy ball glued together, Anonymous 2005b). Phoebe, the
outermost large moon, has diverse surface materials, including ices, organics,
iron compounds, and olivines and other silicates (Clark et al. 2005a,
Dalton 2005, Johnson \& Lunine 2005).
Enceladus must be warm inside and probably has a rocky core (Johnson 2005).

First prize, however, to Titan, mapped to within an inch (well, perhaps a meter)
of its life by \textsl{Cassini} for more than a year and smashed in the face by
the Huygens-probe on 2005 January 14. Interesting results include (a) very rapid
atmospheric rotation compared to the surface (shared, so
far as we know, only by Venus, Porco et a1. 2005b), (b) absence or at least
rarity of much-anticipated  hydrocarbon seas (West et al. 2005, Sotin et al.
2005, Prockter 2005, with possible volcanic release of methane), (c) mesospheric
temperature structure in the atmosphere (Griffith et al. 2005), sparsity of
craters, absence of magnetic fields, and much else (Mahaffy 2005 and next
seven papers).

The gee whiz item (otherwise known as WHEW) was still trapped in press releases
and other non-citable sources as a reference year closed. This was the role of
the Robert A. Byrd Greenbank radio telescope\footnote{The most unprofessional
and disrespectful author---so said the referee of a completely
different paper---takes full responsibility for wanting to call this The Byrd in
the Hand. It was not precisely what most of the radio community said they
wanted, but was surely very  much better than anything obtainable from two
Bushes.} and the VLBA in monitoring the descent of Huygens
 and thus mapping the hurricane-speed winds. \textsl{Cassini} was supposed to
have captured these Doppler data, but was somehow tuned to the wrong frequency.
The ground based telescopes saved the day, or anyhow the data.

 Jupiter's best known moons were discovered by Galileo (meaning the
Florentine, not the satellite, and you can tell because he doesn't walk around
wearing italics)  and probably formed in a disk like planets around stars
(Woolfson 2004), although Amalthea must  have migrated inward since
(Takato et a1. 2004). What surprised us most, however, is how often
 Galilean moons occult and eclipse each other as seen from earth, during the few
months every six  years when sun and earth are, respectively in the moons' orbit
plane. Many of these mutual events  were observed in 1997 and 2003
(Pauwels et al. 2005, Dourneau et al. 2005), something like 21 in 1997 and
15 in 2003.

You already knew that stuff coming out of volcanoes can be pretty noxious, but
Io's volcanic gases are noxiouser than Earth's (Schaefer \& Fegley 2005).

Mars has two moons, and no more need apply, at least none larger than 0.1 km
(Sheppard et al. 2005b).

While moon more or less rhymes with June, spoon, tune, and so forth, luna would
seem to rhyme only with tuna. She (in most mythologies) was formed when some
large, late impacting object hit Earth. Isotopic data have further restricted
this to impact after the earth completed core-mantle differentiation
(Boyet \& Carlson 2005) and the impactor to something that also
formed quite close to 1 AU from the sun (Belbruno \& Gott 2005). The present
lunar surface is mostly basalt (Christensen et al. 2005), as is true for earth,
though we keep most of ours under water; but the moon has been considerably
polluted by solar wind particles (Hashizume \& Chaussidon 2005).
Unlike most of us, the youngest lunar crescent ever seen gets a bit younger
every year. Odeh (2005) reports the new moon of 2004 March 20 seen at age 17h
18min and with only 35 min between sunset and moonset. If you want to do a
decent job of assembly a tag along the lines of ``we were having tea and
missed it,'' you need to check where Odeh was that day relative to
your longitude.

All children are told that stars twinkle and planets don't (a truism already
conditional on location and seeing). In fact, even the moon twinkles, and lunar
scintillation can be used to monitor atmospheric turbulence
(Hickson \& Lanzetta 2004). Does the sun twinkle? Of course it does
(Seykora 1993). It's a star, isn't it?

For the moons of Venus and Mercury, see \S 14 on hens' teeth, flying pigs, and
horse feathers.\footnote{And at least one of the authors would be interested in
feedback from anyone who has recently watched the Marx Brothers film for either
the first or some subsequent time.}

\subsubsection{Asteroids}
As usual the little things appeared in more papers  than the big things of the
Solar System (well there are more of them). We green-dotted for exclamation
(Geewhiz!) the suggestion that the Trojan asteroids of Neptune might have formed
in situ and be the youngest accreted objects in the solar system
(Chiang \& Lithwick 2005). But you are also going to be told that (87)
Silvia has two moons (Marchis et al. 2005), for which the IAU has already
OK'd the names Romulus and Remus (IAU Circ. 8582). They are probably bits
knocked off in some past collision.

 It is, however, future collisions, and with Earth, that one worries most about.
Galad (2005) concludes that 100 years of orbit data are not enough to assess the
probabilities very well, but 99942 Apophis will miss earth in 2029 by rather
more than previously advertized (IAU Circ. 8593). Lyden (2005) suggests what to
do if you see one coming---pull it out to a near
earth orbit or L5 and use the metals.

Binary asteroids, a gee whiz subject a few years ago  are now in the dime a
dozen class, or anyhow at most an IAU Circular each (8483 on 2005AB with a
 period of 17.9 hours and 8526 on a  pair 2.67 arc sec apart, whose period must
be  considerably longer). This has not prevented a certain amount of
unpleasantness about the discovery of 2003 $\mathrm{EL_{61}}$, whose total mass
is probably not all that  much smaller than Pluto plus Charon. We sidestep this
by citing only  IAU Circ. 8577 and by noting that the perusing and purloining of
other folks'  coordinates can be traced back to the early quasar era and
probably to Galileo,  whose solution was the common one of his time, announcing
discoveries in more or less meaningless sentences, whose letters could be
rearranged to say things like ``Venus imitates the phases of the moon.''
 Aw, go on, try it with ``we have seen a really massive binary asteroid.''

Quite a number of
asteroids were mentioned by name as well as number. Here are introductions to
only two: Eros for its complicated cratering history (Thomas \& Robinson 2005),
and 832 Karin, which is perhaps the parent body of the chondritic meteorites
(Sasaki et al. 2004).

Also once whiz-worthy (Ap00, \S 3.4 ) are asteroids that used to be comets (but
whose activity has died away) and the converse (because a coma or tail turned up
after discovery). The Damocloids (with orbits like 5335 Damoc1es) are, says
Jewitt (2005) inactive Halley-family comets, three with comet names and 17 with
asteroid names, all recent discoveries, and not (yet) hanging over anybody's
head. Albedos suggest that nearly half of near-earth asteroids are old comets
(Fernandez et al. 2005) and the class with asteroid names assigned and gas
features caught later reached the point where we gave up recording
anything except IAU Circular numbers (from 8421 to 8622 and a dozen or more in
between). The green dot item here, discussed in IAU Circ. 8582, is an issue of
nomenclature and may not strike you as earth- or even asteroid-shaking. If you
are an asteroid, you can have a permanent number (meaning a well-established
orbit) after being seen at four oppositions, one recent. But a comet needs two
perihelion passages (a much longer time interval for any period longer than
two years). This seems somehow  unfair to the objects concerned, and possibly to
their discoverers and observers. All are Centaurs, and the decision has been
made to let them keep their comet names  but also declare them to be periodic,
with numbers 165P, 166P, and 167P, from which you may get the idea that there
are not really so very many well-established periodic comets.

\subsubsection{Comets}
 Tempel 1 = 9P (well, we told you there weren't very many of them)
became comet of the year when the \textsl{Deep Impact} mission  impacted it
deeply on 2005 July 4.  Its speed changed a bit (giving hope to all who aspire
to knock assorted NEOs into O's  less N to E (Kuehrtt et al. 2005). The actual
scientific results appeared slightly out  of period (A'Hearn et al. 2005, and
the next five papers). Some of the things to be said   are that Tempel 1 came
from the Kuiper (Edgeworth) belt; the material was very loosely
consolidated (more like ashes rising from a fire than even house dust); that the
density was only that of porous ice; the surface was cratered; the dust to water
ratio in the object was larger than expected; and the volatiles included organic
stuff. We suppose this would have gladdened the heart of Prof. Raymond Arthur
Lyttleton (of Cambridge and of the sand bank model) if it were still beating.
Curiously, the comet spat out a couple of jets shortly before impact
(Anonymous 2005c) on 14 and 21 June. If the \textsl{Deep Impact} mission was
the cause, this demonstrates the existence of advanced potentials
with two polarizations and some birefringence.

What else might one say about comets? Well, first you have to discover them.
Messier found a mere 20, a yield of only one for every five of his catalogued
non-comets.\footnote{You are assumed to know that Ml is the Crab Nebula and M3l
 the Andromeda Galaxy. But, quick, without looking them up, what are M2
and M30?} But if the King of Denmark were still giving comet medals, Caroline
Shoemaker would have set him back 32 so far, and images recorded by SOHO 1000,
and counting. Actually there is once again a Comet Medal, named for
Edgar Wilson and intended to recognize non-professional (and human) discoverers.
It was shared this year by R.A. Tucker (C/2004 Ql) and D.E. Machholz (C/2004 Q2,
IAU Circ. 8554),  and Machholz's comet, at least, was briefly a naked eye
object, in January and  February 2005 (IAU Circ. 8484).

Next best to discovering comets is keeping track of them (compare human
relationships). P/18l9 Wl (Blanpied) was recovered in 2003 (IAU Circ. 8485). And
Hale-Bopp has, so far (IAU Circ. 9490) been followed out to 21 AU, where it
 still had an 8.5 arc sec tail. It has a way to go to beat the Halley record
detection distance. When you do follow them
to large heliocentric distances, comets occasionally show unexpected bursts of
activity, which have been attributed to collisions with otherwise invisible
material  (Gronkowski 2004b) and to the solar wind charging up the dark side,
leading to  electrostatic ejection (Gronkowski 2004a).

At the opposite extreme are the sun-grazing
comets (well reviewed by Hoffman \& Marsden 2005). Nearly all the SOHO
discoveries are of this sort, and have short life expectancies. They arise from
the break-up of larger bodies and so come in several kinematic groups, of which
we remember only the Marsden group (because we like him) and the Kreuzer
(because we like his sonata).\footnote{We attempted to normalize our judgement
in consultation. The Faustian Acquaintance assures that the sonata is at least
worth more than the small Austrian coin of the same name, and the Medical
Musician, whose instruments are piano and organ, said ``Find me a violinist and
I'll show you.'' Jack Benny was rejected without audition.}
No comet (even Halley, concerning which we caught only one index-year paper,
Saxena 2004) is worth a lot more than a small coin, because, say
 Neslusan \& Jakubik (2005) there are at least $10^{11}$ more waiting patiently
out in the Oort cloud. Any one that experiences a single passage through the
inner solar system (as a result of encounters with giant molecular clouds, other
stars, or whatever) has a 25--50\% chance of capture or complete removal
(Dybczynyki 2004). Such numbers, plus the inventory of Halley-like comets lead
 Napier et al. (2004) to conclude that the dark Halley-type comets currently
outnumber the luminous ones and constitute a dangerous reservoir of potential
and nearly invisible earth-impactors. The very small albedo comes from a
coating of loose, fluffy organic materials. 3200 Pantheon is well on the way
(Hsieh \& Jewitt 2005), with a surface less than $7\times 10^{-6}$ of which
is covered by freely sublimating water ice. It is the parent body of the
 Geminids.

Just how organic is that organic stuff? Keheyan et a1. (2004) would like to
attribute pre-biotic petroleum in the earth's crust to early comet arrivals
and the petroleum itself to cosmic-ray effects on hydrocarbons. The idea of
pre- (or non-) biological petroleum has, in western countries, been most
closely associated with the name of Thomas Gold (1987), but
even if we were sure he had been right, we would not be sure whether to buy or
sell ChevronTexaco stock. The Medical Musician says ``sell,'' on the grounds
that the company no longer sponsors the Metropolitan Opera radio broadcasts.

Other comets bearing gifts were those of 44 BCE and, conceivably, 4 BCE
(McIvor 2005). The least un-Roman author is inclined to feel that Rome might
have been better off if Julius had not been assassinated that day, but this may
be because the Ceasar she knows best is that of Shaw, not Shakespeare.

\subsubsection{The M's}
Still smaller things are called meteoroids, meteors, and meteorites, depending
on when and where you encounter them, and, we promise,  which is whichy won't be
on the exam. Shower meteors appear to be the most thoroughly studied, perhaps
because it is easier to plan observing runs in advance than for sporadic ones.
They display structure on at least three time scales. First a given meteor trail
can flicker as rapidly as 100 Hz (Babadzhanov \& Knovalova 2004, reporting data
on the Geminids from Dushanbe), due, it seems, to being made of small grains
glued together by stuff of lower boiling point. That the trails are sometimes
heliacal must mean that the grains are irregular enough
for rotation or precession to show, and we mention it primarily for the sake of
noting that the discovery was made on 1 January 1986 (photoelectrically by
J. Westlake) with a pre-discovery by visual methods by W. H. Steavenson on
26 July 1916, conceivably a record for time interval between pre-discovery and
recognition (Sky \& Telescope, 110, No. 3)

Second, there is structure within individual showers (Pecina \& Pecinova 2004 on
the Leonids as seen from Ondrejov in 2000--02; Porubican \& Kornos on the
Quadrantids). The latter consists of five streams, only two of which appear to
share the orbit of comet 2003 EH1, which struck us as a tad odd until
Jenniskens \& Lyytinen (2005) pointed out that the comet was also
$\mathrm{C/1490Y_1}$ so that there had been lots of time for both it and
its debris to shift orbits. The same paper notes that the Phoenicids are
to be associated with a comet that is both 2003 $\mathrm{WH_{23}}$ and D/1819
$\mathrm{W_1}$ (Blanpain), a set of phenomena ripe with opportunities for
mispronunciation.

Third comes temporal structure on scales longer than the annual recurrence time
implied by names like Quadrantids, Geminids, and Lacertids. The Leonids won't
really be back until 2034 (Vaubaillon et al. 2005), and still less should you
plan your career around the Tau Herculids, seen in 1930, but not expected back
until 2022 and 2049 (Wiegert et al. 2005). The parent comet,
73P/Schwassmann-Wachmann, split in 1995, leaving, we suppose,
Schwassmann and Wachmann.

Even sporadic meteors show an annual frequency cycle (Ahn 2005, reporting that
the amplitude and peak time were the same in the years 960-1179 CE as now). This
must be some product of the direction of Earth's orbital motion, rotation, and
day length (or, rather, night length,

day-time meteors being rare). Explanation of why the Leonids have slipped back
about 1.5 days per century is left as an exercise for the student. (No, not you
Mr. H.; your thesis is supposed to be about X-ray astronomy).

Shower particles would not be useful sources of terrestrial petroleum or other
volatiles, even if the particles were big enough to survive passage through the
air. Spectra of Leonids (Kasuga et al. 2005) show evidence of Mg, Fe, Ca, and
Na, but no atoms or obvious molecules of C, H, 0, or N.

The chunks that do reach earth are promoted to meteorites (compare the queening
of pawns in  chess). Individual grains within some of these meteorites preserve
records of the formation  of the solar system and the events immediately
preceding and following. Decoding is more  or less at the stage of completion
where Thomas Young left Egyptian hieroglyphs to take  up light as waves.

So-called pre-solar grains are recognized by isotopic ratios different from
those of general solar system material, especially for isotopes with progenitors
of short half life (extinct radioactivities), though it may not be possible to
exclude inhomogeneities in the pre-solar
nebula of $\mathrm{{}^{26}Al}$ and other parents as an alternative (Boss 2004).

The grains of the year in 2004 (Ap04, \S 3.1.6) were the first nine pre-solar
silicate grains.  Messenger et al. (2005) conclude that they probably came from
supernovae, but picked up their  organic mantles in the general interstellar
 medium. Other grains have been remelted within the  early solar system. They
are called chondrules (from the Greek word for cartilage, and we are
 grateful not to have studied anatomy with whoever chose the name). We caught
during the year  four very localized heat sources that might have done the
melting. McBreen et al. (2005)  suggest lightning due to a nearby gamma ray
burst; Goss \& Durisen (2005) favor shocks when  planets are formed by the
gravitational instability mechanism; Krot et al. (2005) make their
 planets by the accretion mechanism and melt chondrules when planetesimals
collide; and  Aleon et al. (2005) use high energy particles from the young sun.
It is not obvious that they cannot all
be right (unlike multiple hypotheses for, say, the formation of the moon).

\subsubsection{Zodiacal Dust}
At some point we have crossed over into the regime of grains so small that they
scatter sunlight into the plane of the zodiac. This is, after the sun and moon,
the largest \textsl{natural} contribution to the background light of the night
sky (Flanders 2005). The largest non-natural contributions\footnote{The
terminology is perhaps not quite right. To quote one of our favorite defectives
on the police farce, what is more natural than to die when a bullet goes through
your heart?} hardly bear thinking about, but see Cinzano \& Elvidge (2004) for
an update on where the damage has been worst. There is a 2175 \AA\ absorption
feature to be found in interplanetary dust (Brad1ey et al. 2005). And if you
really want to see how the dust moves around, then look at the Doppler shift of
the Mg I 5184 \AA\ line in scattered sunlight. Reynolds et a1. (2004) report
mostly upper limits and are gracious  enough to credit the idea to
Ingham (1963), one of our favorite infrared spectroscopists and
bicycle repair persons.

\subsubsection{Where Did it All Come From?}
Perhaps there have always been two models of how the solar system formed. Long
ago, an encounter of the sun with another star competed with processes occurring
around the sun as it formed (and lost). Distant encounters are not so rare as
close ones and are now blamed for some of the more distant solar system
features, for instance the outer edge of the Kuiper Belt
(Kenyon \& Bromley 2004),
with Sedna as a possible captured member of the other system.

But the ``processes around'' story has now split into two sub-scenarios: a
rapid1y-occuring gravitational instability in the protop1anetary disk, giving
rise to large planets in one gulp (Boss \& Durisen 2005) versus gradual growth
from grains to planetesimals to embryos to planets. Significant numerical
progress seems to have been made on the latter idea during the year
(Go1dreich et al. 2004, who note that the late phases have a good chance of
including impacts like the one responsible for our moon, but that Charon is more
difficult; Leinhardt \& Richardson 2005, also on the ``oligarchic'' phase when
the biggest planetesimals win out over the others). We think it is
within this gradual accumulation model that Matsumura \& Pudritz (2005) explain
why only terrestrial planets are to be found close to the sun. Their model is
 a sort of 3/16'' socket wrench,  since, unlike a variable or  monkey spanner,
it cannot apply to the large number of other systems with
close-in Jovian planets.

After formation comes migration, and we caught precisely a handful of
papers associating migration of the major planets with the phase of late
bombardment on Earth, Moon, Mercury and Mars, and with the
 establishment of the Kuiper belt (and see \S7.3).
\begin{enumerate}
\item Mobidelli (2004) says that the outward migration of Neptune and formation
 of the KBO must have happened after late bombardment or the bombardiers would
not still have been available. \item Murray-Clay \& Chiang (2005) conclude from
the asymmetric distribution of KBO  orbits that Neptune
took at least $10^6$ years to reach its present position.
\item Gomes et al. (2004) attribute late bombardment to migration of
Jupiter and Saturn.
\item Strom et al. (2005) similarly ascribe late bombardment to asteroids
ejected from the main belt by  migrating big planets.
\item Tsiganis et al. (2005) and the two following papers provide a
four-way association, in which Jupiter
and Saturn moving in chased Neptune out, caught the Jovian Trojan asteroids, and
 sent other stuff inwards about 700 Myr after solar system formation.
\end{enumerate}
Some of these considerations are clearly of the variable wrench category and
applicable also in \S6. Others· are not.

\section{DOWNSIZING AND OTHER ISSUES IN GALAXY FORMATION, EVOLUTION, AND
    CONVOLUTION}
Downsizing is not a good thing to have happen to your employer or your space
mission, but it may just be a good minimalist description of the history of star
formation, that is rate versus time (or redshift), metallicity, and site. We
have expended a good many words in previous editions trying to describe the
results of ever-larger surveys of moderate to high redshift galaxies and
ever-larger numerical simulations intended to hindcast the observations, being
sometimes forced to say that the data on stellar populations, colors,
 metallicities, ages, redshifts, locations,and environments just
couldn't be collapsed into anything much shorter than the original paper
(Ap0l \S\S 11.3 \& 11.4; Ap02 \S 10.4; Ap03 \S 9; Ap04 \S 10.6).

It helps to start with the ancient idea of ``primordial galaxies,'' meaning
hypothetical glaring sources of Lyman $\alpha$ emission resulting from the first
major burst of star formation in elliptical galaxies and other spheroids. There
simply weren't any, until the concept was
recast as ``star-forming galaxies at high redshift'' (Ap96 \S 12).

\subsection{A Current Picture}
Under the rubric ``first lights'' the 21st century version of that
search is now a driver for a wide range of upcoming large telescope projects.
Meanwhile, Drory et al. (2005) provide a succinct description of current
knowledge: at every redshift, the most massive galaxies have the oldest stars.
Here are a baker's dozen papers expressing aspects of the same sentiment.
(1) Kajisawa \& Yamada (2005), (2) Jimenez et al. (2005), (3)
Bauer et al. (2005), who note that the specific star formation rate,
$(dM/dt)/M$, declines toward higher masses at all redshifts,
(4) Labbe et al. (2005) noting that the largest $M$'s have the largest $M/L$
ratios at $z = 2$--3 the same as they do now, (5) Smith et al. (2005), pointing
out that fainter galaxies become quiescent at smaller $z$, (6) Yee et al.
(2005), phrasing the process as most of the star formation in a given epoch
declines with time, (7) Tanaka et al. (2005), with things happening first in
both large masses and dense regions, (8) Perez-Gonzalez et al. (2005), saying
that the luminosity of the galaxies with most of the star formation in their
epoch declines with time, (9) McCarthy et al.
(2004) concluding that massive galaxies formed early and fast, (10)
Ferrari et al. (2005), (11) Dola et al. (2005), the largest masses at large
redshift are in clusters with mergers, (12) Hammer et al. (2005), ellipticals
make their stars earlier than spirals, and (13) Holden et al. (2005a).

There were undoubtedly some equally informative papers
that got indexed under some topic other than downsizing and so are missed out
here. Virtually all the cited papers give numbers for quantities like the
fraction of stars formed after $z = 1$, before $z = 5$, or at some intermediate
range, in galaxies of  particular (stellar) masses or luminosities. None of
these much changes the picture  from that of Ap04 \S 10.6. The current star
formation rate is about $0.02\, M_\odot$ $\mathrm{Mpc^{-3}\  yr^{-l}}$. It was a
good deal larger at intermediate redshifts of 1--4 (e.g., Le Fevre et al. 2005),
and smaller again in the more distant past ($5.7\times 10^{-4}$ in the same
units, say Taniguchi et al. 2005). Or to say it all again differently,
ellipticals were half through by $z = 2.3$ (Holden et al. 2005b), while spirals
like the Milky Way dawdled in comparison (Ortolani et al. 2005), and nearly
everybody was about through by $z = 0.35$ (Panter et al. 2005).

 Inevitably, discordances remain. Some of
these arise if, as Caputi et al. (2005) say, massive galaxies were assembled and
stars formed in a two stage process, at $z\ge 4$ and $z< 1.5$, with different
fractions of the two components in different galaxies, and, therefore, in
different samples and surveys. The two stages can be described as the formation
of ``galaxy parts'' and final mergers.

Qualitatively, at least, the items in the preceding three paragraphs are
consistent  with a standard bottom-up, hierarchical picture of galaxy formation.
Less obviously consistent are (a) the conclusion of Lin et al. (2004) that
galaxies now brighter than $L^\ast$ experienced a merger after $z = 1.2$ and
(b) the more quantitative remark of Silva et al. (2005) that the continuity
between SCUBA (ultraluminous, far infrared) galaxies at large
redshift and giant ellipticals now implies that large spheroidal galaxies
formed most  of their stars when they were already single objects. Perhaps this
means that at least some SCUBA galaxies conform to the early definition of
``primordial'' and that their modern counterparts can be recognized.

Well, it may remain true that the details cannot be summarized in any format
shorter than the original papers. But we still like ``downsizing'' as a
first step.

\subsection{Environmental Issues}
 Does the amount or rate of downsizing depend on anything except the mass
of the entity  you are looking at? Because the biggest,
most evolved galaxies tend to live in the
most massive, evolved clusters in the deepest potential wells at any given time,
intrinsic and environmental effects are not easily separable. It is, therefore,
perhaps not surprising that one can find support for, at least, the statements
that (a) environment is unimportant (Treu et al. 2005), (b) environment matters
and, for fixed morphology, mass in stars, and ages of those stars, SFR is
smaller in dense regions (Christlein \& Zabludoff 2005), (c) environment matters
and there is more star formation in clusters (Moss \& Whittle 2005),  and (d)
environment matters now but did not at $z\approx 3$ (Bouche \& Lowenthal 2005).
In contrast, we would not expect to find any contradiction of the conclusion
that dynamical evolution proceeds faster in dense regions
(Einasteo et al. 2005). It is not all that difficult to find hands to
wave at each of the contradictory items (b), well, there is less gas
(Koopmann \& Kenney 2004)
and (c), well there are more mergers (Conselice et al. 2005).

\subsection{Digressions}
 Absolute numbers for star formation rates obviously depend on their being
reliable measurement techniques. Blue light (hot stars), emission lines (HII
regions), X-rays (high mass X-ray binaries and supernova remnants), radio
(electrons and field from SNRs), UV (hot stars), and mid to far IR (blue light
absorbed and reradiated by dust) have all been used. And no they don't always
all agree, but, in optimistic mood, we highlight
Dopita et a1. (2005), who report that absolute calibration of UV, H$\alpha$
emission, near IR,  and mid IR are all in good shape, relegating to a
 dependent (nay, even dangling) clause  the conclusion that even infrared bands
 are a less than gold standard  (Conti \& Crowther 2004, Bose11ie et a1. 2004).
 The difficulties are dust heated other
 ways and new star light escaping unmodified, and one paper focusses on each.

Long ago, the least teachable author was taught to worry about the
``last gasp'' issue, that is, the current gas supply and current star formation
rates in nearly all galaxies cannot last more than a small fraction of the
Hubble time. The factual statement remains true, for example $10^{7-8.3}$ yr
for luminous compact blue galaxies (Garland et a1. 2005), at most $10^9$ yr for
b1ue compact dwarfs (Kong 2004), right on out to a lensed $z = 2.5$ galaxy
(Sheth et a1. 2004). Gas exhaustion has been ubiquitous since about
$z= 0.7$ (Bell et a1. 2005). But she is no longer much worried. Star burst
galaxies (meaning  by definition that they don't do it for very long) are
 naturally over-represented in any  magnitude-limited sample. And, as the
very word ``downsizing'' implies, star formation  really is dying away, due,
say Bell et a1., primarily to gas exhaustion and not to a  decline in the
merger rate. We live in an evolving universe in which no extended epoch
 can reasonably be regarded as typical or untypica1.

\subsection{Galaxy Types: L, S, and D}
 L is for lenticular, otherwise known as S0. Since the
time that Hubble's early/late classification scheme ceased to be thought of as
an evolutionary sequence, astrofo1k have been asking what process(es) can be
reponsib1e for these disk galaxies with little or no remaining gas or star
formation, found commonly in intermediate zones of rich clusters.
Phrased that way, the question almost answers itself (though perhaps with a
wrong answer). Ram pressure stripping is reponsib1e for one well known
transformation in progress in Virgo (Crowl et a1. 2005). Burkert et a1. (2005)
note penetration by smaller galaxies and Vollmer et a1. (2005) other processes
in clusters that remove gas. Burstein et a1. (2005), however, say that gas
stripping is not the answer, and Owen et a1. (2005) blame bursts of star
formation (probably however also connected with entry into clusters).
Another case, we think, where two or more can be correct.

S is for spiral, and they get only sound bites this year, while the Milky Way
has been dispersed among several sections.
\begin{itemize}
\item M82, the quintessential messy galaxy (well, Irregular is the polite term)
has underlying spiral structure (Mayya et a1. 2005), and you will be happy to
hear that the arms trail, at a pitch angle near $14^\circ$.
\item Thick HI layers, whose rotation lags that of the thin disk, are fairly
common  (Barbieri et al. 2005 with a map of NGC 4559).
\item Low surface brightness (LSB if we should need them again sometime)
galaxies have thin disks (Bizyaev \& Kajsin 2004).
\item Thick disks, once nearly the exclusive property of the Milky Way,
are actually common once you know how to look for them
(Tikhonov et al. 2005, Jould 2005). That they don't all share the rotation speed
of their thin disks (Yoachim \& Dalcanton 2005) would seem to rule out puffing
up of a thin disk by gravitational encounters as a universal formation mechanism
(not to mention top-down type processes), but don't worry. There are lots of
other possibilities (Brook 2004 on accretion processes).
\item Barred spirals (SBs) form a continuum with the unbarred (we're saving the
disbarred remark for next year), and the strongest bars have the strongest
spiral arms and shortest lifetimes (Buta et al. 2005). Since significant numbers
of bars exist by $z = 1$, again, once you know how to look for them
(Jogee et al. 2004), we think this must mean that they come and go.
Fadotti \& de Souza (2005) appear to agree, but their picture of thick bars
with large stellar velocity dispersions disappearing quickly into thick disks
won't fit with everything else published this year.
\item Cores decoupled from their surrounding galaxies in kinematics
(Shalyapina et al. 2004)  or composition (Sil'chenko 2005) or both were a
discovery within living memory, but are now ``a well known class of
astrophysical object.''
\item Some spirals have
the most massive (``maximal'') disks that their inner rotation curves permit,
and therefore very little central dark matter (Fuchs et al. 2004). Others do
not (Dutton et al. 2005, Kregel \& van der Kruit 2005).
\item With equal confidence we can say that some spiral disks appear to have
sharp outer edges (truncations, Trujillo \& Pohlen 2005), and others do not
(Bland-Hawthorn et al. 2005 on NGC 300, Tanvir 2005).
\item And how well you can see through the disk of a spiral depends mostly on
just where you look (Holwerda et al. 2005).
\end{itemize}

D is for dwarf. Well, usually d is for dwarf, as in dIrr, dE, dSph (irregular,
elliptical, spheroidal) etc., but we are saving what little clout
we have with the editor and  University of Chicago Press for
 more important battles than being allowed to start a
sentence with a lower case letter!  Whichever case, they are the commonest sorts
of galaxies, making up 85\% of a nearly complete catalog of 362
(Karachentsev et al. 2005), reason enough perhaps for them to appear in more
indexed 2005 papers than any other class. But, in addition, one hopes that they
may preserve (or re-enact) the processes
by which larger, more famous galaxies have formed.

Let's start with a nice contradictory pair of results. According to
Mathews et al. (2004) there are no dwarf galaxies in groups at redshifts
$z\gtrsim 2.5$ (based on a dynamical analysis of the nearby NGC 5044 group).
This implies that they all arrived recently, and indeed, the star ages are close
to 5 Gyr, so they didn't exist at large redshifts. That is, there are no old
dwarfs. On the other hand, according to Momany et al. (2005) there
are no young dwarfs. All have stars at least 10 Gyr old, including Sag DIG which
they present in detail. Can these be reconciled? Of course! What do you think
theorists are for? If, as Corbin et al. (2005, writing about HS 0872+3542)
indicate, dwarfs are assembled from star clumps whose sizes come between those
of globular clusters and those of small galaxies, the stars can be much older
than the dynamical entities. We picked out a dozen or so other dwarf highlights
and present them without specifying which are truly new, lest the ghost of
Fritz Zwicky haunt our proofs.

There are relatively more dwarfs in clusters than in the field (dEs and dSphs,
 not surprisingly, Trentham et al. 2005). Conversely, they are not
 over-represented in voids (Hoyle et al. 2005).

The first truly isolated dwarf (a dSph) appears in Pasquali et al. (2005) and,
say the authors, ``Following IAU rules \ldots is named APPLES1.'' A search for
ORANGES1 with which to compare it is under way.

Some of the Local Group dwarfs began life
before reionization (Ricotti \& Gnedin 2005), meaning, we think, their oldest
stars, not necessarily the dynamical entities.

Some live in real halos with their own dark matter (Piatek et al. 2005 on the
Ursae Minoris dwarf, Wang et al. 2005 on Fornax). Fornax is likely to have
experienced a recent merger (Colemanet al. 2005). Well, we said\ldots

A couple of
dEs in the NGC 5044 and 3258 groups have kinematically decoupled cores
(De Rijcke et al. 2004), as does our own Sextans dSph (Kleyna et al. 2005).

Dwarf S0's coexist in the Coma cluster with dEs (Aguerri et al. 2005). The
dE's they say, are descended from dIrr's (this has been disputed in earlier
years), and the dwarf S0's are harassed former bright, late spirals.

Dwarf galaxies can form from tidal tails, but the conditions are more
restrictive than we used to think (Duc et al. 2004; and there indeed is
Zwicky's ghost; forming
galaxies this way was originally his idea, he said). Duc et al. say that the
torn-out protogalaxies remain bound only when they are still inside the main
dark matter halo.

dIrr's have star formation concentrated toward their centers with older stars
farther out, more or less the opposite of spirals
(Tikhonov 2005, Hunter \& Elmegreen 2004).

Two categories that are more or less always dwarfs are the LSB galaxies
(Sabatini et al. 2005) and the BCDs (Gil de Paz \& Madoie 2005).

And in case you have been entertaining doubts, there was reaffirmation of the
article of faith that dwarf galaxies occupy different regimes of the
correlations of size, dynamical, and chemical properties from those belonging to
globular clusters and to big ellipticals (De Rijcke et al. 2005). Indeed they
derive their blue light from different sources, residual star formation rather
than extended horizontal branch stars (Boselli et al. 2005).

\subsection{E is for Elliptical}
We indexed only about 30 papers of this shape in 2005 (a bad year for the
featureless, perhaps, as political events have shown) and starred for special
attention an ``it's all OK'' paper. Dekel et al. (2005) reassure that the outer
parts of giant ellipticals really are dominated by dark matter. The appearance
of a small velocity dispersion is the result of highly eccentric star orbits,
representing the first stars torn loose during the last major merger. An
analysis of gravitational lensing (Ferreras et al. 2005) concurs,  with
DM dominating outside 20 kpc and rather little inside 4 kpc. The smaller the
galaxy, the further out luminous matter remains dominant.
Padmanabhan et al. (2005), after looking at 29,469 SDSS ellipticals with
spectra, report details of the correlations of $M/L$, dark matter fraction, etc.
as a function of galaxy luminosity,  surface brightness, size, and so forth. In
 quick summary, more dark matter means more everything except starlight.

As your authors age, we become ever fonder of the oldest stars in the oldest
systems, perhaps the globular cluster populations of ellipticals. The Readers
Digest Condensed version mentions only the total number of clusters relative to
galaxy brightness, $S$. Davidge \& van den Bergh (2005) have managed to measure
 $S$ for the heavily obscured Maffei 1, the nearest (field-ish) elliptical, and
find $S = 1.2\pm0.6$, not particularly anomalous for field Es they say (it is
bigger in clusters). What are the units? Um, er. Something like number of
clusters per $10^8$ $L_\odot$ in the B band. And, lest we forget to mention it
elsewhere, another ``it's all OK'' item.
If ellipticals are made from spirals (which have smaller $S$), can you get the
right number in the product without having to wait several Hubble times for all
the young, bright stars to evolve to death? Yes, say Goudfrooij et al (2004)
and Li et al. (2004). You also automatically account for two populations, old
clusters from the parts assembled and the younger, more metal rich clusters
from the assembly process.

A proper characterization of cluster properties is not, of course, limited to
the $S$ value but includes ages, metallicities, and deviations from the solar
pattern of heavy element abundances. Kaviraj et al. (2005) and Puzia et al.
(2005) contributed to that topic during the index year, and yes, multiple
populations exist.

Where are the SCUBA galaxies of yester-$z$ (1--3 or thereabouts)? Gone to
ellipticals  every one, say Takagi et al. (2004) and Swinbank et al. (2004).
No information was provided on when will they ever learn.

Why do some ellipticals continue to form stars longer than others (Cotter et al.
2005)? Surely almost as many reasons as there are ellipticals, but we rather
like the suggestion of Springel et al. (2005) that, when a merger product
includes two  black holes, their orbiting quickly quenches core star formation,
turning accretion of gas into an outgoing wind. And if not, not, and the core
remains blue, with low level on-going star formation for as long as 7 Gyr after
the merger. Does this contradict the statement a few paragraphs above that the
blue light in Es comes from extended horizontal branch stars?

\subsection{X is for XBONG and Many Other Types}
 At least a dozen appeared during the
reference year, some perhaps more interesting than others. X, say
Hornschemeier et al. (2005) stands for X-ray Bright, Optically Normal Galaxies,
and there aren't any, the phenomenon being an artefact of poor wavelength
resolution at large redshifts. In fact, they report, in suitable spectra, [OIlI]
looks as strong as it is in Seyfert galaxies for their X-ray bright sample. Some
other portions of the optical/X-ray/radio ratio space are empty because of other
selection effects (Anton \& Browne 2005), so don't sell all your XBONG stock
(and don't go looking for it in the back pages of the Wall Street Journal
either; 5 letters is too many for a stock ticker symbol; see \S11 at 702)

A category that may or may not be empty is that of HI-filled but starless halos.
Doyle et al. (2005) find that none of 3692 HI sources are invisible in regions
where the Milky Way imposes less than $\mathrm{A_B = 1^m}$. But there were some
fiscal 2005 HI clouds that  rated press releases. The key issue is whether these
clouds  really live in their own dark matter halos or are merely gas pulled out
of some more normal galaxy in a group or cluster (Minchin et al. 2005;
Osterloo \& van Gorkum 2005,
Virgo clouds; Walter et al. 2005 a stray HI cloud in the M81 group).

Existence continued throughout the year for---
\begin{itemize}
\item[(a)] Markarian galaxies (Stepanian 2005),
\item[(b)] clump cluster galaxies (Elmegreen \& Elmegreen 2005), another sort
of pre-elliptical,
\item[(c)] UV excess galaxies, which are rather like Markarian ones, but
found by Kazarian \& Martirosian (2004),
\item[(d)] void galaxies, which are unbiased
(Goldberg et a1. 2005), apparently the nicest thing anybody said about them
during the year,
\item[(e)] assorted blue compact galaxies (Lin \& Mohr 2005, who note that
mergers are commoner in clusters).
\item[(f)] Extremely Red Objects, which are, variously, obscured AGNs
(Brusa 2005); another sort of pre-elliptical (Brown et a1. 2005); a mix of pure
bulge, disk, and interacting galaxies (Sawicki et al. 2005); distant, old
clustered objects with intermediate star formation rates, including about
one-third E+A galaxies (Doherty et al. 2005). We shouldn't dream of saying these
chaps don't read each
others' papers, but we hope they don't try to believe all of them at once.
\item[(g)] E+A galaxies (meaning ellipticals but with spectra displaying the
absorption lines of A stars), which are not the same as IRAS galaxies, another
sort of post-starburst (Gogo 2005); not all the same sort of beasts
(Pracy et a1. 2005); not another sort of proto-elliptical, but arguably a form
of proto-S0 (Bekki et al. 2005); not inhabitants of rich clusters, being found
mostly in small groups and the field now (Blake et ale 2004); not to be confused
with e(a) galaxies, which display nebular emission lines (Balogh et a1. 2005,
who prefer to call E+A's k+a, meaning a mix of stellar K and A spectral types,
which, if you were a star might get you declared symbiotic.\footnote{Symbiotic
stars are a class of cataclysmic variable with one hot and one cool star
and, as a rule, less exchange of bodily fluids than in the novae, dwarf novae,
and so forth.} It probably is easier not to be a number of different things
simultaneously than to be a number of different things simultaneously.
\item[(h)] Lyman break galaxies which also show broad absorption lines, like
some QSOs (Ivison et al. 2005),
\item[(i)] Damped Ly$\,\alpha$ galaxies, which could be
either the edges of big things or the whole of smaller things
(Hopkins et a1. 2005, the latter view, and Chen et al. 2005, the former view;
Weatherley et al. 2005 favoring protogalactic fragments for those at $z> 1.75$;
Okoshi \& Nagashima 2005
favoring LSB galaxies for $z = 0$--1, again not a contradiction),
\item[(j)] Ly$\,\alpha$
emission blobs (Mori et a1. 2004), which are currently petitioning to be renamed
primordial galaxies with lots of supernovae.
\item[(k)] SCUBA galaxies are named for a bolometer (operable under water we
suppose) rather than any specific physical characteristic, and so are allowed to
have a range of underlying energy sources, all of which lead to big and bright
and dusty (Greve et al. 2005;
Chapman et al. 2005; Pope et al. 2005; Houck et al. 2005) all with an assortment
of details, and the last reporting redshifts from the Spitzer Space Telescope
for some objects with no optical counterpart).
\item[(l)] Satellites of spirals, with a color~luminosity correlation extending
over a range of 12.5 magnitudes (Gutierrez \& Azzaro 2004),
\item[(m)] low surface brightness
galaxies, with a new catalogue of 81 of them, versus 18 known before
(O'Neill et al. 2004).  Malin 1 remains unique, say Minchin et al. (2004), but
 a little Irvine bird has whispered in our ears that even Malin 1 isn't exactly
like Malin 1.
\item[(n)] And the favorite of the year, the ultracompact dwarfs, which are
 neither dE's (Mieske et a1. 2005 on a couple of M32 clones in Abell 1689, plus
fainter galaxies bridging the luminosity gap to the UCDs in Fornax) nor
overblown globular clusters (Huxor et al.
2005, on objects in the halo of M31 with colors like globular clusters but
half-light radii of 30 pc like small galaxies). As for what they are, well,
you know the choices. They started out to be that way (Bastian et, a1. 2005);
something bad happened to them on the way to the cluster
(Hasegan et a1. 2005)\footnote{The author who has just acquired her first-ever
device for playing old movies
opines that of the ones that seemed hysterically funny 40 years ago,
\textsl{A Funny Thing Happened on the Way to the Forum} stands up best. Going
back another two decades, it is Jack Benny's \textsl{To Be or Not to Be}.};
or, naturally, more complex scenarios
(Fellhauer \& Kroupa 2005, Clark et al. 2005).
\end{itemize}

\subsection{Galaxies as Families}
 The traditional collective noun is the wastebasket, but
galaxies as a collectivity, are, we learned this year, a one-parameter family
(Coenda et al. 2005), a two-parameter family (Lauger et al. 2005), whose
parameters are central concentration of 1ight and degree of symmetry under
$180^\circ$ rotation in the UV to I bands, at least for $z = 0$ to 1, or bimodal
in distribution over one or more parameters. For instance, Wiegert et a1.
(2004) report that several properties are bimodal on either side of B-V = 0.29
out to $z = 3$, but that there are fewer red galaxies at $z>1.4$.
Nuijten et al. (2005) find a bimodal number versus color relation, with the
red/blue ratio dropping from $z =0$ to $z = 1$, and also a division by whether
bulge or disk dominates. And
Gallazzi et al. (2005) focus on the bimodal distributions of mass, star age,
and metallicity on either side of a stellar mass of $\log{M} = 10\pm0.3$.

Looking back to $z= 2.5$ or more, as you surely do not need to be told, the
commonest class is ``peculiar'' or ``irregular'' (Cassata et al. 2005), though
still with 20\% normal E and S0 galaxies and 27\% normal spirals. One begins
to feel the need of another classification scheme to record changes of type,
luminosity, mass, etc. at moderate to large redshift. Conselice et al. (2005)
 have suggested one.

Collective clustering properties and their changes with redshift are presented
by Ilbert et a1. (2005) and Le Fevre et a1. (2005a,b). A two-word summary is
``bias evolves.'' That is, the brightest galaxies were more concentrated in the
most dense regions at $z = 1$ than they are now. This sounds like a large scale
 manifestation of ``downsizing,'' which is where this section began, and it is,
 therefore, almost time to move on to the universe as a whole.

\subsection{Anthropogenic Downsizing}
 ``Anthropogenic'' these days is normally associated
with ``climate change'' or ``extinctions,'' and you will find it hidden there.
But the Milky Way has experienced a sort of anthropogenic downsizing from the
time of Harlow Shapley (1919) to the present. We remarked upon this some years
ago   (Trimble 1993), but Vallee (2005) has followed the trend down to the
present.  Shapley's galactocentric distance for the solar system of 18.5 kpc has
shrunk to a mere 7.9 kpc. The interarm spacing has also shrunk a bit
(3.5 to 3 kpc over about 20 years) while the pitch angle since 1980 has remained
 steady at $12^\circ$.   Most curious of all, the number of arms reported over
the years has shifted  gradually from two to four, without passing through
three. We think this may  have been organized by the same group that markets
end-of-the-season peaches that go from green to rotten without passing
through ripe.

\section{TAKE MY UNIVERSE, PLEASE}
No, you are too young to remember the stand-up, borscht-belt comedian from
whom this line is borrowed (though two of your three authors and the Faustian
Acquaintance of earlier ApXX's are not). It means that, although patient
readers will eventually encounter updated values of the standard cosmological
parameters and other conventional progress, the more unusual ideas come first.
Oh, and Earth is in the middle, as per Ptolemy, Brahe, and many other
distinguished predecessors in summarizing the cosmos.

\subsection{Old Kosmoi Never Die}
 Well, no, we didn't encounter any earth-centered
models this year, but Grujic (2005) advocates a Newtonian model, with vacuum
energy outweighing the matter and with fractal structure. The universe of
Skalsky (2005) is also Newtonian, but with $\Lambda = 0$. Other old friends
whose age begins to approach ours include---
\begin{itemize}
\item Hoyle-Narlikar conformal cosmology,
discussed by Papoyan et al. (2003), in which the CMB is due to the decay of a
primordial vector boson.
\item Quantized non-cosmological redshifts, supported by
Bell (2004) from structure in $N(z)$ for sources in the Sloan Digital Sky
Survey, and opposed by Basi (2005), who concludes that apparent redshift
periodicities in the distribution of gamma ray bursts arise from selection
effects. If so, then we won't need the mechanism, involving stimulated Raman
 scattering, suggested by Holmlid (2004).
\item Quantized orbits, on the Keplerian scale (vs. Bohrian) would seem to be
even more problematic, but that  appears to be what is intended by Chatterjee \&
Magalinsky (2004).
\item Modified Newtonian Dynamics explains (well, they say predicts) the
observed relationships of scale lengths and accelerations in disks and halos of
spiral galaxies (Milgrom \& Sanders 2005). It could be tested by the behavior of
star streams in the halo of the Milky Way, say Read \& Moore (2005).
\item G varying with separation fails as a model for local peculiar velocity
fields (Whiting 2005). \item The Dynamics of Dinculescu (2005) seems to be even
more Modified, so as to bring
the temperature of the CMB into galactic structure.
\item Tired light cannot explain the time dilation seen in the light curve of SN
1997ex at $z = 0.361$, say Foley et al. (2005). Our only objection is that they
credit tired light to a 1986 paper rather than to Zwicky (1929). Variable
particle masses don't fit either, according to the same paper, with credit this
time to  Narlikar \& Arp (1997), not quite the first team in that race, but at
least coming out of the right stable.
\item An Einstein-de Sitter model (held down flat by matter alone) can fit
supernova data if there is intergalactic metallic dust says Vishwakarma (2005).
He says it also explains ``all other existing observations.''
\item Gravity might be Lyra
(Rahaman et al. 2005), Saez-Ballester (Mohanty \& Sahu 2004) or repulsive
(Raham 2004), though we cannot claim to be wild about any of them.
 \item The spin (vector) driven inflation of Garcia de Andrade (2004) probably
 belongs here,  too, although, given that ordinary inflation is described as
driven by a  scalar fie1d, it is primarily the venue and the September, 2001
submission data  that suggest the conclusion. The related model of a torsional,
Einstein-Cartan universe (Garcia de Andrade 2004a)
may or may not deserve a separate bullet
 \end{itemize}

\subsection{Newer Universes Hang in There Too}
Coles (2005) gave the ``State of the Universe'' address for the year, with
a solid review of the current conventional wisdom. It has, of course, both dark
matter and dark energy (next section). Here are some variants that would seem to
be acceptable within a broad conventional cosmo-church.

None-zero rotation and shear (Jaffe et a1. 2005) provide a possible fit to the
weak quadrupole and octopo1e asymmetries and the strong north-south one in the
WMAP first-year measurements of CMB fluctuations. Will you have heard about the
next two years' data by the time this appears? We hope so! The Jaffe et a1.
model is of Bianchi type VII, with $\omega/H = 4.3 \times 10^{-10}$ and
$\sigma H = 2.4 \times 10^{-10}$ toward $\mathrm{(1, b) = 220^\circ}$,
$60^\circ$. Oh, and this universe is open, with $\Omega = 0.5$.

Branes remain popular and will undoubtedly do so right up to the end of (our)
universe, which will occur in the collision and mutual annihilation of positive
and negative tension 3-branes (Gibbons et a1. 2005).

Loop quantum gravity allows you (well, some of you anyhow) to avoid an initial
singularity in the universe (Mu1ryne et a1. 2005, Boyarsky et a1. 2005,
Bojowa1d et a1. 2005), but the middle one of these also has a big black hole.
Such a universe can bounce, and its volume at the bounce (in Isotropic Loop
Quantum Gravity) reveals the minimum length scale on which one cuts
off modes when calculating things to prevent getting infinite answers. But if
Date \& Hossain (2005) gave a number, we missed it.

Non-trivial topology, like rotation and shear, continues to hover at the edges
of the WMAP data. Aurich et a1. (2005) propose Picard space, which is hyperbolic
in the form of an infinitely long horn of finite volume. The space (Picard 1884)
is older than relativity, and we are not sure whether the author was one of the
ballooning Picards or if he thought of his hyperbolic space while hanging from
a (we hope) positively curved balloon.

There exist, says Lake (2005), positive $\Lambda$ , $\Omega\approx 1$ non-flat
universes, and, what is more, we live in one (WMAP again). The models are not,
according to the author, finely tuned to get $\Lambda$ right.

And, say Alam et al. (2004), the best fit to supernova data
remains a changing $w$ in the equation of state $P = -w\rho$ from $w = 0$ at
$z = 1$ to $w = 1$ now. This is more or less a Chap1ygin gas, whose death
(by liquefaction?) we announced a year or two ago. Alcaniz \& Lima (2005) more
or less concur, on the basis of angular diameters
of radio sources (a very old and frequently misleading cosmological test).
Biesiada et a1. (2005) agree with Alam et a1. that supernovae are the right
test, but conclude that the answer is not yet in.  Incidentally,
Alam et a1. provide a very nice brief introduction to eight
possible alternatives to conventional constant $\Lambda$, some of which we have
 probably  missed in earlier years.

\subsection{Degrees of Darkness}
 As in the previous 10 or more years, there were conventional and
unconventional candidates for dark matter. Traditions upheld include that it is
not mostly in galactic disks (Ciardullo et al. 2004 on M33), not mostly halo
white dwarfs (Creze et al. 2004), not mostly cosmic string
(Jeong \& Smoot, 2005, though maybe a bit (Sazhin \& Khovanskaya 2005), and not
a major constituent of the sun (Kardashev et al. 2005)

Approaching this time from the conventional side, we green-starred the thought
by Boehm \& Schaeffer (2005) that candidates might best be classified not into
three discrete groups of cold, warm, and hot, but rather into a continuum based
on their damping lengths in the early universe and the requirement that they not
wipe out structure on the scale of galaxies. This permits consideration of
candidates with some coupling to neutrinos (superweak, presumably) and photons
(weak electromagnetism?), as well as those that do only gravitation.

\textsl{WIMPs and axions} (reviewed by Zioutas et al. 2004) are the candidates
of longest standing,  and there could be, they say, some of each, though neither
has yet interacted detectably with  laboratory apparatus (Akerib et al. 2004).

\textsl{Kaluza-Klain particles} (meaning the lowest mass ones that conserve KK
parity) get to be number two this year (Bergstroem et al. 2005), and it would be
lovely to be able to say with enthusiasm that the TeV gamma rays coming from the
direction of the center of the Milky Way are their annihilation products. There
were, however, at least three other ``DM annihilation has been seen'' papers
during the year. Zhao \& Silk (2005, neutralinos clustered around
$10^{2-3}\,M_\odot$  black holes), Beacom et al. (2005, the 511 keV line from
the direction of the galactic center),  and Elseasser \& Mannheim (2005, the
extragalactic EGRET background) unfortunately require three
 different mass ranges for the mutually self-assured destructive particles, and
Ando (2005) says  that none of them works wildly well anyhow.

\textsl{Unified DM-dark energy scenarios} entered their fourth
 year (Zloshchastiev 2005) and are perhaps ready for pre-school.\footnote{We
have already discovered that a Ph.D. is not quite enough to understand some of
these papers.} The tensor graviton of  Dubosky et al. (2005) also gives the
Friedman equations an extra acceleration-inducing term, and so probably
belongs in this paragraph.

\textsl{Supermassive black holes} (meaning $\sim10^5\, M_\odot$) have been in
and out of the inventory a number of times. This year, Jin et al. (2005) favor
them for dwarf spheroidal galaxies because they promote shallow (core vs.~cusp)
central density profiles. If, however, they contribute a typical density of only
$1.5\times 10^5\, M_\odot$ $\mathrm{Mpc^{-3}}$ (Mahmood et al. 2005, on
evolutionary scenarios vs.~data), then they are not even a 1\% solution, however
respectable they may be.

And we proceed onward to Fermi balls (Munyaneza \& Biermann 2005); baryon clumps
(Froggatt \& Nielson 2005); neutralino clumps of about one earth mass and one AU in size
(Diemand et al.), one of which should pass through the solar system each
year.\footnote{The most recent having undoubtedly been responsible for the
10 plagues.} Annihilation in galactic ones could be the unidentified gamma ray
sources. Since the clumps are made of a traditional DM candidate, perhaps they
should appear higher on this list.

Here are some more. Droplets of non-hadronic color superconducting phase of very
dense strange\-lets of quarks and antiquarks (Oaknin \& Zhitnitsky 2005). These
produce positrons in the decay process and so account for the 511 keV line from
the galactic center. The charged monopoles of Dubrovich \& Susko (2004) have
masses of $10^{16}$ GeV like all other undetected magnetic monopoles, but they
carry a charge of 68.5 $e$, ratherthan 1/3 or 1 or zero .

Lest you conclude that (at most) one of these can be right, take comfort from
 Karachentsev (2005) whose analysis of motions in the Local Group and four other
 nearby small groups of galaxies implies the existence of two sorts of dark
matter, one  more dissipative (but apparently too abundant to be all baryons)
than the other.

Concerning the nature of the dark energy, Ap04 \S 8.6 noted the remarkable
absence of seriously non-standard DE candidates. With the papers of
Arbab (2004) and Garcia de Andrade (2004, 2004a), this class would seem no
longer to be empty.

Mainstream on the basis of its source, but seriously inflammatory was the
thought (Kolb 2005) that there might be no dark energy, with the appearance of
accelerated expansion due to very large scale structure. Mainstream, but
apparently wrong. Large scale inhomogeneities, at least up to the size of the
horizon, just don't have the desired effect (Siegel \& Fry 2005). Other dark
energy alternatives include:
\begin{enumerate}
\item Extra-dimension gravity on large scales, which modifies the Friedmann
equations (Elgar{\o}y \& Multamaki 2005).
\item A new $M\approx 10^{-5}$ eV particle (Dvali 2005), where some of the DE
resides in the masses of the particles and some in the scalar field potential.
It should be detectable through its interactions with neutrinos, slowly changing
their masses with time. \item ``A scalar field self-interacting through
Ratra-Peebles or supergravity potential'' (Maccio 2005). It predicts more lensed
arcs at the epoch of cluster formation ($z = 0.2$--0.3) than does a pure
cosmological constant.
\item A candidate for phantom energy (Amendola 2004), that is $|w|>~1$
in the equation of state. The field could cluster on astrophysical scales and so
contribute to structure formation, acting like a long-range repulsive part to
gravity (and so, we suppose, making voids rather than clusters if it clumps).
\item A fit with $w = -2.85$ to supernova and other data (Bassett et al. 2004).
\item Another highly-flavored (or anyhow non-vanilla) equation of state that
accounts for the very small quadrupole and octopole moments of the cosmic
microwave background  (Enqvist \& Sloth 2004).
\end{enumerate}

Perhaps in some ways more useful than a list of candidates is the suggestion
that they should be classified in terms of what they ``predict'' for $H(z)$,
$d^2H/dz^2$ etc. (Evans et al. 2005), in the same way that
Boehm \& Schaeffer (2005) suggested classifying DM candidates by their damping
lengths during structure formation. We confess, however, to having indexed
Evans et al. as ``Back to McVittie,'' who was a great fan of $q_0$ and
higher-order perturbative terms for the equations of cosmology.

Are you expecting a big rip? Or have you ever worried that some of your own
expansion  over the years might be cosmological? Then you will forgive our
violating the ApXX rules  and referring you to a paper not yet published
(Price 2005). He has shown that the Earth (etc.) are not expanding and why, and
 described what will happen if acceleration gradually overcomes, first,
gravitational binding, then electromagnetism, and eventually the color force.

\subsection[Is the Universe Full of Stuff?]{Is the Universe Full of
Stuff?\footnote{Frivolous readers may remember this as the title of a paper
parodying the style of the Keen Amateur Dentist of earlier ApXX's. The answer
 was, of course, considerably more complex than yes, no, both of the above, or
none of the above.}}  Oh yes, no fewer than 40 kinds, tabulated by
Fukugita \& Peebles (2004), including negative binding energies, the dark
sector, thermal relics, baryons in many forms, stellar radiation and neutrinos,
cosmic rays,  magnetic fields, and kinetic energy of the intergalactic medium,
all reported as  fractions of the closure density ($\Omega$'s) for $H = 70$
$\mathrm{km\ sec^{-1}\ Mpc^{-1}}$. In particular,  neutrinos contribute 1.1\%
of the total and the CMB only 0.1\%. Other compilations of the numbers of
standard, hot big bang, $\Lambda$CDM cosmology are to be
found in Abazajian et al. (2005 using halo occupation data from SDSS) and in
Rebolo et a1. (2004), also with no surprises, though mild evidence for deviation
from the $n = 1$  Harrison-Ze1dovich spectrum of perturbation amplitudes, in the
direction of less power on smaller scales.

The conventional numbers, you will recall unless you have spent the last three
years in a witness protection program, hiding from press releases, the standard
astrophysical literature, and even these reviews, are a Hubble constant of
 65--70 $\mathrm{km\ sec^{-1}\ Mpc^{-1}}$, a little less than 30\% of the
closure density in all forms of matter (about 4\% baryons), and a little more
than 70\% in cosmological constant, dark energy, or whatever, plus the
Harrison-Zeldovich spectrum (equal amplitude on each scale as it enters the
horizon), a normalization of the matter fluctuations now, called $\sigma_8$,
near unity, and consistent numbers for the
age of the universe, $q_0$ (the second derivative of cosmic length scale),
and bias (the degree to which luminous matter is more (or less) clustered
than dark matter).

Nothing happened in index year 2005 to disturb this consensus. Perhaps one
should not be surprised. In January 2002 an official distance modulus for
the Large Magellanic  Cloud was announced ($\mathrm{m-M = 18.50}$). And the 21
papers published since then all agree within  0.5$\sigma$ with this HST
Key Project value, yielding a $\chi^2$ of only 0.189 (Krisciunas 2005).

$H_0$ is once again attracting more independent researchers. We caught only
eight values  published during the fiscal year, but they ranged from 86.4
$\mathrm{km\ sec^{-1}\ Mpc^{-1}}$ (Bonamente et al. 2004, from the
Sunyaev-Zeldovich effect in Abell 64) to 50 as an lower limit
(Stritzinger \& Leibundgut 2005, from the requirement that
Type Ia supernovae make less than 1 $M_\odot$ of $\mathrm{{}^{56}Ni}$).
The median is 66 (Jones et al. 2005a) from an S-Z measurement in a different
cluster.

The sole $q_0$ for the year is a concordant $-0.7$ to $-1.0$ (John 2004) from a
five-term  fit to SNe Ia data. His third derivative, called $r_0$, is between
0 and +6, while the 4th derivative is close to +10. Oh, and $H = 60$--70,  with
which few would quarrel.

The baryon contribution needed to form structure on the 0.054
$\mathrm{Mpc^{-1}}$ scale (Tocchini-Valentini et al. 2005) falls close to the
$\Omega_bh^2 = 0.22 \pm 0.002$ level needed to agree with big bang
nucleosynthesis. It is however, arguably too early to declare
these numbers off bounds for revision. The observed deuterium abundance remains
a smidge less than the theorists would like (Crighton et al. 2004), and the
 predicted $\mathrm{{}^7Li}$ in old stars is more like two smidges too much
(Fields et al. 2005 on the theory;
Melendez \& Ramirez 2004 on data). Fields et al. put forward a way to lower the
theoretical prediction to match the raw data. Richard et al. (2005) suggest,
contrapuntally, that the theorists are right and the oldest stars have hidden
about  half their lithium via gravitational settling.

The usual number near 0.24 for total matter density can be extracted from data
on gamma ray bursters (Ghirland et a1. 2005), the 2dF survey of galaxy redshifts
(Eke et al. 2004), and the Sloan Digital Sky Survey (Cole et al. 2005, actually
a full, standard set of parameters).

The universe truly needs something like a cosmological constant, even if
flatness ($k = 0$) is not a prior (Mitchell et al. 2005, on lensed QSOs and
radio sources). And all current data are perfectly happy with $w = 1$ in the
pressure-density relation, $P = -w\rho$ (Rapetti et al. 2005) despite all the
imaginative variants in the previous section.

The age of a consensus universe is close to 13.7 Gyr, and the oldest galaxies
and stars ought to be at least a bit younger, since a source with age equal the
total should have $z = \infty$. We caught no disagreements on the actual number,
but only on whether numbers arising from particular methods are meaningful.
Dauphas (2005) says ``yes'' for the $14.5^{+2.8}_{-2.2}$ Gyr for the Milky Way
from U/Th in meteor1tes and old stars. The age comes
from counting backwards til you reach the U/Th production ratio, and we wonder
whether the conclusion remains true if the production ratio, said to be 0.571,
varies with place or time because of different neutron exposures in the
 r-process. In this context, Christlieb et al. (2004) note a subclass of
r-process that can make a range of Th abundances.

On the other side, Yushchenko et al. (2005) focus on the range of barium to
mercury abundances coming out of r-process sites, conclude that the ratio of Pb
peak to actinides will vary, so you can't do it that way. Within the thin disk,
there do not seem to be these large variations in Ba-Hg abundances, and so the
$8.3 \pm 1.3$ Gyr value found by del Peloso et al. (2005) applies at least to
the stars in which Th/Eu was measured.
They are about as old as the oldest disk white dwarfs.

\subsection{More Diffuse Stuff}
The two numbers for which we found the largest variances were the bias parameter
$b$ (the extent to which luminous matter is more, or less, clustered than the
DM) and the mass normalization $\sigma_8$. It has become clear in recent years
that bias simply does depend on galaxy type (Wild et al. 2005,
Conway et al. 2005, larger for early types), galaxy mass
(Seljak \& Warren 2004, big for big galaxies and clusters), length
scale (Myers et al. 2005), redshift (Croom et al. 2005, Le Fevre et al. 2005b,
larger at large $z$), and galaxy luminosity (Zehavi et al. 2005). Worse,
Eisenstein et al. (2005) and Ouchi et al. (2005) conclude that the variables are
 not separable, luminosity and length scale getting mixed up in the first
 analysis and topology and density in the second.

The sad implication is that serious models of large scale structure and
streaming  will be expected to match all these trends as well as the average
value of 1.3 or  whatever you would like it to be.

We wish we could say something equally rational about $\sigma_8$, although it is
supposed to be the rms value of the excess of total matter over the cosmic
average on a comoving length scale 8 $h^{-1}$ Mpc. Thus it could reasonably be
an (increasing)
function of time, but should not depend on the types of objects used as tracers
or the volumes occupied by  those objects, as long as they are large compared to
8 $\mathrm{Mpc^3}$. But, in fact, the values reported during the year did not by
any means all fall within each others' error bars, ranging from 0.6
(Blanchard \& Douspis 2005,
using X-ray clusters) and other small values (Weinberg et al. 2004, otherwise a
standard parameter set) up to 1.11 (Percival et al.
2004, from consideration of the 2dF redshift survey). Blanchard \& Douspis
suggest that baryons may be greatly depleted in the cores of rich clusters,
which are then more than half dark matter. The median value is not terribly
helpful. It is 0.85, but the published numbers cluster in two clumps, one at
0.6--0.7 and one at 0.9--1.1, and a number that nobody found from any data
sample is a funny choice for the ``right'' answer. It may, however, be a
reasonable choice if you merely need to calculate something. Well perhaps we
will have an answer for you, or at least a better posed question in Ap06.

The ``I wish I had thought of that'' green with jealousy star goes this year to
Menzies \& Mathews (2005) for cosmological aberration.
Our earth-based velocity relative to the microwave background is about 370
$\mathrm{km\ s^{-1}}$, more than 10 times the speed around the sun
that James Bradley (looking for heliocentric parallax) found in 1729 (We
remember it well.), and the shift in apparent position is therefore
$0.07^\circ$ or 254 arcsec. This will not, for the most part shift back and
forth every six months (only the Bradley part, as it were),
 but it can matter in some studies of very large scale structure from surveys,
weak   lensing, non-Gaussianity of the CMB, and so forth, and perhaps the
Sunyaev-Zeldovich effect (Chluba et al. 2005). The authors provide a correction
formula and a table of   corrected positions for the 50 most aberrated objects
at $z> 1$. Neither Zwicky nor an   earlier ApXX candidate for the second most
aberrated astronomer appears in the table.

WE INTERRUPT THIS UNIVERSE TO BRING YOU THE EARTH

\subsection{Ptolemaic Cosmology}
 Here, right in the middle of the universe, is the Earth.
Why? Well, we don't want adherents of older models\footnote{The most adherent
author drives a 1980 Toyota and has a deep understanding of what it means to be
an older model.} to feel that we don't give their ideas a fair hearing.
Discussion of the Earth obviously ought to focus on it as a prototype for
planets in general, so the core and mantle come first. The rest of the Solar
System lives in \S 3 and exoplanets in \S6.

\subsubsection{Inside Out}
The inner (solid) core rotates a smidge faster than the mantle
and crust, by $0.3^\circ$--$0.5^\circ$ $\mathrm{yr^{-1}}$, report
Zhang et al (2005), following analysis of the
speeds of earthquake waves versus the direction they are going. That the
rotation is slowing was known to Darwin (George) and comes, among other sources,
from reports of ancient eclipses. The data (Tanikawa \& Soma 2004) are not
 precise enough to  distinguish central from surface rotation. The dynamo that
 makes us so magnetic is   reputed to live in the core, so this must be the
 proper place to record (a) that   the late Triassic field was a geocentral
axial dipole (Kent \& Tauxe 2005, an analysis   of paleolatitudes), (b) that the
last major pole flip occupied the time from   795 to 776 kyr ago (rather slower
than sometimes stated), and (c) that the flips   have been modeled (again) by
Takahashi et al. (2005) with a 5200 yr transition   period, over which there can
be briefly two N or two S poles. They note that we are, given the average time
between pole reversals, badly overdue for the next one, and that the average
surface field strength has declined about 10\% in the last 170 years. The world,
or at least the current polarity episode, is perhaps coming to an end.

Radiogenic terrestrial neutrinos come from inner and outer cores and the mantle.
In a most impressive analysis of antineutrinos ($\overline{\nu}_\mathrm{e}$)
captured by protons in the  KamLAND detector, somewhere between 4.5 and 54.2 of
them have been recorded, where  19 were expected (Araki et al. 2005).

At the core-mantle boundary, you will find a great deal of ``geography,'' though
we think they mean topography---high and low points, rather than features with
names (Post et al. 2005). There is, for instance, a dense, partially molten blob
under Australia.\footnote{In principle, this must have contributed to attracting
the 2003 General Assembly of the International Astronomical Union to Sydney,
though probably less so than the quality of the wine and the charm of the
wombats.} Differentiation of the mantle from the core occurred about 4.53 Gyr
ago, only 30 Myr after formation (Boyet \& Carlson 2005).

Every year, somebody tells us whether the convection in the Earth's mantle
occurs in one zone or two, and so we pass on the information that indeed it
occurs in one zone or two, though not with confidence which. A vote this year
for one from Class \& Goldstein (2005) and White (2005). The issue they address
is whether there is a  need for an undegassed primordial reservoir of
$\mathrm{{}^3He}$. They say not.

Crust formation began 4.35 Gyr ago (Watson \& Harrison 2005) and it must be an
ongoing process, because every other solid Earth paper indexed this year dealt
with cratering
and other processes that destroy crust. Some of them addressed---
\begin{itemize}
\item Possible periodicities,
yes according to Yabushita (2004), though with a 10\% probability of being a
chance result, for cratering episodes.
\item Possible periodicities in the behavior of the fault
that will eventually dump two of your three authors into the Pacific Ocean
(Weldon et al. 2005), though not, you will be sorry to hear, soon enough to stop
the publication of this paper. The next epoch of major stuff is due in 2051, or
perhaps 2000 if the period is getting shorter. Oops. That's now.
\item Erosion at 24 meters $\mathrm{Myr^{-1}}$ for the past 500 Myr, some sort
of average over the land
surface (Wilkinson 2005). If the Earth's surface is 30\% land, this actually
comes quite close to the 3 $\mathrm{km^3\ yr^{-1}}$ of fresh crust material
produced at divergent boundaries that the  rockiest author learned about the
years she taught geophysics.
\end{itemize}

\subsubsection{Air}
 The atmosphere appeared in 56 fiscal year papers, roughly half
pertaining one way or another to climate and climate change. Some of the
following paragraphs are more astronomical than others, so feel free
to read them in random order.

Gamma ray flashes from the upper atmosphere were a 1994 discovery
(Fishman et al. 2004) and were predicted by C.T.R. Wilson (1925) of the
cloud chamber. RHESSI has been seeing them all along, in numbers that
imply rates of something like 50 per day (Smith et al.
2005b). But the global lightning flash rate is, they say, $44\pm5$ per second,
so 50 per day (or even 5000 if the beaming correction is like that
for GRBs) isn't all that many.
The total power is about 40 MW/flash (of typical 0.5 msec duration).
Photons extend up to 20 MeV, corresponding to the potential drop from
cloud tops to ionosphere.

The Van Allen belts have been putting themselves back together after a
nasty accident with a giant solar flare on 1 November 2004 (Horne et al. 2005).
For details of the damage see Surridge (2004, ionosphere) and.
Baker et al. (2004, Van Allen belts). Similar processes must be important for
the radiation belts of Jupiter and Saturn. A nearby GRB would also be hard on
our atmosphere (Thomas et al. 2005).

Co-listed under ``oops'' is a second observation of the wavelength dependence
of the Earth's albedo as measured from earthshine. It does not show the red
chlorophyll edge at 7000--7400 \AA\ that was a highlight of Ap04
(Montanes-Rodriguez et al. 2005). Both could, however, be right, since this
year's data were recorded when the center of the sunlit hemisphere was in
mid-Atlantic, a seriously deforested area.

Not everything that happens on Earth is our fault (``anthropogenic'' is the
polite term). There are solar cycles, with which some aspects of the Indian
monsoons show correlations (Hiremath \& Mandi 2004, Wang et al. 2005a).
The major, long icy episodes on Gyr scales may be causally correlated
with epochs of nearby star formation (de la Fuentes Marcos \& de la
Fuentes Marcos 2005). Something happening in the interstellar medium was
also blamed for ice ages by Yeghikyan \& Fahr (2004), in particular spiral
arm passages (Gies \& Helsel 2005) for a 100 Myr time scale. Coming down to
timescales of $10^{4-5-6}$ yr, there have been seven glacial cycles during the
late Pleistocene (last 700,000 yr), for which there are more than 30 models
(Huybers \& Wunsch 2005). One of them (Haug et al. 2005) was indexed under
``Milankovich lives,'' meaning primacy of changes in Earth's orbital
parameters.   In addition, cycles, or anyhow variability, in the solar flux that
reaches the   surface, is probably responding to anthropogenic aerosols
(Pinker et al. 2005,    Wild 2005). The current broadband albedo, in case you
should need it, is 0.29   (Wielicki et al. 2005).

There was a very early $\mathrm{H_2}$-rich atmosphere according to
Tian et al. (2005), whose eventual replacement by an oxygen-rich one might be
described as floragenic, but was not by Falkowski et al. (2005). They provide
numbers for the Jurassic and Tertiary (thereby falsifying a claim we saw
somewhere else that no one uses ``Tertiary'' any more). There was also the
Permian catastrophe, one of the 30+ hypotheses for which was the flora not doing
their duty, so that atmospheric $\mathrm{O_2}$ dropped (Huey \& Ward 2005). One
of the implications is that organisms of similar sensitivity can live only below
an altitude of 500m. As many of the Permianently extinguished were ocean
dwellers, we are not quite sure how to interpret this, except by noting that at
least one of the authors can do integrals only below
about 6500 feet.\footnote{And, in response to the two most obvious queries,
yes, she can do integrals at sea level, and no, it is not likely that this
particular skill was an essential one for brachiopods. (The junior, but oldest,
author lives at the ragged integral edge of 6400 feet.)}

And, in as due solemnity as is possible to the author who generates the most
$\mathrm{CO_2}$, nearly all the rest is global warming, changes it is likely to
wreck, and human responsibility therefor (Means \& Wentz 2005, correcting an
earlier error in calibration of satellite microwave data for the troposphere).
Interesting tracers include dates of the French grape harvest
(Chuine et a1. 2004, reporting earlier warm periods but 2003 unprecedented);
sediments and tree rings (Moberg 2005, with 1000-1100 CE like the 20th century
before 1990, but ``now'' out of statistics); glacier lengths (Oerlemans 2005,
with coherent warming over the globe since 1850, probed from more than a dozen
sites, one of which we had to look up; Jay Mayen is an island off Greenland).
See also Schar \& Jendritzky (2004) and Stott et a1. (2004), with considerations
 of how much is anthropogenic.

Another aspect of the atmosphere obviously relevant to astronomy is the quality
of various observing sites. A grim thought from a UCLA colleague is that, now
more than ever, you are likely to choose a site that is better during your test
years than it will ever be again, because of climate change. A dozen papers
mentioned a comparable number of sites. The firmest statement came from
Sage (2004), saying that Dome C (Antarctica) is the best spot on earth for a new
telescope. It is also praised by Walden et al. (2005) and
Aristidis et a1. (2005), and is compared with others in a variety of ways by
Racine (2005). He makes the point that surface
turbulence is responsible for much of seeing and can be corrected by adaptive
optics over a larger field of view than higher turbulence. Indeed an
out-of-period  paper indicated that a very large fraction of Dome C seeing
happens  below 35 m, and you could build your dome on a platform thicker
than that.\footnote{Presumably in consultation with the folks who designed
Mickey Rooney's shoes.}

The importance of high altitude winds is stressed by Carrasco et al (2005) and
by Garcia-Lorenzo et al. (2005), who are enthusiastic about La Palma, where the
wine is much better than at Come D, sorry Dome C. American Antarctica is dry not
just in the precipitational sense.

\subsubsection{Air Breathers}
 The non-human biosphere appeared in 33 indexed papers. As
long as we are busy blaming ourselves, recall that earlier generations of humans
did the continents they entered (or anyhow the resident megafauna) no good
(Miller et al. 2005 on Australia).
The current situation is that more species become extinct while waiting to be
listed as ``endangered'' than do afterwards (Fish and Wildlife Service, 2005).
The most artificially augmented author is reminded of a recent series of events
at her home institution, but dare not be more specific or organ-ized.

The evolution of the horse has been declared to be considerably more complex
than the Eohippus-Mesohippus-Equus sequence many of us learned as children. But
the most obvious change (MacFadden 2005) is that Eohippus is now (or rather once
again) Hyracothere.

Your turtle paper for the year concerns the death of Aeschylus (Nisbet 2005).
Funk et al (2005) deny that clipping off their toes is terribly bad for frogs,
though if strict grammatical rules were enforced, so that ``their'' referred
back to Funk et al., we bet they would change their minds.
Bradshaw et al. (2005) have confirmed that elephants don't forget (and we wonder
how frogs feel about people who clip their toes).

Platypi have 10 distinct sex chromosomes (Grutzner et al. 2004). We understand,
we think, that only two sexes are involved but have not essayed
the experiment.\footnote{Chickens??? goes the punchline of a joke we are
absolutely sure the editor will not let us get away with.}
A dead whale can support an exceedingly complex ecology of truly revolting
creatures for about a century (Smith 2005), equivalent, we estimate, to
supporting 100 humans for 10 years, if only calories count. But somebody has to
be willing to eat the bones as well as the wings. Whales don't have wings, you
say? Well, neither do pigs, though they were domesticated many more times and
places (Larson et al. 2005) than chickens.

Some early mammals ate some of the late dinosaurs, report Hu et al. (2005) on
the basis of a fossil of the former with a fossil of the latter inside. Long
ago, mammals eating dinosaur eggs was one of the 30+ hypotheses for the
extinction of the latter. But they didn't eat them all, because
Sato et al. (2005) report finding some fossil dinosaur eggs in pairs, implying
that the oviducts existed in pairs, as they do in modern birds. Chickens???

A few survival issues appeared during the year. ``Reagan lives,'' meaning that
trees really do produce some atmospheric pollutants (Purves et al. 2004; their
arboreal methane  production is out of period). The flora and fauna of Mount
St. Helens are gradually  recovering (Dale et al. 2005). Leeches can live 20
years according to Dobos (2004), who describes them as ``hardworking animals''
deserving of eventual retirement. Armadillos didn't used to get leprosy, until
people started spreading it all over the place (Monot, Honore et al. 2005).

The gerenuk (photo, Science 308, 1040) is the living original of a widely
reproduced gold Ur-statue of a ram, goat, or deer standing on hind legs,
with front feet and horns entangled in some sort of thorn bush. The best
known of these is said to have been among the irreplacible archeological
artifacts lost this most recent time war came to Ur.

The no-man's \ldots no-hominim's \ldots no-hominid's \ldots land between human
and non-human is occupied by the longstanding and definitely unresolved
questions of how many species and subspecies of Homo co-existed at various times
and places and how much, if any, interbreeding there was. The green wow-dot of
the year was, obviously, the meter-high (kneehigh to a giraffe?) Homo floriensis
(Brown et al. 2004, Morwood et al. 2004, Mirazon-Lahr \& Foley 2004), and no,
we are not  voting on the possibility of dwarfism or something else other than
 separate descent from Homo erectus under island pressure for small stature.

The 10 Homo species that appeared within a million years are briefly
presented by Carroll (2005), as part of a piece on adaptive radiation
of lobe-finned fishes. There were 11 of those, one of which eventually
gave rise to land animals (including us). Only one Homo has survived as
well, and we shudder to think what it is likely to give rise to.
We caught no opinions this year on the amount of genetic mixing with
Homo sapiens neanderthalis, but, with a typical male BMI of 28.7, and a
diet with about 30\% more calories than the Inuit, he would have fit right
into modern America (Churchill 2005).

\subsubsection{Hot Air Breathers}
 We recorded 73 papers about humans collectively or
individually, while noting on the fringes that the wizard gene is recessive
to Muggle, though arguably with incomplete penetrance (Craig et al. 2005).
Young Mr. Potter and Cinderella both appeared in the index year literature
as examples of ``reduced parental investment when parent-child relatedness
is low'' (Raymond 2005). We green-dotted the unpleasant fact that foreign-born
postdocs are typically paid rather less for more work than ones whose
relatedness to the PI is likely to be higher (Davis et al. 2005). As for the
saga of Larry Summers (2004), in response to his remarks about Jewish farmers
and white basketball players, the white female Jewish author, who once grew
enough wheat to make one loaf of bread and was the free-throw champion of her
junior high school,\footnote{Lest you doubt these matters, let it be confessed
that everybody in Mrs. Miller's B3 class was required to grow wheat; VT was the
tallest female in her 7th grade class; and something like 1/3 of both groups
were Jewish.} gives him fair warning, based on the recent history of
 women being called to the Torah in reform synagogues, that, if
you once let us in on equal terms, we will quite probably take over!

More computing power is already needed for climate modeling, says Palmer (2005).
This will become even more true if Benford (2005) is correct that the
perfect mate for most women will eventually be a very intelligent robot.
We have not asked him about the state of his own relationships, though his
office is only a few doors down the hall.
A few of you may even remember when computers were human, and very
typically female (Agar 2005).

Science has nothing on the real world in some of these respects. An
advertizement in Time Magazine (21 March 2005, pp. 44--45) shows 26 outstanding
 Ford dealers. These include two women, one black, one Asian-American, and
 22 white males. Oh well, and while we are at it, Shapiro (2005), reviewing
 the \textsl{Oxford Dictionary of Science Quotations}, asks why Simone de
Beauvoir's remark, ``to be a women, if not a defect, is at least a
peculiarity,'' is relevant to science.  Where has he been? In case you had any
doubts, gender and social class both matter  for (or anyhow are statistically
correlated with) Ph.D. receipt and subsequent careers
(Bornmann \& Enders 2004). The X and Y chromosomes arose from autosomes, and the
X has more than its fair share of ``disease genes,'' meaning that common
mutations of them are bad for you (Ross et al. 2005).

Want to get your papers published?
Well, really, why else would you write them? So, when you are submitting, it
pays to make both positive (``X, Y, and Z would be suitable reviewers.'') and
negative (``P, Q, and R might not be entirely fair to this paper.'' )
recommendations (Haynes et al. 2005). But even very high skills at this are
unlikely to enable you to beat the record held by Ernst Mayr (Diamond 2005,
an obituary), for the largest number of books published past age 90. Mayr
himself also credited sitzfleisch. But it is worth noting that the astronomical
pattern of publication (a flood, from thesis and postdoc years; another as
tenure decision approaches) is not universal (Upadhye et al. 2004). Indeed
middle-aged persons also do neat science (Wray 2004).

Making significant money from popular science writing is harder than it used to
be, says Barker (2005). The obvious individual to mention in this context is
Carl Sagan, whose non-election to the US National Academy of Sciences is still
being discussed.  The new president says (Cicerone 2005) that ``there were good
people on both sides of the debate.'' Some of them apparently did nothing, this
being the requirement for the bad people to win, and if you remember the quote
as ``for good men to do nothing,''  you may well have put your finger on part
of the problem.

The only higher honors in science than various Academy memberships are postage
stamps (Feynman of the diagrams, Gibbs of the free energy, von Neumann of
multiple theorems, and McClintock of jumping genes this year) and Nobel Prizes.
Karazija \& Momkauskaite (2004) provide some statistics on the physics winners.
It pays to be a theorist and also to live a long time. The average age of the
laureates and the interval between the  critical work and the prize have both
increased with time. Given that retirement often means no more research grants,
this may not be as dreadful as it sounds. In any case, the records are held by
Ernest Ruska (1986 for work done 53 years earlier) and
S.~Chandrasekhar (1983 for work done 49-53 years earlier). And in case you might
 have forgotten, TNT was  invented by Bernard J. Fluorscheim not Alfred Nobel
(Nature 436, 477, reproducing an item from the 30 July 1955 issue).

An earlier ApXX noted with envy that armadillos sleep about 19 hours a day, much
of it spent dreaming. This year, Roenneberg et al. (2004) have investigated the
mid-point of normal sleep cycles as a function of age, casting, it is widely
supposed, some dawn's early-light on why high school students find it difficult
to get going in time for 8:00 AM classes. The sleepiest author extrapolated the
graphs and concluded that her 1:30 AM midpoint is appropriate for age near 102.
Given that the constellation called by Hooke ``the English rose'' consists of
stars from $m = 6.3$ to 7.8 (Beech 2004), he must have managed to stay awake
through some very dark hours. Twitchiness will help you stay slim
(Ravussin 2005), but we recommend that this be confined to daylight
hours if you share sleeping accommodations .

Levels of literacy: Farmer \& Meadow (2004) opine that the Harrappan script was
not a real written language, but only a record of tribal names and such.
H.G.~Wells was about as literate as folks get, and his version of the future
became less dystopian as he aged
(Nature 436, 785, reprinting an item from the 10 April 1905 issue), quite the
opposite of most of our friends and relations. And if you need to fold a
newspaper for delivery, Petroski (2005, surely the most literate engineer whom
we have never had the pleasure of meeting) is not only an expert but remembers
before rubber bands and plastic bags that tell you not to put them over your
children's head.\footnote{No, the rubber bands do not usually carry text
advising you not to put  them over your children's heads, but only because there
isn't space, and we suspect that the designer of the California raisin label
will soon find a solution to that.} There were many styles, appropriate to
different thicknesses of paper. It cannot be quite true that every kid of that
generation had a paper route, but these days the normal adult American doesn't
even read a paper, let alone have a kid who delivers them.

The World Year of Physics (2005, calendar, not fiscal) will have come to an end
by the time you read this. For some of the highlights, see Stachel (2005, and
eight following papers) and Bennett (2005 and several surrounding papers).
But according to a spring 2005 appeal for donations from AAAS, Einstein doesn't
rank at all, and the donor levels are Da Vinci (low) through Copernicus,
Galileo, Franklin, Edison, and Director (high). And Director hasn't even
appeared on a postage stamp.

 The reference to the Hirsch number (for Jorge at UC San Diego) seems to have
disappeared. It is a way of ranking one's colleagues and such, a tad difficult
to explain,  but here goes. Suppose you have written umpteen papers, some very
frequently cited, some only rarely. Your Hirsch number is then the largest $N$
for which it can be  said that $N$ of your
papers have been cited at least $N$ times. Twelve at tenure decision was said
to be adequate for some disciplines. Impossible not to be reminded of an
Eddingtonian number $N'$, the largest number of days, $N'$, on which you have
cycled at least $N'$ miles. We suspect that 120 might be a pretty high number
for someone at tenure decision (and unlikely to get much larger in later years).
Next to Cambridge, Eddington's home base, is Oxford, whose student-faculty ratio
has deteriorated in recent years from 9:5 to 12:2 (Collier 2005).
This ratio hovers around 20:1 for most University of California campuses, except
UC Merced in spring 2005, where there were already some faculty but no students
and, therefore, that most desirable ratio, zero.

This last could count as either a UC item or as ``yes, but are they on postage
stamps?'' item. ``The new (2004) Laureates' names will rank right up there with
Newton and Coulomb,'' said UC Santa Barbara Chancellor Henry T. Young (2004).
Well, Newton and Coulomb never won Nobel Prizes either, and we think we would
probably have picked Maxwell rather than Coulomb to represent electromagnetism.

\subsubsection{Friends in the News}
Ene Ergma is described as ``an astrophysicist who is now the leading politician
in the Estonian parliament'' (Villems et al. 2004).
Not only that; she still comes to astronomical conferences. Angela Merkel,
new leading politician in Germany, holds a Ph.D. in quantum chemistry
(Science 309, 1471). We don't claim friendship with her, but only with the
quantum, upon long ago advice (``just be happy with the quantum,'' meant
ironically) from Richard Feynman, on whose diagrams see the book review
 by Kane (2005).

Wayne Rosing, director of Las Cumbres Observatory, has joined the LSST group
at UC Davis (Tyson 2005). Michel Mayor and Geoffrey Marcy have shared the
million dollar Shaw Prize (Science 308, 1739). And Vitaly L. Ginzburg (2004)
has invited us all to become familiar with 30 problems that every physicist
should know something about, (not, typically, including the solution, since
controlled fusion and string theory are among them). He adds that
``posthumous recognition is not all that important to me, because I am an
atheist.'' Carlo Rubbia has been removed as head of the Italian agency ENEA
(Science 309, 542) for reasons other than purely
scientific. That same page and the next one record something comparably
unfortunate happening to the former chair of the Committee on Science and
Technology in the British Parliament and the possibility that another Brit,
whose inaccurate testimony put a mother who had lost two babies to SIDS in jail
for 3 years might himself be slightly punished (at the year in jail level, not
at the slaughter of the firstborn level, though if we had been on the
jury\ldots). Other arguably disproportionate punishment attended the case of
gene therapist J.~Wilson (Anonymous 2005f), compared to that, say, typically
imposed on a driver whose vehicle kills someone while he is supposedly in
control of it. Probably the less said about J.~Robert Sehrieffer (who, unlike a
couple of those intermediate
people, really is an acquaintance for many of us) the better.

A method due to Sherlock Holmes (who is not cited) is noted by Muscarello \& Dak
(2004) as valuable. ``Many more crimes might be solved if detectives were
able to compare the records for cases with all the files on past crimes,''
they say. And a result claimed long ago by Thomas Gold (also not cited) that
``all positive quantities are correlated'' has been confirmed for lQ and health,
even after correction for socioeconomic status (Deary et al. 2005).
Padmanabhan (2005) quotes Raman\footnote{Who received his Physics Nobel Prize
the same year, 1930, that his nephew
began the work on degenerate stars that would result in his 1983 Prize.}
as having declared that there would be no astrophysics within miles of
Bangalore. No, we didn't actually
meet Raman, but we do know his son, the astrophysicist of Bangalore.

Money. Now that we have your attention, the states that get the most federal
science dollars per capita are Massachusetts and Maryland, with the least
going to South Dakota and West Virginia (Anonymous 2005g). Federal dollars
per scientist in the various states might be more revealing, but the numbers
were not provided. Zebrowitz \& Montepare (2005) have found a signature for
hard times: the popular actresses have more mature faces than those belonging
to more prosperous decades. That more prosperous countries do more science and
more science per capita remains a fact of life (e.g. Nature 436, 495). But if
you look for actresses with more mature faces in poorer countries, we think
you will be disappointed. Our own personal sample is limited to
 Hollywood/Broadway actresses who were fellow graduates of Hollywood High School
(Swoosie Kurtz, Linda Evans, Stephanie Powers), all of whom now have mature
faces, owing to membership in the class of 1960.

There is already a whole generation of astronomers who were not there when
SN 1987A went off (Gaensler 2005), let alone, we suppose, able to remember
Jack Benny in \textsl{The Horn Blows at Midnight}, the Kennedy assassination, or
Pearl Harbor. Remembering the Maine, Plymouth Rock, and the Golden Rule is
now reserved for cast members of \textsl{The Music Man}.

Astronomers are a bit better than other
species of physical scientists about participating in outreach activities
(Jensen 2005). There is even an lAU Working Group on the topic. And some other
things being accomplished with a little help from our friends:
The US now leases icebreakers as well as rocket launches from Russia (Erg 2005).
Nepalese (Sherpa) porters really can carry heavier loads than the rest of us,
 more efficiently (Bastien et al. 2005). It seems sadly probable that humans
 are included in the statement ``Many animals may spend most of their time at
 or above the carrying capacity of their ecosystems'' (Science 309, 609).
 Are the depressed more likely to walk into door frames, have their computers
 both fail and fall on their toes, and so forth, or is it that, as
 Smiley (2005) says, ``\ldots proneness to report minor injury can be added to
 the list of other known signs of emotional distress.'' We are currently
 seeking a suitable pseudonym for someone who appears to demonstrate this
 syndrome in spades. No one was ever tangled in so many traffic jams,
 airport delays, sexual harassment cases, and assorted violence and
mayhem. He appears here as the Medical Musician on a pro tem basis.

Progress comes from dissatisfaction (Nettle, 2004, a book review, and
G.B. Shaw said from irrational people) but the scales of stellar magnitude
used by Ptolemy, Al Sufi, Ulugh Beg, Tycho, and some early telescope builders
(Beyer, Flamsteed) were actually quite close to the official semi-modern Pogson
scale (Fujiwara \& Yamaoka 2005).
And we rather like the units of time in which the day was divided into 100 ke
(Soma et al. 2004). You will have had an extra 0.001 ke for your activities on
31 December 2005, owing to the leap second. We hope you made good use of it.

And a comparable number of items have fallen off the edge of the typewriter,
some deserving cheers (to ICTP Trieste for its hospitality to East African
Ph.D.'s and to ArXiv for holding the line on who can post\footnote{Your least
posted author hastens to report that she is not in the privileged group.}),
some  wows (for neat new information on tracing history of languages and on
primates as sprinters and endurance runners),
and other expressions of earthly engagement.

WE RETURN YOU TO OUR STUDIOS WHERE THE UNIVERSE IS IN PROGRESS

\subsection{The Backgrounds}
 The 18 CMB papers recorded in academic year 2005 failed to
outnumber the 23 pertaining to all the other backgrounds, but have nevertheless
been forced into a separate section while we start with the shortest
wavelengths.

The gamma ray background can nicely be accounted for as the sum of blazars, once
the Milky Way contribution has been removed (Strong et al. 2004). The Milky Way
part itself can be modeled with proton and electron spectra much like the local
ones, and the $\gamma$'s derived from $\pi^0$ decay, inverse Compton scattering,
and Bremsstrahlung. Dark matter
decay or annihilation also raised its candidential head
(Elsasser \& Mannheim 2005), requiring the DM particle mass to fall near 515
GeV, not otherwise a popular choice.

The X-ray background is either slightly older than the CMB, if you think of its
discovery in 1962, or slight younger, if you think of it as the sum of emission
by active galaxies at moderate redshift. A double handful of papers considered
the situation, and we cite only one that expresses content with that model
(Civano et al. 2005, because they have finally found a sample of sources harder
than the background at 2--8 keV) and one that expresses discontent
(Mainieri et al. 2005, though with abiding faith that the missing sources will
eventually be resolved, if not in the Chandra era, then by some later, greater
collecting area).

At 90--265 \AA\ there are only limits to be had, and less than you would expect
just from our local bubble of ionized interstellar gas (Hurwitz et al. 2005).
Your choice whether
these photons should be called soft X-rays or hard UV.

Most of the far UV we see comes from the Milky Way and is quite patchy
(Murthy \& Sahnow 2004), though obviously there must be an extragalactic sea
as well, unless all galaxies are perfectly opaque shortward of, say,
912 \AA. No,
say Shimasaku et al. (2005) who find that lots of photons get out of galaxies,
with the UV background brightest near $z = 3$ and declining on either side.

The less harsh UV is typically called ionizing radiation, and the main
disagreement over the past decade or so has been the relative contributions from
QSOs and from star-forming galaxies, and how the ratio varies with redshift.
Some of each at $z = 2$--4 said Bolton et al. (2005), not quite the conventional
answer.

The optical background has been close to $10^8$ $L_\odot$ $\mathrm{Mpc^{-3}}$
since Oort estimated it \textsl{many}
years ago. There are two ways to attempt the measurement: add up all the
galaxies you can see and divide by the volume they occupy; or get above the
atmosphere and attempt to look between the galaxies. Some folks did each during
the past year. Driver et al. (2005) added up galaxies and hit very close to
the Oort value, at $1.99\pm 0.17\times 10^8 h$ $L_\odot$ $\mathrm{Mpc^{-3}}$
in a band they call $b$. So did
Baldry et al. (2005) using a different galaxy survey and the AB band, but
reporting in different units so that $\rho_L$ is $10^{19}$
$\mathrm{W\ Hz^{-1}\ Mpc^{-3}}$. Minowa et al. (2005)
compare the results of the two methods, and conclude that the sum of Subaru
galaxies (that is, a third survey) is, at 9.43 $\mathrm{nW\ m^{-2}\ sr^{-1}}$
only about half of what they see peeking between galaxies. It is left as an
exercise for the student (Not you, Mr. H: back to that thesis) to show that the
numbers reported in these three different ways are at least approximately
consistent. We gave up somewhere around the time we realized that you can see
only the steradians above the horizon. That Oort came reasonably close says,
among other things, that, while little galaxies greatly outnumber big ones
(\S4.4), most of the light comes from the big ones.

The 1--20 $\mu$m background has a QSO component, though a modest one, say
Silva et al. (2004) and Franceschini et al. (2005). Galaxies dominate and
include a new class much brighter in the 15 $\mu$m ISO band than at 2.2
$\mu$m (Johansson et al. 2004). The confirmation from looking between the
galaxies has proven over the years quite  difficult, but
Matsumoto et al. (2005) say that they and the IRTS (Infrared Telescope
in Space) have done it, seen more IR than they expected from the redshifted
UV of Population III stars, and conclude that Pop III star formation ended at
$z = 9$ (which acquired a green dot in the process of being redshifted from
UV to IR).

\subsection{The 3K Microwave Relict Microwave Cosmic Background}
Now about the CMB, or 3K or relict radiation, depending on our mood. You have
the choice of an ``all is well''
school, e.g. Eisenstein et al. (2005a) on detection of correlation between large
scale structure in the 2dF and SDSS galaxy surveys and in the CMB,
Barkats et al. (2005) on the $\theta = 4'$ polarization structure seen from
Crawford Hill, and Readhead et al. (2004), somewhat similar results from
the Cosmic Background Imager,
or, if you prefer, a ``problems remain'' school. Items on that slate include (a)
failure to detect gravitational lensing of the radiation by groups and clusters
 of galaxies (Lieu \& Mittaz 2005), though the effect, called magnification
 dispersion, is seen for distant QSOs, (b) evidence for non-trivial topology
 from Aurich et al. (2005), and (c) evidence for non-Gaussian distribution of
  amplitudes (Land \& Magueijo 2005a, McEwen et al. 2005 ).

Here are three items that are meant mostly for fun. First,
 Amendola \& Finelli (2005) note that the spectrum of primordial fluctuations
must have included decaying modes as well as growing ones. While these won't
have contributed to present large scale structure (\S 5.10), they might still be
detectable at less than 10\% of their strength when recombination occurred.
Second, Perjes et al. (2005) opine that the CMB fluctuations on the sky
``underwent reversals approximately 2 Gyr ago,'' so that we now see a negative
image of the last scattering surface. We suspect that, given the very extensive
analyses of COBE and WMAP data and everything in between, that this must be
(a) well known to everybody except us, (b) trivial, or (c) not true.
And third yet another source of spectral distortion of the black body radiation,
 the ``gradient-temperature Sunyaev-Zeldovich,'' which would measure the
electron  conductivity of the gas in X-ray clusters (Hattori \& Okabe 2005).
Incidentally,  most of the S-Z distortions that are seen can be tied to clusters
already known  or, in the case of the zone of avoidance, suspected
(Hernandez-Monteagudo et a1. 2004).

And, finally, the aspects of the observed 3K radiation that most of us find at
least a little worrisome. First is the north-south symmetry, which has been more
fully described, but not understood (Hansen et al. 2004). This is a fluctuation
larger than expected. Second are the fluctuations smaller than expected, the
quadrupole and next couple of umppoles. We caught four viewpoints.
\begin{itemize}
\item A mere statistical fluctuation say O'Dwyer et al. (2004, arising from our
choice of vantage point, well, not a whole lot of choice perhaps available).
\item The small $\ell = 2$ and 3 amplitudes are OK, but the correlation of
azimuthal planes of $\ell = 2$, 3, and 4 is ``uncannny''
(Land \& Maguerio 2005).
\item Meaningful and a potential source of information about the dark energy
equation of state according to Enqvist \& Sloth (2004).
\item More complicated than you thought say Schwarz et al. (2004), because the
quadrupole plane and two of the three octopole planes are perpendicular to the
ecliptic and normally aligned with the dipole and with the equinoxes, while the
third octopole plane is perpendicular to the supergalactic plane (all at more
than 99\% confidence). This is all odder than it sounds, for if the solar system
is a source or sink of some of the radiation, there should be an annual term in
the data beyond the $\pm30$ $\mathrm{km\ sec^{-1}}$ Doppler swing due to our
orbit around the sun, and there is not.
\end{itemize}

\subsection{Large Scale Magnetic Fields}
Is the intergalactic magnetic field a background?
Well, sort of, we suppose. It is, anyhow, going to get no more attention that it
does here, because there doesn't seem to have been much 2005 progress over the
previous few years. The situation remains that you can start with fields made in
small things (pulsars, quasars, gamma ray bursts, or whatever), push them out,
and stir them around. Or you can start with very week seed fields ($10^{-19}$
G is perhaps enough, Takahashi et a1. 2005a) on longer length scales and amplify
them with spiral galaxy rotation (Schekochihin et a1. 2005a), with
Ba1bus-Hawley instabi1ities (Kitchatinov \& Rudiger 2004, building on an idea
from Ve1ikhov 1959, where $10^{-25}$ G might be enough to start with), or
turbulence from supernovae (Ba1sarsa et a1. 2004). And a mechanism we truly do
not understand, even by the standards of the previous
ones, by which Siemieniec-Ozieb1o (2005) gets primordial field emerging on all
length scales simultaneous1y.

Dynamos have been operating in laboratories for more than a century (starting at
Siemens in the 1880s). Schekochihin et a1. (2005b) compare more recent ones to
the dynamos of planets, galaxies, and clusters in a space defined by magnetic
and ordinary Reynolds  and Prandtl numbers.

Did you know how far back the observations can be pushed? Yamazaki et a1 (2005)
report that the mean field on 1 Mpc scales must have been less than 3.9 nG at
$z = 1000$, or we would know about it from, yes, the faithful old CMB. At
redshifts less than that but larger than zero, the fields are less ordered than
they are here and now (Goodlet \& Kaiser 2005, a Faraday rotation result). For a
review of observations since $z = 1000$ and on many 1ength scales,
see Vallee (2004).

\subsection{Very (and Not So Very) Large Scale Structure and Streaming}
A potentially very stringent test of the consensus $\Lambda$CDM cosmology is
its ability to match observations of structure on scales from the largest
superclusters of galaxies  down to the cores of individual galaxies and the
satellites around them,  when theorists start with the initial conditions of a
$\Lambda$CDM universe and evolve them forward in time to $z = 0$.
Our magenta star paper (the green pen was hiding under the newspaper that
day) from the theory side is Springe1 et a1. (2005), reporting the largest ever
calculation of this sort. They begin at $z = 127$ with $(2160)^3$ particles in a
box that is today 684 Mpc on a side (Gnedin 2005). There is postprocessing to
pick up baryon-induced features, and they do, for instance, get enough halos of
$10^{13}$ $M_\odot$ by $t = 850$ Myr to host the high redshift QSOs being found
by SDSS and other surveys, and in place by $z = 5.7$ (Ouchi et a1. 2005). An
envelope back (well, we used a form letter from the NSF) will show you that the
mass per particle must be $1.3\times 10^9$ $M_\odot$, so that the calculation
cannot be expected to resolve objects smaller than that. The largest things are
sheets, filaments, and cores where the sheets and filaments cross.

The starred observational paper is Miller et al. (2004) reporting that the
biggest  structures in the distribution of 2dF QSOs are some 200 $h^{-1}$
comoving Mpc between $z = 0$ and 2.5. Jones et al. (2004) concur that, while it
may get better than this, it doesn't get any bigger. It would be improper to
proceed without allowing Ribeiro (2005) his word.
His word is ``you guys are all wrong,'' and you are failing to find fractal
structure on still larger scales because you have chosen to use the wrong
definition of distance in analyzing the observations. Only the luminosity or
redshift distance is appropriate, he says, not the comoving distance and not
another sort whose source he does not cite.
Angular diameter and parallax-type distances are different yet again, and also
inappropriate in models more complex than Einstein-de Sitter, where several are
degenerate (Mattig 1958). Well, it may be so. A good friend once remarked that
the fact you can't get bread to rise doesn't mean there is no yeast effect.

Balancing back the other way, a few more ``all is well'' papers pertaining to
big things, before we start with small ones and come back up. (1)
Jena et a1. (2005) report that they get a good match to the statistical
properties of the Ly$\,\alpha$ forest (of QSO absorption lines) with the usual
universe and plausible values of ionization
and heating (e.g. Madau  et a1. 1999). (2) Weinberg (2005) concludes that all is
well with galaxy surveys, the red galaxies being in the cores and the blue ones
in filaments. (3) The north/south asymmetry in galaxy distribution is merely a
local hole, underdense by about 25\% and at the upper end of the normal range
of big things (Busswell et al. 2004). Frith et al.
(2005) find that we are within that hole and that it could be as large as
430 Mpc ($z = 0.1$), not easy to get out of $\Lambda$CDM; and then they
partially back off again because the samples don't extend far enough beyond
this distance to be sure of the normalization. (4) It is definitely good news
that various codes for evolving the early universe
down to $z = 0$ more or less agree (O'Shea et a1. 2005, Heitmann et a1. 2005).

\subsubsection{The Smallest of the Large}
The general idea is that standard models predict
more small scale structure than is seen. Two manifestations of the problem are
called core/cusp (meaning predicted central density profiles of galaxies and
clusters are steeper than the observed luminosity distributions) and missing
satellites (meaning less substructure in large halos is seen than predicted).
The pendulum has swung between ``problem'' and ``OK'' several times in previous
ApXX's.

This year we will merely report, first, that the data may not be so unambiguous
as generally advertized (Metcalf 2005 on substructure from gravitational
lensing; Mashchenko et al. 2005 on satellites of the Milky Way as massive and
largely dark, our couple of dozen being seriously outnumbered by the 160
belonging to NGC 5044, Faltenbacher \& Mathews 2005), and, second, that the
models may not be so unambiguous as generally advertized, (a) because  of
insufficient mass resolution (remember those $1.4 \times 10^9$ $M_\odot$
particles in even the most  extensive simulation),
including an explicit statement from Xiao et al. (2004) about the
importance of being resolved for cusps, and (b) because of the enormous
complexity in tying the mass patterns calculated with the light patterns
observed (Gao et al. 2004).

You will have to decide on your own emotional reaction to the following.
For decades there has been a deficiency of satellite galaxies in the planes
of disk primaries (Ap00, \S 7.2), and it was called the Holmberg effect.
Early this year, that distribution was confirmed for the Milky Way's tribe
(Kroupa et al. 2005), and there was an explanation in terms of satellites
falling in along filaments (Benson 2005). But, about the time Christmas
ornaments began appearing in the stores (August), the Holmberg effect was
replaced by an anti-Holmberg effect in the data (Brainerd 2005), and there
were also, as it were, anti-predictional calculations (Knebe et al. 2004,
Zentner 2005) saying that satellites should be found preferentially in the
disk plane.

\subsubsection{Medium}
 Are the shapes or angular momenta of individual galaxies
typically aligned parallel or perpendicular to the enveloping large scale
structure? Yes, say the data (Aryal \&  Sauer 2005, and a handful of additional
index-year papers). And, curiously, both are
predicted, or, anyhow, calculated (Bailin \& Steinmetz 2005).

The Local Group is our own particular medium sized structure. It should be
regarded as consisting of subgroups belonging to the large members
(us, Andromeda, and perhaps M33) says Karachetsev (2005). This reflects the way
it probably formed, from an off-center collision between proto-M31 and a
similar galaxy (Sawa \& Fujimoto 2005). Both subgroups continue to grow, not
so much in cosmic time as in observing time. The addition of AND VIII and AND
IX began \S 10.9 of Ap04. And this year we welcome a Milky Way satellite in
the direction of Ursa Major, at $\mathrm{M_v = -6.75}$ the faintest yet
(Willman et al. 2005b).   Hardly more than a faint overdensity of stars in SDSS,
it probably also has the lowest surface brightness seen to date. A
characteristic radius near 250 pc makes this a real, if feeble, galaxy. The
same group (Willman et al. 2005a) report also a 23 pc sized,
$\mathrm{M_v = -3}$ entity about 45 kpc from us that could be described as a
very faint dSph galaxy,  a diffuse globular cluster, or an intermediate sort of
object.

We indexed 40 some other papers about individual members of the Local Group,
roughly half concerning M31, which, you must certainly be tired of being told,
is rather less like the Milky Way than it was when it (or at least we) were
younger (Hurley-Keller et al. 2004 on the planetary nebulae;
Fusi Pecci et al. 2005 on
globular clusters; Mould et al. 2004 on the history.of star formation;
Williams \& Shafter 2005 on which is larger). This is not what is generally
meant by galactic evolution.

 The SMC, LMC, NGC 6822, And IX, Sculptor,
Fornax, NGC 185 and NGC 147 are all to be left hanging alone around the
church door, except the last two which are bound to each other
(McConnachie et al. 2005) at least since 1998 (van den Berg 1998), while
we elope with M33. Quite remarkably, both its overall proper motion and
its rotation have been seen, using VLBA positions of water masers
(Brunthaler et al. 2005). Its galactocentric transverse velocity is
$190\pm 59$ $\mathrm{km\ s^{-1}}$, and the authors have
derived a distance  near 730 kpc and a mass for M31 of at least
 $1.2\times10^{12}$ $M_\odot$ if M33 is bound to it. We indexed this under
``van Maanen revisited,''  because his (incorrect) reports of spiral rotation in
the plane of the sky before 1920 retarded the recognition of the existence of
external galaxies for decades. His spirals had leading arms (like leading
questions always suspect). M33 today has trailing ones.

If you care to go looking for other tribes like ours,
Karachentsev \& Kasparova (2005) provide advice that only galaxies bigger than
$10^9$ $M_\odot$ can have two or more companions and only those more massive
than $10^{10}$ $M_\odot$ can have more than three.
And, being kinder than we, you will not entirely direct your hunt
toward trying to disprove this.

Shakhbazian compact groups (Tovmassian et al. 2005) have the same mass to
light ratio (37 on average), sparsity of radio sources, and occasional
discordant redshifts as  Hickson compact groups. The latter consist largely
of old galaxies and so must either last a long time or have just formed out
of previously existing units (Mendes de Oliveira et al. 2005). The two classes
appear to differ primarily in discoverer name. Comparable compact groups exist
at slightly larger redshift according to de Carvalho et al. (2005) who, by not
citing Shakhbazian, perhaps hope to have the class named for themselves.

\subsubsection{A Little Bigger}
 As a special treat, we shall refrain from telling you
the ghastly joke of which the punch line is ``The baby is a little Bigger'' and
confess immediately that this subsection deals cursorily with clusters of
galaxies, apart from a topic or two (like cooling flows) which live in \S10.

The least-bound author cherishes a long-standing affection for what are
variously  called intergalactic or intracluster stars and starlight. Such must
actually exist since 10 indexed papers reported properties, and no one claimed
not to be able to find the stuff (see ``yeast effect'' above). It begins to seem
probable that there are at least two types. One of these is traced by planetary
nebulae, whose distribution in the Virgo cluster is clumpy and associated with
large galaxies in both position and velocity space (Arnaboldi et al. 2004,
Feldmeier et al. 2004, Aguerri et al. 2005a). The other is bluer, somewhat more
diffuse, and associated with infalling galaxies
(Willman et al. 2004), disrupted spirals (Adami et al. 2005), and even on-going,
in-situ star formation in gas filaments in the clusters (Crawford et al. 2005).
Clusters as far back in time as $z = 0.25$ already have some intracluster light,
find Zibetti et al. (2005) by stacking SDSS images.

A calculation designed to put 20--40\% of light between galaxies also says that
those liberated stars should have velocity dispersion about half that of the
galaxies in the same location. The calculators (Sommer-Larsen et al. 2005) were
not surprised, so probably we shouldn't be either. Data for stray planetary
nebulae in the Coma cluster concur (Gerhard et al. 2005).

\subsubsection{That Last S}
VLSSS is Very Large Scale Structure and Streaming, and indeed deviations from
homogeneous mass distribution are necessarily accompanied by deviations from
uniform Hubble expansion, as long as Newton, Einstein, Galileo, or somebody
like that was roughly right. Whether the structure or the streaming is
dynamically and logically prior, we will let you know as soon as we find out why
the egg crossed the road.  Meanwhile,
the closest thing to a paradox lying around is that the local velocity field is
quite cold (small deviations), while the larger scale includes items like the
600 $\mathrm{km\ s^{-1}}$ dipole (Karachentsev 2003) and the Great Attractor
(Mieske et al. 2005a).

By analogy with ordinary bias (that is, luminous stuff is more clustered
than  dark stuff) there is also velocity bias. That is, objects made of
baryons can have different velocity distributions from the dark matter particles
in the galaxies (etc.) that they share. Which way does this go? Well, we think
that Faltebacher et al. (2005) conclude that galaxies in clusters move faster
than the dark matter and Kim et al. (2005) conclude that gas moves slower. Both
of these are necessarily calculations, as there are no direct measurements of
DM particle velocities. Not only do those two theoretical conclusions sound
contradictory, Whiting (2005) reports that, while there are indeed local
deviations from Hubble flow, the peculiar velocities don't actually point toward
the light concentrations. That is, either the peculiar velocities are not a
response to gravitational tugs of massive lumps, or light
doesn't trace mass, or both.

But at least we have good measures of the sizes of the peculiar velocities,
n'est pas?  Well, not entirely. Errors in distance determinations can amplify
them (Gibbons 2005,   on use of the fundamental plane, but the phenomenon would
seem to be general). As always, we collected a couple of dozen papers addressing
various standard distance indicators, most of them expressing
reservations.\footnote{Expressing reservations should not be confused with
making reservations. The two verbs are essentially synonymous only for very
special cases like olive oil. In any case, the author who belonged for
$28\: 1/2$ years to the ethnic group maligned in the joke for which
``reservations'' is the punch line wishes to record a strong preference for
under-recognized restaurants where they are not necessary.}

\subsection{Formation of Galaxies and Clusters}
 There used to be two models---top down (or monolithic) and bottom up
(hierarchical,  with mergers). The closest to anti-merger statements we found
this year were (a) well, all right, but it has been a \textsl{long} time since
some big, early-type galaxies experienced a major merger and had
significant star formation (Fritz et al. 2005), and (b) well, all right, but
sometimes the process gets a little out of hand and leaves you with a single
overluminous elliptical (Sun et al. 2004, who provide the name fossil group
for these).

The pro-merger ideas and data are sufficiently numerous that we provide only
capsule summaries of three favorite subtopics.
\begin{itemize}
\item The host cluster of Cygnus A, with 118
members, 77 new, is really two clusters of an Abell richness of 1 in the
process of making an Abell 2 (Ledlow et al. 2005).
\item The product depends on the mass ratio of the input disk
galaxies---up to 3:1 you get an elliptical; 4:1 to 10:1 yields an SO or
something similar; while greater than 10:1 merely disturbs the larger disk
(Bournaud et al. 2005). Apart from the precise numbers  this seems obvious
enough not to mention, except that there is a counterclaim that disks
can survive even 1:1 mergers (Springle \& Hernquist 2005), provided that a good
deal of gas was there to begin with and can settle back down into a flat layer.
\item All the processes at once can be seen in the gradual change of
populations in the calculations of Martin et al. (2005) and the observations of
Conselice et al. (2005).
\end{itemize}

\section[EXOPLANETS]{EXOPLANETS: PLEASE, SIR, MAY I HAVE
             SOME LESS?\footnote{The alternative
title for this section was ``Waiting for a phone call
from Stockholm,'' because it is the opinion of the only one of your authors to
have danced with a Nobel Prize winner that the discovery of this whole new set
of astronomical objects is the single most exciting event of the last 15 years
or so and fully prize worthy. The actual phrase comes from the late Howard
Laster, whose daughter, living in Sweden, was expecting her first child during
that critical October week many years ago.  Oh. It was Eugene Wigner, who used
to cut a mean Viennese waltz.}} Yes, more, more, more
remains a major theme (McCarthy et al. 2004a with two doubles,
Moutou et al. 2005  with a set providing modest support for the idea that the
smallest orbits go with the most metal rich hosts, and many others uncited). But
also more discovery and detection methods are beginning to prove themselves.
Udalski et al. (2005) reported the second clear microlens case from OGLE.
Alonso et al. (2004) found the first non-OGLE transit planet. And there are
another 40 OGLE transit candidates (Udalski et al. 2004) to be followed. They
only had to look at 230,000 stars to find them, and much of the hard work
remains to be done to get confirming (or falsifying) radial velocity data for
them all. Pont et al. (2004) say yes for OGLE Tr-111, with $P=4$ days. This is
the first transit planet with a period in the standard ``hot Jupiter'' range,
rather than near 1.5 days. The systems don't actually pile up at $P\approx 1.5$
days but are just passing through, say Patzold et al. (2004). The transit method
is not, incidentally, a good way to find brown dwarfs (Bouchy et al. 2005).

\subsection{More Observations}
 Shkolniket al. (2005) say they have seen the Io effect
(activity enhancement) in the form of Ca II H and K emission synchronized to the
orbits of two hot Jupiters orbiting Upsilon And and HD 179949. The
Rossiter-McLaughlin effect (distortion of line profiles of rotating host stars
as planets transit across) has not been seen by Ohta et al. (2005). We mention
it partly for the pleasure of being able to cite Rossiter (1924) and
McLaughlin (1924) who first thought of it.

Now about ``direct detections.'' Jura (2005) notes that the tail of a
Hale-Bopp-ish comet will reflect as much light as an Earth, and
Griessmeier et al. (2005) conclude that LOFAR might be able to separate the
radio emission due to a Jupiter from that of the parent star (this is already
well done within the solar system).

There was, we were told during the year, an honest to gosh image of a planetary
companion (Chauvin et al. 2005). It is called 2M1207b and orbits the brown
dwarf 2MASS J1207334-393254 (never mind; it won't come when you call it anyway)
in TW Hya. The pressier releases, however, attended the announcements of TrES-l
(Charbonneau et al. 2005) and HD
209458 (Deming et al. 2005). In each case, the authors started with a known
transit system, measured its brightness when the planet was off to one side
somewhere, measured it again when the planet was in back, and then subtracted
the smaller number from the larger one. The difference is then very approximate
 photometry of the planet in 1--2 colors (so far). In principle this could
 presumably be pushed to spectroscopy, though variability of the Earth's
 atmosphere will be a problem. HD 209458b observed this way is larger than you
  would expect for its mass and equilibrium temperature, as is TrES-l.
The problem may be a generic one (Laughlin et al. 2005a) for reasons that
are not entirely understood (Laughlin et al. 2005b).

What is ``more'' buying us? A second M dwarf host (Butler et al. 2004);
additional ``hot Neptunes'' (Marcy et al. 2005), which conceivably descend from
hot Jupiters via evaporation (Baraffe et al. 2005); and the first triple star
host (Konacki 2005).  The system is currently stable, but forming it must have
been a bit tricky  (Hatzes \& Wuchterl 2005).
The close stellar pair has had an orbit since Griffin (1977) published it in
a series of papers still in progress.

The class ``not many more'' appeared
in a search of (a) the open cluster NGC 7789 (Bramich et al. 2005) and (b)
of the globular cluster 47 Tuc (Weldrake et al. 2005), each of which yielded
fewer transits than expected if the incidence of planets is like that in the
solar neighborhood. On the positive side, each team now has a nice new set of
variable stars to study.

The preference of planets for metal-rich stars is familiar enough to rate only
 one mention this year (Fischer \& Valenti 2005). That the hosts have a fairly
 uniform distribution of ages across 3--12 Gyr is less familiar, but still gets
  only one citation (Karatas et a1. 2005), because that is all we found. As the
  stars age to red giants, their habitable zones move out and, for solar type
  stars, the Gyr duration may be long enough for
life to evolve (Lopez et al. 2005).\footnote{Although not of the index year,
we do like Stern (2003) with his ``Delayed Gratification Habitable
Zones'' to describe this situation.}

The possibility of living with a red giant was number one on our SETI list. It
also includes (a) the search for transits by non-spheroidal objects
(Arnold 2005) and (b) a search for at most 1 ns optical laser pulses from
13,000 solar type stars (Howard et al. 2004). The one possible candidate was
HIP 107395, and the authors note that we could outshine our sun by a factor
$10^4$ in a sufficiently narrow cone and narrow wavelength band.

Remember Epsilon Eridani and Tau Ceti? These were the first stars ever asked
to produce SETI-type radio signals, almost 50 years ago. They failed, but
remain the only G to early K single dwarfs within 5 pc. Both have debris disks,
say Greaves et al. (2004).
 Lynden-Bell \& Debenedetti (2005) ask whether there might be life without
water. That depends, we feel, on the quality of the wine available. The
observation, however, comes from editor Ball (2005) who notes that physicists
and astronomers are more likely than biologists and chemists to ask these
``what if'' questions. We think the genre was probably invented by historians,
but note that author  Lynden-Bell has lived with a mathematical physicist who
often appears in these  pages for many years. It's all right. They are married
(and serve very good wine).

``More'' also means you are allowed to do statistics. The minimum required
is $N = 3$ to define the direction of a linear correlation and the dispersion
around it. Working with a somewhat larger sample, Mazeh et al. (2005) deduce
that, within the hot Jupiter class, there is an inverse correlation of planet
mass with period, and Halbwachs et al. 2005) find that planets and binary
companions occupy different zones in a period-eccentricity diagram, even after
allowance for migration, circularization,
and so forth. They believe this implies different formation mechanisms.

\subsection{More Theory}
 Clever planning has brought us to the end of the observations
with a paper that just cries out to proceed from page 16 of the index (exopl
date/search/SETI) to p. 17 (exopl calc/dyn). Here live a couple of dozen papers
about formation, migration, and  orbit stability. The six topics
following  were chosen for microhighlighting more or less for their
discouraging words.

Beer (2004) says that none of the exosystems formed the same way our endosystem
did and none will have earth mass planets. They use a Box-Cox transformation
without citing Gilbert or Sullivan.

There is an on-going worry about whether proto-stellar disks last long enough
for any planets to form. This is obviously silly; they must, or it wasn't done
that way. Hernandez et al. (2005), Brice\~{n}o et al. (2005),
Calvet et al. (2005), and Carpenter et al.
(2005) explore some of each of those possibilities. So you are not allowed to
let worry about this problem keep you awake until after you have read all these
 papers. Afterwards you will be too tired to stay awake.

Agnor \& Asphaug (2004) report that more than half of planetesimal collisions
during a supposed planetary growth process actually break things up rather than
accumulating larger masses. This does not, of course, matter in the
gravitational instability and hybrid formation scenarios (Boss 2005,
Currie 2005).

As for migration, note, say Cody \& Sasselov (2005) that it does not lead to
much planetophagia, because the resulting changes in mass, composition,
convection zone depth, temperature, and age do not lead to the patterns seen in
real hosts.  They trace the idea of accreting stars just a little further back
than the paper that often results in the name Bondi accretion to
Lyttleton (1936) and Hoyle (1939).

It is comforting to know that, where we see two or three planets around the same
star, the orbits are stable for reasonable lengths of time, but less so to
realize that this has been achieved by theorists revising the observational
data to produce resonances (Gozdziewski et al. 2005 for $\mu$ Arae and
Ferraz-Mello et al. 2005 for HD 82943 b and c).
The original data for the latter system appear in Mayor et al. (2004).

Poor lone, lorn planets and brown dwarfs can be left behind when a core that
aspired to stardom is photo-eroded by a nearby OB star
(Whitworth \& Zinnecker 2004). The
Hollywood equivalent is not having your option picked up.

\subsection{More Disks}
 In an ideal world, there are two sorts---protoplanetary, before
the planets have formed, and debris or exozodi, some combination of leftovers
and broken up comets, asteroids, etc.
after planets form. In practice, the two phases are likely to overlap, and the
words do too. Smith \& Bally (2005) attribute a debris disk to IRC9 in Orion,
which will someday be an AV star and which they describe as a young analog to
Vega, likely still to have protoplanetary stuff. On the other hand, they provide
an ideal introduction to the green dot on this topic, the conclusion
(Su et al. 2005) that the Vega debris disk must itself be transient.
Observations with \textsl{Spitzer} led Su et al. to a calculation that a
production rate of $10^{15}$ g/sec is needed to maintain the supply of small
grains. The alternative to a sporadic event is a truly enormous reservoir of
asteroid material, much larger than $3\times 10^{30}$ g.

A not quite random selection of other debris disk and exozodi items during
the year includes (a) first examples of stars with both planets and 70 $\mu$m
excess disks (6 of 26 stars examined by Beichman et al. 2005 with
\textsl{Spitzer}), (b) the pre- to post-transition around the Be star
51 Oph which has a warm inner dust disk and gas (Thi et al. 2005), the
only Be star that can make this claim, (c) partial clearing and asymmetry
of the $\beta$ Pic disk, the very first discovered (Telesco 2005,
Weinberger 2005), and (d) the relative rarity of collisionally-produced
grain disks around main sequence stars (Song et al. 2005 on BD $20^\circ$ 307
which has amorphous and crystalline silicates around it but no PAH).
Rieke et al. (2005) conclude that the transition from the lost
protoplanetary disk to secondary stuff is largely complete for stars
150 Myr old, and that there is thereafter considerable variety in disk
sizes, central holes, and so forth.

\section[ASTROBIOLOGY]{ASTROBIOLOGY\footnote{The Gold Star
for this section is awarded to United States District Judge
John E.~Jones III for his decision in favor of the plaintiffs in
the Dover PA case concerning the teaching of ``intelligent design''
 versus Real Science. Every budding scientist (and lawyer) should
consult Jones  (2005b) for his opinion.}}
It has taken us 15 editions of ApXX to recognize that a comparatively new
discipline has joined our traditional two of astronomy and astrophysics,
and it is one deserving its own section.
You may perhaps excuse our tardiness by noting that astrobiology draws on
such seemingly disparate fields as chemistry (organic and inorganic),
geology (terrestrial and other solar system bodies, if we play fast and
loose with meaning of ``geo''), biology (molecular,
traditional, evolutionary, etc.), or what about \textsl{Organic Geochemistry}
(as a subject and the name of a journal), biogeochemistry, and just about
any subdiscipline in astronomy  and astrophysics you care to name. One section
in Ap05 is barely an introduction to the subject and we shall concentrate on
but two related topics that were chosen by the junior, but oldest, author
(who was on his way to becoming a dipterologist
in his youth but found he couldn't remember or pronounce the Latinate names).
Even these two, however, call on a number of journals (and books) not often
referenced in, for example, PASP, AJ, ApJ, or MNRAS. One of these is
\textsl{Astrobiology}, the namesake journal for the field,
which is a mere child whose first issue appeared in March 2001.
(Compare this to  AJ ``Founded in 1849 by B.~A.~Gould\ldots .") Going through
one issue of \textsl{Astrobiology} we counted references to 39 different
journals, not all of which did we peruse for relevant papers. Perhaps this
is why we waited so long for a try at the subject.  In any case, since this
is our first shot for a full section
in this field, we shall often call upon papers from outside the index year.

For a recent review in the usual astrophysical
literature, see Chyba \& Hand (2005).
Another useful resource is the 2005 National Research Council study \textsl{The
Astrophysical Context of Life}, which may be downloaded (free!) from
\texttt{http://www.nap.edu/catalog/11316.html}. It is a critical study, with
recommendations, of the status of the field. Also check out NASA's Astrobiology
Institute website (see NAI 2006).

And now, first things first.

\subsection{Life Is Where You Find It, Or Not}
What is life? Sure we know the answer: ``I can't define it but I know it when I
see it.''---quoting a Supreme Court Judge's opinion on an entirely different
subject. But, sorry, Your Honor, life is not that simple, so to speak. How
can we define it so that we can recognize it when we do see it, or think we
 detect it---and not just on Earth? Conrad \& Nealson (2001), who happen
to be the  authors of the first research paper
published in \textsl{Astrobiology}, put it this way: ``Elimination of
Earthcentric biases from life detection strategies
thus increases the probability that we will not only know life when we see it,
but have the statistical acumen to prove that we have seen it, as well.''

	   Schulze-Makuch \& Irwin (2004)  list three ``fundamental
characteristics'' that they deem necessary to distinguish life from non-life
(and see Irwin \& Schulze-Makuch 2001). Whether these make up a ``definition''
 rather than a ``check-out'' list is a matter
we shall not go into. You might wish, however, to read the long discussion by
 Ruiz-Mirazo et al. (2004)
on how a proper definition should be posed and what issues it should address.
Cleland \& Chyba (2002)  go further and argue that we must understand life
at a deeper level before we make up definitions; e.g., in defining ``water''
without knowing what $\mathrm{H_2O}$ is on the molecular level we may be
spinning our water wheels. Worse yet,
we may even miss recognizing strange microbial life on Earth that
doesn't fit our preconceived notions---as discussed by
Cleland \& Copley (2005) and Davies \& Lineweaver (2005).

The first characteristic for life from  Schulze-Makuch \& Irwin
 is that it be ``composed of bounded microenvironments
in thermodynamic disequilibrium with their external environment." Since,
it is supposed, that the external environment consists, at least partially,
of a solvent that can contain accessible nutrients, the ``bounded'' makes
sense. Otherwise, the organism soon becomes indistinguishable from its
external environment because of diffusion driven by
gradients. ``Disequilibrium''
is better than ``in thermodynamic equilibrium'' because the latter is a
fancy way of saying that you and your  environs are one; i.e., you're dead.

The second characteristic is that life is ``capable of transforming energy
and the environment to maintain a low entropy state.'' This defines the
interaction of what is in, or on the
surface of, the bounded microenvironment with its surroundings. With no
interaction the parcel of life would, by the second law of thermodynamics,
``move spontaneously  toward a state
of maximum entropy,'' thus leading to an adverse result as in the above.

Finally, life is ``capable of information coding and transmission.''
Were this not so then the ``organism''---and we might just as well use that
term---would be incapable of
passing on information that could be used to create a duplicate or
near-duplicate of itself.\footnote{``Organism'' need not imply ``organic''
in the chemical sense. Some argue that silicon could form the basis for
life instead, although not as efficiently.
See, for example, \S5.3 of Schulze-Makuch \& Irwin (2004).}
Non-duplication seems like a dead end, although this may be overly picky.
 Note that lateral gene transmission may (and does) occur between different
organisms without replication or reproduction.

How does what we find on Earth conform to the above conditions?
Leaving aside viruses, which appear to be a special (and probably degenerate)
case, Earth teems with microorganisms.
The simplest organisms are the prokaryotes.  Each consists of a
membrane---with often an outer protective cell wall---that surrounds the
cellular cytoplasm and its contents. Inside resides a (usually circular)
free floating chromosome containing the cells genetic DNA. (Extrachromosomal
DNA may be present in plasmids, which have various functions.) Ribosomal
inclusions in the cytoplasm are involved in
protein synthesis. The prokaryotes are subdivided into the Bacteria and
the Archaea based on distinct
differences in DNA and cell wall composition and structure. (See the pioneering
efforts of Woese 1997.) All the rest of terrestrial life are eukaryotes,
which have a distinct nucleus containing most of the genetic material,
energy modules (mitochondria and, in plants, chloroplasts), and other
material.\footnote{There are some exceptions to
this statement. \textsl{Giardia lamblia}, an intestinal parasite that
can infect campers drinking water from pristine looking streams, has
neither chloroplasts nor mitochondria. Its, and its cousins', place in
evolutionary biology is problematic.} Since eukaryotes are most likely chimera
composed of prokaryotes who decided to combine forces in the distant past,
this year's review  will let them be. (See, e.g., Margulis 1992, 1999, and
for a discussion of many of the topics gone into here we recommend the
splendid book by  Knoll 2003.) In any case, terrestrial life fits the
above definition of life---which is no surprise.

Modern prokaryotes are doing very nicely. D'Hondt et al. (2004) estimate that
the Earth contains 4--6$\times10^{30}$ cells (finally an astronomical number!),
mostly in open ocean, soil, and
in deeper oceanic and terrestrial subsurfaces. (See also the oft-cited paper
by Whitman, et al. 1998.) They used samples from the Oceanic Drilling Program
retrieved from depths down to 420 m from Pacific Ocean sites. Typical cell
concentrations were $10^6$ cells $\mathrm{cm^{-3}}$.
Schippers et al. (2005) and Teske (2005) have verified that such samples contain
live bacteria and archaea (rather than just inactive or dormant cells), with
the latter perhaps being more abundant. In any case,
prokaryotes seem able to survive trying conditions, to say nothing of
hyperthermophiles who enjoy basking in the hot springs of Yellowstone or around
midocean ridge vents.

How small are the smallest terrestrial prokaryotes? Prompted in part by possible
organic remains in Martian meteorites, but also for identification of
terrestrial life dating back perhaps nearly
four billion years, the National Academy of Sciences organized a workshop to
address this very question  (Nat.~Acad.~Sci. 1999). The consensus of the
participants was that modern---give or take
a billion years or so---terrestrial cells have a lower size limit of
$250\pm50$ nm. There are exceptions, and possible exceptions, to this limit.
Huber et al. (2002) report a novel (perhaps representing a new phylum) member
of the Archaea plying its trade in a hot submarine vent off Iceland. It tops out
at 150 nm. It appears to be, however, a symbiont that attaches itself to an
archaean host. Further down the scale, Kajander et al. (on page 50 of the NAS
report) find organisms (``nanobacteria'')
of size between typical viruses and bacteria in animal serum that can be
cultured in suitable media. Whether these organisms are really self-sufficient
was a matter of contention. For now we shall stick with
the consensus view. This is not to say, however, that life operating under a
different set of molecular rules could not be
smaller and still function. Note that there are also some Sumo wrestler
sized bacteria. \textsl{Thiomargarita namibiensis,} a colorless sulfur
bacterium, has a diameter of 750,000 nm.
As  Schulz \& J{\o}rgensen (2001) point out, this means that the range of
prokaryote volumes exceed $10^6$
(about ten times more than the span between mouse and elephant).

\begin{figure}[ht]
\plotone{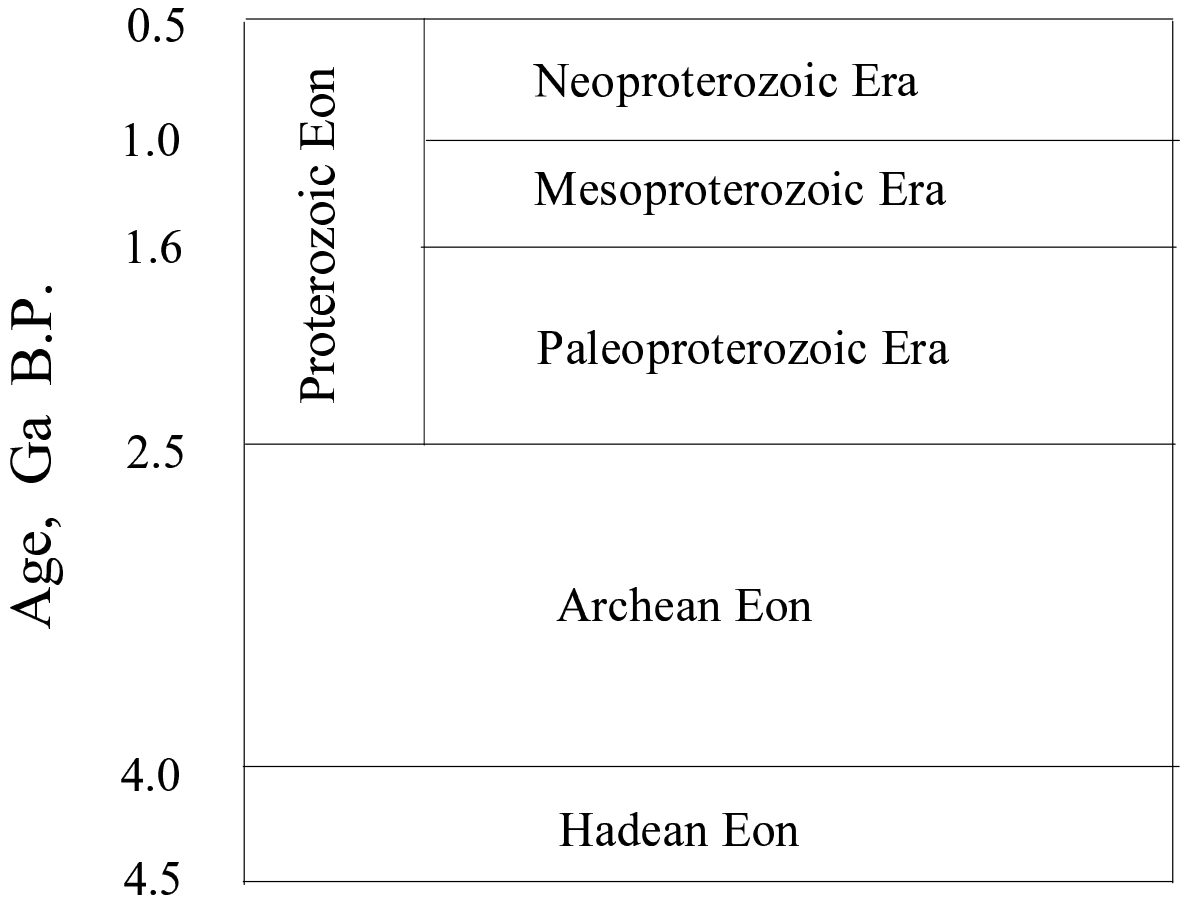}
\caption{The eons and eras of Earth's distant past from 0.5 to 4.5
Ga as measured before the present (B.P.). Adapted from Knoll (2003).}
\end{figure}

\subsection{Life in the Old Country}
If we are to detect and identify life on Mars, for example, we should ask how
it is done for the very early Earth when life was in its infancy. The earliest
identifiable microfossils of good
pedigree appear to be those in cherts from the Transvaal Supergroup
(South Africa) and date from about $\mathrm{2.6\,Ga}$ B.P. (See, e.g.,
Altermann \& Schopf 1995. We use Ga, Gyr, and
Gya interchangibly, as all appear in the literature.) They are in the form of
rods, spheroids, and filaments that bear close resemblance in shape and size to
 well-attested prokaryotic microfossils
found in later geological formations---although, to our untutored eyes, they
could be anything. They were part of an assemblage that formed components of a
community that formed stromatolitic
reefs in an ancient sea. (Stromatolites are domed, candelabra, and
wavy-laminated shapes that are common in some fossil beds, although they may
be distorted by later geological processes. You can
still find them in select places, such as coastal Bermuda and Western Australia.
See Knoll 2003 for a good selection of photographs.) This gets us back to the
very early Proterozoic or late Archean
eons. (For a snapshot of where we are in Earth's history, see Fig.~1.)
How much further back can we go? And here is where things get
murky---and controversial.

On an optimistic note Chyba \& Hand (2005) opine ``It is broadly agreed that
robust and abundant fossil
evidence is present in $\approx$3 Gya rocks, and that substantially
controversial isotopic evidence
exists in 3.8 Gya rocks.'' This puts us smack dab in the Archean.

One promising example (not mentioned by
Chyba \& Hand) is Rasmussen (2000) who reports on the ``probable fossil remains
of thread-like organisms'' in 3.235 Ga Australian rocks of deep sea origin
formed in a  hydrothermal setting. The threads are 550
to 2000 nm in diameter and up to 300 $\mu$m long, and are of uniform thickness.
If they are the remains of living organisms, then thermophilic prokaryotes are
the likely suspects. (Fossil remains from deep-sea hydrothermal
systems formed prior to the Cambrian [i.e., prior to $\sim600$ Ma B.P.] are
more then rare. Rasmussen claims his are the first found.) It seems that no
further work has been done on these deposits but Brasier et al. (2005, and see
below) suggest they are ``promising and worthy of re-examination.''

Schopf, et al.~(2002, and see the many earlier references therein) discuss
putative microfossils from the Apex chert, Chinaman Creek, of the Warrawoona
Group in Western Australia.\footnote{Compared to this, astronomers sorely lack
romantic names for their objects. Consider a new IAU Commission XXX,
\textsl{Astronomical Nomenclature, With Panache}.}
They are 3.5 Ga old and resemble modern (and ancient) cyanobacteria
(also known as ``blue-green algae,'' but are not always of that color and they
are prokaryotes, not eukaryotes). Most modern cyanobacteria make their living
 through photosynthesis and, in the past, must have played a
crucial role in the oxygenation of Earth's atmosphere (see Knoll 2003, Chap.~6,
and his color Plate 2; and the review of Canfield 2005).
The possible microfossils look, to our eyes, so very much like later accepted
cyanobacteria fossils that we smile in appreciation. However---

The most complete discussion we found of the c.~3.5 Ga Apex cherts is
Brasier, et al. (2005). They combine a detailed study of the geology of the
Warrawoona Group with a suite of microscopic studies
of the chert (including optical and electron microscopy, digital image analysis,
etc., and they include many informative figures).\footnote{Their description of
the geology of this ancient land, while not poetry,
almost lets you hear and smell that part of Australia as it was assembled. For
more information on the geology of the region see the website for the
Geological Survey of Western Australia (given as GSWA 2006).} Their conclusion
is  that the Apex microfossils are ``pseudofossils'' that resulted from the
incorporation of carbon-rich material into recrystallizing silica.
The debate continues.

What other kinds of evidence point to biological activity in ancient rocks?
Terrestrial organic organisms  must deal with carbon as a vital part of being
``organic.'' Isotope-wise, however, they seem to show a preference.
Samples of biotic material show an underabundance of $\mathrm{{}^{13}C}$
compared to the more common isotope
$\mathrm{{}^{12}C}$ (more common terrestrially by about 100 to 1).
It's not that our life ``likes'' the lighter isotopes better but
rather the slightly lighter, and thus  faster atoms (or carbon containing
molecules) collide more frequently with their targets.  Hence the small
statistical preference for the lighter atoms in the reaction product is the
result of reaction kinetics. The process is referred to as ``fractionation.''
(See, e.g., Hayes 2004 for a complete,
though difficult for us, discussion.) The underabundance of the heavier isotope
compared to the lighter (and this is not restricted to carbon) is expressed as
the difference of the ratio $\mathrm{{}^{13}C/{}^{12}C}$ in a biological sample
compared to that in an accepted laboratory standard.
The difference is denoted by $\delta\mathrm{{}^{13}C}$ in parts per thousand
(\ppthou); that is
\begin{displaymath}
\delta\mathrm{{}^{13}C}=1000
      \frac{\left(\mathrm{{}^{13}C/{}^{12}C}\right)_\mathrm{sample}}
       {\left(\mathrm{{}^{13}C/{}^{12}C}\right)_\mathrm{standard}}-1.
\end{displaymath}
Typical values for biotic remains are --20 to --30\ppthou.

Brasier, et al. (2002) report fractionations in the Apex cherts
of --30 to --26\ppthou\ entirely consistent with a biological origin. But they
then point out that such fractionations may be produced abiotically by
Fisher-Tropsch synthesis and have nothing to do
with biology.\footnote{The Fisher-Tropsch process(es) was designed to produce
liquid hydrocarbons and was used extensively by coal-rich, but oil-poor, Germany
in WWII for synthetic oil production.} To test whether
abiotic fractionation can take place in hydrothermal geological environments,
McCollum \& Seewald (2006) performed some neat laboratory experiments and found,
alas, that fractionations typical (or exceeding) those associated with biology
can indeed be produced. A general review is due to Holm \& Andersson (2005).

This leaves us in a quandary. As amateurs, the evidence, for or against, the
presence of life in the deep Archean seems to be up for grabs. Sniffing between
the statements in many papers, however, we get the distinct impression that
most investigators in the business do believe that life started well before
rock solid (so to speak) positive evidence makes its appearance in the
geological record. How else then to explain the presence of prokaryotes in the
very early proterozoic or late archean? Those little bugs were, and are,
complicated creatures.

It is now a little over 50 years since Miller (1953) demonstrated that organic
compounds, including amino acids, can be made by zapping, spark-wise, a glass
container filled with methane, ammonia, hydrogen, and water vapor.
(See Lazcano \& Bada 2003, for a brief history
of the experiment.) A true it-can-be-done experiment, and there have been more
done like it since. As usual, however, questions remain about the state of the
Earth's ancient atmosphere, and other sites may have been where life originated,
such as hydrothermal vents, as just one example.

One, among many, fundamental question yet to be answered is why  certain
biological molecules are essentially either all left(L-)-handed (amino acids
in proteins) or right(R-)-handed (sugars in RNA and DNA).
This is the ``homochirality problem.'' To force  homochirality requires either
a left- or right-hand bias in radiation influencing the chemistry, for example,
or an initial bias in handedness as life chemistry
starts its thing. Jorissen \& Cerf (2002) review several mechanisms by which
this might be accomplished. Examples are circularly polarized solar UV, or
unpolarized UV acting in concert with a magnetic
field (e.g., the Earth's) that is not perpendicular to the light beam. Both
seem marginal, but possible. That Earth in its early history could have acquired
homochiralic material from outside remains a possibility.
Among suggestions of how such material could have acquired a L- or R-hand bias
in space we have, for example, Lucas et  al. (2005a) who invoke circularly
polarized UV in star formation regions.

\subsection{Life in the Really Old Country}
We have had little to say about the origin of terrestrial life but if we think
it was around in the middle-aged Archean, then might it have started in the
earliest Archean or in the Hadean (from ``Hades'' itself)?\footnote{Earth may
 have been seeded with life from the outside, as in ``panspermia,'' but we
won't touch that one.} The Hadean eon is well-named because early on the
evidence is for intense meteoritic bombardment. When, how intense, and of what
duration is a matter of controversy. Looking at exposed terrestrial rocks
is tricky because we lack a continuous record prior to some 3.7 Ga B.P., which
is just a little after the Hadean (see Fig.~1). Our nearby Moon may serve as
a surrogate as its geology settled down long before Earth's. Hartmann (2003)
argues that the rate of lunar impacts has decreased exponentially and rather
smoothly with time from 4.0 Ga (and perhaps back to 4.2 Ga). As an example
(using his Eq.~1) we find that for each $\mathrm{km^2}$ of lunar surface there
was roughly a one in ten chance that a crater of diameter greater than one km
would have been formed by an impact in the interval
4.0--4.2 Ga. Not a healthy environment.

Another model for lunar bombardment is the ``Late Lunar Cataclysm'' or
``Late Heavy Bombardment'' (LHB), which may have taken place around $3.9\pm0.1$
Ga B.P. The evidence for this, as an event involving a goodly
part of the solar system, include: dating of the Martian meteorite ALH84001
(of which more later, see Turner et al. 1997); dating  of lunar melt samples
showing a lack of ages older than 3.9 Ga, although earlier lunar impacts may
have been covered (Cohen et al. 2000); and some evidence
preserved in Hadean zircons (which will appear again later, see
Trail et al. 2006). This model turns the exponential drop-off of the above on
its head with relatively few  impact events prior
to $\sim3.9$ Ga (excluding the early assimilation of solar system bodies) and
then all Hell breaks loose in the late Hadean or early Archean.  One suggested
cause of the LHB is a major reshuffling of the positions
of the gas planets after the dissipation of the solar nebula.
Gomes et al. (2005a, and see the related papers Tsiganis et al. 2005, and
Brunini 2006) performed numerical simulations of planetary and planetesimal disk
interactions (and orbits) starting from what they consider to be a reasonable
initial configuration. They find that after some
0.7 Ga the orbital periods of Jupiter and Saturn came into a 1:2 resonance
($\mathrm{P_S/P_J=2}$) and their orbits became eccentric. The result was a
dance involving Saturn,  Uranus and Neptune that destabilized the planetesimal
disk (and the asteroid belt) thus delivering a flood of material into
the inner solar system. Hence the LHB at about the right time.

And now the possible good news. Zircons are remarkably tough crystals and can
survive even after their parent rocks have been eroded or broken up.
Wilde et al. (2001), and its companion paper Mojzsis et al. (2001), report
examination of zircons from Western Australia that have ages ranging
back to 4.4 Ga B.P. (from U-Pb-Th dating). An updated examination with more
detail is discussed in Cavosie et al. (2005, and see the popular version in
Valley 2005). Harrison et al. (2005) trace the crust to perhaps 4.5 Ga,
or a little before, as in Watson \& Harrison (2005,
from Hf isotopic ratios) although most, or all,  of it was rapidly recycled
back into the mantle. The better news from them is that the evidence suggests
that supracrustal rocks were around as long ago
as 4.2 Ga  and liquid water oceans lapped (if that's the appropriate word)
 their shores. (``Cool'' water is implied by the ratios of
$\mathrm{{}^{18}O}$ to $\mathrm{{}^{16}O}$ in the zircons.)

Considering all the above, a mental picture of the ancient Earth still escapes
us, as does the beginning of life. Perhaps we should go along with Knoll (2003)
who concludes that ``Origin-of-life research resembles a maze with many entries,
and we simply haven't traveled far enough down most routes to know which ones
end in blind alleys.''

\subsection[ALH84001]{An Extraterrestrial Visitor to Antarctica\footnote{We
recommend the long review chapter by Jakosky et al. (2006) for much more
information about Mars than we go into.}}
Observational astronomers don't necessarily have it easy in Antarctica but they
don't have to rise out of sleeping bags to begin their daily trek across rock,
snow, and ice looking for meteorites during the two months or so when the
climate is bearable.  A prime location
are the Allan Hills off the Ross Sea about 150 miles from McMurdo Station (a
convenient distance for a helicopter). The hills are mostly free of ice but
there are several fields where meteorites stand out. A description of one of the
fields and its finds may be found at Schutt (2006, website for the
\textsl{Antarctic Meteorite Location and Mapping Project}).

On 27 Dec 84 a 4.2 kg, $15\times10\times8$ cm, meteorite was found at Allan
Hills and now has the designation ALH84001 (ALH for the hills). It is of a class
called ``shergottites'' named after a similar one found at Shergotty, India,
in 1865.\footnote{This class is often referred to as the
SNC class from the names of the type specimens Shergotty, Nakhla, and
Chassigny.}
Of the some 30,000 known meteorites that have been found on the Earth, ALH84001
and Shergotty are only among 16 that are known to
have arrived from Mars. (That number seems to fluctuate over the years in the
literature.) How do we know their origin? The most convincing evidence we have
seen is summarized in Fig.~5 of McSween (1994, and see Bogard \& Johnson 1983)
adapted from the review article by Pepin (1991). It shows
the correlation between ground level
concentrations of gases ($\mathrm{CO_2}$, $\mathrm{N_2}$, $\mathrm{{}^{40}Ar}$,
$\mathrm{{}^{36}Ar}$, $\mathrm{{}^{20}Ne}$, $\mathrm{{}^{84}Kr}$, and
$\mathrm{{}^{132}Xe}$) in the Martian atmosphere sampled from \textsl{Viking}
landers and gases trapped in glass inclusions in the shergottite EET79001.
Would that every scientist be blessed at some time
with such a straight line with small error bars.

There are several classes of shergottites and ALH84001 is an
orthopyroxenite---and now you know. Bridges \& Warren (2006), in reviewing the
properties of the shergottites, report that ALH84001 is 97\% orthopyroxene,
which is the commonest silicate in meteorites and is the major constituent of
most chondrites. ALH84001 was not recognized as a shergottite until nearly ten
years after its discovery when Mittlefehldt (1994) took a good look at it. He
found that in amongst the dominant orthopyroxene were
carbonate inclusions (as interstitial grains) of about 100--300 $\mu$m size.
He suggests that these inclusions were formed by ``multiple infusions of
fluid.'' The material was then subject to shock, after which
much smaller ($\sim10\,\mu$m) carbonates were deposited in fractures.
Borg et al. (1999), using Rb-Sr and Pb-Pb dating,
conclude that the carbonate inclusions (or ``globules'' or ``flattened
pancakes'') were formed at 3.9--4.0 Ga B.P., an age
when Mars was subject to heavy bombardment (Neukum \& Wise 1975, and shades
of LHB). It may be, however, that water did infiltrate  ALH84001 at a time near
enough to confuse the issue (of which, more later). The
initial crystallization age of ALH84001, from Sm-Nd and Rb-Sr analysis, is given
by Nyquist et al. (1995) as $4.5\pm0.13$ Ga, which makes it the oldest of the
known shergottites.\footnote{Our eyebrows lifted
when we read this. We are back to the earliest Solar System days!} Measurements
 of the radiometric exposure age indicate that it was ejected from the Martian
surface by an impact a mere $15\pm0.8$ Ma ago according to
Nyquist et al. (2001).\footnote{This long review paper is a great resource
with many clear figures. On reading it, you will find out that some shergottites
crystallized only $\sim180$ Ma ago (during our Jurassic period)
implying that magmatic activity on Mars is probably an ongoing affair. EET79001,
discussed earlier, was ejected  about 0.7 Ma ago---nearly yesterday. For a neat
website giving updated geological period designations, with colorful charts, go
to the site \texttt{www.stratigraphy.org} maintained by
the International Commission on Stratigraphy.}

Inside the carbonate globules, and especially concentrated around their rims,
are small magnetite crystals (McKay et al. 1996) some 5--100 nm in size. A
subgroup of these are identical (or nearly so) to those made by a terrestrial
magnetotactic bacterium called MV-1 (Thomas-Keptra et al. 2001).
The crystals  are not terrestrial contaminants picked up by ALH84001 while it
sat in Antarctica for 13,000 years before being picked up.

Also present are traces of organic compounds such as polycyclic aromatic
hydrocarbons (PAHs).\footnote{PAHs are molecules of linked benzene rings.
They are nothing new under the stars as Yan et al. (2005) have detected them
in dusty ultraluminous IR galaxies at $z\sim2$ using \textsl{Spitzer}.} Some of
these appear to have been produced by
terrestrial bacteria, but not all---especially in the carbonate globules.

\subsection{The Leap to Life}
McKay et al.~(1996), in a paper that has generated a great deal of healthy
controversy and study, proposed that ALH84001 contains the remains of Martian
life forms, albeit on the nanometer scale. As a result, no Earth-born rock has
received as much attention as has ALH84001. McKay et al.
took the above (of \S7.4) into account but also pointed to ovoid particles of
size 10--100 nm in the carbonate globules
plus rod-like structures a mere 100 nm long. If they are biotic, then they
qualify as some sort of nanofossils by the standards of \S7.1.

Herein lies the first difficulty. The argument that
independent terrestrial life has a lower size limit of $250\pm50$ nm means that
if the Martian nanofossils are biotic, then they must have functioned in a mode
different from what we are accustomed to.\footnote{We leave the serum
nanobacteria out of this because they are probably not
independent.} There is nothing intrinsically wrong with this idea but, at
present, it isn't testable. Of course the ALH84001 ovoids and rods may be the
desiccated shrunken remains of what were
originally more robust organisms. But we don't know.

The magnetite crystals as  biomarkers fair no better.  Golden et al. (2004)
conclude that---a) the ALH84001 crystals do not have the same structure as those
made by the bacterium MV-1 and, b) the crystals could just as well have been
 produced abiotically by hydrothermal processes early on in the
life of the meteorite.

The presence of organic molecules such as PAHs has somewhat the same problem.
Zolotov \& Shock (2000) calculate (but no with experiments) that these could
have been synthesized by a combination of the cooling of magmatic and impact
generated gases. And, talking of impacts, Treiman (1998) says there were
four to five impact events in the life of ALH84001 spanning roughly 3 Ga
(his Table 1) with question marks as to
the timing of some of them. So when were the nanofossils formed? We don't know.

Direct evidence for biotic markers in ALH84001 seem ambiguous at present.
However, the good news is that early Mars was both wetter and warmer than it is
now (or in the comparatively recent past). \textsl{Opportunity},
in its travels, has come up with evidence for deposits that seem to demand water
at some time, to say nothing of the channels seen from orbiters. The story is
too long to tell here---and there are some detractors---and we thus suggest
Bullock (2005) and Christensen (2005a) for more popular reviews,
Jakosky \& Mellon (2004),  Jakosky et al. (2005), and Bibring et al. (2006) for
 more of the hard science.

But hope springs eternal, as it should. For example, Gibson et al. (2001)
discuss the much younger shergottites Nakhla and Shergotty  and conclude they
have the same (perhaps) biotic virtues at ALH84001. Stay tuned, as they say.

\section{BETWEEN AND BEFORE THE STARS}
 It is always a little difficult
to decide the precise particle density at which interstellar material
becomes star formation, and so, not wanting to have to say that
\S8 differs from \S9 only in Section Number, we have run
them together here.

\subsection{Interstellar Gas Compositions}
 Just which molecules occur diffusely
in space? Glycine, no (Snyder et al. 2005, and the main problem remains
uncertainty on which wavelengths to look for, though earlier ApXX's
have trumpeted its discovery more than once). The first ketone, yes
(Widicus Weaver \& Blake 2005). It is $\mathrm{CO(CH_2OH)_2}$ the simplest
ketose monosaccharide. They believe it is part of a pattern of formation of
``pre-biotic'' compounds in hot pre-stellar cores. And yes for
$\mathrm{CH_3CCD}$ in the Taurus Molecular Cloud
(Markwick et al. 2005). The least musical author had greenstarred this in hopes
of being able to report it as  MethylMusic,\footnote{The Medical Musician
declined to be drawn on this one. Even our friends have some
taste!} but suspects it must somehow belong to the propyl family.

Molecular oxygen remain elusive, according to Wilson et al. (2005), who examined
the Small Magellanic Cloud with the Swedish satellite \textsl{Odin}. And all is
not  PAH that smells, though it is likely to have multiple carbon bonds
(Ruiterkamp et al. 2005 on the diffuse interstellar bands).

Only senior molecules will remember when $\mathrm{HCO^+}$ was called X-ogen, but
it remains true that negative molecular ions are much less common than
positive ones (Morisawa et al. 2005).
Well, it is generally said to be more blessed to give than to
receive (Take my Chancellor, Please!). Indeed all of $\mathrm{H_3^+}$,
$\mathrm{H_2\,D^+}$, and $\mathrm{H\,D_2^+}$ are wandering around out there
(Flower et al. 2004). No news yet of $\mathrm{D_3^+}$, but the authors report
predictions for ortha/para/meta  state ratios just in case. Also calculated
and expected to exist are $\mathrm{He\,H^+}$ and $\mathrm{{}^3He\,{}^2H^+}$
(Engel et al. 2005), the latter  of which at least must be rather rare.

Deuterium is, obviously, much overrepresented in ISM molecules compared to
atomic abundance, which sloshes around 1--$2.5\times10^{-5}=\mathrm{D/H}$,
reduced from the Big Bang production number of 2.5 by both astration (nuclear
processing in stars) and depletion onto molecules and grains
(Williger et al. 2005). The umpteenth first detection of the 92 cm
spin-flip transition of neutral deuterium came from Rogers et al. (2005),
looking toward the galactic anticenter with a purpose-built array at
Haystack and finding emission at the $6\sigma$ confidence level. They plan
to look for absorption toward the galactic center soon.

Additional new molecules of the year include $\mathrm{HC_4N}$ in IRC +10 216
(Cernicharo et al. 2004). This is only the second such molecule with even
numbers of carbon in the middle (the first was $\mathrm{HC_2N}$, not
surprisingly), while odd ones, up to and beyond 7 Cs in the middle are known,
the H and N representing the continents.

The familiar formic and acetic acids, methyl formate, methyl and
ethyl cyanide, and  methanol were traced out in a high mass star
formation region by Remijan et al. (2004).
Might one possibly ever find purines and pyramidines? Peeters et al.
 (2004) say they would last only hours in a solar nebula at 1 AU,
10--100 years in the diffuse ISM, but the cloud lifetimes in dense clouds.
This is, you may notice, a good, firm, positive maybe.

Comito et al. (2005) report so many molecules in a single paper (929
transitions of 26 identified species plus some unclaimed features) that
one would almost suppose the authors had to pay their own page charges
personally, like the least grant-worthy of your ApXX authors.

\subsection{Interstellar Dust}
 The oldest and most stable fact about Galactic dust
is that it absorbs about one magnitude per kiloparsec in the plane
(Trumpler 1930, Amores \& Lepine 2005). Establishing chemical and
structural properties and how these vary within and among galaxies
has taken a little longer. Here are a few one-word descriptions---

\begin{itemize}
\item Alphatic, meaning carbon chains as well as the rings of PAHs
(Mason et al. 2004), and rather similar in the Milky Way, ultraluminous
infrared galaxies, and Seyferts.
\item Fluffy (Cambresy et a1. 2005),
meaning that it emits better than it absorbs, they say, leaving us
in hopes that this doesn't violate one of Kirchhoff's laws or even one
of the laws of thermodynamics.
\item Hollow (Min et al. 2005), an idealized calculation.
\item Frozen, a new component of hydrocarbons, reported by
Simonia (2004) and seen via photoluminescence.
\item Greyer, around ultracompact $\mathrm{H\,II}$ regions (Moore 2005) with
implications for grain sizes more complex than we expected.
\item Heated, in mergers, with details
requiring the power of ALMA and \textsl{Spitzer Space Telescope}
(Xilouris et al. 2004).
\item Greyer also around active galaxies and QSOs (Gaskell et al. 2004),
 but denied by Willott (2005).
\item Deficient in metal-poor galaxies (Galliano et al. 2005) and in damped
Lyman alpha clouds (Junkkarinen et al. 2004).
\item Chemisorbed to account for the 2175\AA\ feature (Fraser et al. 2005a).
\item Multicomponented in both the Milky Way (Rawlings et al. 2005) and
other spirals (Stevens et a1. 2005).
\item Dangerous, in supernova remnants, owing to the presence of iron
needles (Gomez et al. 2005 concerning the Kepler remnant). Well, haven't you
been warned not to stick your hand casually into public wastebaskets, lest
you encounter unshielded, unclean needles?
\end{itemize}

\subsection{Gas Phases and Their Motions}
 Would you like it hot or cold first?
Well, some like it hot, which means the coronal phase now probed with
Ne IX, O VII, O VIII, and such (Yao \& Wang 2005). The idea that there
should be such a phase, made up of partially overlapping supernova
remnants, belongs to Spitzer (1956, the person not the space mission,
though we would be the last to deny him the right to appear in italics).
Other galaxies also have some (Doane et al. 2004). The filling factor is
 more than 10\% but how much more remains unclear.

Our very own local superbubble consists largely of this hot, ionized,
tenuous stuff, though with neutral clouds embedded (Welsh \& Lallement 2005;
Oergerle et al. 2005; Witte 2004; Vallerga et al. 2004). Welsh \& Lallemant
add that they have seen, for the first time, the $10^5$ K interfaces between the
clouds and the $10^6$ K bubble medium and at the bubble surface.
Redfield \& Linsky (2004) reported that the local ISM is subsonically
turbulent with mean velocity 2.24 km/sec.

\textsl{Voyager 1} may actually reach this medium in another 10--20 years.
Meanwhile, 95 AU out in December 2004, it crossed The Terminator (shock) where
supersonic solar wind gives way to subsonic solar wind (Stone et al. 2005 and
the next three papers). You will live to see the final crossing; we will live
to see it; Ed Stone et al. will live to see it;
\textsl{Voyager 1} we are not so sure about.

Less hot ionized gas is called $\mathrm{H\,II}$ (except by a few territorial
members of the Division of Plasma Physics who think it should all be called
plasma, along with stellar interiors, QSO emission line gas, and so forth). The
ionized phases were responsible for most of the very fine scale structure
recognized in the ISM (Lehner \& Howk 2004 on O VI), because the recognition
normally comes from variations in electron density along the line of sight to
pulsars (Hill et al. 2005, Zhou et al. 2005, for instance). But it now seems
fair to say that the neutral hydrogen is at least as particulated
(Brogan et al. 2005 on 10 AU scales).

$\mathrm{H\,I}$ was obviously the phase of the year. It came (a) warped
(Revas \& Pfenniger 2004), (b) cold (Givson et al. 2005), (c) en route to
molecules and with cosmic rays (Giammanco \& Beckman 2005), and, perhaps
most important for the future, (d) increasingly well surveyed
(McClure-Griffiths et al. 2005, writing, or anyhow observing from
the Dominion Radio Astrophysical Observatory) and characterized
(Heiles \& Troland 2005), though this last paper scored 2.5 on our
 ``um'' scale for the remark that ``observed quantities are only indirectly
 related to the intrinsic astronomical ones,'' especially, they say,
 magnetic field strength, for which they report 6 $\mu$G.

Then $\mathrm{H_2}$ forms (more easily on rough grain surfaces,
 Cuppen \& Herbst 2005), and where there is enough of it, you get giant
molecular clouds (Stark \& Lee 2005), dark clouds (catalogued by
Dobashi et al. 2005), and Bok globules, which rotate,
say Gyulbudaghian \& May  (2004) with periods near $10^7$ years.
This seems long, given that characteristic time scales for formation,
destruction, and conversion to stars of GMCs are also a few times
$10^6$ to 1--$2\times 10^7$ years (Bergin et al. 2004, Monaco 2004,
Tassis \& Mouschovias 2004, Goldsmith \& Li 2005).

Most authors during the academic year described the cloud motions
and internal structure as turbulent (Lohmer et al. 2004, one of many),
though Tarakanov (2004) held out for generalized Brownian motion, and he
 blamed the fractal structure (with $n = 2.35$) on clouds bouncing around
 off each other after ejection by stars.

The most puzzling velocity structure remains the high velocity clouds,
and not everything that was written about them during the year can be
simultaneously true of all of  them.
If we are allowed two votes, one will go to the general idea that some are
left-overs from galaxy formation, analogous to the clouds responsible
for certain QSO absorption lines (Maller \& Bullock 2004). A second,
even more diffuse, vote goes to the conclusion that the HVCs associated
with M31 and M33 are not all the same sort of beast (Westmeier et al. 2005).
 The principle alternative to left-overs falling in is gas expelled from
  galactic disks in fountains falling back down.

\subsection{Star Formation}
Subtopics one might reasonably worry about include
time scales, efficiencies, turn-off, triggering,
the special problems of making massive stars (which probably exceed their
Eddington luminosities en route),  effects of turbulence
and magnetic fields, and accounting for the distribution of
masses of single and binary stars (aka IMF) that must result.
Notice that some of these could be answers to some of the others.

The most attractive idea of the year is called ``collect and collapse,'' for
which only the name is new, the idea going back to Elmegreen \& Lada (1977).
It is that a massive star and its $\mathrm{H\,II}$ region can sweep up a great
deal of gas, which will then fragment. Examples are given by
Deharvang et al. (2005), Hosokawa \& Inutsuka (2005), and Oey et al. (2005 on
the W3/W4 region).  Of course that first massive star had to come from
somewhere,  and disagreement persists on whether the normal mechanism is
accretion onto a single core from its surrounding disk and envelope or merger of
several smaller protostars. In lieu of voting this year, we green-dotted
two papers that indicate observational signatures of the two processes
(Lintott et al. 2005, Bally \& Zinnecker 2005).

Pushing the problem back in time to ``cores in molecular clouds'' brings
us to the SCUBA map by Kirk et al. (2005), revealing cores with flat
centers, sharp edges, and molecular masses about the same as their
virial masses of 0.4--4.8 $M_\odot$. Some cores fall apart without ever making
stars (Vazquez-Semadeni et al. 2005a); others are just about to make stars
(Crapsi et al. 2005, some portions of whose argument are not totally obvious);
and in between comes the Balbus-Hawley instability (Padoan et al. 2005), which
deposits stuff onto both the incipient star and its disk. All stages from
starless cores to young clusters can co-exist in a given star formation region
(Teixeira 2005).

The bigger questions received no global answers this year. We think these
are a few incremental steps. Local fields in star formation regions are
in the mG range (Fish et al. 2005), and the efficiency with which gas is
turned into stars declines as the field strength goes up, down to 5\% when the
magnetic pressure exceeds the thermal pressure (Vazquez-Semademi et al. 2005,
a calculation), presumably because the field has to leak out by ambipolar
diffusion (Boss 2005a). This also renders the process somewhat spasmodic
(Tassis \& Mouschovias 2005), a choice of words endearingly reminiscent of
some of the people working in the field.

The main trigger for assembling large molecular clouds remains, we think,
passage of gas through a spiral arm, though if anybody said so this year,
we missed it. On the next scale down, molecular gas is set into contraction,
and there were votes for passage of a globular cluster through the Galactic
molecular gas layer (Kobulnicky et al. 2005) and an intergalactic cloud
hitting a galactic one (Wang et al. 2004), at which point the former ceases
to be a high velocity cloud. Within clouds, important phenomena include
collisions of subclumps  (Koda 2005), plane parallel shocks
(Urquhart et al. 2004), and Type II supernovae (Salvaterra et al. 2004),
whose behavior takes us more or less back to collect and collapse.

Binary star formation is, perhaps, half as well understood as the single sort,
but the most divided author has private reasons for liking the tendency of
Ochi et al. (2005) or at least their model to produce binaries with mass ratios
near one. It should be noted  in this context that turbulent star formation
(Krumholz \& McKee 2005) tends fairly naturally to establish the range of clump
masses seen in NGC 7538 (Reid \& Wilson 2005) and to produce
multiple fragments close enough together to be relevant both to binary formation
(Machida et al. 2005, Clark \& Bonnell 2005) and to ejection of occasional
stars from clusters, perhaps explaining why these runaways tend to have
smaller masses than the stars left behind in the clusters (van den Bergh 2004).

How did most of the globular clusters form? Um, er has been a traditional
answer, since they aren't making them any more, at least in our Galaxy.
But the answer closest to home is that of Kravtsov \& Gnedin (2005), who say
that the immediate parents were giant molecular clouds in the gas disks
of disk galaxies, with baryon mass equal to $10^9$ $M_\odot$, which later
merged to make  the bigger galaxies we see now. The peak formation epoch via
this mechanism   was $z = 3$--5. Big, young non-globular clusters were typically
put together   from a bunch of smaller clusters (Homeier \& Alves 2005;
Chen et al. 2005a).

\section[STARS]{STARS OF STAGE, SCREEN, RADIO, AND Ap05\footnote{The
absence of television and blogs from the list of things to be
stars of dates the section heading to, roughly, pre-1952 and originally
described Jack Benny, whom we still join a few times a year in travelling
to Washington to visit our money.}} As recently as 2001, ``optical observations
of stars'' was still the largest single class of astronomical paper published
world wide. The rule that our own papers are not highlights of the year
precludes citation, but a reprint-preprint package will be sent in plane brown
wrapper to anyone who requests it. Stars also made up 15
of the preliminary topic classes (from YSO to aging neutron stars) for Ap05,
plus four more for binaries and two for star clusters (out of 76 total). The
ordering of topics is the least imaginative possible, from young stellar
objects onward.

\subsection{Young Stellar Objects}
Three  classes, 0, 1, and 2 (or 0, I, and II, well we
didn't say our lack of imagination was unusual in the field) are distinguished.
The class zero objects are supposed still to derive most of their energy from
accretion (Groppi et al. 2004), and the envelop in waiting remains more massive
than the core (Froebrich 2005). Class I (the traditional YSOs) and Class II
(the T Tauri stars) are already dominated by nuclear energy, and only those
with the larger outflows can power Herbig-Haro objects
(White \& Hillenbrand 2004). Vorobyov \& Basu (2005) say that the transition
from Class 0 to Class 1 represents the exhaustion of the reservoir of material
available for accretion. Doppmann et al. (2005) disagree, saying that accretion
and outflow can coexist in Class I.

Models of these phases have improved to the point where it is sometimes
possible to get the same age for a cluster from lithium depletion and from
pre-MS isochrones (Jeffries \& Oliveira 2005 on NGC 7547). Many YSOs are
X-ray sources, and it can be a good way to pick them out. Ozawa et al. (2005)
report Types I, II, and III, post T-Tauri's, as X-ray sources in the Rho
Ophiuchus region. If you care to ask whether the X-rays are produced primarily
by magnetic processes or by accretion shocks, the answer is yes (Preibisch 2004
on magnetic processes, Swartz et al. 2005 on shocks). YSO X-ray sources do not
show conspicuous activity cycles (Pillitteri et al. 2005).

A slightly modified classification scheme says that $\mathrm{0 = all}$ accretion
energy; $\mathrm{1 = accretion}$ + nuclear;
$\mathrm{2 = all}$\ nuclear + outflow; and $\mathrm{3 = junk}$ cleared out of
the way, starting at the inside of the disk (Barsony et al. 2005). Notice that
the energy sources are not directly observable, so that YSOs are generally
studied and classified using somewhat different criteria, for instance growth
and then dissipation of disks over millions of years (Rodriguez et al. 2005),
with faster dissipation at larger masses, dust evolving chemically and settling
to the plane of the gas disk, and small grains disappearing first.
Schutz et al. (2005), Hernandez et al. (2005), Brice\~{n}o et al. (2005a),
Calvet et al. (2005), and Carpenter et al. (2005) are by no means the only
papers on these processes, but their clustered publication makes them
easy to consult.

Conservation of angular momentum from an interstellar blob will
inevitably yield a protostar rotating faster than break-up. Indeed young stars
tend to be fast rotators, but the YSO rotation papers indexed this year all
focussed on spin-down and took three points of view (a) magnetic coupling
to the disk is not how it happens (Matt \& Pudritz 2005, who suggest wind
on open field lines as an alternative), (b) the disk is not the whole story
(Littlefair et al. 2005, comparing NGC 2264 and IC 348), and (c) extraction
of angular momentum by magnetic coupling to a disk is quite a likely mechanism
(Covey et al. 2005).

Which are our favorite YSOs? T Tauri itself, of course,
a bound triple star (Skinner et al. 2004), whose variability on time scales
from 2.68 days to 40 years is explored by Mel'nikov \& Granking (2005). Both
the naked and the dead, sorry, the naked and the classical T Tauri stars, have
magnetic fields in the few kG range, with the sample not, we think, large
enough to tell if there is a systematic difference (Johns-Krull 2004,
O'Sullivan et al. 2005, Symington et al. 2005).

FU Ori and its ilk, for   which the conspicuous flaring is attributed to
accretion disks, but for   which the evidence for companions (planetary or
stellar) is piling up (Malbet et al. 2005, Clarke et al. 2005,
 Grinin et al. 2004).

And anything named for George Herbig, whether Ae, Be, or Ze. None of the
last class turned up during the year, but there did appear (a) the second
Herbig Ae star with a magnetic field near half a kG (Hubrig et al. 2004), and
no decision on whether this is smaller than the T Tauri fields, (b) the second
Herbig Be with an $m = 2$ spiral in its disk
(Quillen et al. 2005),\footnote{Is there also a Hirsch or Eddington number
for the maximum number $N''$ of  astronomical disks with at least
$M = N''$ arms? If so, it must be close to $N'' = 4$.}  (c) 10 Ae  and Be stars
with X-ray luminosities either about the same as the less massive T Tauri YSOs
(Skinner et al. 2004), or, on the other hand, brighter (Hamagushi et al. 2005),
and (d) Ae/Be's with disks either centrally puffed up (Eisner et al. 2004,
reporting resolution with the Palomar Test Bed Interferometer) or
flaring outwards  (Acke et al. 2005). The former are dust disks, the
latter gas, so both could be correct.

\subsection{Brown Dwarfs}
``Brown dwarf'' is widely held to be a good name because (a) brown is not an
(additive) color, (b) the spectra are conspicuously non-thermal,
so that cooler = bluer in some infrared colors, and (c) they are not stars.
Not surprisingly, they are very like stars in some ways, different in others.
Let's start with the idea that BDs can do just about everything that stars can
do.
\begin{itemize}
\item Live alone (Luhman et al. 2005a)
\item Have BD companions (Zapaterio Osorio et al. 2004, Burgasser et a1. 2005,
the latter on brown dwarf pairs orbiting M dwarfs)
\item Be companions (Forveille et al. 2004, Pravdo et al. 2005), with smaller
separations for smaller masses
\item Have planets (Chauvin et al. 2004)
\item Be subdwarfs (Burgasser et al. 2004, the second case)
\item Exhibit resolvable outflows (Whelan et al. 2005), with the
method, called ``spectro\-astrometry'' (line centroid offsets vs. velocity),
explained in detail and fine print
\item Pulsate (Palla \& Baraffe 2005) but
with deuterium fusion as the driver rather than hydrogen ionization and,
admittedly, a calculation rather than an observation
\item Gravitationally lens stars behind them (Jaroszynski et al. 2005,
observations)
 \item Emit X-rays (Stelzer 2004, the second example)
\item Form the same way as low mass main sequence stars (Muzerolle et al. 2005,
 Mohanty et al. 2005, Luhman et al. 2005b), which is to say that the young ones
have accretion disks
\item Be triples (Bouy et a1. 2005)
\item Sustain magnetic fields despite being fully convective.
Berger (2005) reports radio emission from a 2MASS source, arguably
gyrosynchrotron despite the absence of H$\,\alpha$ and X-ray emission. There is
a 3 hour period, which could be rotation or orbit, and the L3.5 object is
J00361317+1821104
\item Have both magnetic fields and accretion when young (Scholz et al. 2005a on
a couple with variable emission lines)
\item Rotate with more or less the same range of periods when young
(Caballero et al. 2004).
\end{itemize}
And now some differences, with the most notorious last.

More weather
(Maiti et al. 2005) in the form of variability due to dust clouds moving
around in the atmosphere.

More complex atmospheric structures that need a
fifth fitting parameter (Tusji et al. 2005), the usual temperature, $\log{g}$,
chemical composition, and microturbulence, plus the thickness of the cloud deck.

Absence of binary X-ray sources. We don't mean that there are not some neutron
star XRBs and ``black widow pulsars'' with $M_2$ in the brown dwarf mass range,
but that close double BDs are not Chandra X-ray (or ACTA radio) sources, where
similar M dwarfs would be (Audard et al. 2005).

Additional formation mechanisms that involve ejection from triples and
clusters before they have had a chance to grow to proper stardom. Compare
Deanna Durbin with Shirley Temple and see Lucas et al. (2005),
Umbreit et al. (2005), and Luhman et al. (2005c), who all also address the
issue of---

The Brown Dwarf Desert. You must not imagine either a water-based organism
 crawling desperately across the surface of a brown dwarf or the BD itself
 calling plaintively for ammonia in the midst of a terrestrial desert,
 but, rather, a distressed observer in an otherwise perfectly nice desert
 that just happens to have very few brown dwarfs in it.

 This doesn't even mean few, total, compared to low mass stars and
 orphan planets in regions where you expect all three (Lucas et al. 2005).
 Instead, if you plot numbers of spectroscopic and visual binaries in
 which $M_2$ = low mass star, brown dwarf, or (hot) Jupiter, there is a deep
 dip in the BD mass range (Kouwenhoven et al. 2005, Chauvin et al.
2005a, Bouchy et al. 2005, Umtreit et al. 2005, Luhman et al. 2005c).
Part of the reason, say Matzner \& Levin (2005), is that their formation
by fragmentation in disks is inhibited by radiation. According to
Padoan \& Nordlund (2004) very large density fluctuations due to supersonic
turbulence are needed to overcome this inhibition

\subsection{A Few Favorite Stars (Besides Jack Benny and Carole Lombard)}
 Some of
these are binaries and probably belong in another section, but we suffer
from the delusion that you can deny the consequent by denying the major
premise, so that the absence of foolish consistency is a hobgoblin of
large minds. Indeed we begin with some newly declared or affirmed binaries.

\textsl{Arcturus} (Verhoelst et al. 2005)

\textsl{FK Comae} (Kjurkchieva \& Marchev 2005)

\textsl{AB Doradus}. Well, it has been a triple for a long time, but the newly
measured mass and luminosity of component C (SM Close et a1. 2005, Reid 2005)
provide a revised calibration of the bottom end of the luminosity-mass
relation. It is fainter and cooler than had been expec£ed for its
mass of 0.09 $M_\odot$ and age of 50 million years. This then casts some doubt
upon the identification of brown dwarfs from luminosity and temperature
alone in open clusters of known age.

$\alpha\,$\textsl{Cen A} has also been accompanied for many years, but it has a
new Zeeman magnetic field of 247 Gauss (Kordi \& Amin 2004). Well, newly
measured anyhow.

\textsl{Fomalhaut}, resolved with the VLT interferometer in what was mostly a
stability test (Davis et al. 2005a). Is it inappropriate to claim as a
favorite telescope something that shares a set of our initials?

\textsl{Altair}, which is not only $\alpha\,$Acq but also the brightest
$\delta\,$Scuti star, displaying
seven modes at amplitudes less than a millimagnitude (Buzasi et al. 2005).
Well, at least most people can pronounce that one.\footnote{Improbable
as it may seem, the most acquisitive author this year actually
turned down money to make a recording of acceptable pronunciations of the
names of a number of stars and constellations on the grounds (a) there are
real differences between the amateur and professional communities even
within the USA and (b) she could think of no honorable purpose
(for either community) for which some of these would be needed.
The potential employer declined to explain what the purpose was.
 But if you should be offered such a recording for a price, stand
 assured that none of us will profit from it.}

\textsl{The most massive stars} (who have our Atkinsonian sympathy) galumph
in between 120 and 200 $M_\odot$, and this is not just a matter of running
out of statistics in a Salpeter  (power-law) initial mass function
 but a real cutoff (Oey \& Clarke 2005, Figer et al. 2005,
Kroupa 2005a).

 \textsl{The most dense stars}, on the other hand, are those of
smallest mass, with as ponderous an average density as  75 $\mathrm{g/cm^3}$
for a 0.092 $M_\odot$ star
(Pont et al. 2005). Surprise at its being this large means it is
time we taught stellar structure again, preferably using the text
book that differs only in middle author from the present paper.

\textsl{The nearest stars}. Remarkably, these continue to proliferate. This
year, the solar system was nearly and newly assaulted by the 16th nearest star
(Deacon et al. 2005, SCR 1845-6357), a white dwarf, at 4 pc probably
closer even than van Maanen 2 (Scholz et al. 2004---and we are a little
vague on the nature of van Maanen 1), and at least a gaggle of others
less than 10 pc away according to Golimowski et al. (2004),
Costa et al. (2005 whose Figure 3 $\mathrm{[Fe/H] = +0.5}$ isochrone should,
we think be --0.5) and Scholz et al. (2005b).

 A bunch of peculiar
 chemical compositions for CP stars, of which we record only the very
 high abundance of tantalum in Chi Lupi (Ivarsson et al. 2004),
 to express sympathy with the original Tantalus crawling over the desert
 looking for a brown dwarf. Or something like that. Oh all right.
 The metal cannot absorb water, and who knows how it does with fruit trees.

Additional items that surprised us:
\begin{itemize}
\item The discovery of a number of faint DY Per stars in the LMC led us to
wonder a few years ago whether DY Per was a DY Per star. (Alcock et al. 2001).
It is (Zacs et al. 2005).
\item More than half of all known yellow hypergiants live in the young
open cluster Westerlund 1 (Clark et al. 2005b).
\item Oe stars exist, but are rather rare compared to Be stars
(Negueruela et al. 2005).
\item The 65 stars originally classified as B[e] are actually a mixed
bag (Miroshnichenko et al. 2005). Mixed heritage is also
characteristic of the sdB and EHB stars (Maxted et al. 2004, the
proceedings of a conference), and of the blue stragglers (Clark et al. 2004),
though it sounds as if Porter \& Townsend (2005)
and Sills et al. (2005) may be advocating their respective mechanisms
(rapid rotation and main sequence binary collisions and mergers)  for all blue
stragglers.
\item Am stars are considerably more common than plain old A's, especially
in binaries (Yushchenko et al. 2004).
\item Pre- and post-main-sequence stars cross the same stretches of the HR
diagram above the zero age main sequence, and it is not always that
easy to tell them apart. Miroshnichenko et al. (2004) have reclassified
HD 35929 from a Herbig Ae, pre-MS star to post-MS.
\item Spectral types by integer steps, for instance from A0 to A9 were enough
for Morgan, Keenan, and Kellman (indeed some of the steps are almost
never used), but Luhman (2004b) has found some stars that require
further subdivision at the level of M2.25, for instance.
\end{itemize}

\subsection{The Sun}
 The Sun?!?! Out, out. Down Bowser. Back to \S2 where you
 belong. But perhaps two small items can creep through the pet door here.

First, the Sun has been losing metals, not primarily to the solar wind
but to theorists. With the new, lower $Z$, it comes exceedingly close to
the average of nearby disk stars, both early (Lyubimkov et al. 2005)
and late (Luck \& Reiter 2005, Taylor \& Croxall 2005). In light of the
well known correlation between high metallicity and hosting planets,
you might wonder whether the Sun has also been losing planets. Not unless
you count Pluto. The (other) disadvantage of reduced metallicity is a
poorer fit to the spectrum of solar oscillations, via effects on the
depth of the convection zone and such (Bahcall et al. 2005).

Second there is the activity cycle. The Sun has been, of late, very
spotty, and, if you can trust tree ring data as a proxy, has not been
this active since about 8,000 BP (before 1950, you may recall from Ap04).
There has been a corresponding increase in solar luminance at earth of
about 1 $\mathrm{W/m^2}$ since the Maunder minimum (Wang et al. 2005b, who are
not completely clear about what data go into their record of solar luminosity
and magnetic field since 1713). And, predict two of the papers that call
attention to the present vigorous activity, the future will be less so,
perhaps by 2010 (Ogurtsov 2005), perhaps not until more like 2100
(Solanki et al. 2004, Reimer 2004).

Carrington's numbers for his elements (of the solar rotation and spot poles)
were very nearly right (Beck \& Giles 2005), a confirmation which would have
excited the fastest rotating author more if she had known they were
in doubt. Carrington, you may recall, was the first to see a white
light solar flare and not blame it on the beer produced by the family firm.

\subsection{Pulsating Stars}
Twenty-some classes of pulsating variables appeared
in the 2005 literature? Should keeping track of them be aspersed as
botany? No! It's stamp collecting, and we're proud of it. In addition,
detailed matching of pulsation properties is often the best handle on
equations of state, opacities, and convective energy transport.
The Cepheids are perhaps the most important since they
anchor the extragalactic distance scale. In this
context it would be more comforting if---

1. There were general agreement about just how the
period-luminosity-color relation depended on composition (Perssen et al. 2004
vs. Gieren et al. 2005). Age comes in there somewhere too  (Bono et al. 2005).
The good news on this front is confirmation of the number near 1.3 used to
convert observed radial velocities to photospheric motion in the
 Baade-Wesselink method of measuring Cepheid  luminosities provided by
Merand et al. (2005) who used CHARA to measure $R(t)$.

2. The relationship between masses from evolutionary tracks and
from pulsation analyses  were more like an identity than it is.
The pulsation masses always come out smaller,
which, since we all believe that massive stars shed at various times
should be fine (Brocato et al. 2004), apart from the
tiresome detail that the required mass loss is smallest for
the biggest stars (Caputo et al. 2005, and noted in our summary as ``odd'').

3. More of them showed increases in period as expected for a
first crossing of the Hertzsprung gap, instead of
the negative $dP/dt$'s found by Moskalik \& Dzimbowski (2005).
Polaris and a few others do have large positive period derivatives,
and Polaris itself,
as has been much advertized, also has a smaller pulsation amplitude
than in the good old days. Turner et al. (2005) conclude that it is
at the red edge of the fundamental instability strip for first
crossers, where convection is winning over pulsation. For more of the
effects of convection-pulsation coupling see Grigahcene et al. (2005),
Dupret et al. (2005), both focussed on Gamma Dor and Delta Scuti stars,
and Munteanu et al. (2005) on LPVs.

If you aren't massive enough to be a Cepheid, you can be a Pop II Cepheid,
an anomalous Cepheid, a Type II Cepheid, or a short period Cepheid
(Caputo et al. 2004). Pritzl et al. (2005) are clear that at least
three of these are physically distinct classes (with Type II Cepheids
approximately equal to the old class of W Virginis stars), but less
clear on how you know what to call one when you meet it. Adopting the
technique that has, over the years, led to friendly relations with a
number of outstanding senior graduate students, we will stick with Dr. Cephei.

If you are even more mass challenged,\footnote{Don't you wish.}
you can be an RR Lyrae star, about whose masses (Cacciari et al. 2005) and
period changes (Derekas et al. 2004) much can be said, as in the case of the
Cepheids. The OGLE samples of RR Lyrae stars in the Large Magellanic Clouds and
in the Milky Way are now large enough that the manifestations of the Blazhko
effect can be correlated with other properties (Smolec 2005). The differences
are not primarily a metallicity effect, though the metal-poorest stars are
brightest. The shortest known Blazhko period, 7.23 days, belongs to RR Gem
(Jursik et al. 2005).

Each year a few stars get promoted from mere low-amplitude pulsation to
asteroseismology. This year we caught $\theta$ Oph (which is also a Beta Cephei
star) observed by Handler et al. (2005) and Briquet et al. (2005), who
are polite enough to cite each other. And Procyon has been restored to the
pantheon, by Claudi et al. (2005, radial velocity data, not to mention a whole
conference on the subject, Kurtz et al. 2005 and 21 following papers, some of
which report data from places where even Jay Pasachoff has never
been).\footnote{We held a small, informal competition for this naming
opportunity, and the eclipse chasing and other activities of Jay Pasachoff were
deemed to have taken him to more, and more difficult, places than even the
outreaching of Edward Sion, the site testing of Jan Erik Solheim, and
the crescent moon viewing of Brad Schaefer.}

Every other pulsation class is rudely confined to one paper each.
LSIV $-14^\circ$ 116 is the first pulsating, helium-rich sdB
(Ahmad \& Jeffery 2005). The non-He rich sort are now called V361 Hya stars if
they have periods of minutes and PG 1716+427 stars if they have periods of
hours (Ramachandran et al. 2004). GW Vir stars are the same as pulsating
PG 1159 stars, from which you can deduce that there are at least two of them;
in fact about 10 say Nagel \& Kwerner (2004),  with the blue edge of the
instability strip at 160,000K. Of the 10, four are type DOV white dwarfs
($\log{g} \ge 7$) and six are nuclei of planetary nebulae ($\log{g} = 1$--6).

Pulsation of Be stars is responsible for
 some of the periodic optical variability of Be X-ray binaries (Fabrychy 2005).
 Like the wheels on the stuff that is tall and skinny and
green and grows around houses, this was just added to make it more difficult,
though not by us. What? Oh. Grass.

Delta Scuti stars and Gamma Doradus stars live in the same part of the HR
diagram (near the A main sequence). Does any star do both? Candidates come and
go (Chapellier et al. 2004, Henry et al. 2005), and HD 8801
(Henry \& Fekel 2005) is this year's candidate. It is an Am star without known
binary companion (relatively rare). In fact, many chemically peculiar stars
pulsate, or, if you prefer, many Delta Scuti stars are spectral types
Am-Fm (Yushchenko et al. 2005a). RV Tauri stars are known primarily for
their minima of alternating depth. Their chemical peculiarities are best
described as dust/gas separation. (Giridhar et al. 2005).

The semi-regular (asymptotic giant branch)
variables occupy multiple ridge lines in HR diagrams, period-luminosity,
period-temperature, and other diagrams. Papers on the subject this year
(Schultheis et al. 2004 and several others) left us no wiser than Ap04
\S 4.12, but of the opinion that ``chaos'' would be a good description,
 except that it has to be saved for a handful of large-amplitude stars
 whose behavior is chaotic (not stochastic and not the sum of a bunch of
 constant-amplitude modes) in the technical sense (Buchler et al. 2004).

 A star approaching the main sequence is quite likely to pass through one
 or more instability strips, and indeed a few of them
 show periodic, low-amplitude variability (Kwintz et al. 2005), though
 they seem to avoid the center of the instability strip in NGC 6383.

Whether the radii of Miras vary through their light cycles depends on
whether you use infrared interferometry (yes, Boboltz \& Wittkowski 2005)
or the SiO maser emission (no, same paper), though both sets of photons
come from 8--11 arc-sec from the centers of the stars.

Some of the elliptically modulated variables in the LMC also show a
``long secondary period.'' Soszynski et al. (2004) do not claim this as a
pulsation phenomenon, but it's new this year and so has to go somewhere!

Despite the conciseness of the paper, Kopacki (2005) managed to mention
SX Phe stars, Pop II Cepheids, BL Her stars, red giant tip stars, and
RR Lyraes, all in M13, all variable, and all apparently pulsating.

Like each other class, the Beta Cephei stars get only one paper
(Davis et al. 2005) of a handful recorded, but you get two stars, because
it is a $P = 557$ day binary (with both interferometric and radial velocity
data) consisting of two 9 $M_\odot$, $\mathrm{M_V = -3.8}$-ish stars, both of
which are Beta Ceph variables and Bl giants. Several earlier papers have
reported Beta Ceph masses, and they always seem to be very close to 9
$M_\odot$. The variability of Beta Cephei itself was discovered by Frost (1902),
who selected the name Beta Canis Majoris stars for the class, a confusion
which has persisted down to the present time.
The paper by Daszynska-Daszkiewicz \& Niemczura (2005) on them is cited for
the shear challenge of spelling the names correctly.

\subsection{Some Other Favorite Stars}
Many of these are variable. Some probably
pulsate. But they appeared in the index year papers for some other reason.

Pugach (2004) points to some variable stars whose correlation of B-V color vs V
magnitude has the opposite sign to what you would expect from temperature
variations.

Vogt et al. (2004) have discovered that if you wait long enough
(34 years in their case) you find variability on time scales up to
8000 days or more. It will presumably take another 34 years or so
to determine periodicity if any.

R CrB stars, though known firstly for fading and secondly for
pulsating, also have mass outflow from both disk and bipolar structure,
like other AGB stars (Rao et al. 2004).

Eta Carinae, the prototypical luminous blue variable, gets two papers
because it is generally advertized as a binary. In one, it displays a
new sort of pumped $\mathrm{Fe\,II}$ emission (Johansson \& Letokhan 2004),
and in the other, the Homonculus Nebula around it is reflecting X-rays for
the first time (Corcoran et al. 2004). Some other LBVs are also pulsators
at periods longer than the expected fundamental (Dziembowski \& Slavineka 2005).
The implication is that we are seeing strange modes driven by an iron opacity
bump. The stars must be in the helium core burning phase and have lost a good
deal of mass.

Though we have cast nasturtiums at the multiple ridge lines of
 semi-regular variables (for more, see Fraser et al. 2005), some day
 they are all going to have individual names, though the largest class
  may remain ``other'' as it currently is for 6616 of the 10,311
stars catalogued by Pojmanski \& Maciejewski (2005).

Asymptotic giant branch stars lose mass. That's what they are best at,
and the details depend on metallicity (Marshall et al. 2004), or only
on luminosity and temperature, not on composition (van Loon et al. 2005),
or on luminosity, radius, mass, temperature, and surface gravity
(Schroeder \& Cuntz 2005), which is probably enough parameters to
take care of what might really be extra radiation pressure on grains when
there are more metals to condense.\footnote{Yes, there are five, and you
were expecting Gamow's elephant, weren't you?} Garcia-Segura et al. (2005)
believe  that magnetic fields are also important in driving the higher-speed
collimated winds, though we and Schoenberner et al. (2005) endorse a
prolonged, low velocity superwind as the main mass stealer. In any case,
stripping goes so deep that the layer seen can briefly be as hot as 200,000K
(Werner \& Drake 2005 for H1504+65 and a number of new candidates).

And pretty soon, if they are very good, they get to be planetary nebulae.
 Actually the key parameter is probably not goodness but some combination of
 mass and mass loss (because, as remarked in earlier years, some stars go
 from extended horizontal branch to white dwarfs without ever being nuclei
 of PNe). Definitive distance scales for Galactic planetaries have been
 established every couple of years. Phillips (2005a) has done it again.
His is on the short end of the existing range. We are long folks ourselves,
but as Phillips' references extend back only to 1992, it is inevitable that
our own great works on the subject, as well as those of Josef Shkovsky and
Michael Feast, go uncited. The observations also say that his set of PNNs
 have no main sequence companions earlier than K0.
The various observed PN shapes are not primarily an evolutionary sequence
but reflect the initial stellar mass, hence, presumably, the amount of
stuff available (many papers, of which Phillips 2005b can be the
representative). The bipolars come from the biggest stars.

What you see is not always what you get. The systematic difference
between element abundances found from collisionally excited and
recombination lines (recombination is bigger) implies large quantities of
cold, hydrogen-poor, gas hiding somewhere (Wesson et al. 2005 the last
of a number of papers on this during the year). Liu et al. (2004) suggest
that vaporized planets might be responsible, making this the 29th way of
detecting exoplanets.

Given the wide range of PN progenitors (all stars up to $8\pm2 M_\odot$ and
some merged binaries, Ciardullo et al. 2005), lifetimes must vary a
good deal, but a thousand years will do as an average.
Sabbadin et al. (2005) report on NGC 6741, which, at age 1400 yr
has already been recombining since year 12 of the revolution.
And then they get to be---

\subsection{White Dwarfs}
 Fifty-one papers on WDs were indexed (plus some on
possible Type Ia progenitors and cataclysmic variables exiled to
other pages). A very old issue is the DA/DB (H vs. He atmosphere)
dichotomy and how individual stars decide which to be when. No, the
answer didn't come in this year, but if you have only one green dot
to give out, it should probably go to Kalirai et al (2005a) for the
 remarkable discovery that there are no DBs in young open clusters.
 An out of period result extends unexpected WD populations to older
 open clusters and a globular or two. That WDs with M-dwarf companions,
 with 5\% DBs, fall half way in between single field WDs (10\% DBs) and
 cluster stars (0\% He) must be trying to tell us something about this
 phenomenon (van den Besselaar et al. 2005). The paper reports numbers
 3--15 of DB + MV binaries and numbers 2 and 3 of DC + MV. A good many
 of their DBs live in the temperature gap at $T = 30,000$--45,000K where there
 are few or no single  white dwarfs with  helium atmosphere.

The correlation between main sequence mass and white dwarf mass matters
for figuring out how much stuff they give back to the interstellar medium,
for deciding how many core collapse supernovae you should get from a
given initial mass function, and other issues in stellar populations.
A critical part of the calibration is study of white dwarfs in young
open clusters. This year, Williams et al. (2004) and M35 told us that
stars up to at least 5.8 $M_\odot$ make WDs. The biggest main sequence stars
make the biggest white dwarfs. Low metallicity yields more massive WDs
from a given MS mass (Kalirai et al. 2005b). And the magnetic WDs are
on average a good deal more massive (0.93 vs. 0.66 $M_\odot$) than the others
 Wickramasinghe \& Ferrario (2005).

As long as the magnetic fields have crept in, this is probably the place
to note that one implication of the Wickramasinghe \& Ferrario results is
that WD fields are fossils left over from their main sequence lives rather
than the products of ongoing dynamos.
Tout et al. (2004) concur, and Jorday et al. (2005) report that 4 of 4
nuclei of planetary nebulae have field strengths intermediate between
those of Ap/Bp stars and WDs, though their statistical conclusions about
flux loss should perhaps await a slightly larger sample.

White dwarfs in cataclysmic binaries more often than singles have strong fields.
Townsley \& Bildsten (2005) say you can make sense out of the pattern
if the birthrate of strong fields is 8\% for both, but the ones in CVs
last longer. If so, then the total absence of detectable fields in WDs
paired with M dwarfs that are not CVs is remarkable (Liebert et a1. 2005).
The authors suggest it is a selection effect against high mass, small
radius (faint WDs) in their sample, which comes from SDSS. The strongest
white dwarf field reported to data is a GG (GigaGauss,
Vanlandingham et a1. 2005) and intrudes on the neutron star range.

Some of the DB fields are strong enough to prevent hydrogen accretion via a
prope11er mechanism, while allowing interstellar metals to get in as grains
(Friedrich et a1. 2004).
Does this at last solve the problem of how there can be helium-atmosphere
WDs with significant metal abundances? No, because other DB fields are weaker.
But having landed in the heavy e1ement swamp, let's look around a bit.

It is possible to find (holding our ear trumpets up to try to catch your
question) astrofo1k who remember when white dwarfs came with H or He
surfaces plus van Maanen 2 with some iron. Then arrived the PG 1159
stars, with mostly C+O, which they retain, until gravitational settling
changes them to DOs and DAs (Gautschy et a1. 2005, a pulsation calculation).
They are also allowed a bit of H or He if they want (Vauc1air et a1. 2005,
also on pulsations). But the advent of high resolution UV spectroscopy made
clear that heavy elements are actually quite common and that, for
hydrogen-dominated atmospheres, accretion of interstellar material
provides a reasonab1e exp1anation (Koester et a1. 2005a, 2005b,
Gianninas et al. 2004 on L157, a 1.24 $M_\odot$ star). The heaviest, rarest
element seen so far is germanium (Vennes et a1. 2005), in three DAs, and the
Ge abundance is nearly solar.

The chiefest puzzle is presented by helium + metals. Petitc1erc et a1. (2005)
reviewed four possible mechanisms: radiative levitation, mixing or dredge up
from the CO (etc.) interior, ISM accretion with the hydrogen batted away, or
left overs from the PG 1159 phase. They vote for this fourth for the stars
they studied with FUSE. Where hydrogen and helium coexist, the surface ratio
 can be inhomogeneous (Pereira et al. 2005), and $\mathrm{He\,H^+}$ can be a
significant opacity source (Harris et al. 2004).

White dwarfs are, in general, slow
rotators (Karl et a1. 2005 and other papers stretching back 40 or more
years). So are at least some of the sdB progenitors (Charpinet at a1. 2005).
Ferrario \& Wickramasinghe (2005) suggest that WDs are not really any slower
than neutron stars in relation to what is possible, and that the magnetic
fields are also analogous, but their fast rotator has a period of 700 sec.
They propose that it is a merger product (one of three possible initia1
conditions for white dwarfs), and the faithful envelop back says
this is ``fast'' in the same way that a 0.7 sec pulsar is ``fast,'' given
the respective break-up periods near 1 sec and 1 msec. Their ``slow''
is 50-100 years, corresponding to weeks or months for a neutron star
(not in the observed, or observable, range).

White dwarfs can pulsate. You have already met the GW Vir
(pulsating PG 1159) stars. The ZZ Cetis are the ones with hydrogen
atmospheres, and, in contrast to Ap04 (\S4.10), which declared the
instability strip to be free of
non-variables, this year it is impure (Mukadam et al. 2004,
Mullally et al. 2005).

White dwarfs necessarily cool and fade as they age, and, if we
fully understood the underlying physics, the process could be
used quite independent of main sequence turn-offs to measure the ages
of stellar populations. Alternatively, one can try to decide whether
one understands the processes by seeing whether the ages come out the same.
A major glitch occurs when the interior nuclei settle into a crystal and
lock up  in zero point
fluctuations much of the energy that could otherwise be radiated
(Mestel \& Ruderman 1967). For only one star is the crystallization
expected within the ZZ Ceti strip, and its period spacing suggests
that half or thereabouts of the interior has indeed crystallized
(Kanaan et al. 2005, Brassard \& Fontaine 2005).

Beyond this point, the white dwarf sequence in globular cluster M4
yields a perfectly reasonable 12.1 Gyr (Hansen et al. 2004).
Less satisfactory is the case of the old open cluster NGC 6791, with
a main sequence turnoff age of 9 Gyr and a white dwarf age (from the
shape of the luminosity distribution) of 2.4 Gyr (Bedin et al. 2005).
It is also somehow difficult to assign reliable effective temperatures
to the cooler WDs (Farihi 2005) and so to decide which cooling curve they
should be compared to (Jao et a1. 2005).

Given the local density of $\mathrm{0.0l/pc^3}$ (Pirzkal et a1. 2005), there
should  anyhow be an adequate number to study, and the range in masses in the
least biased samples available is wider than some thinkle peep
(Liebert et al. 2005a, Ferrario et al. 2005).

\subsection{Active Stars}
 Well, they all are, at some level, but the section
addresses manifestations, correlations, cycles, causes and such.
First, the broadest sort of definition: stellar activity is taken to
mean chromospheres and coronae, star spots, flares at whatever
wavelength you happen to be looking at, (and) other evidence for
strong magnetic fields and winds. Solar activity so defined
results in the Sun controlling space out to the  edge
of the heliosphere. We caught for the first time this year
``astrospheres'' used to mean the same thing for other stars, plus
evidence for their existence (Wood et al. 2005a).
Detection requires there to be some neutral gas around the star
for the wind to interact with. Wood et al. picked out 10 of 17 stars
 within 10 pc and only 3 of 31 further away via Ly$\,\alpha$ absorption at the
 interface. And since they absolutely had to look through our
 heliosphere to see the other stars, that registers in many absorption
  features as well. Some of the systematics arise because very active
  stars have mostly polar spots leading to bipolar winds, while for
  the Sun (Tu et al. 2005) the wind starts out from coronal
  holes at a range of latitudes.

\subsubsection{Magnetic Fields}
Now it is a truth universally acknowledged that a single star
in possession of a good field must be in want of a dynamo. Many are to be found
in Peter \& Stix (2005, the proceedings
of an October, 2004 conference). And this is in truth relevant
because it is also generally acknowledged that stellar activity
is largely driven by magnetic processes. Stellar dynamos in turn
require both rotation and convection, the former being as a rule
the easier to measure independently and so the topic of more papers.

Dynamos, it seems, can exist in stars ranging all the way from
brown dwarf masses (\S 9.2) up to O stars, though $\theta^1$ Ori C is
the only O with a measured field (Gagne et al. 2005), and the change
from a shell dynamo (in stars with radiative core and convective
atmosphere) to a distributed one (in fully convective stars) changes
spot numbers or distributions (Scholz et al. 2005). At the other,
high mass, end Mullan \& MacDonald (2005) say they can get up to
100--200G, approaching the observed range for OB stars. The dynamo
fields are not always nice, tidy dipoles. The example that jumped
 out this year (Petit et al. 2005) is the binary component Sigma Boo A,
 a G8V star with both poloidal and toroidal field and rotation
 period = 6.4 days. The orbit period is much longer, and you would
 not expect synchronization unless the system were a few Tyr old
 (see Abt \& Boonyarak 2004 on the range of periods for which
 synchronization does occur).

The dynamo process must also in some sense be self-defeating.
A strong dynamo and field will drive out more wind (at least up
 to some limit) and keep it co-rotating further out, thus slowing
 the rotation, as is generally said to have happened to the Sun.
 The two papers noted on this general topic, however, said that
  magnetic braking is not the reason that many chemically peculiar
  stars are slow rotators (Glagolevskiu 2004) and that the expected
  faster rotation at a given mass for LMC and SMC stars compared to the
  Milky Way (fewer heavy elements = less radiative pressure wind
  driving = less slowing) is not actually seen (Penny et al. 2004).

\subsubsection{Rotation}
 Rotation does, however, in general slow with age (after the
   initial spin-up during accretion mentioned above, and see
   Strom et al. 2005a on h and $\chi$ Persei spin-up compared to field stars).
   The slowing can be both calculated and measured and the expected
   die-off of activity observed (Telleschi et al. 2005, Ribas et al. 2005).
   Both papers compare solar mass stars over a range of ages, and part of
   the purpose is to ``predict'' what the Sun must have been like in the past
   and plausible effects on the solar system as a habitable environment.

Just how slow can rotation get? The 521 days for HD 2453 discussed
by Glagolevskij (2004a, including a Mercator projection of the surface)
is a more direct measurement than the lower limit of 37 years deduced for
one roAp star by Byabchikova et al. (2005).

Differential rotation is a topic likely to appeal to track runners,
especially if phrased as the number of rotations required for the
equator to lap the poles by one.
This happens in just 3 or 4 periods for some A stars with shallow
convection zones (Reiners et al. 2005), and also in a few periods for
the Sun (but this is 120 days rather than the 40 hours for the A stars),
and takes more than 400 of the 0.424 day rotation periods of LO Per
(Barnes et al. 2005). Yes, of course there are models that redistribute
internal angular momentum to produce the differential rotation seen
(Charbonnel \& Talon 2005). We worried a bit that the dynamos might
suffer, but the effect seems to go the other way---the dynamos disturb
the differential rotation (Covas et al. 2005, Itoh et al. 2005).

To summarize again, yes, activity does die away (Gondoin 2005 on NGC 188
vs. younger clusters), but you will not be any more surprised than we
 were to be told that both the early spin up/down processes
 (Lamm et al. 2005) and the later evolution of activity are more
 complicated than generally supposed. Indeed Pace \& Pasquini (2004)
 despair and say that at least the Ca II K flux cannot be used as an age
 indicator at all. Silvestri et al. (2005), however, rode, well,
  published to the rescue with a calibration that extends beyond
  4 Gyr (based on old open clusters and MV + WD pairs dated with
   WD cooling ages) on out to 10 Gyr. The Vaughan-Preston (1980) gap,
    meaning the absence of nearby stars with activity levels between
    those of the Sun and Hyades is just more of that tall, skinny,
    green stuff with wheels growing around observatories.

\subsubsection{Cycles and Other Periodic Behavior}
 What about stellar activity cycles?
 Well, they are bound to appeal to those who have always resented Paris
  being a movable feast.\footnote{No, the least cosmopolitan author does not
understand what this means,
 unless that different people will most appreciate Paris at different
 times in their lives. If so, she hasn't reached hers yet. Traditional
movable feasts drift around the calendar (like Easter, and unlike Christmas).}
That AB Dor (single young star
near 1 $M_\odot$) has two cycle periods of 20 and 5.5 yr (Jarvinene et al. 2005)
goes some way to make up for the stars found by Hall \& Lockwood (2004)
which have none. For some of their  10 not-even-unicycles,
the H and K emission is larger than the flux at solar minimum. To decide whether
the stars are experiencing the equivalent of a Maunder minimum, one really
needs to wait patiently and see the transition back to cyclic activity
in a few to a few hundred years. In and out of Maunder (etc.) minimum
is a long sort of aperiodic cycle, we suppose. There are also short
ones. UX Ari (a sort of RS CVn binary) for instance joins the Sun in
 having a Rieger cycle (294 days) which Massi et al. (2005) attribute
 to trapped Rossby waves.

``Active longitudes'' is the idea that spots, excess magnetic flux, and
chromospheric  emission may tend to pop out at the same place
repeatedly over very long periods of time (e.g., Alekseev \& Kozlova 2004 on
LQ Hya, a single dMe star). The behavior of $\mathrm{K^1}$ Ceti
(Rucinski et al. 2004) may mean that both an active longitude and
the pattern of differential rotation have been stable for 30 years.
It is, however, one of the five dwarf novae (of which WZ Sge is the best
known) whose orbit period has bounced back from the minimum possible
and is now increasing and so should probably not be taken as a model
for anything else. Should one actually believe in active longitudes?
The evidence in the case of the Sun is an artefact (Pelt et al. 2005).
This does not mean that the phenomenon isn't there, only that the current
data don't demonstrate it (compare, again, the yeast effect).
The authors do not address whether other stars might also be misleading us.

A few details of how the magnetic, rotational, and convective energies
are fed to chromospheres and coronae so that we can see
them remain to be worked out. This surely has a better chance of happening
for the Sun than for stars where observations have little or no angular
resolution, though sometimes stellar astronomers rush in where solar
ones fear to tread: (a) a review of coronal X-rays (Guedel 2004), (b)
the first extra-solar flare X-ray oscillations, on AT Mic (MVe,
Mitra-Kraev et al. 2005), (c) an assortment of possible correlations
 with magnetic field, mass, age, and metallicity (Lyra et al. 2005),
 not, we suspect, separable correlations.

Some observations whose authors surely meant to be helpful: (a) EV Lac,
a dMe, flares at radio, optical, and X-ray wavelengths, but not all at
the same time (Osten et al. 2005), (b) normal single stars behave as if
they have acoustic flux plus a uniform distribution of magnetic flux
tubes (Cardini 2005); RS CVn, BY Dor, and similar classes do not.

And one of our favorite paper pairings from the year. The ergodic theorem
applies to single AB Dor and binary V471 Tau (that is, only rotation and
$T_\mathrm{e}$ matter, not how the stars got that way,
Garcia-Alvarez et al. 2005);
and the ergodic theorem does not apply to HD 283572 (a weak lined T Tauri)
and 31 Comae (in the Hertzsprung gap) and EK Dra (ZAMS), all in the same
region of the HR diagram, but their coronae know the difference
(Scelsi et al. 2005).

\subsection{Stellar Physics and Evolution}
The physics contemplated here is largely
that which goes into the three auxiliary relationships of the four equations
of stellar structure (radiative opacities, equation of state, nuclear
reaction rates) and the convective version of the equation for
temperature equilibrium and energy transport.

\subsubsection{Opacities}
 The standard remark about opacities is that real stars always
seem to need more than the theorists can find. This year, not true for A stars
in the UV (Garcia-Gil et al. 2005) but otherwise fairly pervasive
(Ramiez \& Melendez 2005). Depending on the context, water
(Jones et al. 2005), other molecules and dust (Ferguson et al. 2005),
and magnetic line broadening (Kochukhov et al. 2005) are likely
to be contributors .

Most disquieting, the problem of insufficient opacity has spread to
the Sun, now that its CNO abundance has dropped to the local stellar
average (Bahcall et al. 2005, Turck-Chieze et al. 2004). Two recent
 major compilations of calculated opacities  (OPAL, Opacity Project)
 do not differ enough for it to matter which one you use
 (Badnell et al. 2005). One can just about restore equilibrium by
 assuming a neon abundance also equal to the average of nearby stars
 and gas (Drake \& Testa 2005) and by pushing your choices of initial
 heavy element abundances, opacities for them, and diffusion effects
 all in the same direction (Guzik et al. 2005). The required equilibrium
 is being able to calculate observed helioseismological frequencies
 from the same model that reproduces the observed solar neutrino flux.

\subsubsection{Equation of State}
 We caught only one positive statement, that
the standard model for the Sun is good enough (Young \& Arnett 2005),
and no negative ones.

\subsubsection{Nuclear Reaction Rates and Cross Sections}
 Nuclear astrophysics
has advanced to the stage where uncertainties in the major energy-producing
 reactions do not dominate our (mis)understanding of anything, with the
  exception, it seems of triple alpha building of $\mathrm{{}^{12}C}$.
  Fynbo et al. (2005) and
  El Eid (2005) report a change in laboratory data for a strong
  resonance at 11 MeV above the ground state of the product nucleus,
  whose effect is to increase the triple alpha rate at AGB temperatures
  and decrease it in supernovae.
Because of the competition between $3\alpha\rightarrow\mathrm{{}^{12}C}$ and
$\mathrm{{}^{12}C}(\alpha,\gamma)\mathrm{{}^{16}O}$,
 one effect should be a somewhat larger C/O ratio both for massive stars
  heading into later nuclear reaction phases and for the white dwarfs
  that are Type Ia supernova progenitors. The final C/O ratio in the
  universe at present is somewhat less sensitive to the triple alpha
   reaction rate than has been suggested in the past (Schlattl et al. 2004),
    a result of potential importance in anthropic considerations (\S 7, above).

Among other reactions, the NeNa and MgAl cycles, high temperature analogies
of the CN cycle, must occur because the right mix of products is seen in
assorted red giants (Antipova et al. 2005). Authorities disagree about
whether (Werner et al. 2005) or not (Lugaro et al. 2004) expected reactions
can produce as much fluorine as some RGs have. This is, however, primarily
 a matter of getting sufficient mixing between the H and He burning shells
  rather than of reaction rates. The dominant process is
\begin{displaymath}
    \mathrm{{}^{14}N}(\alpha,\gamma)\mathrm{{}^{18}F}(\beta^+)
  \mathrm{{}^{18}O(p,}\alpha)\mathrm{{}^{15}N}(\alpha,\gamma)\mathrm{{}^{19}F}.
\end{displaymath}

Other nucleosynthetic issues nearly neglected here include the
existence and source of primary nitrogen and which sorts of stars
make most of the neon, carbon, and various isotopes of oxygen. This
leaves us with lithium, as frequently seems to be the case. It is required that
some stars produce it (the Li-rich red giants) and, arguably, that others
destroy it (those on the lithium plateau in globular clusters). In both cases,
rotationally induced mixing is helpful (Steinhauer \& Deliyannis 2004 for main
sequence stars, Denissenkov \& Herwig for red giants), and recall
Cameron \& Fowler (1971) concerning transport of $\mathrm{{}^7Be}$, the form in
which the $\mathrm{{}^7Li}$ must be made in stars, which leads us to---

\subsubsection{Extra Mixing}
Rotationally induced mixing and, perhaps, other sorts in excess of that expected
from the simplest  mixing length theory (which carries energy and stuff just to
the point where the radiative temperature  gradient drops to
the adiabatic value and then stops) has been declared necessary
in a number of contexts. Pasquini et a1. (2004) discuss lithium
yet again and Origlia et al. (2005) the ratio $\mathrm{{}^{12}C/{}^{13}C}$
in a couple of open clusters. As for how it happens, Mathis et al. (2004)
address shear-induced turbulence from differential rotation and
Young \& Arnett (2005) convection-induced mixing into radiative
regions in close binaries. A test of the extent of overshoot, via
its effect on the C/O ratio in white dwarfs, should be possible
from changes in ZZ Ceti pulsation frequencies when their cores
crystallize (Corsico 2005), provided of course that one has understood
all the other physics that enters into determining C/O.

Within mixing length theory, mild surprise was occasioned by (a)
values of the mixing length to pressure scale height ratio less than
 one for a couple of exoplanet hosts (Fernandez \& Santos 2004)
 because it is 1.63 for the Sun and (b) the existence of no fewer
 than four sorts of scale length in three-dimensional stars
 (Kaplyla et al. 2005). If your reaction is that three-dimensional
 stars are the only sort you have seen, then probably we
should back off and say three-dimensional models of stars.

 Although the Sun is the only star on which we can
 resolve small convective elements as a test of theory, rising and
 falling gas distorts the line profiles of other stars in ways that
 are now more or less understood (Gray 2005).

\subsubsection{p, s, r, and $\nu$ processes}
Well, these have to go somewhere, and, since they happen in stars, here
they are in with the Marx Brothers (about whose amusingness your authors
disagree) and all. The p-process makes proton-rich isotopes of heavy elements
and, say Hayakawa et a1. (2004), what really happens is knocking loose of
neutrons by energetic photons. The s-process happens in stars less metal poor
than $\mathrm{[Fe/H]} = -2.6$ (Simmerer et al. 2005), and its products have now
been seen in a globular cluster and the nearly-merged galaxy IGI
(Caffeu et al. 2005, firsts for both, say the authors).

The r-process makes the isotopes with more neutrons than the most tightly
bound ones at the bottom of the valley of beta stability. Where does it
happen? We caught one vote for NS + BH mergers (de Donder \& Vanbeveren 2004)
 and two votes for proto-neutron star winds in core collapse supernovae
 (Suzuki \& Nagataki 2005, Kohri et al. 2005).

A few rare isotopes, e.g. $\mathrm{{}^{45}Sc}$ and $\mathrm{{}^{49}Ti}$,
can be made only by neutrino interactions.
Pruet et al. (2005) recommend hot bubbles with p/n ratio $>$ 1 in
supernovae. As long as
this is not a travel recommendation, we are perfectly happy to take it.

\subsubsection{Real Time Stellar Evolution}
We take this to mean significant changes in a time less than, say,
the sum of the ages of the authors, with a courtesy extension to
events rather longer ago than that, which, however, happened at a
sharply enough defined times that they could have been observed as RTSE
events if astronomers had been looking with suitable tools. Examples
of this extension include the Becklin-Neugebauer object, which must have
started its escape from the $\theta^1$ Ori region about 500 years ago
(Rodriguez et al. 2005a) and HD 56126, which stopped being an AGB
star 1240 years ago and is now a proto-planetary nebula (Meixner et al. 2004).

The 2005 examples of evolutionary changes in our combined lifetimes
 have nearly all appeared in previous editions of ApXX and so get
 only one paper each this time around.
FG Sge was first, and it would seem that the fun is now over,
since it has looked like a typical AGB star for the past decade,
with constant luminosity, appropriate mass loss rate, and so
forth (Gehrz et al. 2005). Two other (probable) examples of very
late helium flashes and associated major changes in surface temperature,
 composition, and luminosity are V605 Aql (Lechner \& Kmeswinger 2004)
  and V4334 Sgr (Sakurai's object) which should loop back to the red
  again in about 2250 (Hajduk et al. 2005), with special recognition
  for the commentary by Asplund (2005), who mentioned Our Book.

The status of McNeil's nebula is less certain. Ojha et al. (2005) this
year claim it as an EXOr or FUOr (young, variable accretion disk) and
three papers uncited concur (two) or at least don't discur (one).

V838 Mon appeared in the most notebooked papers (six) and with the
least certain status. Several make firm statements, though none so firm
as the colleague who told us insufficiently privately at a conference
that Ap04 was full of banana chips for doubting the planetary
companion hypothesis, so the pale green star goes this year to
Banerjee et al. (2005) for their firm ``cause and energy source still
unknown.''

Groenewegen (2004), however, received the coveted pink
  blot for having noticed that 3 of 2277 known S/M/C stars have
  switched from O rich to C rich in the last 20 years (a comparison of
OGLE, 2MASS, and DENIS data sets). If you will get
out your abacus\footnote{The darkest-haired author owns at least four
of these and remains ashamed to admit to her Uncle Hugo and
 Aunt Hepzibah, a.k.a. Mitsunori and Nobuko Kawagoye, that she has
 still not learned to use any of them.}
you can verify that this means one such switch per star in 15,000 years.
The author posits last thermal pulses of helium burning that just
happen not to have been seen as FG Sge-type events. The sample stars
are in the Magellanic Clouds, where producing C stars is relatively
 easy because of the smaller initial amount of O to be overcome.

\subsubsection{Slower Evolution}
 This is the sort calculated from the standard
four coupled non-linear differential equations or deduced from the appearance
of Hertzsprung-Russell diagrams and their ilk. Important public service in this
context consists of calculating and publishing large numbers of evolutionary
tracks like those of Claret (2004) for 0.8--125 $M_\odot$ solar composition
stars, using the best available opacities,  nuclear reaction networks, and
equations of state, and reasonable amounts of mass loss and convective
overshoot. The larger masses are followed up to the end of carbon burning.
Nearly everything turns out well.

Now comes the less happy part. There are discrepancies between the
masses of stars estimated from spectroscopic criteria and from evolutionary
tracks. The spectroscopic masses are larger for FGK dwarfs
(Valenti \& Fischer 2005a) and smaller for early O stars (Massey et al. 2005).
And there are stars like Nu Eri, with a number of Beta Ceph type modes, whose
full range of properties cannot be
fit with the same model from any one set (Ausseloos et a1. 2004).

If you want to attempt stellar  pulsation synthesis to compare with
the integrated spectra of a bunch of stars, you need to connect the $L$,
$T_\mathrm{e}$, and composition of your models with the observed quantities
color, magnitude, and 1ine intensities using a reference set of
individual stellar spectra, observed or calculated (Martins et al. 2005).
Detailed questions someone might want to ask or answer include:

 a. Do evolved stars sometimes loop back blueward of the red giant region? Yes,
 sometimes say the calculations of Ventura \& Castellani (2005) and the
 observation by Southworth et al. (2004) of V621 Per.

b. Do stellar collisions and mergers ever really matter? Yes, say
Laycock \& Sills (2005) in a paper that got a multicolored
dot,\footnote{Remember we also own a collection of raisin labels.}
because, by including pre-main-sequence
stars in their repertoire (for instance pre-MS + WD merger) they
were able to produce high mass horizontal branch stars and other
curiosities.

c. What is the minimum mass a star must start with to
survive the superwind phase and  carbon ignition and evolve on to
core collapse? Well, it is different for single stars and primaries
of close binaries which are likely to lose lots of mass to their companions.
So the mass cut is larger in close binaries, right? Wrong, say
Podsiadlowski et al (2004). Typical numbers are 10--12 $M_\odot$ for single
stars and 6--8 $M_\odot$ for close binaries. The cut also comes at smaller
masses for small metallicity and larger overshoot, again backwards
from what at least one of us would have guessed. And to top it off,
the neutron stars they liberate from close binaries go off with
smaller kick velocities than do the singletons, as a result of
differences in the supernova onset process.

At this point, nothing could surprise us, and so we end the
subsection with a few thoughts on the initial mass function,
which may remind you of the few words said by Prof.~Dumbledore
before the opening-of-term dinner in Harry Potter I.\footnote{These were
``nitwit, blubber, oddment, tweak'' and it has only recently occurred to us
that the purpose may have been to elude a magic spam filter, confirming evidence
being the absence of Spam \copyright from the extensive menu listed for the
dinner. The next notebook line is a description of the first camel
race run with robotic jockeys.}

What is the IMF?
The distribution of stellar masses, $N(M)$, on the ZAMS at some
formation event. Are binaries as a rule included properly? No.
What does it look like? A power law or right hand side of a
Gaussian at large mass, a flattening or turn over somewhere
below 1 $M_\odot$, and a real turn down into the brown dwarf desert.
Is it everywhere the same? The traditional answer, pertaining
largely to the upper mass end, is yes, and this persists at least
approximately for big young clusters  that will eventually be
globular clusters in two dIrr galaxies (Larsen et al. 2004). But
there were also some no's, with star burst clusters having a
larger maximum star mass than the field in the same galaxies
(Chandar et al. 2005), though this could mean simply that the
biggest stars don't live long enough to be liberated.

Variations on the low-mass end of the IMF are better established.
The peak is near 0.8 $M_\odot$ in the Taurus star formation region vs.
0.1--0.2 $M_\odot$ in Orion (Luhman et al. 2004). The Cha I
region also has fewer 0.1--0.3 $M_\odot$; stars than Orion
(Feigelson \& Lawson  2004). And three young c1usters
studied by Barrado y Navascues et al. (2004) all have some brown
dwarfs but a gap in $N(M)$ near 0.05 $M_\odot$.

Is the IMF understood? Yes, by several groups, though it is not quite certain
that they all have the same understanding. Bate \& Bonnell (2005)
emphasize your old friend the Jeans mass (for the peak) and
stochastic processes (for the slope of the high mass end),
while Larson (2005) and Jappsen et al. (2005) emphasize
other properties of the gas, cooling processes and turbulence respectively.

\subsection{Binary Stars}
To say that binary stars don't get no respect
would be an exaggeration as well as a double negative. But they don't
get as much respect as they deserve, making up more than half of all
stars, at least outside the M dwarf range, but only 4 of 19 pages in our index.
Neglect of binaries in a stellar population will make you think
it is younger than it really is (Zhang et al. 2005a, Xin \& Deng 2005)
and will lead you astray in matters of chemical evolution as well
(de Donder \& Van Beveren 2005). Many interesting astronomical
entities are found only in binaries, including cataclysmic variables,
X-ray sources, and recycled pulsars. Nelson et al. (2005) have
provided a very extensive set of evolutionary models that include
white dwarf cooling, shell flashes, and much else and lead to no
disastrous contradictions with observed populations.

Evolution of individual systems was historically done by assuming
conservation of both mass and angular momentum. This is no longer
the norm. Petrovic et al. (2005) is a fairly random sample relevant
to Wolf-Rayet binaries. And, in a nod to another very old problem,
 Williamon et al. (2005) have found that the dumpee during the first
 phase of Roche lobe overflow mass transfer settles down to become
 a normal star again fairly soon.
According to folklore, the first person to try to follow the mass
recipient was a Berkeley graduate student named Benson who, in
despair, abandoned the problem and astronomy.

On the observational side, a whole handful of green dots, reproduced
 here in black and white (because we pay our own page charges)---

Really good numbers (measured and derived) for mass, luminosity,
temperature, and age of the Sirius AB system (Liebert et al. 2005b).
Given a system age of 225--250 Myr and a white dwarf cooling time
of $124\pm 10$ Myr, the initial primary must of had a pre-WD
lifetime of 91--136 Myr and so an initial mass of $5.056\pm0.3$
$M_\odot$. And theory with calibration says that a 5 $M_\odot$ star should
leave a 1 $M_\odot$ WD, which is just what you find there. That
Sirius may not be a member of the Sirius supercluster
(King et al. 2003) is  just one of those things that are sent to try us.

The largest dynamical masses we've seen in a long time, 80 + 80 $M_\odot$ for a
pair of Wolf-Rayet stars (Raux et al. 2005).

In another echo of the distant past, the revived suggestion
that population II stars have a smaller fraction of initial
binaries than do population I stars. Carney (2005) says 10\%
for halo stars on retrograde orbits vs. 28\% for Pop I stars in
the same range of binary periods etc, though to end up with the
10\% binaries now seen in the core of 47 Tuc, Ivanova et al. (2005)
say the initial incidence must have been very close to 100\%.

 Binary Cepheids, of which there were zero when we were children,
now add up to at least 18, of which 8 or 9 are actually triples
(Evans et al. 2005a).

 On the subject of triples and quads in
general, about 1/3 of the weak hierarchical systems examined by
Orlov \& Zhuchkov (2005) will not last more than a million years,
suggestive of recent capture, perturbation, or exile from their
parent clusters. Apparently lots of now single and double stars
 have come up (or down?) through this route. Data for a range of
 young star samples imply that a typical formation event yields
 2--3 stars per core (Duchene et al. 2004, Ratzka et al. 2005),
 while calculations yield 5--10 (Goodwin \& Kroupa 2005), a
  configuration not often observed.

 And now the green dot you faithful readers have both been waiting
for, another study of the distribution of binary system mass ratios
with a second peak at $M_2/M_1\approx1$ as well as the peak at small values
(Fisher et al. 2005). We find ourselves cited but not credited with
the once anathematic idea that the distribution is bimodal.
If you would like to look for yourself, the largest sample
available is the 9th (mostly on-line)
catalog of the orbits of spectroscopic binaries (Pourbaix et al. 2004).
The previous, 8th, catalog appeared on paper in 1989.

Because stars are not actually point masses or rigid spheres, members
of a pair drag on each other, gradually rotate the line of apsides
(Wolf \& Zejda 2005), and, in due course, circularize  the orbits
(Meibom \& Mathlieu 2005, who find that older populations are
circularized out to  longer orbit periods), synchronize rotation
and orbit periods (Abt \& Boonyarack 2004), and bring together into close
orbits star pairs that must have  been wider when they formed
(Guenther et al. 2005 on BS Indi, at about 0.435 day and 30 Myr the
shortest period young binary). Ivanov \& Papaloizou (2004) give
an improved model for tidal circularization and synchronization
for the case of low mass stars, which are fully convective with
turbulence providing the viscosity.

Ah, but wipe that smile off your face (or wherever you customarily
carry your smile). Many of the calculations and often the
interpretation of data assume that stellar rotation and orbit
axes are parallel. This is wrong for about half the systems
examined, with apsidal motion data, by Petrova \& Orlov (2004).

And please wave goodbye to the following sorts of binary system,
since we will not mention them again, at least until Ap06---

\textsl{Cool Algols} (Mader 2005 on AV Del, the 7th with an orbit, the
first having come from Popper 1976). Also Algols with superhumps,
which, like CV superhumps are caused by a beat frequency between
the orbit and disk precession frequencies (Retter et al. 2005).

\textsl{RS CVn} systems with apparent abundance anomalies, for which
Morel et al. (2004) say that non-LTE effects are a better bet.

\textsl{A Beta Lyrae} system with material still flowing from M1 to M2
(Djurassevic et al. 2004).

\textsl{A W UMa} star on the way to becoming an
FK Comae (rapidly rotating giant) or blue straggler (Qian et al. 2005).

\textsl{V471 Tau}, the prototype of the V 471 Tau stars, now fading and
perhaps perturbed by a 3rd star (Ibanoglue et al. 2005). And,
since V 471 Tau stars these days are more often called pre-CVs,

$\circ$ The cataclysmic variables. What? A whole notebook page with
30 highlighted papers gets one lousy hollow bullet? Yup, and 4 papers.

\textsl{One} for the single nova in a local dwarf galaxy (Neill \& Shara 2005),
which is a lot for that observation, we promise.

\textsl{Two} for GK Per (nova 1901), which now looks a good bit like Cas A and
might be called a CNR (no, not a French sponsoring agency,
but a classical nova remnant by analogy with
SNR (Anupama \& Kantharia 2005, Balman 2005).

\textsl{Three} for a fixed amount of accreted hydrogen as the trigger for
nova explosions (Schaefer 2005) from a very clever examination
of the three recurrent novae that have exploded at least three
times in recorded history (another version, we suppose, of
the Hirsch citation or Eddington bicycle number).

\textsl{Four} for R Aqr, which these days is a mere symbiotic star, but
which experienced two outbursts in 1073 and 1074 CE (recorded
as guest stars in Korean history) plus two more in 1350 and 1814
that launched expanding shells (Yang et al. 2005).

\subsection{Population III, First Lights, and Reionization}
These topics were belabored in Ap04 (\S 7) and Ap03 (\S 3.8)
and not another word do you get until another year, two years, three
years\ldots of WMAP data have been released and interpreted (or,
perhaps, interpreted and released). But perhaps just a few numbers.
Most of the early metals came from Pop III stars of 10--50 $M_\odot$,
while most of the reionizing UV and black holes came from 50--140 $M_\odot$
stars (Tumlinson et al. 2004, Diagne et al. 2004). But the metals turn
off Pop III star formation when only 1/3 of the necessary photons
have been generated (Matteucci \& Calura 2005).

Less than half the early UV comes from QSOs say
Dijkstra et al. (2004). Or is it more complicated, with up to
$z = 15$ dominated by soft X-rays from accretion on black holes,
then stars taking over to wipe out remaining high density gas at
$z =6$--7 (Ricotti et al. 2005)? And we all know that the current
intergalactic ionization is largely maintained by AGNs.
0\% QSO light at the beginning would suit Malhotra et al. (2005).

And there we will leave the issue until someone wins the
Vera \& Robert Rubin Prize for spectroscopic confirmation of a
$z>7$ source, WMAP speaks again, or the Messiah comes
(on whichever number visit you are anticipating).

\subsection{Stars in Clusters---Open}
The smallest cluster contains one star and is reached via ``the cluster
richness distribution  [which is] continuous down to the smallest cluster,
consisting of  one'' (de Wit et al. 2005). This way of looking at things
accounts  for the distribution of O stars amongst clusters, runaways
(at least half), and 4\% non-runaway field stars.
Clusters with mass less than 25 $M_\odot$ exist and cannot contain any
early O stars. Their demise is also not the main source of field
stars (Soares et al. 2005). Mechanisms for producing runaway stars
were first put forward by Poveda (1967), which is relatively easy to
find, and Ambartsumyan (1938, which we put forward as a challenge to
search engines). If, as above, a single star can count as a cluster,
then each of the  following topics is described in a cluster
of papers.
\begin{itemize}
\item The closest moving group, centered   literally if not
physically on AB Dor (Zuckerman 2004), contains about 30 other stars.
\item The TW Hya moving group really consists of two associations
of different ages (Lawson \& Crause 2005).
\item There are moving groups that are not remnants but the products of
transient spiral waves sweeping existing stars into dynamical groups
(Famaey et al. 2005).
\item The Pleiades is part of one of these sweep-up operations
(Quillen et al. 2005).
\item In h and $\chi$ Persei, only the former shows mass segregation
(Bragg \& Kenyon 2005), another dynamical process unless it
is primordial, and watch out for the use in this paper of
``Hubble law'' to mean the density distribution in the cluster.
\item Other moving groups have escaped from larger clusters
(Chumak et al. 2005).
\item Open groups do fall apart, concerning
which we highlighted 8 papers. You get only the one that
points out the cluster can still be identified for a long
time after it ceases to be bound (Fellhauer \& Heggie 2005).
\item Gould's belt seems to be an example of this process,
currently expanding but still identifiable (Bobylev 2004).
\item The star formation efficiency determines whether a cluster
is ever really bound, if you start with a virialized cloud.
Hard1y ever, say Tilley \& Pudritz (2004).
\end{itemize}

If you would like to initiate a project on open clusters yourself,
Karchenko et al. (2005) provide a handy catalog of new ones, typically richer
than the 520 previously known. All clusters contain some
binaries, and we think the range found (15--54\%,
Bica \& Bonatto 2005) is too large to be just statistical
fluctuations and selection effects. There must be real differences, and if you
ask whether these are primordial or the result of dynamical evolution in
the cluster, the answer will surely be ``yes.'' Some real differences among
clusters in their initial mass functions inc1ude substructure and breaks
(Pollard et al. 2005 on M10 and Balaguer-Nuneq et al. 2005
on the sum of 5 clusters).

Stars tend to huddle (cluster may be too technical a term)
around the centers of galaxies. There are clearly two or more
types of such huddles. Sarzi et al. 2005 report stars mostly $\ge$ l Gyr
old at centers of 23 spirals. Walcher et al. (2005) discuss
younger ones with $5\times 10^5$--$6\times 10^7$ $M_\odot$ within 5 pc of the
centers of late spirals which, if stripped, would then look
like ultra-compact dwarfs or big globular clusters.
Some ellipticals also have central clusters of $10^{6-8}$ $M_\odot$,
$10^7$--$2\times 10^8$ yr, blue stars (Elmegreen et al. 2005). A mechanism for
forming the young stars very near the center of the Milky Way was proposed by
Christopher et al. (2005) and one for forming young star
clusters from galactic accretion disks by Nayakshin (2005).
And you will surely agree that the subject of nuclear star
clusters deserves more serious reviewing than it gets here
(ARA\&A new CEO, are you listening?).

\subsection{Stars in Clusters---Globular}
Perhaps one ought to start by distinguishing globular clusters from other sorts
of astronomical objects. This was easy as long as only Milky Way
populations were under consideration, but is no longer
entirely possible. For instance, M31 has some collections
of stars that are intermediate between dSph galaxies and
globular clusters (Huxor et al. 2005). And a question almost as old as the most
pack-rattish author's Toyota\footnote{We are indebted to the Faustian
Acquaintance for the information that the Maxwell, Jack Benny's transport of
choice, were manufactured until the year of FA's birth  (though not
in the same country), so that Saxon, the 1980 Toyota, may well
be about the same age as the Maxwell, which vanished when Benny
made the transition from radio to television, for the excellent
reason that it was played by Mel Blanc, who did not look much
like J.C. or any other Maxwell.} is whether the clusters with very large numbers
of massive stars forming today are ``really'' young globulars. This obviously
asks for a prediction of what will still be there in 10 Gyr or so.
The answer, according to our predictors of the year, is ``yes''
for some (de Grijs et al. 2005), but by no means all (Eggers et al. 2005). Less
extrapolation is needed as the clusters age, and the case is firm for
intermediate age globular clusters (1--5 Gyr) in early type galaxies
(Hemple \& Kissler-Patic 2004). A confirmed hierarchist would say that they are
the product of the last major merger during assemblage of the  galaxies. Don't
wait too long to go back and check, though, even globular
clusters can die (Koch et. al. 2004).

If you've seen one globular cluster, you've seen them all? This
was once nearly true (and probably still is through
Messier's 4-inch refractor). But if you also measure
luminosity, characteristic radius, location and velocity in
the host galaxy, age, metallicity, and $[\alpha/\mathrm{Fe}]$ or some other
deviation from solar heavy element mix (or even a sort of
product of age and composition called color), the clusters
of nearly every galaxy separate into two or more populations.
These can be studied either for their own sakes or as guides to
the galaxy formation process. The Milky Way now has three
populations (Mackey \& Gilmore 2004) representing about
7 merger events and a major initial collapse. A similar
monolith + sacrifice of dwarf spheroidals scenario is discussed
by van den Bergh \& Mackey (2004) and by de Angeli et al. (2005).

Great big elliptical galaxies have great big cluster systems,
with metallicity extending up to solar and a wide range of all
properties (Woodley et al. 2005 on Cen A, Forbes et
al. 2004 on M60, and Brodie et al. 2005 on NGC 4365, with probably
three populations of different composition).

The Galactic globular cluster of the year was, once again,
$\omega$ Cen, with papers referring to its unsavory past as a dwarf
elliptical galaxy, to the remarkably large value of helium
abundance in some of its stars, and its unfittable white
dwarf population (Ideta \& Makino 2004, Piotto et al.
2005, Monelli et al. 2005). And the maximum complexity
star goes to Sollima et al. (2005) for identifying five
separate populations of red giants in $\omega$ Cen with different
ages, metallicities, and kinematics.

The multiplicity of pulsars in 47 Tuc (Ransom et al. 2005, one with a mass as
large as 1.68 $M_\odot$) and in Terzan 5 (Ransom 2005, the current
record-holder) pail by comparison (you need a bucket to carry all the
preprints), and fade into the general, long-standing problem of whether there
are enough X-ray binaries in the clusters to give rise to all the
recycled and binary pulsars seen (Ivanova et al. 2005a).
In case you aspire to settle this observationally, the
field of 47 Tuc actually contains about 300 X-ray sources
(Heinke et al. 2005), but 70 are background sources, about 25
the msec pulsars themselves, and most of the rest are
cataclysmic variables and chromospherically active binaries.

The better you get to know globular clusters, the more
types you find, and we allude cheerfully to the
multiplication of entities beyond the two original Osterhof types,
called I and II curiously (Contreras 2005 on M62
with its 200+ RR Lyraes; Castellani et al.
2005 on M3) without aspiring to tell you the cause.

The second parameter problem means an attempt to assign
cause(s) to the range of horizontal branch morphologies
among clusters with the same overall metallicity.
This year, the discussions should probably be described
as ``presented'' rather than ``voted for'': Caloi \& D'Antona (2005)
on helium abundance, which requires a population of stars that
produce  $\Delta\mathrm{Y}/\Delta\mathrm{Z} = \infty$; Cho et al. (2005)
on CNO/Fe  variations; Zhao \& Bailyn (2005) on fraction of
close binaries; and  Smith (2005a) on deep mixing.

\section{BETTER MOUSETRAPS, SQUARE WHEELS, AND DOGS' DINNERS}
The first of these are generally regarded as good (assuming you want
mice to beat a path to your door), the second as bad (unless you have
misinterpreted the consequences of the kinetic coefficient of friction
being smaller than the static), and the third as rather a mix (at least
in cultures where dogs were fed table scraps\footnote{The Faustian Acquaintance
has recently acquired a dog, who, being named Pele, eats, we presume,
Brazil nuts.}). This section contains
some of each, and your authors feel that, by \S10, they are already
in enough trouble without stating which is which.

\subsection{Widgets}
One ought to be able to distinguish widgets that actually exist from plans
(and we will try to do so), but there are borderline cases. A contract has been
 signed and casting begun for the first of the 8.4 m mirrors needed for the
Giant Magellan (he was only 5'4'' you say?) Telescope, but you should not try to
apply for observing time just yet (Schechter 2005, Anonymous 2005e). OWL, the
OverWhelmingly Large telescope, received another official blessing
(Gilmozzi 2005), but no mirror segments have been cast yet.

Among existing devices, and starting with the longest wavelengths,
we welcome the increasing productivity of the Giant Meter Radio Telescope
(BASI 32, 191, and following papers, the proceedings of a conference
honoring Govind Swarup's 75th birthday). In the submillimeter regime,
first results came from a new array (Ho et al. 2004 and the next 17 papers)
and from a portable submillimeter telescope (Oka et al. 2005),
meaning the wavelength, not the size, which is 18 cm. It has been
used at the Atacama site at 4840 m (15,880 ft, at which height your
most oxygen-challenged author can't even do derivatives).
MINT has been observing the cosmic microwave background at 2.1 mm
from Cerro Toco (Fowler et al. 2005). And Motohara et al. (2005) have
carried out a submm study of a $z = 2.565$ dwarf galaxy that will end up
with less than $10^{10}$ $M_\odot$ of stars when all the gas is gone.
They used a Zwicky telescope (gravitational lensing). A submm telescope
made with four parabolic cylinder reflectors is in the planning stages
(Balasubramanyam 2004).

Also in the transition from planning to
 construction is the lower-frequency LOFAR which will, it seems,
 go ahead on more than one site (Kassim 2005), including The
Netherlands plus Germany, Western China, and NW Australia. The plural,
we think, is LOFARIM. The square kilometer array isn't that big yet,
but has already defined a number of key projects (Carilli \& Rawlins 2004,
proceedings of a conference).

Optical astronomy remains more than half of the observational
total, and mirror coatings last longer if purged with very dry air
(Roberts et al. 2005). We experienced an out-of-period, back to the
future moment in reading that the Gemini north mirror will be coated with silver
rather than aluminum next time around.

Partial adaptive optics (unlike half an eye, according to the
intelligent design folks) is actually useful (Tokovinin 2004).
New methods of wavefront sensing were proposed by Bharmal et al. (2005)
and by Oti et al. (2005). A laser guide star is now in use at Keck
(Melbourne et al. 2005).
The superconducting tunnel junction detector (whose advent we
hailed a few years ago) is approaching routine use at the
William Herschel Telescope (Reynolds et al. 2005a with the
AM Her nature of V2301 Oph among its discoveries).
CHARA is now using all six telescopes (McAlister et al. 2005)
to measure shapes of rotating stars, gravity darkening, and such,
while the VLT tries to fool mother nature with a new sort of
coronagraph, a four-quadrant~phase mask (Boccaletti 2004).

If robotic telescopes are still not quite routine, conferences on them have
become so (Strassmeier \& Hessman 2005, and the following papers). A sort of
giant speckle interferometer with a balloon-borne  focal camera called Carlina
has seen fringes on Venus (Coroller et al. 2004). A 1500 m effective aperture
with adaptive coronagraph could image planets like Jupter out to a few parsecs.
Carlina lives with the robotic telescopes because live volunteers for ascending
to the focal plane are likely to be scarce.

GALEX, an ultraviolet survey instrument, launched in April 2003,
reported back in a set of 31 papers (Martin et al. 2005a and the next 30).

At the highest gamma ray energies (where photons reveal themselves by doing
horrible things in the upper atmosphere), HESS in Namibia yielded so many papers
this year that it hardly feels new. All can be recognized in our reference list
and elsewhere as Aharonian et al. (200$\aleph$).

VERITAS was the subject of a conference (Swordy \& Fortson 2004),
but with the site  somehow permanently under attack, perhaps
it is time to fill it with wine and declare
``in veritas vino.''\footnote{This thought somehow involved
us in an extended discussion with our
advisory committee on the identity of the best vintage ever.
The Faustian Acquaintance  voted for Haut Brion 1964.
A name-tagless bearded AAS participant advocated the 1961.
The Keen Amateur Dentist doesn't drink, making him a marvelous person
 to sit next to at conference dinners where the glasses are filled
 automatically. And the Medical Musician responded with an incomprehensible
  anecdote concerning a very elaborate meal served in a Paris penthouse by
  an Enron executive. Mr. H. had been consuming some less prestigious
  vintage, making him unavailable for comment.}
(Northcutt 2005 on the difficulties of using an O'odham site.)

The search for gravitational radiation soldiers on. Frossati (2005) has
designed a spherical detector to operate at 0.068 K, at Leiden, and at
a cost of 3 milliLIGO. The one that cost 1 LIGO reported an upper limit on flux
from pulsars  (Abbott et al. 2005), while the AURIGA bar set a limit on emission
from the giant flare of SGR 1806-20 of $10^{-5}$ $M_\odot c^2$
(Baggio et al. 2005). And an award for subtle courage goes to
\textsl{Physics Today} and \textsl{Nature} this year.
The volume by Collins (2004) devotes much of the 2nd half of its 864 pages
to how LIGO won out over all other technologies, to the sorrow, distress,
and sometimes damage to careers of their proponents. The greatest loser was
arguably Ronald
Drever of Caltech. \textsl{Physics Today} invited him to review the book.
The first half consists largely of unkind (and sometimes untrue)
remarks about Joseph Weber and the searches he carried out
(for astrophysical neutrinos as well as gravitational radiation).
And \textsl{Nature}
invited his widow to review the book.\footnote{By analogy with Winnie the
Pooh, who lived under the name of
Sanders, she has always lived under the name of Trimble, and
yes, like \textsl{The Horn Blows at Midnight} and SN 1987A
as current events, there must be a whole new generation who will never know.}
Both reviews are a good deal more restrained
than you might have expected.

Time standards get ever better (Diddems et al. 2004 and several
following papers).
They start somewhere around the klepsydra era, as do Davis (2004)
on photographic emulsions and Taylor \& Joner (2005) on photometry
of the Hyades.

Widgets designed for use in the laboratory have made or imitated
(1) aurorae (Pederson \& Gerken 2005), (2) Herbig-Haro objects
(Lebedev et al. 2004) and other sorts of magnetic jets
(Lebedev et al. 2005), and (3) grain alignments analogous to the
Davis-Greenstein alignment of interstellar grains. Abbas et al. (2004)
 used micron-sized non-spherical grains illuminated by lasers and
 rotating at 1--22 kHz. But they need to use much stronger ambient
 magnetic fields than are present in the ISM for the grains to align
 this year or decade. The purpose of the laser illumination is to
 permit measurement of the rotation rate. And foots are aplan
 (should this be feet?) to produce ``Hawking radiation in an
 electromagnetic waveguide'' (Schutzhold \& Unruh 2005).

\subsection{Forces Majeures}
Each year there are, of course, people who want to
improve human understanding of the universe by abolishing relativity in
favor of Newton or even Galileo, quantum mechanics in favor of
diceless play, and thermodynamics in favor of free lunches. But most of them
do not publish in the journals we read. Thus the time machines of 2005
(or 1905 or 2105?, Ori 2005), as well as the violations of Lorenz invariance
(Alfaro 2005), and the entities that might challenge the laws of thermodynamics
(Barnich \& Compere 2005) and improvements of the Michelson-Morley experiment
(Antonini 2005) in our notebooks were relatively innocuous.

The four forces were all alive and well during the year, at least
at low energy and redshift. Gravity always wins, both in numbers
of papers and in dominating the structure of large things made
spherical (authors and readers excluded) and so comes at the end.
The nuclear shell model is still useful (Caurier et al. 2005),
though the magic numbers are different for nuclides whose neutron
numbers are either much larger or much smaller than that of the
most stable nuclide of the same atomic weight
(Fridmann et al. 2005, Janssens 2005). And we had never noticed
that no stable nucleus has exactly 19 neutrons (plus
vice neutrons etc.).

The weak interaction remains sufficiently weak
that one is always glad to hear that anything has been detected.
Believe it or not, $\mathrm{GdCl_3}$ dissolved in water makes a good Cerenkov
detector for antineutrinos (Beacom \& Vagins 2004, with a special
thanks to the second author for the vial of this very sour salt
that lives on our bookcase; it is not noticeably poisonous).
The neutrino flavor switching seen by the MINOS experiment
between Fermi Lab (source) and the Soudan mine in Minnesota
(detector, Anonymous 2005k) does not seem to have sour as any
of the available flavors, which have been coupled by a new,
bestever value of $\sin^2\theta_\mathrm{w}= 0.2397\pm 0.0017$
(Czarnecki \& Marciana 2005).
This is said to be  the last experiment that will be
performed at SLAC. A second nuclide has exhibited double
beta decay, $\mathrm{{}^{54}Zn}$ (Blank et al. 2004). The
first was $\mathrm{{}^{45}Fe}$ ending
up as Cr. This is not the magical sort of double beta decay that
would imply some neutrinos are their own antiparticles (majorons)
but the plain old difficult sort, in which two neutrinos must be emitted.

 On the electromagnetic front, the unanswered (or multiply answered)
question of the  decade or thereabouts is whether
$\alpha=\mathrm{e^2}/\hbar c$ was
different in the part of the past explored by QSOs at large redshift.
Six papers voted during the year, of which we cite only
Levshakov et al. (2005) riding both horses in midstream to report
that their VLT sample of spectra between $z = 1.88$ and now shows no
evidence for change, but the earlier Keck sample does. Other things
that didn't change much during the year were the proton-electron mass
ratio (Ivanchik et a1. 2005), the charge on the photon, less than
$3\times 10^{-33}$ of that on the electron (Kobychev \& Popov 2005), and the
sizes of the electron orbits herded by Maeda et a1 (2005). Their whip
is applied radiation at the 13--19 GHz that would be the orbital
frequency if lithium electrons were classical horses.

Gravity being the weakest force required the largest number of
indexed papers (24) to keep it together. If you are having only
one thought on the subject this year, it should probably be that
general relativity continues to triumph over its enemies (Williams et a1.
2004 on lunar laser ranging, Stairs et al. 2004 on PSR 131534+12).
Some GR effects that appear as expected include Lens-Thirring
precession (Miller \& Homan 2005, from a BHXRB not Gravity Probe B),
gravitational radiation (Espaillat 2005, from the CV ES Ceti, not LIGO),
 ergospheres in Seyfert galaxies (Niedzwiecki 2005, from spectrum fitting,
  not visits), and black holes in higher-dimension supergravity
  (Eluang et al. 2005, Gibbons et al. 2004,
from calculations, not measurements).

Nemiroff (2005) suggests that one might be able to see the
gravitational lensing of the gravitational force itself (if you
should happen to find yourself 24 AU from a transparent sun).
Various limits were set to secular changes in $G$, the gravitational
coupling constant (Pitjeva 2005). We suspect that the number,
$\dot{G}/G \le-2\pm 5\times 10^{-12}$/yr probably applies to $GM$ of the sun,
rather than $G$ alone. Thus it might suffer a glitch when the kilogram is
redefined in terms of the Planck constant or Avogadro's number
(Mills et al. 2005), instead of in terms of a chunk of metal in Paris
(which has always
been at risk of small additions or subtractions during those
movable feasts). At least five non-GR descriptions of gravity
also appeared, of which loop quantum gravity (Mulryne et al. 2005)
appears to be the most conventional, and inhomogeneous gravity
(Clifton et al. 2005) the least.

\subsection{Physics of the Early Universe}
 There must once have been a
quark-gluon plasma. Whether this has been recreated in accelerator
experiments remains to be determined (Wisczek 2005, Aronson 2005).
Baryogenesis obviously also happened and has definitely not been
duplicated in the laboratory, so that we are all made of
13.7 Gyr year old baryons. Four possible mechanisms appeared in the
reference journals of which the most mysterious is that of
 Davoudiasl et al. (2004), in which there is a gravitational interaction
 between the derivative of the Ricci curvature scalar and the baryon
 number current in the expanding universe. This breaks CPT
 (charge-parity-time-reversal) invariance and, with baryon number
 violation, can make the  observed baryon to photon ratio of
 $6\times 10^{-10}$.

An expert has assured us all that ``\ldots the world is a multi-colored,
multi-layered'' superconductor of Higgs condensate (Wilczek 200Sa).
 Whether this contradicts the earlier conclusion that all the world's a
 stage remains to be determined. And, as for what you have to look
  forward to, another expert opines that ``\ldots a physical theory of
  everything should at least contain the seeds of an explanation
  of consciousness'' (Penrose 2005).

\subsection{The Forces at Work}
 Gravity comes first in this round, since it
always wins one way or another. One way it has won in galaxies over
the years is called violent relaxation (Lynden-Bell 1967) and green
stars and stripes for the recognition that a new dynamical theory is
needed, because the existing one is not, as it were, transitive, for
successive processes, $\mathrm{A+B}\neq\mathrm{B+A}$
(Arad \& Lynden-Bell 2005). Readers will
 perhaps have noticed that dots, stripes, and other graffiti are often
 awarded to authors who change their minds or correct their own
 mistakes. This is probably not an adequate motivation for
 deliberately publishing a wrong paper.

Where gravity meets electromagnetism, you find some of the
traditional instabilities and also the extraction of energy
from black holes. The two-stream or Kelvin-Helmholtz instability
may happen for the two interpenetrating superfluids in neutron
stars if the relative velocity of the two fluids is large enough
(Andersson et al. 2004). Numerical simulations of ordinary
Bondi-Hoyle-Lyttleton accretion show instabilities whose underlying
physics is unclear (Foglizzo et al. 2005). Such accretion is
reviewed (stably we hope) by Edgar et al. (2004). Accretion disks
for galaxies, young stellar objects, black holes, and
all are the topic of Greaves et al. (2005 and five surrounding papers).
 A new mechanism for producing quasi-periodic oscillations in neutron
 star X-ray binaries (Rezania \& Samson 2005) probably also belongs here.
 It can also, they say, make drifting subpulses in pulsar radio emission.

  Energy extraction from black holes via the Penrose process, the
  Blandford-Znajek process, and perhaps others (Wang et al. 2004a)
  has had its good years and bad years. Komissarov (2005) seems to
  be saying that 2005 will not be remembered with the Mosel
  wines of 1972, let alone the haut brion of 1961 (or 1964).
  Production of astrophysical jets through the Balbus-Hawley
  instability giving rise to hoop stresses was new this year
  (Williams 2005) apart from conference proceedings (Massaglia et al. 2004).

Does MHD require gravitation to work, or is it purely electromagnetic?
In either case, Blackman \& Field (2005) conclude that the Zeldovich
relations are not applicable to real cases with large magnetic
Reynolds number. If this is true, we are sure that Zeldovich would
have been first\footnote{Well, second. In Russian, Z
comes between B and F, and he was a great believer in alphabetical order.}
to sign on the paper, apart from the very small difficulty
associated with being dead. We count it only a small difficulty
because Krisciunas et al. (2004) provides an example of a paper with
two deceased authors, and one who has disappeared. RC, please,
phone home (or Kevin).

Electromagnetism left to its own devices tends to radiate.
Nineteen radiation processes went into the notebook this year,
many of which appear elsewhere in company with the sources that
use them. You will surely be thinking of electrons, so we begin with
a TeV flare mechanism in which relativistic protons excite $\Delta$ resonances
(Boettcher 2005, interpreting data in Daniel et al. 2005).
The charmingly named ``striped wind'' process is a possible source for
optical radation from the Crab pulsar (Petri \& Kirk 2005, properly
crediting the idea to Pacini \& Rees 1970 and to Shklovsky 1970).
The name was even more charming in the first draft, when temporarily
displaced fingers dubbed it dytiprf einf yo rmiy bidinlr lihy grom
the Crab pulsar. The ``Carousel of sparks'' for drifting subpulses in
pulsars has a certain charm too (Janssen \& van Leeuwen 2004).

Quantum electrodynamics matters for the magnetar radiation process proposed by
Heyl \& Hernquist (2005), in which MHD waves are modified by polarization of the
vacuum (not the one in the closet, the one in the equations).

Gyrosynchrotron radiation has been around for a long time (well,
probably very close to 13.7 Gyr) but Burgasser \& Putman (2005)
may well be the first to inflict it upon M and L dwarfs (for their
radio emission). The coherent cyclotron maser process
(Begelman et al. 2005) is one way to produce radio
temperatures in excess of $10^{12}$ K (an inverse Compton
limit that applies to incoherent, single electron radiation,
Kellermann \& Pauliny-Toth 1969, Readhead 1994). A simpler trick
is beaming (Horiuchi et al. 2004).

 And two more putatively new
processes this year challenge Ehrenfest's theorem. First is optical
Cerenkov line radiation (Chen et al. 2005b), which happens when
thermal relativistic electrons hit gas and drive  its refractive
index above $n = 1$ close to the frequency of a resonance line.
 And there is the inverse Faraday effect, in which a circularly
 polarized laser pulse changes spin states in a magnet in 200 fs
 (Kimel et al. 2005). Ehrenfest's theorem? Ah, we mean the one
 that says it is difficult to explain something even when you
 understand it, and almost impossible when you don't.

A few others of the processes of 2005 defy assignment to a specific
 force except  perhaps the force and road of
casualty,\footnote{This is an FSQ (Famous Shakespearean Quotation),
making no sense out of context and not much more in
[\textsl{The Merchant of Venice}, Act II, Scene 9, Line 30].}
 including anthropic reasoning (Livio \& Rees 2005);
the rediscovery process for $P(D)$, this time for the CMB
(Herranz et al. 2004); and Fourier transforms in which you either
 throwaway the phase information and keep only the amplitudes,
 or conversely (Singal 2005). The intention was to improve analysis
 of radio interferometric images, but the test photos shown are
 pictures of people at an India-New Zealand test match. You still
  see faces if you keep only the phase information, but not if you
  keep only the amplitudes. Many folk at test matches (we think it
  is a form of spectator sport) see faces best before the third beer.

\subsection{Cooling Flows}
 The phrase is short hand for X-ray-emitting
 clusters of galaxies whose central gas temperatures and densities
 imply the gas should radiate away most of its energy in much less
 than a Hubble time. They are common enough that the ``last gasp''
 picture won't do.
What has been done over a number of years (with 22 papers this time around)
is to reheat them somehow or otherwise evade the problem. Among the
more or less discrete (meaning separate, not modest, like many of our
colleagues) ideas were---
\begin{itemize}
 \item Try looking at it as gas flows going with ways,
 with central heat input from Type Ia supernovae etc, and the problem
 disappears  (Mathews et al. 2004a), plus a bunch of specific heating
 mechanisms.
\item Turbulent scattering plus thermal conduction (Chandran 2004),
\item Core oscillations (Titley \& Henriksen 2005),
\item Radio lobes (Reynolds et al. 2005 with viscosity as an important
transport mechanism,  comparable with conduction; Nulsen et al. 2005 with a link
to energy deposition   far from the ``cooling core'' otherwise known as
pre-heating),
\item Gas flow through   or near an accretion disk (Soker \& Pizzolato 2005),
\item Dynamic friction (El-Zant et al. 2004),
\item Intergalactic supernovae (Domainko et al.),
\item Active galaxies plus conduction (Fujita \& Suzuki 2005).
\end{itemize}

And the green sources of the year were (1) bubbles driven by jets
from AGNs (McNamara et al. 2005, the first of many papers on the
 general idea) and (2) the Tsunami model
(Fujita et al. 2005, with the publication schedule of ApJ
such that they must have called
it that before 2004 December 26). Despite all this energy input,
 classic cooling flow clusters continued to exist in the index year
 (Morris \& Fabian 2005) and hadn't changed much since $z = 0.4$
(Bauer et al. 2005a).

\subsection[Milky Way]{The Milky Way Swollaws\footnote{Occasionally
a word spelled backward remains marginally
pronouncible and so can be used to indicate the inverse operation.}
and Other Unsettling Issues}
The star most anxious to leave the Milky Way was clocked at
853 km/sec (heliocentric) and 709 km/sec relative to the Galactic
rest frame (Brown et al. 2005b). It must have read some of the same
 papers we did, including the one on the Paranago (1959) effect,
 (Drobitko \& Vityazev 2004, the general idea being that the disk
 kinematics are different for O-F stars and F-M stars), and the ones
 about crystal-like structure in the nearby interstellar medium
 (Anisimova 2004), the co-existence of leading and trailing density
 waves (Mel'nik 2005), and the presence of two pattern speeds for
 SiO masers (Deguchi et al. 2004).
Either the Milky Way has been swallowing another satellite to make
the Monocerus Ring, or its disk is warped. Conn et al. (2005) conclude
 that we cannot currently tell the difference.

The Cartwheel galaxy shows non-thermal radio spokes as well as optical
 ones, but they are not the same spokes (Mayya et al. 2005a).

 Is the
 Universe a WHIM? For all that operators of a submillimeter telescope
 opine that the bulk of the visible universe is at about 10 K
 (Ho et al. 2005), the majority view is that the single largest pool of
 $z = 0$ baryons resides in a Warm/Hot Intergalactic Medium at about $10^6$ K
 (Nicastro et al. 2005). Shull (2005) said it first during the index year
 and McKernan et al. (2005) were the last to say that some issues still
 need to be resolved, in their case whether the hot gas emitting O VII
 and O VIII lines near the Milky Way is real WHIM versus the North Polar Spur,
  SN outflows or something else. And there were about 10 related papers that
appeared in between. We will invoke the principle of the excluded middle by
saying some of our best friends are made of baryons; some of our best
friends are slightly missing (or a few pickles short of a sandwich as
Ann Landers would have put it);
and, therefore, at least a few baryons are still missing
(Sembach et al. 2004 on observations; Kang et al. 2005 on
theory of the various phases).

\subsection{Unusual and Alternative Histories}
 Most of these pertain one
way or another to the history of astronomy (etc.), but a couple
belong to a history of the universe in which young galaxies have
significant intrinsic metallicity which decreases as they age
(Harutyunian 2004). In close association, of course, the abundance
of hydrogen increases with time (Harutyunian 2003). The Milky Way,
whose oldest stars are the least metal rich, whether you examine the
field or globular clusters (Cohen \& Melendez 2005), either missed the
boat or caught one going the wrong direction. And a datum you could
 attach to either point of view is that QSO absorption gas
has an Fe/H ratio which grows from $z\approx 3$ to $z\approx0.3$
(Prochaska et al. 2004). If large red shift means long ago and
far away, then heavy elements have been created over the years.
If large redshifts belong to sources recently expelled from
nearby galaxies, then heavy elements have been destroyed in
the expulsion process.

The chief historical green dot is not, perhaps, science,
but the statement (Time Magazine, 1 August 2005, p.~39),
``there have been some 525 nuclear explosions above ground since
 Hiroshima; not one of them has been an act of war.'' Try telling
 that to the people who were at Nagasaki in August 1945. A more
 useful factoid is that 45\% of the Hiroshima and Nagasaki initial
 survivors are still alive (Land 2005), as were precisely 25\%---114 of
 456---of the 1940 graduating class of the US Naval Academy, as of
 2005 September 30. A sizeable fraction of that class did not
 survive 1941 December 7. And, if you can bear to look a gift
 tooth in the mouth, dates of birth of corpses that began life from
 about 1943 to nearly the present can be determined from
$\mathrm{{}^{14}C/{}^{13}C/{}^{12}C}$
 ratios in teeth (Spalding et al. 2005) because of the range of ages
 at which various teeth sprout\footnote{Presumably in the absence of
attention from the Keen Amateur Dentist.}
and the radioactive input from those not-in-anger above ground bomb explosions.

In more traditional scientific oopsery, Stevenson (2005) told us
that WKB stands for Eugene Wigner, Hendrik Kramers, and Leon Brillouin.
The ghost of Gregor Wentzel would rise to protest, but he is busy dancing one of
those Viennese waltzes with Graffin Maritza. Sir Harold Jeffreys was apparently
unknown to the author but would  surely have volunteered the information that
tsunami is its own plural, like zori and sheep.
We are always very careful to say WKB-J method. And some others:

``Scientists have known since the 1950s that they were seeing too few solar
neutrinos'' (Science 306, 1458), but Ray Davis didn't start looking
until the 1960s.

``He [Fred Whipple] was the Leonard Medalist (1970) and the Bruce Medalist
(1986) of the Meteoritical Society.'' (Hughes 2004.)  But the Bruce is given by
the Astronomical Society of the Pacific, whose advisory committee one of us and
a Las Cumbres colleague  have joined so recently that we do not yet know whether
our advice will be taken.

 ``\ldots Which enabled Joseph [Barclay] to announce in 1856 the
discovery of a companion star to Procyon'' (Barclay 2005). This
would have surprised both Alvan G.~Clark, who did not discover
the companion to Sirius until 1862 (and is generally credited as
the first to see photons from any white dwarf) and even more
Schaeberle who thought, and said, in 1895 that he was the
discoverer of Procyon B.

``\ldots firm evidence that the universe is expanding'' credited
to V.M.~Slipher in 1971 (Heavens 2005). Not quite. It was the
redshift-distance relation, and, while Slipher measured the first
set of redshifts (and Milton Humason the second) the distances and
the publication of the correlation came of course from Hubble.
Granted that Slipher and Humason sometimes get too little credit
and Hubble too much, this is, nevertheless the sort of over-correction
of steering that sometimes afflicts cyclists with small Eddington numbers.

``Fermi won a Nobel Prize in 1938 for his discovery of the properties
of slow neutrons'' (Maltese 2005, in a book review). Well, the
citation, which mentions new elements first, was of course wrong,
and we fly swiftly back in memory to the moment when, perusing
the aging pages of \textsl{Comptes Rondue}, we discovered that the
French Academy has hastily revised the
citation of the LeConte Prize to award it to Blondlot ``pour le corps
de ses ouvres'' rather than for the discovery of N-rays.
It was a talk by Philip Morrison called ``The N-ray dosage and
protection problem'' that sent us to the library to see what had
 been done about the citation. About half the talk, like the
 Collins (2005) book, consisted of unkind remarks about Joseph Weber.

``1955 \ldots there was almost no television'' (Nature 433, 785).
We cannot speak from personal experience about the situation in London,
 but in Los Angeles there were 7 channels (3 network, CBS=2, NBC=4, and
 ABC=7, and 4 local,  5, 9, 11, and 13, other even numbers belonging to
surrounding communities like San Diego and Santa Barbara) and the first live
 remoted coverage of an on-going news event (the search for Kathy Fiscus,
  a little girl who fell down a disused well shaft) was already
  more than five years in the past.

``Happy hundredth birthday to \ldots Dippy, the giant Diplococus \ldots
that took up residence on 12 May 2005 \ldots" (Nature 435, vii).
Either they mean 1905, or postal copies of Nature have been
coming even later than they used to.

``Eddy dubbed this the Maunder minimum, after E.~Walter Maunder (1851--1928)
who had called attention to this aberration in the 11-yr sunspot cycle''
(Robinson 2005). William Herschel and Gustav Sp\"{o}rer had actually
noticed the prolong minimum earlier. The name clearly obeys
Stigler's law (things get named for the last person to notice them
and not credit his predecessors), but Sp\"{o}rer gets his minimum for
the one around 1450, and Herschel almost got a planet (as well as a
recording of his symphonies in the ``contemporaries of Mozart'' series
along with Salieri, V\'{a}clav Pichl, Fran\c{c}ois-Joseph Gossec, and
nine others of mostly comparable musical obscurity).

Radick (2005) addresses alternative histories, wondering what would
have happened if Darwin had stayed home. Far more credit to Wallace,
he concludes. And if Einstein had given up on math, then, said
Einstein himself, Langevin, though we are inclined to favor
Lorentz and Fitzgerald for special relativity and suspect
that general relativity would have had a long wait

\subsection{Oops, Being a Compendium of Undiscoveries and Other
Unfortunate Events}
Supernova SN 2002kg was a brightening of luminous blue variable V37
in NGC 2403 (Weis \& Bomans 2005). SN 1954J in the same galaxy can
make the same claim (Van Dyk et al. 2005).
And 2003qw has been promoted to an AM CVn star (Nogami et al. 2005).

Poor PN H2-1 (where H stands for Haro) has been recognized so often
 as a bright knot in the Kepler SNR, forgotten, and rediscovered
 (Riesgo \& Lopez 2005) that the poor thing is almost raw from
 being dragged in and out of catalogs.

Since all stars vary at some level, some day we will cease to
sympathize with photometrists whose standard stars vary
(Viehmann et al. 2005, IRS7 in the core of the Milky Way in this case).
Very similar bad luck afflicted the search for planet transits in
NGC 6940 by Hood et al (2005), nearly all of whose observed stars
were non-members. And they didn't have any transits either.

DO Dra and YY Dra are both variable, and are in fact the same star
(Hoard et al. 2005), which may be useful to it if it wants to be
observed from two countries who don't honor each other's passports.

FH Leo is a common proper motion pair (late F plus late GV), not a
cataclysmic variable (Dall et al. 2005). The cause of its outburst,
caught by \textsl{HIPPARCOS}, remains unclear, though the authors suggest
engulfment of a planet or scattered light from Jupiter.

\textsl{HIPPARCOS} itself is still digging out of the difficulties with its
non-uniform sky coverage, coordinate system, and so forth.
The coordinate system rotates (Boylev 2005).
The Pleiades will never quite forgive it for pulling them into
119 pc versus the correct 132--138 pc (Soderblom et al. 2005). And while
there will some day be a new catalog (van Leeuwen 2005),
 don't sit up all night waiting for it, unless you are the sort of
 astronomer who normally sits up all night anyhow. In the day time,
 of course, one sits down (or so said Victor Borge).

The Palomar-Green catalog of QSOs is only about 50\% right near its
magnitude limit because the color cut fell near the peak of N(U-B)
and the $2\sigma$ error bars on colors were about equal to the FWHM of the
real distribution (Jester et al. 2005). Thus, about half the real
 QSOs are missing, half of those included don't belong, and the
 colors are perilously close to random. It is, however, doing very
 well compared to the USNO-B1 catalog, in which 99.9\% of the listed
 objects are not real, and only 47\% of those that belong are present
 (Levine 2005). It is a catalog of objects with proper motions of
 1--$5''$/yr, and the comparison sample is the Luyten half-second, LHS, catalog.

``Saturn is in the Southwest after sunset, south in midevening''
Planetary Report 25, No.~2, p.~1. Well,
Velikowski said the direction of the Earth's rotation had reversed at some time.
As Huygens descended upon Titan, only one-half of the intended
700 pictures were sent back, because a controller forgot to
send the command to switch on the right side of the hardware (Anonymous 2005d).

At least two periodicities were so odd that apparently they
 aren't true, 246 days for  $\mathrm{CH_3OH}$ in a star formation region
 (Goedhart et al. 2005); and midinfrared counts of galaxies from
 ISO have spectral features passing through the wavebands, not
 quantized redshifts (Pearson 2005).

``Einstein, who had no formal scientific training beyond a qualification to
teach high school physics'' (New Scientist, 20 April 2005, p.~46).

``If a star is greater than about 3 solar masses, it ultimately evolves into a
 black hole'' (Sky \& Telescope 110, No.~3, p. 103--104).

``Young stars
 are made mostly of hydrogen \ldots The tremendous heat inside them
 turns some of the hydrogen into other gases. Older stars also
 have helium and even carbon. Even the Sun has some helium'' (Levy 2005).

\section{FIRST, LAST, ALWAYS, AND OTHER ORDINAL NUMBERS}
Here reside
 assorted astronomical extrema, things of which there are many,
 things of which there is at most one, and candidates for the
  Lincoln's doctor's dog prize.

\subsection{Countdown}
 Numbers in the index year literature ranged
from 1--$2\times 10^{11}$ (not, as you might have supposed, the number
of stars in the Milky Way but the number of objects in the
Oort cloud required to keep up the supply reaching us,
Neslusan \& Jakubik 2005) down to $10^{-25.7}$, the cooling of
the interstellar medium provided by excited CII, in
erg/sec/H atom (Lehner et al. 2004).

Green stars went to 44,000 and 702, each of which requires
a bit of explanation. The NSF plans to improve the success
rate in its proposal application process by making its RFPs
 more narrowly targeted, so as to receive fewer proposals
 per year. The current number is 44,000 (Bement 2005).
 And 702 is the number of chemical elements that could
 receive symbols under the current system of one or two
 letters each. Only 111 are in use so far (Hayes 2005),
 including a few of which the news hasn't come to Harvard.
 Of various other such systems, airport codes are
 the fullest, at 10,678 out of $17,576=26^3$. Chemical
 elements are the least heavily utilized, while radio
 stations beginning with K (west of the Mississippi) and W,
 internet country codes (242/676, with a small prize for the
 correct identification of ``to'' and ``tv'' which puzzled us in
 reviewing a paper a couple of days ago), and stock ticker
  designations (3928/10,278 probably the most rapidly
  variable of these numbers) come in between. And an
  assortment of other numbers, only about half of which can
   truly be described as odd.

$141\times 10^6$ sources in the third SDSS data release
(Abazajian et al. 2005a, and the last author is Zucker,
close to a record in itself).

$11\times10^6$ observations of
variable stars logged into the AAVSO system (Waagen 2004).

693,319 galaxies, QSOs, and stars in the NYU Value-Added
Galaxy Catalogue (Blanton et al. 2005, and if the value added
is more than about 2 cents apiece, we can't afford it).

473,207 graduate students in technical fields in the US last
year, an NSF report quoted in Science 309, 1169.

365,xxx days
in each of the past two and upcoming millennia. Yes, no two
are the same because of the changing rules for leap years,
and we are irresistably reminded of G.B. Shaw's attempted
calculation (\textsl{In Good King Charles' Golden Days}) of the
number of mistresses per day needed to achieve a particular
lifetime total. He got it wrong, and the first 10 readers
to send in the correct version will receive an item from
our celebrated collection of envelop backs.

193,123 QUEST1 objects in a variability survey from QSOs,
using a 1-meter telescope in Venezuela (Rengstorf et a1. 2004).

110,563 UV-excess candidates for QSOs from SDSS
(Richards et a1. 2004). About 95\% of them really are.

61,977 stars with proper motions exceeding $0.15''$/yr from the
POSS digitized catalog (Lepine \& Shara 2005). They recover all
the LHS and NLTT stars from surveys originally carried out by
Luyten.

 20,000, the approximate number of astronomy papers
published per year, according to Colless (2005), including
meeting abstracts, conference proceedings, and all.
The number read for this review is smaller by a factor
about 2.6.

11,788 sources in the DIRBE point source catalog (Smith 2004).

10,862 light curves of eclipsing binaries in a fully
automated analysis from OGLE II (Devor 2004).

3000 nuclei
with known spin, parity, and half lives (NUDAT at BNL 2005).

2980 isolated galaxies (Allam et al. 2005), where the
definition involves both separation and relative size of
nearest neighbor.

 2204 gamma ray bursts in a BATSE catalog
that includes 589 non-trigger events (Schmidt 2004).

 2200 white dwarfs in the globular cluster $\omega$ Cen (Monelli et al. 2005).

1319 EGRET photons remaining as the background when 187
of 1506 belonging to sources have been accounted for
(Thompson et al. 2005). This is a considerable improvement
over the first report of the gamma ray background, ``and the
remaining 22 photons\ldots" (Kraushaar \& Clark 1962).

1095 the largest number of ADS papers attributed to a single author at the
moment when A.V. Filippenko surpassed the 1094 of Ernst Opik (whose oevre is
unlikely to increase further). On the other hand, Opik also has 16 musical
compositions to his credit.

899 groups of
 galaxies near $z = 1$ (Gerke et al. 2005), but the authors
 say that both type 1 and type 2 errors are close to 50\%,
  meaning that something like 450 of the supposed groups
  aren't, and a comparable number are missing.

871 Herbig-Haro objects up to the time of Phelps \& Ybarra (2005).
Their new one is on the edge of the Rosette Nebula and has
outflow on 1--3 pc scale.

827 point X-ray sources in Chandra images of 11 spiral galaxies
(Kilgard et al. 2005).

NOTE: we are leaving out another six
items in the 700-800 range to keep you from noting that this
year's numbers truly do not favor initial digits of 1 and 2.

748 days in orbit reached by Sergei Krikalev in ISS on
16 August (distributed among 6 flights).

 555 emission lines
in a single VLT spectrum of Orion (said to be the maximum
number anywhere in that wavelength range, Esteban et al. 2004).

520 open clusters cataloged by Kharchenko (2005a).

 388 GRBs
seen by INTEGRAL up to the submission time of Rau et al. (2005),
 an average rate of 0.3/day, very close to what was forecast from
 BATSE data. INTEGRAL is sensitive to smaller fluxes but does not
 see as much of the sky for as much of the time.

231 radio supernova remnants in the Milky Way (Green 2005).

225  the number of papers written by \textsl{Observatory}'s most prolific
author in the 1971-2000 period. He is D.J. Strickland (also one of the current
editors).  Second place would seem to be held by The Most Underappreciated
Astronomer (see \S11.3) with 220. VT scores a mere 5.

213 items in a table in Science 308, 943, for which the
text says ``nearly 200.'' Well we suppose 200 is near to 213,
or conversely, but the word surely carries an implication of fewer.

115 protons in the nucleus of the most recently confirmed
 element (Dmitriev et a1. 2005).
The story is second hand from \textsl{Nature}, who appear to have
simplified somewhat. They say $\mathrm{{}^{243}Am}$  zapped
by $\mathrm{{}^{48}Ca}$ makes element 115 which decays to 5 alpha
partic1es plus $\mathrm{{}^{268}Db}$.
Now 5 $\alpha$'s is 20 particles; 268+20 = 288; but 243 + 48 = 291,
so we suppose three somethings (probably neutrons) must
spray off at the first step.

104 lensed arcs (plus 12 more radio ones) in 128 HST clusters
(Sand et a1. 2005). Notice that the average is close to one per c1uster,
though we don't suppose that is how they are really distributed.

79 pulsation frequencies in FG Vir (Breger et a1. 2005).

MYSTERY NUMBERS: A\&A 435, 1173 lists observatories by
MPC number, with Greenwich = 0, Heidelberg = 24,
Hamberg = 29, He1wan = 87. Not by longitude, since
Palomar = 261 and 675, while Mauna Kea = 568. And not
by foundation date, since Mt.~Wilson and Yerkes come
after Palomar. The paper (Emelyanov 2005) is really about
emphemerides of 54 outer satellites of Jupiter.

63 total moons of Jupiter to spring 2005 (Sheppard et al. 2005),
and 50 for Saturn.

48 high mass X-ray binaries in the Small Magellanic Cloud
(Coe et al. 2005). Most are BeXRBs with the X-ray variability
 partly tied to the Be star variability.

40 new DQ type white dwarfs (Dufour et al. 2005).

32 leap seconds added since 1972 (Anonymous 2005h).
You will have had another one by the time this appears, and
 we hope you used it well.

 31 short period RS CVn stars
 (Dryomova et al. 2005), which they say are the same as
  pre-contact W UMa stars.

22 lensed quasars (the radio sort) in the CLASS survey
(York et al. 2005).

 14 micromagnitudes, the precision of the
photometry reported by Kurtz et al. (2005a).

13 AM CVn stars (Nogami et al. 2004).

 11 magnetars,
 McGarry et al. (2005 reporting the seventh anomalous
 X-ray pulsar, which is also the first magnetar in the SMC).

 10 eigenvectors to fit 95\% of the variance of 16,707 QSO
 spectra (Yip et al. 2004).

 10 planets (Brown et al. 2005a)
 counting 2003 UB313, whose radius is probably rather
 larger than that of Pluto (and a brief nod to the
 Abraham Lincoln story about the number of legs on a horse).

  9 catalogs of spectroscopic binaries (Pourbaix et al. 2004).

   8th supernova in NGC 6946 (Li et al. 2005).

 7th cool Algol
    with an orbit (Mader et al. 2005). Definition of the
     class and the first orbit came from Daniel M.~Popper
     (IAU Symposium 151) in 1992. There are times when
he would seem to be a candidate for second most
underappreciated astronomer.

6 accretion-powered millisecond pulsars (Galloway et al. 2005).

6d search under way (Jones et al. 2004a). No, they are
not covering two 3d universes, but only most of the
southern sky to $z = 0.15$. It means 6 degree field.

5 periods in a CV (Araujo-Betancor et al. 2005). They
are orbit, superhump, rotation, nonradial pulsation,
and a 3.5 hour period of unknown cause.

 5 often the maximum
 number of authors given credit for a paper on ADS. One
 wonders whether Thoralf J.~Aaboen (the first real name
 in the Orange County phone directory) might be prepared
 to offer adoption, in much the same way that a
 mathematician with an Erdos number of five offered to
 sell ``six'' a year or two ago.

 5, not, the pentaquark has fragmented (Close \& CLAS 2005).

 4 different mixing lengths
  in a 3d calculation of convection (Kapyla et al. 2004).

 4, the number of terrestrial hemispheres in which \textsl{Sky \& Telescope}
   is published (Fienberg 2004). There must be some analog of the
   Hirsch citation number and the Eddington cycling number to be
    found here, $N'''$ for a publisher, the maximum number of
    hemispheres $N'''$ in which he publishes $N'''$ magazines.

 3 white dwarfs in CVs with fields in excess of 100 MG
    (Gansicke et al. 2004). They are not, say the authors,
    particularly rare, just very faint and so difficult to
    find.

 3 asteroid orbits that fall entirely within that of
    the earth (Meeus 2005). The third is 2002 $\mathrm{JY_8}$. It actually
     has a semimajor axis a smidge larger than 1 AU but with
     its orbit somehow oriented to stay inside ours.

 3 Virginias
     in astronomy, the largest percentage increase among any of
     our small numbers, with Virginia McSwain whose thesis
     abstract appears in PASP 117, 309 and Virginia Kilborn of
     Swinburn University, who co-authored a paper on dark HI
     galaxies in \textsl{ApJ Letters} during the year. Both were (also)
     named for much older relatives.

 3 intermediate polars with
      periods below the gap (de Martino et al. 2005), number 3 being
HT Com.

 3 black South African co-authors on
Jerzykiewicz et al. (2005), a paper we also indexed under
 ``bad luck'' because it reports that their first comparison
 star proved to be a slowly pulsating B variable and their
 second a $\delta\,$Scuti star.

 2 subdwarf L stars (Burgasser 2004,
 adding 2MASS 1626+3926).

 2 the number of arms you would
 think you were seeing in the inner 150 pc of the X-ray
 gas of cluster Abell 2029 (Clarke et al. 2004).

 2 (through 5)
  the QSOs in which the big blue bump has been confirmed as
  an accretion disk by seeing the Balmer edge only
  in polarized light (Kishimoto et al. 2004).

 2 EGRET sources
  which are not Blazars (Guera et al. 2005 adding 3C111 to
  Cen A).

 2 hybrid PG 1159 stars, meaning there is some
  hydrogen in the atmosphere as well as C, O, and He,
  affecting their pulsation periods (Vauclair et al. 2005).

  2 pre-main-sequence stars in which X-rays come entirely
  from accretion, not magnetic activity (Swartz et a1. 2005).

 zero used to be the number of Cepheid
 variables in star clusters. It is now 20, with the most
 recent in the LMC cluster NGC 1866 (Brocato et a1. 2004).

  Non-integer, the cosmic abundance of holmium is twice
  that of hafnium (Wallerstein 2005).

\subsection{Firsts}
 Dozens of these were recorded, a good many of
which have crept into the object-oriented earlier sections.
Of those that did not, we cannot resist (say ``the first'' in
front of each, or somewhere in the middle).
\begin{itemize}
 \item Natural source
of transition radiation (Nita et a1. 2005). What is
transition radiation? Oh dear, we were afraid you would ask that and have a
copy of the new \textsl{Oxford Dictionary of Physics}
on order, but it was predicted by Ginzburg  \& Frank (1946) and found
in cosmic ray detectors by Yodh et al. (1973).\footnote{And why, you will
ask, didn't we just tunnel through the 20 yards to
 Prof.~Yodh's office and ask? Well, only if all else fails does your
 least penetrating author try following directions. But here is what
 he said: transition radiation is the electromagnetic radiation that
 is emitted when a charged particle traverses the boundary between
  two media of different dielectric or magnetic properties. Like
  Cerenkov radiation, this process depends on the velocity of
  the particle and is a collective response of the matter
  surrounding the trajectory. Like bremsstrahlung, it is sharply
  peaked in the forward direction if the particle is
  ultrarelativistic, in which case most of the energy is
  in X-rays. TR from particles traversing successive boundaries
  exhibits interference and diffraction patterns. We suspect
  that stage, screen, and radio are indeed media of different
  dielectric properties.}
\item Detection of the hyperfine splitting of HI radiation
(aka 21 cm) is one of our favorite
 stories of predict, discover, and confirm, and everybody
 behaving like a community of scholars and publishing the
  three detection papers together. This time around,
  hyperfine splitting of deuterium, in the form of the
  $\mathrm{DCO^+}$ molecule in the ISM (Caselli \& Dore 2005).
\item Astrophysical masers remain the best buy explanation of
strong, variable emission of OH and  a number of
other molecules, especially with anomalous line ratios.
But Weisberg et a1. (2005) have seen the driving directly as
 pulsed OH maser emission on source vs. off, when looking
 at the pulsar B1641-45 (P = 0.455 sec). The on-off
 subtraction is necessary because there is a good deal of
 diffuse OH emission in the region.
 \item The Lyman alpha forest
 in spectra of distant QSOs (indeed some nearby ones if you have
 a UV spectrograph above the atmosphere) has been around for
 30 years, but Nicastro et a1. (2005a,b) have reported the first X-ray forest.
It consists of two lines and is, therefore, unlikely to be a
forest that you are unable to see because of the trees.
\item All-sky survey at short radio wavelength (1.6 cm)
underway at ATCA (Ricci et a1. 2004).
They found 221 sources in the first 1216 square degrees,
about half each galactic and extragalactic. And we refrain
from saying that, if this is to be truly all-sky, they will
need either an additional, northern site, or more penetrating
radiation.
\item Gustave Arrhenius estimated greenhouse warming at
the end of the 19th century (and came very close to modern numbers).
 But the idea can be traced back to Fourier in 1827
 (Pierrehumbert 2004), and concocting the best remark about
 decomposition as a solution is left as an exercise for the
 student (No, Mr. H. X-ray clusters).
\item Five dimensional space. No, not an observation, but as a
way of tackling unified theories. The first was Nordstrom (1914)
rather than Kaluza (1921). Nordstrom also owned half a solution
to the Einstein equations (with non-zero electric charge; the other half
belonging to Reissner) and a whole bunch of department stores.
\item Successful use of long baseline optical interferometry to measure
polarization  (Ireland et al. 2005).
\item X-rays from an AGB star, a flare in Mira (Karovska et al. 2005).
\item And our candidate for the 2005 Lincoln's Doctor's Dog's
Favorite Jewish Recipes award is the first 3D spectroscopic
 study of H$\,\alpha$ emission in a $z\approx 1$ field galaxy with an integral
  field spectrograph (Smith et al. 2004). Indeed it is very
  probably the first galaxy to have such an instrument,
  qualifying it also for inclusion in \S13.
 \end{itemize}

\subsection{Extrema}
 Some of these are human (indeed occasionally
all too human) and some astronomical.
They are mixed this year as they were in the literature.

The longest time interval between parts I and II of a series
is 26 years (Ahmed et al. 2005). This is, however, dwarfed by
the 40+ years ``in press'' for a paper cited by Chumak et al. (2005).
 The paper, by Olin J. Eggen was intended for volume IV, on
 clusters and binaries,  of the \textsl{Stars and Stellar Systems}
 compendium. This has not (so far) been published, though
Chumak assigned a 1965 date to it.

The shortest half-life of an element possibly present
in stars? Well, you know about Tc and perhaps even Pm,
but Bidelman (2005), examining a high-resolution
spectrogram of HD 101065 = V 816 Cen, has found lines
possibly due to transitions of Po, Ac, Pa, and the
transuranics Np through Es. Of the 11 known isotopes of
Einsteinium, $\mathrm{{}^{254}Es}$ has the longest half life, 276 days,
comparable perhaps with the interval between the writing of
this (Christmas 2005) and your first chance to read it.

Largest redshifts are easy to tabulate because somehow authors
always mention the point in some conspicuous place in their
papers. The 2005 ones were: (a) supernova, $z = 1.55$
(Strolger et al. 2004), (b) radio galaxy $z = 5.2$
(Klamer et al. 2004), (c) ``overdensity'' meaning something
that might evolve into a cluster or supercluster,
$z = 5.77$ (Wang et al. 2005a), consisting of 17 Ly$\,\alpha$
emitters stretched across 70 Mpc (comoving we assume),
(d) X-ray selected clusters, $z = 1.39$ (Mullis et al. 2005
who say that others should be easy to find), (e) giant
radio galaxy, $z = 3.22$ (Mack et al. 2005). It is
B3 1231+397B and is a compact, steep spectrum (young, recurrent)
core source, (f) PAHs, $z\approx 2$ (Yan et al. 2005), to be seen
again in \S7.4,
 (g) $\mathrm{H_2O}$ maser, $z = 0.66$ (Barvainis \& Antonucci 2005),
 (h) cluster detected with weak lensing,
  $z = 0.9$ (Margoniner et. al. 2005). This is the lens
  redshift not the lensee. The lowest $z$ galaxy functioning
  as a lens is ESO 325-G004 at $z = 0.0341$ (Smith et al. 2005d).

 Largest error this year may be the $2\times10^{45}$ $\mathrm{cm^2}$ cross
 section for neutrino capture mentioned in AJP 73, 495.
 It is also described as belonging to
 ``carbon tetrachloride $(\mathrm{C_2Cl_4})$.'' Well no. Carbon tet
is $\mathrm{CCl_4}$. That other cleaning fluid is perchlroethylene.
What's the use of having had a chemist father if you can't
tell these apart by sniffing
them (cautiously, of course). The most chlorinated author
spent measurable parts of her childhood joyfully sniffing
$\mathrm{CCl_4}$, and we know at least one referee who will not be at
all surprised to hear this.\footnote{The junior, but oldest, author
recalls the glorious days of yester-year when such sniffing was  part
of the pleasure of revealing watermarks in stamps. And we know the
``most chlorinated'' author will not be surprised to hear this.}

Some stellar extrema. The hottest main sequence star has $T = 48,000$ K say
Massey et al. (2005). The largest radius is 1500 $R_\odot$ for KW Sgr, V354 Cep,
and KY Cyg (Levesque et al. 2005). All are about 25 $M_\odot$ and
$3\times 10^5$ $L_\odot$ and probably in the double
shell burning phase. The shortest M dwarf binary period is
0.1984 day for BW3 V38 (Maceroni \& Montalban 2004). Candidates
for the most massive star, the shortest Blazhko period
RR Lyrae, the hottest post-AGB star, etc. hide in other sections.

The smallest space telescope at present is the
15 cm, 60 kg MOST (Matthews 2005). It is used primarily
(you thought we were going to say MOSTly, didn't you) for
astEroseismology (we lost that one several years ago).

Among neutron stars, the slowest rotation period is 9600 sec
(Bonning \& Falanga 2005) for LS I +65010. The period has been
shrinking at $-8.9\times 10^{-7}$ s/s, so it will someday cease to be
the slowest. The fastest moving pulsar has $V = 1083$ km/sec
in the plane of the sky (Chatterjee et al. 2005, reporting a
VLBA proper motion). The fastest rotation hovered near
1.55 msec for so long that we were beginning to think the
limit was trying to tell us something (except that there is
now an out-of-period counterexample). And the youngest
neutron star seen as a source of thermal X-rays (Gonzalez et al. 2005) has
$P/2\dot{P} = 1700$ yr.

The fastest obituary into print was probably that for
John Noriss Bahcall, who died on  17 August and was
remembered in the 1 September issue of \textsl{Nature} (Ostriker 2005).
The piece describes Bahcall as a hedgehog (not, given his
expertise in galactic structure, nuclear physics and
astrophysics, and science politics, perhaps entirely fair)
and the author as a fox.

The largest cities (Nature 437, 302) are currently Tokyo,
Mexico City, New York, Sa\~{o} Paulo, Mumbai, Delhi, Calcutta,
Buenos Aires, Shanghai. and Jakarta. Interesting perhaps to
compare with the list from 1950 that the most populous author
memorized in Miss Munro's 7th grade social studies class,
when she wasn't busy growing wheat or shooting free throws.
New York, Shanghai, Tokyo (so far so good), Moscow (oops),
Chicago, London, Berlin, Leningrad, Buenos Aires, Paris.
The lists of longest rivers and largest islands seem to
have held up better and are occasionally still useful.
The countries around the Mediterranean  and their capitals
fall somewhere in between (but quiet rejoicing that the
bit of the list that says ``Israel, Palestine'' is going
to be true again, something she had not hoped to see in
her lifetime, along with one Germany). Oh, and would
someone please clarify just how many pieces of the
former Yugoslavia actually touch the sea?

Some cosmic biggies. The largest structure is still the
SDSS Great Wall, about 80\% larger than the Harvard Great Wall
(Gott et al. 2005). The widest (radio) gravitational lens
sprawls over $41''$ behind Abell 2218 (Garrett et al. 2005).
Mind you, $41''$ is the apparent
height of a middle-sized author seen from a distance of
5.5 miles. Too close, we can hear you exclaiming.

The brightest radio BAL quasar (Benn et al. 2005) probably
isn't very bright, but in the good old days, broad
absorption line sources were all radio quiet. This one has
$z= 3.37$ and a broad line region velocity width to
--29,000 km/sec. The most extended HI disk
reaches to $8.3\times$ the Holmberg diameter for NGC 3741. The
galaxy has $M/L = 107$ in solar units, mostly because of
 small  luminosity. The dynamical mass is in fact only
 about eight times the baryon mass (Begum et al. 2005).
 And what is the Holmberg diameter?
In previous years we would have told you that is twice
the Holmberg radius. But this year it is twice the
Hafberg diameter.

On scales between the cosmic and the comic we find
(a) the most distant star stream of the Milky Way
(Clewley et al. 2005) out at 70 kpc in the halo and
consisting so far of six horizontal branch stars and
three carbon stars sharing an orbit, (b) the largest
supernova remnant in the SMC (Williams et al. 2004a),
$98\times70$ pc in extent, 45,000 years old, and perhaps a
Type Ia since there is no pulsar and no OB association nearby,
(c) the darkest GRB, 001109, meaning the one with the
smallest ratio $L_\mathrm{opt}/L_\mathrm{X}$ in the early afterglow
(Castro  Ceron et al. 2004).

The last photographic survey, UKST H$\alpha$+[NII], ended
in late 2003 (Parker et al. 2005).
They used Kodak Tech-Pan film. A publication called
\textsl{Kodak Plates and Films for Scientific Photography}
still sits on our bookshelf. It is a souvenir of the
1975 AAS meeting in Bloomington (which also featured a
concert of Scott Joplin music) and cost \$2.50. And
(from another AAS/Kodak publication of the same vintage)
 we note that 2006 marks the 35th
anniversary of the retirement of William F.~Swann whose
30 years of working with the astronomical community
included the production of the $14\times14$ inch plates for the
Palomar Observatory Sky Survey. The article about him was
written by William C.~Miller, then the photo-maven at what
were then the Hale Observatories. But Kodak even then
measured wavelength in
nanometers.

The most under-recognized astronomer? Contemplate
Robinson (2005a) in the usual 50 and 25 years ago column,
and notice that, amongst all the credit, the chap who
developed the radial velocity spectrometer (a descendent of
which caught the first exoplanet) is not named.
The pickiest reader of course wrote and complained, and a
correction appears in \textsl{Sky \& Telescope} (109, No.~5, p.~13).
It may or may not be significant that shortly thereafter
(a) Robinson stopped writing that column and (b) your
present author's subscription vanished. Who? You think
Ap05 should be more meticulous than S\&T? So do we.
It was Roger Griffin of Cambridge, whose orbits of
spectroscopic binaries are rapidly soldiering on toward
200 papers, leaving him feeling rather smug at having
adopted arabic rather than Roman numerals from the beginning.

\section{BIG BLACK BLANGS}
Sorry, no. Make that Blig Back Blangs.
No. Blig Black Bangs. Never mind. Horizontal bars.\footnote{With
apologies to Spike Jones, Doodles Weaver, and the Man on the
Treezing Tri-flap.} Well, we told you last year that the
one twist tonguer we've never been able
to manage is the big black bug bled bad blood (yeah,
it's even hard to type). Anyhow, this section deals
with supernovae and their remnants (compact and diffuse),
active galaxies and other manifestations of black
holes in galaxies, and a few other incendiary strays.

\subsection{Supernovae}
 There used to be two sorts of supernovae,
Type II (with hydrogen lines in their spectra) and Type I
(without). There are once again two types, core collapse
(Type II plus Types Ib and Ic) and nuclear detonation/deflagration
(Type Ia). The former happen to stars
initially heftier than about 8 $M_\odot$ (dependent on
composition, presence of a companion, rotation, and
probably other things). The latter happen to degenerate,
white-dwarf-ish, stars or cores driven by accretion or
merger above the Chandrasekhar limiting mass.

Some galaxies have lots: 2004et was the 8th in NGC 6946
and was perhaps imaged as a yellow supergiant beforehand
(Li et al. 2005). And some people discover a lot---39 up
to the end of 2003 for the Rev.~Robert Evans in Australia,
and 100+ by amateur groups coordinated by Guy Hurst in the UK
(Evans 2004). How many total? Well, index year 2004 ended
with 2004es, and 2005 began with 2004et (IAU Circ 8413),
continuing on up to 2005eo (IAU Circ 8605), making, we think,
223 events in the fiscal year (minus the invariable few that
turn out to be well known variable stars and such).
The system whereby faint SNe get preliminary designations
and move into the mainstream only when/if properly confirmed
(CBET) has now been made permanent (IAU Circ 8476).

What about typical numbers per galaxy per year? The unit
is the SNU (SuperNova Unit) of one per century per $10^{10}$
$L_\odot$,
and, like all good units, it leads to typical values near 1.
Three groups keep catalogs (CfA, Asiago, and Sternberg) and
Tsvetkov et al. (2004) provide web-dresses so you can run the numbers for
yourself. Typical results (Petrosian et al. 2005,
Mannucci et al. 2005) say that (a) only S and Irr galaxies
have core collapse events, while Ia's can happen anywhere,
(b) even the nuclear events are much commoner in late type
 galaxies, with rates of all types varying by factors 20--30
from earliest to latest galaxies,
Scannapieco \& Bildsen (2005) attribute the Ia statistics
to there being both prompt (0.7 Gyr after star formation
and including the brightest ones) and delayed (up to
10 Gyr and including the faintest ones) events, (c) star
forming and star burst galaxies have more than their
fair share, (d) the nuclear explosions are less common
by factors 2--3 in magnitude-limited samples than the
core collapse events and by larger factors in
volume-limited samples, and (e) you find larger rates
(again in SNU) looking back in redshift to 0.25--0.7,
by factors 3--7 (Dahlen et al. 2004, Cappellaro et al. 2005).
And we don't know quite what to make of the discovery that
SNe Ia are commoner by a factor of about four
(0.43 $h^2_{75}$ vs. 0.10 SNU) in radio loud than in radio quiet
galaxies (Della Valle et al. 2005).

But the green supernova of the year is 2003gd
(Hendry et al. 2005), for which pre-need images
establish the progenitor as a red supergiant, the
first thus directly confirmed, though we all firmly
believe that RSGs should be commoner than the BSGs
seen for SN 1987A and 1998A (a somewhat indirect
argument for the latter, Pastorello et al. 2005).
Several other papers recorded limits on SN II
progenitors from images taken fortuitously  in
advance, of which the most interesting is probably 2004dj
in NGC 2403k discussed by Maiz-Apellaniz et al. (2004) and
Wang et al. (2005c). It appeared in a young star cluster
previously examined by Sandage (1984). Which star is
missing is not yet certain and,
therefore, whether it was a red or a blue SG whose
core collapsed.

And now a set of standard answers to three remaining
standard questions about progenitors and mechanisms of
nuclear explosions and mechanisms of ejection from core
collapses.

In the search for SN Ia progenitors, there were votes for
all the traditional ways of forcing a white dwarf up
in mass: (1) supersoft X-ray binaries (Lanz et al. 2005
on the Cal 83 system of white dwarf plus normal star),
(2) novae (Kato \& Hachisu 2004, though they note that
even helium accretion and explosion on a WD will remove
material, (3) WD + AGB (Kotak et al. 2004), and (4) most
obvious of all, merger of two white dwarfs
(Tovmassian et al. 2004), apart from the detail that
they have failed the existence test for decades.
Morales-Rueda et al. (2005) present yet another sample
that has no WD binaries of sufficiently small orbit
and sufficiently large total mass to work.

As for the explosive mechanism, detailed calculations
of subsonic deflagration converting to supersonic detonation,
typically with off-center
ignition, continue to improve (Gamez et al. 2005,
Wunsch \& Woosley 2004), even though the first
ignition point may
fizzle. We've  built some campfires like that and, so,
worrisomely, have the controlled
burn folks at the forest service.
A green star to Spuromilio et al. (2004) for an
examination of infrared lines in the first
year of light decline of a Type Ia that revealed the
gradual conversion of cobalt to iron (a total of
about 0.4 $M_\odot$). This confirms that the
production of lots of $\mathrm{{}^{56}Ni}$ and its subsequent
decay energy really does power these events.

As for the mechanism by which the gravitational potential
energy released in core collapse ejects SN II (etc.)
outer layers and makes them shine, we caught 8 papers
saying that, whatever you might have thought before,
rapid rotation and strong magnetic fields in the parent
star, the collapsing core, winds from the core, and
disk of continuing accretion material are a Good Thing,
and at least 3 expressing doubts about whether the
problem has yet been solved. Let Wilson et al. (2005a) stand for the
``it's all going to be all right,
guys'' camp (since he has been working on the problem
longer than just about anybody) and Ardeljan et a1. (2005)
stand for the ``hold your horses, because you are going to
need more''  camp, because they get an explosion (after
including the Balbus-Hawley instability, a compression wave,
and a shock) but it is a puny one that ejects only 0.14 $M_\odot$.
Ought everybody to get the same answer? Apparently not.
Branch et al. (2004) point out that different explosion
patterns work for different events.

\subsection{Supernova Remnants}
 Most of the obvious ones got at
least a nod. Let's start with SN 1987A and work backwards.
It now consists largely of stuff previously ejected and
now illuminated by the light flash and collisions with SN ejecta.
 There are, so far, no contradictions, say Sugerman et al. (2005),
 but we were frankly unable to identify
the features called the hourglass and Napoleon's Hat
(North and South) in their figures. The limit on optical
emission from a central compact remnant has been pushed
down to 4 $L_\odot$ (with allowance for 35\% absorption,
Graves et al. 2005), and it will not be easy to do better.

Cas A has less than 1.5 $M_\odot$ of dust (Wilson \& Batria 2005).
That sounds large for a limit, but the supernova of
1685 $\pm$ something did eject lots of new heavy elements now
in gaseous form (Hwang et al. 2004). The much smaller amount of
dust present in the Crab Nebula (Green et al. 2004) is equally
unsurprising, given that it is also not a habitat of
new gaseous metals.

Kepler (V 843 Oph = SN 1572 = 3C 358) and Tycho ( = B
Cas = SN 1604 + 3C10) each received a paper focussed
on using the full range of original observations to
confirm that all those = signs are true (Green 2004, 2005a).

The Crab Nebula has become slowly more massive over the years,
 this year reaching at least 6.4 $M_\odot$ for the visible nebula
plus pulsar, according to Negi (2005), almost enough to
account for the entire progenitor star. That less than 20\%
of the pulsar spin-down energy goes into relativistic protons
(0\% would fit the data, Aharonian et al. 2004) confirms a
deduction many years ago by the crabbiest author.
She was also much pleased to see an explanation of
radial fingers of gas as twisted magnetic filaments
(Carlqvist 2004) because she was, long ago, required
to deny their existence and entitle her thesis
``Motions and Structure of the Filamentary Envelop of the
Crab Nebula.'' A fading green star for the gradual fading
of the Crab radio flux, by 9\% since 1948
(Stankevich \& Ivanov 2005), interrupted by two cm/mm
outbursts following the pulsar glitches of 1975 and 1989.
The poor thing used to be a standard source, and we wonder
whether one would now have to say that Cas A faded faster
than reported in the past, when it was compared to 3C 144.
Meanwhile, such radio emission as there still is at the
latter location shows ripples and wisps, much like the
optical ones reported long ago by Walter Baade
(Bietenholz et al. 2004). The relative phasing is
complicated and implies two in situ acceleration mechanisms.

The remnant of SN 1006, supposedly a Type Ia and the
brightest of modern times, was treated rather badly
this year. Not only could Winkler et al. (2005)
find no more than 0.06
$M_\odot$ of iron hanging around (vs. 0.5 $M_\odot$ or so
required to power a bright SN Ia), but also HESS
saw less than 10\% of the TeV flux previously
reported from the HEGRA and CANGAROO facilities
(Aharonian et al. 2005g).

 The Vela supernova remnant(s)
received a whole week's worth of papers, largely
considering whether the bright bit in the corner is
a separate younger SNR. We will weasel by saying just
that the radio structure of the whole region is very
complicated (Hales et al. 2004) and that the
corner bit is a gamma ray source
(Katagiri et al. 2005).

SNR RX J1713.7--3946 = G347.3--0.5 was reported as the
first TeV source among the shell-type remnants
(Aharonian et al. 2004a), but we find difficulty in
imagining it famous under either of those names, which
 pertain to the X-ray and radio emission.

\subsection{Single Neutron Stars and Black Holes}
 These
include the pulsars, isolated (cooling) neutron star
X-ray sources, almost certainly the anomalous X-ray
 pulsars (AXPs) and soft gamma repeaters (SGRs), and
 isolated stellar mass black holes (except there weren't
  any this year). The classic questions about neutron
stars include the initial values of temperature, magnetic
field, rotation rate, and space velocity; how these change
with aging; and the use of the observed or deduced
properties to decide about birthrates, equation of
state, and whatever else you might like to know.

The index holds 63 relevant papers from fiscal 2005,
and the colored dot is attached to the giant flare of
SGR 1806--20 on 27 December 2004. It saturated INTEGRAL
with about 100 times the flux of any previous flare
(Mereghetti et al. 2005) and had various precursors
(within 100 s), tails to 3000 s, a 7.56 s pulsation period
as it faded (presumably the stellar rotation period), and
60 msec QPOs (Hurley et al. 2005, Gaensler et al. 2005,
Terasawa et al. 2005, Lazzati 2005, Palmer et al. 2005,
Cameron et al. 2005).
A mechanism associated with crust instability appears
likely, and considerable importance may attach to the
fact that it would have looked like a short-duration GRB
if we had been observing from a not-too-distant external
galaxy. The burst was powerful enough to affect ionization
of the upper atmosphere, and it was seen as a Sudden
Ionospheric Disturbance by radio amateurs; probably
also by various defense installations, but they don't
report in Sky \& Telescope (109, No.~5, p.~32).

 On the
subject of pulsars, we refer you to a review of observations
(Seirdakis \& Wielebinski 2004) and as many of the following
as you can fit on your buffet plate without spilling.

Timing noise nearly always dominated the second derivative
of pulsar periods (Hobbs et al. 2004, 2005), but one whose
fax number we neglected to record is so quiet that even its
third derivative is meaningful
$(-1.28\pm 0.28)\times 10^{-31}$ $\mathrm{s^{-4}}$
(Livingtone et al. 2005) from 21 years of glitchless data.
It implies $n = 2.839\pm 0.003$ for the parameter whose value
is 3 for pure magnetic dipole radiation (and different
from 3 for the few other pulsars with measured $n$).

Geminga seems to have fled from an OB association
(Pellizza et al. 2005). The details are a bit model
dependent, but the general idea is that it came from
a moderately massive star, and, since the association
is still there, cannot be very old.

The velocity required by the previous point is nothing
like a record for pulsar proper motion (1083 km/sec for
B1508+55, Chatterjee et al. 2005).
Mdzinarishvili \& Melikidze (2004) conclude that pulsars
found from Australia reflect two separate populations
with initial velocity and initial field positively correlated.
Hobbs et al. (2005), on the other hand, conclude that
the young pulsars have a single $N(v)$ distribution and
so are a single population.

In general, there seems to be reasonable accord between
observations and ``predictions'' for both initial field,
centered somewhere around $2.5\times 10^{12}$ G, and initial
rotation period, near 15 msec, (Vranesvic et al. 2004,
Walder et al. 2005, Loehmer et al. 2004a).
Lovelace (2005) concludes that magnetic field is reduced
by accretion and recovers slowly.
Normal processes can slow the rotation period to at
least 8.39 sec before death intervenes (Kaplan \& van Kerkwijk 2005).

Giant radio pulses have become common enough that
Kuzmin \& Ershov (2004) advocate two classes (with
emission arising from outer gaps and from poles).
The Crab pulsar does it,  but, for once, does not
hold the record. $T_b \ge 5\times10^{39}$ K belongs to the msec psr
B1937+21 (Soglasnov et al. 2004), the  temperature
reduction mechanism proposed by Gil \& Melikidze (2005)
for 0532 (beaming and other
relativistic effects) probably also applies to B1937.
You can reduce the implied $T_b$ by a factor $10^8$, which
still leaves it hotter than hell (or heaven, using
a traditional description of the illumination there).

 Temperature evolution
is generally probed with the isolated neutron star
X-ray sources. To quark (so as to accelerate cooling)
or not to quark is the question. On the conservative
(``our'') side, Page et al. (2004) look at the various
possible enhanced cooling
 processes (quark stars, direct URCA, pion or hyperon
 condensate), and conclude that the minimal extension
 to Cooper pairs and modified URCA is sufficient for all
 cases where thermal X-rays are actually seen, while two
 upper limits appear to need enhanced cooling. Additional
 ``other physics'' appears in Gusakov et al. (2004,
strong proton superfluidity) and Khodel et al. (2004, multi-sheeted neutron
Fermi surfaces to activate direct URCA cooling).

``Magnetar'' is shorthand for the class of neutron stars
with very strong magnetic fields, including the soft
gamma repeaters, the anomalous X-ray pulsars, and perhaps
some others. Does everyone agree with this definition?
No. At least three papers during the year held out for
fields near $10^{12}$ G,
like ordinary pulsars, rather than $10^{14-15}$ G, and other
effects contributing to the AXP/SGR phenomena.
Malov \& Machabeli (2004) favor an electric cycle,
Istomin et al. (2005) a field strongly concentrated
toward the poles so that rotation periods can slow to
about 10 sec (with magnetic dipole radiation continuing), and
Mosquera Cuesta \& Salim (2004) propose significant
effects of strong gravitational fields.
We don't entirely
understand these, but it is impossible to dislike
a paper that begins by citing Born \& Infeld (1934).

 The now-conventional strong field view is
upheld by Halpern \& Gotthelf (2005) and Gaensler et al. (2005a).
 Figer et a1. (2005) conclude that the progenitors were
 at the upper limit of the mass range, 30--50 $M_\odot$ of stars
that can make neutron stars rather than black holes,
and readers with long memories may remember from \S9.5
the probable correlation of large white dwarf masses (hence
hefty progenitors) with strong fields for them. Kaspi \&
McLaughlin (2005) may have seen some faint thermal X-ray
emission from neutron stars that correspond to AXPs in
quiescence; and Woods et al. (2005) record what they
indicate is the
third example of a new class of burst peculiar to AXPs.
Sedrakian et al. (2003) would like all pulsars eventually
to evolve to AXPs or SGRs, but how the fields strengthened
as the rotation slows was not obvious.

\subsection{Binary Neutron Stars and Black Holes}
 Are they really
black holes? Well, they are astrophysicists' black holes
anyhow, that is, entities with (1) masses too large for
neutron or quark stars (10.65 $M_\odot$ for V404 Cyg,
Cherepashckuk 2004); 13--14 $M_\odot$ for GRS 1915+105,
Fujimoto et al. 2004); and a distribution through the
mass range 4--15 $M_\odot$ more or less what the theorists
expect, Borgornazov et al. (2005), (2) discernible general
relativistic effects of spin close to the maximum allowed
(Aschenbach 2004), and (3) evidence for a horizon, when
the luminosities and spectra of the BHXRBs are compared
with NSXRBs in quiescence (McClintock et al. 2004).
Indeed allowing for the smaller scales of everything,
they are a good deal like the centers of active galaxies,
including our own feeble Sgr $\mathrm{A^\star}$ (Jester 2005) in several
respects including, probably, 3:2 resonances in QPO
frequencies (Torok 2005, Homan et al. 2005).
Among other analogies, a good deal more than half of the
available energy should and does come out in jet kinetic
energy (Jester 2005, Gallo et al. 2005). The QPOs for
standard AGNs are going to be a tad difficult to observe,
unless the TAC gives you really long observing runs
(Vaughan \& Uttley 2005).

Can you always tell an NSXRB from a BHXRB? As in many
previous years, the accreting, compact member of SS 433
was firmly established as a probable neutron star of 2.9 $M_\odot$
(Hillwig et al. 2004) and as a definite black hole of 30 $M_\odot$
(Cherepashchuk et al. 2005). The donor, whose spectrum we
see, is in either case an A supergiant of 9--10 $M_\odot$.

 Binary
neutron star can mean NS + something else or NS + NS, and
binary pulsar can mean pulsar + something else or
pulsar + pulsar. All exist. The first pulsar + pulsar was a
highlight last year, and theorists have since been
beavering away to interpret all that has been seen.
We note, arbitrarily, two papers from the ``all is well'' camp,
on the X-ray light curve (Campanaet et al. 2004, with the stars
illuminating each other) and on the radio eclipse as
synchrotron absorption (Lyutikov 2004), plus one, we think,
rather odd evolutionary scenario (Piran \& Shaviv 2005)
in which the initial masses of the stars were each only
about 1.45 $M_\odot$. Oh, if you need to phone, the number
is J0737--3039. The galactic center transient radio source,
GCRT J1808.4--3658 could, say Turolla et al. (2005) be
another pulsar pair. The discoverers, Hyman et al. (2005)
suggest several other possibilities.

Faulkner et al. (2005) have caught the fifth binary,
meaning pulsar + another NS, that will merge in less than
the Hubble time, thereby increasing the predicted rate of
short-duration Galactic GRBs by 25\%, we estimate.

In the neutron star plus something else category,
Galloway et al. (2005) report the 6th accretion-powered
millisecond pulsar (with the shortest rotation period yet
of 1.67 msec) and they draw attention to the puzzle of
why one sees the rotation in these six and not in the other
750 LMXRBs, although sometimes the accretion turns off,
and you can then see the rotation period that is otherwise
powering the source (Campana 2004 on SAX  J1808.4--3658
in quiescence). Also in the NS + other bin lives the first
millisecond pulsar to experience a glitch (while anyone
was watching, Cognard \& Backer 2004); the first HMXRB
with a neutron star whose rotation period does not show
in its light curve (Blay et al. 2005);
more of those superoutbursts that are flashes of carbon
 burning on the NS surface (in't Zand et al. 2004); the
 first Type I X-ray burst outside the Milky Way, naturally
 in M31 in a globular cluster (Pietsch \& Haberl 2005); and
 Rossby waves on the surfaces of neutron stars
as an explanation for the decrease in QPO
frequencies when X-ray bursts fade (as an alternative
to the radius of the photosphere shrinking back). Of many
papers, we cite only Heyl (2005), because it seems to
have been his idea last year, and we were having tea and missed it.

The opposite case, of something else not plus a neutron
star, is exhibited by a bunch of OB runaway stars, none
of which is a ROSAT source, implying that none has held onto a
close NS or BH companions (Meutrs et al. 2005), though
others have recorded runaway XBRs on other occasions
(Sepinsky et al. 2005).

\subsection{Ultraluminous X-ray Sources (ULXRS) and Intermediate
Mass Black Holes (IMBHs)}
 Do they or don't they color their
no hair?\footnote{Black holes have no hair and it must, therefore, be
wigs rather than Clairol hair coloring about which only
their hairdresser knows for sure.}
They are X-ray sources, mostly in other
galaxies, bright enough to exceed the Eddington luminosity
for the sort of 5--15 $M_\odot$ black holes found
in the previous section, extending up to about $10^{41}$ erg/sec,
assuming isotrop1c emission. This is the Eddington limit for
a $10^3$ $M_\odot$ accretor. Most definitions just say brighter than
$10^{39}$ erg/sec, and the faint end of the distribution is
commoner than the bright end. The choices then are
(a) sources other than accreting compact objects,
(b) beaming, (c) intermediate mass black holes of
100 to perhaps as much as $10^4$ $M_\odot$, and (d)
accidental projections of very bright, much more distant
sources. The alternative to (d) is ULXRSs with non-cosmological
redshifts (Galianni et al. 2005). We and
Gutierrez \& Lopez-Corredoira (2005). are voting for
the conventional wisdom here, they because the areal
density of the ones with large redshifts is that of
random sources, that is indeed accidental projections.

As for the rest, there is no general agreement on
whether the ULXRSs constitute a class separate from
the general run of high-mass X-ray binaries. Yes, say
 Miller et al. (2004a) on the basis of X-ray temperatures
 less than 0.25 keV for the brightest (vs. 0.3--2 keV
 for galactic BHXRBs) and also the absence of optical
 identifications; and no, say Swarz et al. (2004) from the
 absence of discontinuities in spatial, spectral, color, or
 variability distributions with luminosity.

And here are the cases for some combination of (a), (b),
and (c). Liu \& Bregman (2005) have provided a catalog
of the 109 brightest ROSAT sources in about 65 galaxies.
These preferentially inhabit late type galaxies and star
formation regions. Some are supernova remnants,
HII regions, and compact groups of young massive stars.
A few coincide with old globular clusters (and could be
IMBHs in those). Liu \& Bregman have deferred dividing the
rest between BHXRBs and IMBHs until Paper II. And galaxy
mergers can make shock features potentially confusable
with our target class (Smith et al. 2005c). A popular view
is mostly BHXRBs with a few IMBHs (Fabbiano 2005;
Liu \& Mirabel 2005, whose catalog of 229 in 85 galaxies
has some background AGNs and SNRs mixed in).

The best cases during the index year for ordinary though
massive BHXRBs seem to be (a) M101 in which
Kuntz et al. (2005) have shown that one has a mid-B
supergiant as its optical counterpart, (b) M74 where
Krauss et al. (2005) have recorded spectrum and variability
like those of an end-on microquasar jet, (c) N4559
with a similar source (Soria et al.
2005), and (d) the Milky Way some of whose sources
reach 10 times the Eddington luminosities in cases where
the black hole mass has been established (Okuda 2005).

And the best cases for accretors in the $10^{3-4}$ $M_\odot$ range
include M82, assuming its 55 mHz oscillation frequency
is a low frequency QPO like those of BHXRBs of smaller masses
and higher frequencies (Fiorito \& Titarchuk 2004),
(b) NGC 628 with a 2-hour quasi-period, scaled in the
same way (Liu et al. 2005), (c) NGC 5204 because of its
very cool (0.2 keV) inner disk (Roberts et al. 2005a),
and (d) Holmberg II because the optical and radio emission
from the surrounding nebula argue against significant beaming
(Lehmann et al. 2005, Miller et al. 2005a).

We now hand over even further to the theorists
and ask ``Can you account for these things and put them
someplace where they will have material to accrete at a
rate at least large enough to support the Eddington
luminosity?''

Pas de probleme (we think this sounds less
rude than ``no problem'' as a substitute for ``you're welcome'')
says one school of thought. Intermediate mass black holes
should be left
from the collapse of population III star cores and other
events in the early universe. Indeed the Milky Way might
well have a supply of them, contributing a bit to its dark
matter (Tutukov 2005, Islam et al. 2004, Zhao \& Silk 2005).

Feeding may be more difficult. Baumgardt et al. (2004)
consider the case of a $10^3$ $M_\odot$ black hole grown from a 100 $M_\odot$
seed in a globular cluster, and point out that all that will
be left nearby will be small black holes, whose accretion
will emit very few X-rays and butter no parsnips. Two papers
indicate that, to be an URXRS, an IMBH must have a biggish
star in orbit close enough for Roche lobe overflow
(Portegies Zwart et al. 2005 on M82 X-I and
Tutukov \& Fedorova 2005). Volonteri \& Perna (2005) say point blank that
IMBHs can be left wandering in galactic halos from hierarchical
galaxy formation, but they simply must carry their own baryons
around with them to reach even $10^{39}$ erg/sec.

A green star, therefore, for the idea that at least a few ULSRSs
in M82 and elsewhere may well be the nuclei of captured
satellite galaxies, for whom $10^3$ is the ``right'' mass (\S12.7,
though none has been seen), and who indeed will be toting baryons
(King \& Dehnen 2005).

\subsection{Sgr $\mathrm{A^\star}$ and Its Environment}
 Sgr $\mathrm{A^\star}$, at the center of the
Milky Way, is not even an MLSRX (moderately luminous X-ray source)
at a bit less than $10^{34}$ erg/see in X-rays and less at other
wavelengths, though the radio emission, its spectrum,
variability, and so forth have been very extensively studied
ever since its prediction by Lynden-Bell \& Rees (1971) and
discovery by Balick \& Brown (1974). The proper motion of Sgr $\mathrm{A^\star}$
(after removal of the amount due to galactic rotation) is less than
1.5 milliarcsec/yr, and uncertainty in its three-dimensional location is
the largest source of error in measuring its mass from the
velocities of the stars around it (Ghez et al. 2005).

 The current mass accretion rate is very small, less than a
few$\,\times 10^{-7}$ $M_\odot$/yr, recognized because no gas was lit up when star
S2 passed close in 2002 (Nayakshin 2005), and the  amount that
actually fuels radiation is still smaller, more like $10^{-11}$ $M_\odot$/yr.
Most of the rest that gets as far as the Bondi radius is lost
in a wind, at least this year (Bower et al. 2005). The source
doesn't just sit there, however. A near infrared flare was
caught for the first time this year (Eckart et al. 2005,
who have fit a synchrotron self-Compton spectrum). We think
Clenet et al. (2005) have detected a quiescent counterpart,
for which two green ears and a tail should probably
be awarded. The X-rays display
quasi-periodic oscillations, with frequencies in 3:2:1
resonance (Abramowicz et al. 2004, Aschenbach 2004a).
The periods are 692, 1130, and 2178 sec, which require
the Sgr $\mathrm{A^\star}$ black hole to have a spin parameter $a = 0.996$,
shared with three microquasars. This is very close to the
maximum permitted by general relativity, and Aschenbach (2004b)
suggests this  may be the true maximum, with the estimate
by Thorne (1974) a smidge too large.
The millimeter emission varies on similar time scales, but
not with any obvious P or QP (Mauerhan et al. 2005).
The flux harder than 165 GeV either varies within a year or is
very poorly determined (Aharonian et al. 2004b). Our Galactic
feeble center is by no means unique. Totani et al. (2005) have
found at least one, and maybe six other very faint AGNs, of
which, they say, the Milky Way is typical.

A faded green star (because the idea was already out there
last year) goes to the thought that one can account for
conditions in the gas surrounding Sgr $\mathrm{A^\star}$ (mapped by
INTEGRAL) if the central X-ray source was brighter by a
factor near $10^5$ a few hundred years
ago (Revnivtsev et al. 2004). This is
still only a small fraction of $L_\mathrm{edd}$, which is nearly
$10^{11}$ $L_\odot$ for a $3\times10^6$ $M_\odot$ black hole.
Is there a theorist in the house to explain it all?
Of course. But the most relevant related point may come
from observers. Stark et al. (2004) present CO maps of the
galactic center region implying, they say, that gas piles
up around 150 pc and, every $2\times 10^7$ yr, collapses inward
making a bunch of giant molecular clouds, followed by massive
stars and the dumping of $4\times 10^7$ $M_\odot$ of gas into the central
region. The present moment would then be the end of such an
event, and we are perhaps lucky
to have caught the instant when the gas disk has nearly all
been turned into stars, leaving very little central gas fuel
for the black hole (Nayakshin \& Cuadra 2005).

What is around Sgr $\mathrm{A^\star}$? Well, that gas disk for starters,
whose 119 km/sec HI rotation speed (Dwarkanath et al. 2004)
must mean that it is about a parsec in radius, for the density
and temperature that radiates 21 cm, plus much denser
molecular gas that could continue to form stars
(Christopher et al. 2005). Lots of stars, which are generally
held to be young and massive and to constitute an example of a
nuclear star cluster, or a bunch of  NSCs
(Stolte et al. 2005). Still, there are people who would doubt
existence of the
tooth ferry even if they had ridden on it themselves, and
Davies \& King (2005) counterpropose tidally stripped
low and intermediate mass (hence older) stars, formed much further
from the galactic center and recognized as interesting only
for the subset whose orbits bring them in close.

In the X-ray regime, there are diffuse sources (Muno 2004),
the 511 keV line (Parizot et al. 2005), and Sgr A East, which
is a supernova remnant, indeed just possibly the only known
remnant of a Type Ia event arising in a white dwarf of less
than Chandrasekhar mass, compressed to ignition temperature
and density as a result of passing close to a black hole
(Dearborn et al. 2005). A bunch of faint, mostly hard compact
X-ray sources (Belczynski \& Taam 2004, Muno et al. 2004)
include a number of intermediate polars (moderately magnetized
cataclysmic variables), Wolf-Rayet stars, OB stars, RS CVn
binaries, young pulsars, BH and NSXRBs, and millisecond
pulsars with accretion from winds in some cases and RLOF
in others, Most of the common zoo and domestic species in
other words.

A new sort of radio beast is GCRT J1745--3009 (of unknown
distance), which in 2002 exhibited a string of 10 min bursts,
1.27 hours apart, captured in 330 Hz data by Hyman et al. (2005).
Observations of the region in earlier and later years show
nothing down to 15 mJy (vs. 1 Jy bursts). What is it? A nearby
brown dwarf; a nulling pulsar, magnetar, or
coherent microquasar; example of kinds of beaming and
beacons predicted long ago? Or something that hasn't
been though of yet.

\subsection{The Black Hole Bulge Connection}
 If every galaxy
has one, why do people talk about them so much? Well,
the same could probably be said about human private parts,
which also have in common with black holes a central
location and, as a rule, concealing material around.
And here we had better let the analogy go, and proceed to
outstanding questions. Every biggish galaxy (with a
spheroidal  component) has a black hole whose mass
is somewhere around $10^{-3}$ of the spheroidal stellar mass,
and the correlation is somewhat tighter if you choose
spheroidal velocity dispersion rather than mass for your abscissa. Data
revealing this count as one of the triumphs of extended programs
with the Hubble Space Telescope, whose angular resolution was
required to separate
the dynamical effect of the central black hole from that of
other stuff you find growing
around galaxy centers. Questions not yet fully answered
(or at least not everybody offers the same answer) include,
(a) how far down in mass does the correlation extend?
(b) which came first, the black hole or the stars,
or (c), if (b) is the wrong question, how did they co-form
to end with the ratio they now have (notice that this accepts
that bulge star formation is way past its prime and so are
QSOs and all)?, (d) what was the situation like at moderate to
large redshift, and do the standard $\Lambda$CDM scenarios of structure
formation deal with it well?, and (e) what still needs to be
asked before it can be answered?

The small mass end remains mysterious. Modelers
(e.g. Kawakatu et al. 2005) predict that there should
still be black holes (though perhaps with smaller ratios
to the total) down at least to $10^4$ $M_\odot$, outside the range
currently accessible to observations (e.g.
Valluri et al. 2005 on NGC 205; Barth et al. 2005 on dwarf
Seyferts).\footnote{Whether it is OK to compare Seyferts with normal
galaxies requires a small digression.
AGNs in general are clearly not normal in that they are
making better use of their black
holes than average, but what about the BH to bulge mass
ratios? Silge et al. (2005) say that Cen A hosts a black
hole 5--10 times more massive than average;
Wilman et al. (2005) and Captetti et al. (2005)
say normal for Perseus A and the Seyfert NGC 5252
respectively.
And Mathur \& Grupe (2005) and Collin \& Kawaguchi (2005)
find a number of Seyferts with inferior black holes and
predict that, when these reach the average mass for the
hosts' velocity dispersion, accretion will drop to well
below the Eddington rate and the central sources turn off.}

As for which crossed the road first, the possible answers
are stars first, black holes first, and co-formation
(with two subanswers, constant ratio and BH/bulge ratio
dropping with time). All four appeared during the year.
Begelman \& Nath (2005) predict (if that is the right word)
that feedback from BH accretion into the protogalaxy should
keep the BH/$\sigma_v$ ratio the same for all redshifts and all
halo masses, even small ones. Cai \& Shu
(2005) make the same prediction, from a magnetic
feedback mechanism, provided that all sources begin
with 46 mG. One observed sample concurs,
Adelberger \& Steidel (2005) finding that the
BH/bulge ratio at $z = 2$--3 is the same as now over
the black hole mass range $10^6$--$10^{10.5}$ $M_\odot$. They are,
however, outvoted by samples with BH/bulge larger at
moderate to large $z$ (Akiyama 2005, $z = 2$--4 data;
 Merloni et al. 2004a, synthesis of many kinds of data;
  Bonning et al. 2005 pointing out that both
accretion and star formation are small now, but the stars
have been gaining on the black holes for some time).
And there are probably also more theorists on the side
of BH/bulge larger in the past, including Wyithe \& Loeg (2005)
and Koushiappas et al. (2004), who say that black holes can
grow only a factor two in mass since $z = 15$ when seeds
stopped forming because of reionization. They also
conclude that the smallest seeds will have the Jeans
mass during the dark ages, $10^5$ $M_\odot$, so that galaxies with
bulge star masses less than $10^8$ $M_\odot$ cannot have BHs in the
proper proportion. True et al. (2005) report Seyferts with
large BH/bulge ratios at $z = 0.37$, which counts as BH first.

Alexander et al. (2005) belong to the stars first camp,
although their main point is that quite a lot of the
black hole mass growth occurs behind obscuration.
Martinez-Sansigre et al. (2005) concur, saying that a
complete sample would have obscured QSOs (type 2)
outnumbering the unobscured (type 1) by a factor three.
They present a correctable sample of Spitzer sources
with $z= 1.4$--4.2. The opposite conclusion, that most
black hole growth by accretion is by unclothed
accretion, is reached by Barger et al (2005)
using statistical arguments.

 Is accretion the only way
black holes can grow? Obviously not. When protogalaxies
merge, their central black holes must also merge or go
whirling around each other forever (such binaries exist
but are not terribly common). Ap04 reported a majority
view against  mergers as an important process in
BH mass growth. This is probably still more or less true, firmly so
according to Shankar et al. (2004), but a significant
role for mergers is advocated by Saitoh \& Wada (2004),
Yoo \& Miralda-Escude (2004), Di Matteo et al. (2005,
making the point that outflow from the BH eventually stops
both accretion and star formation),  and
Hao et al. (2005, discussing violent mergers which
preserve a standard ratio because the star formation
rate is a few hundred times the
accretion rate).

Semi-finally, three groups have worried about the most
distant QSO in current catalogs. It has $z = 6.42$ and a black hole of at least
$10^9$ $M_\odot$. Walter (2004), Yoo \& Miralda-Escude,
and Shapiro (2005a) all make the point
that nature and theorists have to work so hard to make
that big a black hole so quickly that they just don't
have a chance to make  all the stars as well.
Walter also notes that the expected $10^{12}$ $M_\odot$ of stars
isn't actually seen either.

And the concept of ``downsizing'' seems to apply to
black holes as well as to their host galaxies (\S4).
Heckman et al. (2004) report that most accretion is
now occuring on black holes of less than $10^8$ $M_\odot$ (just
as most star formation is now occurring in galaxies
of $10^{10-11}$ $M_\odot$), and that this is smaller than the
average accreting black hole of the past.

\subsection{Active Galaxies and their Nuclei}
 Of the 139 papers,
read (R), precised (P), and Indexed (I) on this topic
(where $\mathrm{R> P> I}$), only one ended up with a star,
Nipoti et al. (2005) on the ancient question of why
some are radio loud and more radio quiet. Their answer
is that true, radio loud, quasars and the quiet QSOs
are merely two modes of the same population. A number
of other very old questions received at least one
answer during the year, and for those questions that
seem to invite yes or no, typically both appeared.

\textsl{Confinement} summarizes the puzzle that the extended
clouds of relativistic particles
and magnetic field required to emit synchrotron radio
don't just expand freely at the speed of light. X~ray
gas confines radio jets on 130 kpc scales say
Evans et al. (2005a). This is one of the classic candidates,
 though at one time the pressure was thought to come from
a hot intergalactic medium with density close to the
critical cosmological density.

\textsl{Equipartition} between magnetic field and relativistic
electrons seems to apply in the contexts where it was
first supposed to, the lobes and hot spots of
Fanaroff-Riley II radio sources (the sort with two large
lobes and hot spots on either side of a galaxy with
jets feeding them). So say Hardcastle et al. (2004a) and
Croston et al. (2004). The latter also note that it would be
odd to find field and electrons in equilibrium if the energy
in protons were larger, so it probably isn't.
Beck \& Krause (2005) rediscuss energetics in the case
where the protons are winning (by 40:1 or so), as they
expect from certain kinds of shock acceleration.

The microquasar GRS 1758--250 further justifies its
name by also having equipartition
in its radio-emitting lobes (Hardcastle 2005). The
jets themselves (FR II, microquasar, or what have you)
will generally not be in equipartition
(Tyul'bashev \& Chennikov 2004), but it is the wrong
issue to investigate anyhow, because nearly all the
energy is in bulk, mildly relativistic flow
(Nagar et a1. 2005, Jorstad \& Marscher 2004, semi-randomly
out of half a dozen papers that made the point during
the year). The case of the Milky Way core is slightly puzzling.
We think LaRosa et al. (2005)
are saying that the field wins, although the equipartition
particle density, at 2 $\mathrm{eV/cm^3}$, is like cosmic rays here.
A ha! In this context the protons probably have 100 or so
times the electron energy density (as in cosmic rays), and
so they are perhaps in equipartition with the observed mG
field. Abacus time, guys.

\textsl{Binary black holes} in orbit would seem to be an inevitable
result of mergers of galaxies each of whom had one.
Quasar 0957 with a period near 12 years is the
longest-discussed and probably best established example.
Cases were made this year for 3C 345 with a period  of
480 yr (no, Lobanov \& Rolant 2005 haven't seen more than one), 3C 273
(Zhou et al. 2004 using a model for jet acceleration), NGC 4716
(two variable nuclei, 60 pc apart, Maoz et al. 2005),
Pks 1510--089 (P = 336 days and the 4th minimum arrived
on schedule, Wu et al. 2005).
You might draw two different statistical conclusions
from these examples. If some of the
best and brightest AGNs have binary BHs, they must be
common (or that is why the sources are bright, not we
think claimed by anyone this year); or, conversely,
since the literature isn't totally overflowing with
examples, they must be rather rare. Theorists can, of
course, explain both. On the one hand, once a binary
BH clears out the loss cone (stars available for
disruption and accretion), a Hubble time is needed to
repopulate (= rare among observed sources,
Merritt \& Wang 2005). And, on the other hand if there
are any stars available at all, the binary will be much
better at the tearing  apart than a mere single monster
(Ivanov et al. 2005). The process is a tad complicated,
but our old friend Kozai (1962) of the resonances comes in
somewhere. And here we had thought it was just for asteroids.
Binary black holes even in a vacuum will lose angular
momentum and energy to gravitational radiation and merge after
\begin{displaymath}
    t_0=\frac{5}{512}\,\frac{c^5}{G^3}\,\frac{a^4}{M^3}
\end{displaymath}
where $a$ is the current semi-major axis and $M$ is the mass
of each of two equal black holes, perhaps $10^8$ $M_\odot$ each.
Enjoy the calculation says your server.

\textsl{``Alignments''} means various optical phenomena, including
star light and emission lines, with radio jets, and we can
all think of several possible causes. At one time the
phenomenon was thought to be limited to $z\gtrsim 1$, and it remains
true that more alignment is seen at large redshift and high
luminosity (Inskip et al. 2005 on 6C galaxies). But
NGC 1068 and Cen A show, respectively, aligned line
emission and star light (Gratadour et al. 2005,
Osterloo \& Morganti 2005). We found support for at
least two of the standard mechanisms:
photoionization of gas near the jets (Whittle et al. 2005)
and jet triggering of star formation (Klamer et al. 2004a,
reporting that the first stars and first metals in a
$z = 4.7$ galaxy formed along the radio jet).

\textsl{Super-Eddington luminosities} are apparently rare, say
Paltani \& Turler (2005), though
they discuss in detail only 3C 273, for which they find a
real black hole mass of 6--$8\times10^9$ $M_\odot$ from reverberation
mapping, 10 times the number from emission line velocity
widths, implying a more or less face-on disk.

\textsl{Type II AGNs} have as their prototype the Type II Seyfert
galaxies, whose broad line regions are obscured by their
accretion tori and so visible only in scattered (thus
polarized) light. You might suppose that the brighter
sorts of AGN would be harder to hide. Indeed Type II
QSOs were announced as a highlight a few years ago.
They have become common (Zakamska et al. 2005, a bunch
more from SDSS samples). Grindlay et al. (2005) announced
the first Type II blazar, GRS 1227+035, recognized in
balloon X-ray data.

\textsl{Unification} is the classic yes and no issue, where
the angle from which we view a system is proposed as a
major discriminant among AGN types and subtypes. It is
part of the story, say Varano et al. (2004) who have
compared FR II radio sources with quasars and find that
the opening angle of the torus gets larger as the
luminosity of the accretion disk
gets larger. But not the whole story. There are also
strong correlations of properties whose observed
values  will depend on orientation
with properties whose observed values should be
orientation-independent, for instance Marcha et al. (2005)
 on optical emission lines (independent) versus core/jet
 radio ratio (dependent) and Shi et al. (2005) on radio
 flux (beamed) versus 70 $\mu$ emission (isotropic). And we
 think that Imanishis \& Wada (2004) are proposing strength
 of nuclear starburst proportional to AGN luminosity as
  the cause of some of these correlations.

\textsl{Both please}. Are the best and brightest AGNs also vigorous
star formers? Yes for samples reported in at least seven
reference-year papers, of which only Storchi-Bergmann et al. (2005)
on NGC 1097 get cited. But not always. 3C 31 has
$10^9$ $M_\odot$ of molecular gas
within 1 kpc of its center but is not forming stars
(Okuda et al. 2005). How are we supposed to know it is
an AGN? Well, it carries its 3C around with it.
In contrast, the Wolf-Rayet galaxies like NGC 6794 do
it all with stars (O'Halloran et al. 2005). And every
year someone points out that there is or ought to be
an evolutionary sequence, with mergers yielding a star
 burst and jets forming
later. This year it was Tadhunter et al. (2005).

\textsl{ADAF, ADIOS, and all}. Very many accreting black holes are not
luminous in proportion to the amount of available gas.
The two major competitors for the answer to ``what becomes
of it?'' are down the tubes, taking energy along (known as
ADAF or advection dominated accretion flow) and blow back
(of which ADIOS is one sort). Two not quite random papers of
many, (1) blowback of various sorts as the explanation for
poor correlations of black hole masses, Bondi accretion
rate, and X-ray luminosities (Pellegrini 2005), and
(2) ADAF as a picture of broad line emission regions in
AGNs (Czerny et al. 2004, with a special color-changing
star for their having voted this way over their own
previous hypothesis).

\textsl{Radio loud/quiet}. The basic dichotomy was not
challenged this year (though the exciting class of
radio intermediate galaxies exists). One must begin by
distinguishing correlations from causes. That the
 radio-louds have more supernovae (Della Valle et al. 2005)
 and more microvariability (Jang 2005) are presumably side
 effects of mergers and jets, correlated but not causal. That
 radio-quiets have relatively feeble jets that can neither escape
the host galaxy (Barvainis et al. 2005) nor provide powerful
 X-ray emission (Ulvestad et al. 2005) comes closer to
 sounding like a cause, but why the weak jets? The largest
 sample examined during the year (6000 SDSS QSOs and quasars,
 McLure \& Jarvis 2004) reveals that the black hole masses and
 host properties overlap far too much between the loud and
 quiet groups to be the dominant cause (and three papers that
declare big black bludges,
sorry, holes, to be the determining factor remain trapped
on pp.~49 and 67 of the notebook).
McLure \& Jarvis suggest an evolutionary sequence (cf.
Nipoti et al. 2005) or black hole spin, an idea hallowed
by multiple presentations over the years.
Bachev et al. (2004) voted for spin this year, while
Ye \& Wang (2005) said that there must be a second
parameter, which they describe as ``power law index of
variations of magnetic field on the disk.''
Depending on the extent to which the disk is
magnetically coupled to a (spinning or not spinning)
black hole, this could be an indirect causal connection.

And an observation that is new this year, at least
to us: all radio loud hosts have central optical
profiles that are cores rather than cusps
(Capetti \& Balmaverde 2005, de Ruiter et al. 2005).
The authors propose that galaxies with cuspy centers
will be forever silent, while currently quiet cored
galaxies are the radio sources of the past and future.

\textsl{Lifetimes of AGNs} are another of the truly old questions.
Do a few galaxies do it all their lives, or most for
1\% of the age of the universe each? The recognition
that central black holes are nearly ubiquitous would
seem almost to have settled the issue, and all the votes
we caught this year were for $10^7$ yr of being really bright
and $10^8$ yr of significant accretion on to the BH, much of
it in hiding (Hopkins et al. 2005a, Bonning et al. 2005,
Adelberger \& Steidel 2005a, Croom et al. 2005). The
turn-off comes for lack of gas, and Hawkins (2004)
calls attention to a class of Seyferts and QSOs
(face on and with no broad absorption line features)
 where we can see it happening. An occasional unwary
 star venturing too close can allow a brief resurgence
 (Tremain 2005, Gomboc \& Cadez 2005), and a black hole
 that gets too massive turns itself
off (Ann \& Thakur 2005).

\textsl{The hosts?} Well, they are big
(meaning massive) and fat (meaning spheroidal) and old
(meaning that star formation at least started a long time ago).
 A baker's dozen or more papers provided parts of that
 description, which could equally well apply to a good
 many of our friends and relations, so the coveted green
dots go to two other related ideas. First, there is a strong
positive correlation between accretion rate and metallicity
(Shemmer et al. 2004). And, second, downsizing (\S4)
applies even to nuclear activity, in the sense that, at
$z = 1$, a source about as bright as a modern Seyfert lived
in a much more massive galaxy as a rule (Gilli et al. 2005).

\textsl{Evolution?} There were more in the past, as you have known since
Sciama \& Rees (1966) used the redshift
distribution to refute steady state cosmology. Soon after,
observations were being urged to reveal whether the fraction
of galaxies with nuclear activity had declined (density
evolution), or was it the luminosity per source
(luminosity evolution), or both? A moment's thought, or a
decade of the literature, should persuade you that there
is no way to tell the difference if $N(L)$ is a pure power
law. Thus it is
structure in the luminosity function that now enables
Barger et al (2005) to say luminosity
evolution scaling as $(1+z)^3$ out to $z= 1.2$ for optical evolution and
Wall et al. (2005) to say density evolution for Parkes radio
sources since $z = 1$. Wall et al. also note that the
number density (in comoving coordinates) turns down again at
$z\ge  3$. Silverman et al. (2005) report the first X-ray selected
sample that also shows this turnover (or anti-evolution if
you must). Their second point, that most of the X-ray
luminosity today and back to $z\approx1$ comes from relatively
low $L_X$ sources, is echoed by Merloni et al. (2004). This is
presumably yet another aspect of downsizing (\S4),
though it also presumes that
the evil selection effects of magnitude limited
samples are not overwhelming.

\textsl{Non-cosmological redshifts?} Among many other challenges
nearby large-redshift QSOs would have is that of
passing their light through many (sometimes very many)
absorbing clouds of smaller redshift between them and us.
Ejected gas is one way to do this. Most of the astronomical
community agrees that there is such a class of
``associated'' lines, distinguished by line width, gas density,
temperature, and composition from intergalactic gas clouds;
that they extend only to about --1000 km/sec shift relative
to the QSO rest frame;
and that the velocities are even that large only from the
brightest (assuming cosmological distances) QSOs
(Aoki et al. 2005, Gallagher et al. 2005, Benn et al. 2005).
Probably no set of observations can move members of one
 camp into the other, but Zackrissen (2005) has an idea.
 If QSOs are ejected from the nuclei of nearby galaxies
 carrying large intrinsic redshifts (which decrease with time)
 in their pockets, then the recently-ejected should consist
 only of ionized gas and/or young stars with no proper host.
 Unfortunately, this comes perilously close to what you
 expect for the largest redshifts in the standard picture.

\section{TO RER IS HUMAN}
 Hubristically emulating the
High Priest on Yom Kippur\footnote{In barest outline, he confessed
first his own sins, then
those of his family, and finally those of the whole nation.
A slightly expanded version appears in Leviticus 16.6, 16.11,
and 16.21, but for the full version you must
go to the commentary Mishnah Yoma,
Ch.~3. Mishnah 8; 4:2; and 6:2. The customary tune is
said to be the only one in current use that can be
traced back to the time of the Second Temple. Thus this
last section begins with our own errors of earlier ApXX,
goes on to some family failings, and ends by drawing
on the entire nation of astronomers.},
we begin by confessing our own sins in (mostly) Ap04,
ordered by section number. These are of two sorts,
class A (rigid) where there is no doubt that we were
wrong and class B (limp, or blimp, a folk etymology),
where a correspondent appears to be disgreeing with a
paper cited as a highlight as well as with our taste
in choosing it, rather than one of theirs. Many of the
items made us say, ``urr,''  and occasionally ``um'' or ``duh.''

\subsection{Our Urrs, Class A}
 Sect.~3.2.1: 10--100 $M_\odot$ planets
 should have been 10--100 $M_\oplus$ planets even in these
 days of growing BMIs.

Sect.~4.8: Should $V_r$ in the formula $V_r= 4.74 \mu d$ have been $V_t$ (for
transverse, rather than radial)? Actually not
in context. You cannot measure $V_t$
separately from $\mu$ and the expression applies only
to spherical expansion.

Sect.~4.12: Mira variables with confirmed period
changes include R Hya and R Aql.

Sect.~4.13: The star with the changing Blazhko
period is XZ Cyg, not XZ Cam, which our correspondent
describes as ``a perfectly lovely star in its own right,
 but I believe it is an eclipsing variable rather than
 an RR Lyrae.''

 Sect.~10.4: The galactic ring [of stars]
 is not called Canis Majoris chides a Reliable Correspondent.
 Indeed surely not by its mother, though we and the
 authors being cited meant it in the non-rigorous sense of
 ``the constellation you look through to see something.''

 Sect.~11.4: The telescope with the shortest interval as
 largest, at least in modern times, was the
 Dominion Astrophysical Observatory's 1.8 m, completed
 two years ahead of the Mount Wilson 100-inch, because of
 wartime delays in the US.

Sect.~13.1: The second brightest supernova? How about 1972
 event in NGC 5253 says George (and you must guess which
 one since at least four have contributed comments on
 ApXX reviews over the years).

 Sect.~13.1: The correct
 central wavelength of the 4430 \AA\ diffuse interstellar
 band was first pointed out even earlier, in Zeitschrift
 f\"{u}r Astrophysik 64, 512 (1964) also by George.

\subsection{Our Urrs, Class B}
 Ap03, Sect.~6.4: A Correspondent
casts doubts on the reported detection of a magnetic
field in Beta Lyrae.

Ap04, Sect.~4.14: B Correspondent suggests that any
attempt to identify the Egyptian lion constellation
with Leo is likely to cause heart attacks among
(other) experts.
Before making up our minds, we want to know who they are.
(Compare the issue of whether blowing up the houses of
parliament is a good or bad thing to do.)\footnote{Especially
since the Keen Amateur Dentist sometimes now sits there.}

Sect.~8: One of
the non-standard cosmological models yields a concordance
age of 14.1 Gyr, if $H = 48$ km/sec/Mpc.

Sect.~9: The (proof?) reader who objected to our
consideration of ``who to cite in Ap05'' in favor of
``whoM to cite'' says that he is ``just being objective,
not accusative, and, since he is married, not dative
and certainly not genitive.''

Sect.~9.10: Falsification of the Chamberlin-Moulton
hypothesis and when it occurred.
The first strong theoretical line of argument came from
Lyman Spitzer in 1937. He concluded that gas pulled from
the sun would dissipate not condense. If you think
falsification requires an observation or experiment,
then it is probably the presence of deuterium in the
planets and its absence in the solar atmosphere.

Sect.~11.5: Concerning spectral types of stars in the
SMC. We had said ``4161 spectral types\ldots well, spectral
types for 4161 stars, but apparently only about 10 types.''
 The authors respond, ``there are 4161 spectral types,
 not 10, but a lot of them are the same. On the other hand,
 they're in \textsl{Monthly Notices}, not the SMC.''

 Consciences cleansed, we proceed to the usual assortment of um-provokers.
  No names are mentioned, but the references are real.

\subsection{Numerical Urrs}
``Corresponding to a uniform enrichment to a few hundred
thousand solar'' (ApJ 629, 615, abstract). Hey,
can I have the gold?

``\ldots model uncertainties make the accuracies of these values
 at least twice the magnitude of the precision''
 (AJ 128, 2826, abstract).

``\ldots an estimate of the star
 formation rate at redshift $3.1\times 10^{-2}h^3 M_0 /yr/Mpc^3$"
 (A\&A  430, 83, abstract).

``Finally, in Section 5, we draw our conclusions'' (MNRAS 356, 157).
But there are only
four sections, and we honestly did not understand whether
the 2.2--2.4 cm height was the thickness of the C ring or
amplitude of its warp.

``\ldots a fading of the characteristic luminosity by a factor
1.35 because $z = 0.2$'' (MNRAS 355, 767, conclusions).
Apparent brightness drops as (l+z) to various powers, $x$,
 for various cosmological models, but $x = 1.65$
 isn't any of them.

``several'' is anyhow larger than seven (MNRAS 354, L7,
abstract).

``nascent Trapezium'' with five stars
(ApJ 622, L141, abstract). Oh, all right, there
is also $\theta^1$ Ori E.

``\ldots due to an error in the conversion to SI units, \ldots the
densities are all too small by a factor $10^6$''
(A\&A 435, 339, erratum).

 $H = 71$, $q = 0.5$ still in use
(A\&A 435, 863) to analyze a radio ga1axy at large redshift.

``January temperatures were in the range --25 to $35^\circ$ C''
(Science 308, 397) in an article on the possibility of
recreating Pleistocene ecosystems in Siberia which would
seem to require organisms tolerant of a very wide
temperature range.

16\% of US youth in 1999--2000 were above the 95th percentile
for 2000 in CDC sex-specific BMI growth charts.
(MMWR 54, 203, and no it won't help a bit to know that
CDC = Centers for Disease Control and MMWR = Mortality
and Morbidity Weekly Report, one of our more cheerful
 bits of regular reading). They do better in
 Lake Woebegone.

 909 women report how they spent their day
 in Science 306, 1776. The sum of the mean hours per day
 devoted to all activities was 27.2 hours, of which 6.9
 were spent working and none sleeping. Perhaps medians
 would have been better.

``\ldots [scientists] descended on San Diego. Even many
of those based in the US flew in'' (Nature 432, 257).
A UK author can perhaps be forgiven for not knowing quite
how long it takes to get from Bethesda MD or
Cambridge MA to San Diego CA by bus, train, or car.
The article was about reducing carbon dioxide production.

Eros as described in A\&A 433, 371 has a mass given
in kg and volume in $\mathrm{km^3}$, but density
$\mathrm{g/cm^3}$.

 "\ldots 0.2 mag, plus or minus a factor of two''
(ApJ 627, 634, abstract). Like the adders deprived of their
slide rule, we find it difficult to go forth and
multiply under these conditions.

``$L_X= 28$ erg/sec for class 0 protostars in Rho Oph''
(ApJ 613, 410, summary).

A graph in ApJ 613, 521
(Fig.~5) has axes labeled 1,5,10,15,20 and 1,2,3,4,
but no units or names of the quantities.

``The speed of light, or $c$, is a rea11y big number,
186,282 mi/sec. Multiply it by itself, and the result is,
well, a really big number, 37,700,983,524.''
(Smithsonian, June 2005, p.~11). Now, in what set
of units is this useful? Yes, your least metric author
did the same thing at age 8, but she rounded off to
186,000, much shortening the computation and then
asked her father what to do with it. Publish
immediately was not the answer.

Contrary to Morrison's dictum, it is possible to
waste \$$10^8$ (Nature 436, 14) on a conservation project
for Stellar sea lions (named for their discoverer,
not their sparkling appearance) which requires the
grantees to avoid making the most informative measurement.

``\ldots the leading experiments are still sensitive enough
 to set limits 1--2 orders of magnitude less stringent
 than those traditionally presented''
 (PRL 95, 101301, abstract).
Not, we were going to say, the best argument for
funding, but in light of the previous item\ldots

\subsection{Ur-People}
 From the APS/CSWP Gazette, Issue 24, No.~1, p.~9.
 A very nice picture of Yuri Suzuki, winner of the 2005
 Maria Goeppert-Mayer Award. But the caption on the picture
 says Agnes Pockels. And somewhere it has to be recorded that
  Goeppert-Mayer's daughter (wife of astronomer Donat Wentzel)
  died this past year after a long, painful course of scleroderma.

``The dynamical problem of Henon-Heiles hardly needs any
introduction'' (ASS 295, 325) so they don't give it
one. Another paper read later the same day described it as a
``well known potential,'' which doesn't help as much as they
probably meant it to.

"\ldots members of the Review Committee from universities,
 schools, the oil industry, the shallow geophysics communities\ldots''
 (A\&G 46, 3.7). Gee, we don't say things like that
 about astronomers.

``Luna Hill's limits case" (ASS 293, 271). No, no citation,
but we are reasonably sure that Luna is not a first name.

``The Paloma-Green quasars'' (ASS 295, 397, abstract).
Paloma is the Spanish for dove, and you can make your
own gentle comment.

``I don't have any particular reason to think he is not
up to it, but\ldots'' (Nature 437, 610) is one astronomer
describing another, newly chosen for High Position.
 With friends like. these\ldots

``India's Atomic Energy Commission says\ldots that his country,
on considering\ldots'' (Science 309, 365). The temptation is
to say something about limited democracies, but we found
 ourselves this year participating in an organization
 where not even the Electoral College is allowed to vote.

A plea for contributions to keep the papers of R.~Franklin,
M.~Perutz, etc., together (Science 307, 519)
gives absolutely no indication where to send a check,
who to contact, or, for that matter, whether it
would be deductab1e.

``P.P.J. thanks [two names] for fruitful discussions''
(A\&A 430, 56, acknowledgements). But the authors are
HC, PP, RS, and BA.

``Vandervoort 1983, 1984, 2004, hereafter $03^3$, LM, and $\mathrm{M^2}$
respectively'' (MNRAS 354, 601, introduction).

``\ldots known
as the Leri method\ldots" (MNRAS 358, 397). Leri is one of
the authors, and if you wonder about self-bestowed eponyms,
contemplate what Feynman called the diagrams
(``the diagrams'' of course; we asked him).

``Holmberg relation'' (A\&A 434, 893). This one is
absorption in a galactic disk, $\mathrm{A_v}$ versus blue
magnitude. No citation, of course.

Figure 10 of A\&A 434, 176 would seem to represent
science fiction creatures, chess persons, or a pudgy
pinhead presenting a pear to a princess.

``Stuff works when the repairman is available''
(New Scientist 11 December, p.~64) is so obviously
true that it must be Enoemos's Law.

\subsection{Where-ur and When?}
``\ldots the Indian Ocean tsunami event of Sunday 2005
December 26'' (Observatory 125, 202). Well, December 26, 2005
 was a Monday, but you had to be reading this with us in
  October 2005 to wonder whether it might be a
  ghastly prediction.

``\ldots mutual phenomena of the Galilean satellites in Romania.''
(A\&A 429, 785, title). The most we can say is that,
Romania, currently holding the record for most times
in and out of the IAU, has a better chance for mutual
phenomena than most.

``\ldots quasar host galaxies with adaptive optics" (A\&A 439, 497,
abstract) and if that's where you want to keep your AO system,
you have a perfect right, if you can get it there.

``\ldots assembly of stars and dark matter from the
SDSS'' (MNRAS 356, 495). Well it was a very massive survey.

``\ldots in the sun and in silico" (A\&A 429, 1093,
title). Like Spike Jone's phone call,
they don't say who it is, but, we suspect, not the same as ``in vitrio''
though many glasses have significant silicon content.

``\ldots the Needles in the Haystack Survey'' (MNRAS 354, 123)
would have been a great name if it had been
done from Haystack Observatory. But it wasn't.

\subsection{Nonce Words}
 These are new ones\footnote{The etymology is Middle English
``for then ones''  misunderstood as ``for the nones,'' and go ahead,
admit it, you were expecting Ur-words.}  made up for
a single occasion and, one might hope, never to
be heard from again. Some of these are words, some
authentic (pronouncable) acronyms, some not even that.

Gasoline is a parallel N-body gas dynamics code
(New Astron.~9, 137, title).

Decretion disks are the outer parts where angular momentum is
 transported outward
(Astron. Rep.~48, 800, title). In earlier years,
excretion disks have been mentioned,
and we can understand a non-English speaking author
recoiling from the first dictionary definition.

``Quaternary is a hangover from a previous naming system,
the rest of which has been discarded''
(Nature 435, 865). Yeah, like the Cretaceous-Tertiary boundary.

A caterpillar that eats snails in Hawaii
(Science 309, 575) is a phytophagous species.
If you have eaten snails only in France, you
don't know what you are missing.

The quipu we grew up with are now khipu
(Science 309, 1065), but they arguably carry more
information than they used to, justifying a change.

COSMIC = Continuous Single-dish Monitoring of Intraday
variability at Ceduna (AJ 129, 2024).
CHIPS = Cosmic Hot Interstellar Plasma Spectrometer
(ApJ 623, 911). BEAST = Background Energy Anisotropy
Scanning Telescope (ApJS 159, 93), described as a cosmic
 microwave background ``experiment'' (well, wouldn't you
 like to make some changes?). IMF = interplanetary
  magnetic field (ApJ 625, 525). CHORIZOS
  (PASP 116, 859) is a chi-squared code.
ALIVARS = Algol-Like Irregular VARiable Stars
(formerly anti-flare stars, ASS 291, 123).
HINSA = HI Narrow Self Absorption (ApJ 622, 938), and
 practitioners of ahinsa will have to decide whether
 it is the absorption or the neutral hydrogen
 that is being denied.
Or, perhaps, the narrow self. RASSCALS (ApJ 622, 187)
is a category of groups of galaxies. The GOY model
(A\&A 432, 1049) need not have been, since the earlier
 papers are Gledzer and Yamada \& Ohkitani. Reflections
 on Reflexions (MNRAS 357, 1161) means there will
 always be an England, or anyhow an English
 different from American.

The ``maser mechanism of optical pulsation''
(MNRAS 354, 1201) is suffering from wavelength disorder.
 ``The optical and
electronic regions of the electromagnetic spectrum''
(Science 308, 630) have the related Lambda's disease.
``Losanges" (ApJ 424, L32) might be flat and sweet or
might come from Los Ange(el)s. If the latter, they
might well have been found in the ``Sedentary Survey''
(A\&A 435, 385).

``\ldots despite fulsome opposition\ldots'' (Nature 433, 682)
 may not be precisely what the authors were thinking,
 and ditto for ``the polemic and long-sought correlation''
 (ApJ 629, 797). In contrast, ``photon tiring''
 (ApJ 617, 525) was intended, but it is not the same as
 tired light and is not likely to be found from Sedentary Surveys.

``Standard Sirens'' (ApJ 629, 15) come from the
inspiral of black hole binaries. And now that you
know what song the Sirens sang, have you any thoughts
on what name Achilles  adopted when he hid among
the women? And how did he ever find shoes to fit?

``The nearest clusters with larger $\sigma_\mathrm{los}$ are the
idoneous ones to discriminate models'' (A\&A 424, 415),
 but we hate to think what they must be models for.
 If we had to guess, it would be a tossup between
 ``ideal'' and ``obvious.''

``ugwz-type subclass of ugsu dwarf novae''
(PASP 116, 117, though we hate to bite the journal
that accepts us).

Concerning Auger and the need for enhanced computing power,
 ``all agree that as it gobbles up data\ldots''
 (Science 309, 687). Selbbog is marginally pronouncable
 if you wish  to convey the opposite idea
  by going backwards. ``Spews out'' is also available in the
  realm of graphic metaphor, indeed perhaps excessively graphic.

The green dot for names of the year had to be divided.
Candidate one is Edasich for Iota Draconis
(Sky \& Telescope No.~6, p.~74), reducing us from two
pieces of information (where/when you can see it and
an approximate brightness) to zero. And candidate two
(ApJ 620, 948) is the recommendation to distinguish things
that will produce planetary nebulae from things that will
produce planets by calling them ``pre-planetary nebulae'' and
``proto-planetary nebulae.'' And we managed to remember
which was supposed to be which almost long enough to
tell our class about it.

\subsection{Ur-Symbols}
$Z$ is used to mean metallicity relative to
the sun (A\&A 430, 1133) and so takes on values
0.1 to 2.5 or 3.0 (and we still want the gold).

``\ldots $\rho_Q$ and $P_Q$ being respectively the pressure and
energy density'' (A\&A 436, 27) providing the possible
definition of ``respectively'' as ``bass ackward.'' $M_v$ does
double  duty (ApJ 622, 938) for virial mass as
 well as absolute visual magnitude.

\subsection{Ur-Duh}
The most surprising trait here is that anyone was
 surprised by them. For instance ``complex models are
 more frequently required for sources with higher
 S/N'' (A\&A 433, 1163). ``Faculty members are assessed not
 only on the quality of their teaching or even their
 research, but on how fundable their research is''
 (Nature 434, 10~9). ``A cooling flow does exist in the
 moderate cooling flow model\ldots'' (ApJ 622, 847, abstract).
  ``Students believe that batteries get light as they run down''
   (Science 308, 191). Well indeed they must, though we would
    not want to be in charge of measuring the $\Delta M$ that goes
    with this $\Delta E/c^2$.

``Should large organic molecules be found in extraterrestrial
 samples, it would be interesting to check the handedness of
 their optical activity'' (Nature 435, 437). Amino acids were
 promoted from racemic to a slight excess of the terrestrial
 rotatoriness a few years back. And yes it made ApXX
 at the time. ``Textbooks still parrot the conventional
 thinking that no fossil sharks are found before the
 Devonian'' (American Scientist 93, 248) provides another
 example of the difficulty of keeping on top of things

\subsection{Ur-No}
 This comes as close as we can get to the words
uttered by S.W.~Hawking at a conference many years ago
when Dennis Sciama asked him, ``Is that right, Steve?''
We hope they would both vote with us on most of the following.

``Planets have their elegant circular orbits yanked into
ugly distended ovals'' (Sky \& Telescope 109, No.~1, p.~45),
presumably by that notorious Yank Johannes Kepler.
The primary meaning of ``oval'' is not elliptical but
egg-shaped (have some pity for the  poor hen) and
you might also want to rethink resonances in general
before adopting the advice in that article on how to
push children in swings.

``\ldots the two disks are chemically well separated [and]
they overlap greatly in metallicity'' (A\&A 428, 139,
 abstract).

``The deuterium enrichment (in molecules like $\mathrm{N_2H^+/N_2D^+}$)
is presently ascribed to the
 depletion of CO in high density cores'' (ApJ 619, 379).
 It may well be so, but no
explanation of the mechanism was provided.

``With topspin, the velocity of air relative to the
surface at the top of the ball is higher\ldots if the ball
has topspin, the thin layer of air in contact\ldots is
travelling faster at the bottom than at the top\ldots''
(New Scientist 11 December, p.~65). Could this be why
we have never been able to pitch decently?

``Ecologists have established that nitrogen and carbon
isotopes are heavier in marine organisms''
(Science 306, 1466). Not the same author, but kin to
``\ldots increasing atomic weight (or the positive charge of
the nucleus)'' (Nature 433, 401). Well, Mendeleev had
trouble with that one among the heavier elements.

``All our helium is left over from when the planet first
formed [and] leaks out of the middle of the Earth''
(Nature 433, 906). Well, arguably, but a good deal of
it was initially left over in the form of Uranium and Thorium.

\subsection{Ur-Phrases}
It is possible that some of these
 would have done better in a primordial language.
But not probable.

From a large envelop mailed by Duke University,
``Contents inside.'' Like the closed box model
 of cosmic chemical evolution last year, consider
 the alternatives. Or have your received a shipment in a Klein
  bottle  from them lately?

From a student essay, ``A week ago, Hans Bethe died
and contributed to the field of science.'' Of all the
people who might be suspected of continued, significant,
 posthumous scientific contributions, Hans would
 surprise us least.

From a mailing by an organization we really like,
``If you move, be sure and let us know your new address,
 and the post office won't always forward your
 (Publication Name) to you.''

``You show me your OVI and
 I'll show you mine'' (ASS 289, 469, title, selected
 undoubtedly by authors who also have a large central
 black hole, cf. \S12.7).

 ``High Energy Density Laboratory Astrophysics''
 (ASS 298, No.~1, title of conference proceedings).
 The energy density in our lab depends a lot on
 whether the postdocs are there (the sign is up to you).

``The International Chicken Genome Sequencing Consortium''
(Nature 432, 695), and ``Opportunities in researching disaster''
(Science 309, 983, to show that we are equal opportunity journal
disparagers).

A few cases where the authors seem to have
done it deliberately, but you would have had second thoughts;
 ``In spite of the wealth of data, or perhaps because of it''
 (A\&A 433, 305, abstract). ``\ldots assuming spherical symmetry
 [for the baryons] we find that the halo mass is rounder than
 the baryons'' (ApJ 623, 31, abstract). ``Fresher than fresh''
 was a frozen foods slogan; ``rounder than round'' belongs
  perhaps to genetically engineered apples. ``The introduction
  of a new physically meaningless parameter'' (A\&A 428, 545),
  ``Assuming that all the detected X-ray radiation is either
  non-thermal or thermal'' (ApJ 616, 460, footnote to table).
  Yet again, consider the alternative.

``The shower was not bright meteor and lower activity in
comparison with other meteor showers''
(A\&A 417, L35, introduction).

``Uncertainties in the nebular geometry and the degree of
dust coupling are most likely responsible for the blue rise"
(ApJ 616, 257, abstract). A confused source in all
senses.

And some cases where even we would have had second thoughts.
``\ldots two bodies getting entangled in thin layers of
dynamical chaos'' (MNRAS 360, 401). ``\ldots despite some
philosophical differences with us about the passbands\ldots''
(PASP 117, 502, acknowledgements).
``\ldots send (expendable photocopies of) papers to one of the
following\ldots referees\ldots, and then inquire of him by phone in
40 days'' (Dio 13.1, p.~19) and presumably 40 nights.

``Quod erat demonstrandum; Latin for which was to have been
proved,'' a footnote to QED (A\&A 436, 554). Aw gee.
We thought they meant quantum electrodynamics.

``Local Universe'' has begun to appear all over the place,
though we started cringing only at, arbitrarily, ApJ 624, 155.
 Do they mean the Virgo superc1uster? Redshift less than
 0.01, 0.1, or what? Or perhaps just a region within which
 one can find ``a large dwarf galaxy sample'' (AJ 129, 2129).

``Animals like autists concentrate on details''
(Nature 435, 147), from a review of T.~Grandin \& C. Johnson,
\textsl{Animals in Translation}. If you have never before read
anything by or about Temple Grandin, now is the time to
start. If you don't occasionally have a ``hey I was like
that'' experience occasionally, how did you get to be a scientist?

The same colleague has supplied our closing quote for
many years, and you will hear from him in just a moment,
but first a word from a colleague who was actually pleased
at something in Ap04: ``We are encountering difficulties in
having our [topic mentioned in Ap04] proposals allocated in
larger than 2-meter telescopes. I will use your reference
in the next proposal\ldots and see if I am allocated.''

And, finally, with only the ethnicity pseudochauved,
``Uzbanian people says: wishes to live before next bearthday in order to all
would assemble around the ce1ebra1  table newly and say wishes one more!''
Your authors are not sure what will happen in the
next year, but hope indeed to say wishes one more to you all.

\acknowledgments
Thanks, as always, to our libraries and librarians,
real, virtual, electronic, and biological.
V.T.'s share of the page charges is being paid out of
honoraria from the Peter Gruber Foundation and the
NSF Pre-doctoral Fellowship Peer Review Panel.
C.J.H.'s incredibly generous contribution has been in the form of alphabetizing
everybody's references and keyboarding
V.T.'s typed text.\footnote{What VT  has yet to realize are the tremendous
advantages of being an Emeritus Professor.  CJH}

A few colleagues each year continue to take the risk of
being quoted in these pages,
and we are deeply indebted to Faustian Acquaintance,
Jackie Beucher,  William P.~Bidelman,
 Alain Blanchard, Stephen Elliott,
Tom English, Donald V.~Etz, Mr.~H., Ethan Hansen, Petr Harmanec,
George Herbig, Ian Howarth, Bruce Jakosky,
Vicky Kalogera, Andy Knoll, Kevin Krisciunas,
Harry Lustig, Tom McCollum, Medical Musician, Bohdan Paczynki,
Lord Rees of Ludlow, Michael Rich,  Alexander P.~Rosenbush, Wayne Rosing,
Brad Schaefer, Horace Smith, John Stull, Karel van der Hucht,
George Wallerstein,  Gaurang Yodh, and Ben Zuckerman.

MJA acknowledges the NASA Astrophysics Data System
(ADS)---as does CJH---and thanks the numerous colleagues who provided preprints.
The work was partially supported by NASA contracts from the
TRACE, RHESSI, STEREO, and LWS TRT (Living With a
Star--Targeted Research \& Technology) Programs.


\end{document}